\documentclass[10pt,a4paper]{article}
\usepackage[T1]{fontenc}\usepackage[utf8]{inputenc}\usepackage[english]{babel}
\usepackage{lmodern,microtype,amsmath,amssymb,amsthm,mathtools,graphicx}
\usepackage[gray]{xcolor}
\usepackage{tikz}\usetikzlibrary{arrows.meta,positioning}
\usepackage[margin=22mm]{geometry}\usepackage[hidelinks,hypertexnames=false]{hyperref}
\hypersetup{linktoc=all}

\numberwithin{equation}{section}\numberwithin{figure}{section}\numberwithin{table}{section}
\newtheorem{theorem}{Theorem}[section]\newtheorem{proposition}[theorem]{Proposition}
\newtheorem{lemma}[theorem]{Lemma}\newtheorem{corollary}[theorem]{Corollary}
\theoremstyle{definition}
\theoremstyle{remark}
\newcommand{\be}{\begin{equation}}\newcommand{\ee}{\end{equation}}
\newcommand{\bse}{\begin{subequations}}\newcommand{\ese}{\end{subequations}}
\newcommand{\del}{\partial}\newcommand{\dd}{\mathrm d}\newcommand{\Ric}{\operatorname{Ric}}\newcommand{\Hess}{\operatorname{Hess}}
\newcommand{\Scal}{\operatorname{Scal}}\newcommand{\Ocal}{\mathcal O}

\newcommand \bei {\begin{itemize}}
\newcommand \eei {\end{itemize}}

\usepackage{titlesec}
\titlespacing*{\paragraph}{0pt}{.6ex plus .2ex minus .1ex}{1em}

\titlespacing*{\section}{0pt}
  {2.ex plus 1ex minus .2ex}
  {1.5ex plus .2ex}
\titlespacing*{\subsection}{0pt}
  {2.ex plus 1ex minus .2ex}
  {1.5ex plus .2ex}

\let\originalsum\sum
\renewcommand{\sum}{\originalsum\nolimits}

\title{Schwarzschild--de Sitter space is not rigid}
\author{Philippe G. LeFloch$^1$ and Luc Rozoy$^2$}\date{November 2026}
\begin{document}\maketitle
\footnotetext[1]{Laboratoire Jacques-Louis Lions, Centre National de la Recherche Scientifique, Sorbonne Universit\'e, 4 Place Jussieu, 75252 Paris, France. Email: \texttt{contact@philippelefloch.org}.}
\footnotetext[2]{Universit\'e Grenoble Alpes, 38058 Grenoble, France. Email: \texttt{rozoy.luc@gmail.com}.
}

\begin{abstract}
Starting from any compact subextremal Schwarzschild--de Sitter region with positive cosmological constant, we develop a gluing method to prove the existence of a \emph{new family of static vacuum spaces} with two additional black-hole horizons of prescribed equal area $16\pi\mu^2$. All of the horizons and the symmetry axis are regular, and the metric and lapse are real analytic up to the boundary. As the mass parameter $\mu$ tends to zero, both (metric, lapse) fields converge smoothly to the Schwarzschild--de Sitter reference solution \emph{away from two antipodal points} on its sphere of maximal lapse. After rescaling near either added horizon, the spatial metric converges smoothly to the Schwarzschild metric. For small $\mu$, the lapse attains its maximum on a \emph{unique equatorial circle}. The transverse Hessian is negative definite, and the smaller of the two eigenvalue magnitudes is asymptotic to a positive constant times $\mu$. Nearby regular level sets are tori. The proof uses cutoffs at separated radii and elliptic estimates uniform as $\mu\to0$. An integral condition excludes the infinitesimal displacement of each Schwarzschild center, while a scalar multiple of a localized tensor restores solvability. Matching the Schwarzschild correction to the Green tensor for the two point sources gives the leading term of this scalar coefficient in terms of the radial derivative of the background lapse. Reflection makes the coefficients at the two centers equal, and varying their common radius makes them vanish. A radial Green function and the heat equation on the sphere determine the sign of the first angular variation of the lapse.
\end{abstract}


{\small
\setcounter{tocdepth}{1}
\setcounter{secnumdepth}{2}
\begingroup
\makeatletter
\renewcommand{\l@section}{\@dottedtocline{1}{0em}{2.8em}}
\makeatother
\tableofcontents
\endgroup
 }

\section{Introduction}\label{sec:introduction}

\subsection{Formulation of the problem}

We are interested in the static vacuum Einstein equations with positive cosmological constant. Schwarzschild--de Sitter, de Sitter and Nariai are the classical models with compact static regions. These models motivated a longstanding expectation of \emph{geometric rigidity} under suitable regularity and boundary assumptions. Let $h$ be a Riemannian metric on a compact three-dimensional manifold with boundary, and let $N$ be its \emph{lapse function}. For a fixed cosmological constant $\Lambda>0$, the static vacuum Einstein equations for $(h,N)$ are
\be\label{eq:static}
\Hess_hN=N(\Ric_h-\Lambda h),\qquad \Delta_hN=-\Lambda N.
\ee
We seek solutions for which $N$ is positive in the interior, vanishes on the boundary, and has nonzero gradient there. Each boundary component then represents a nondegenerate horizon. The system \eqref{eq:static} is nonlinear and elliptic in the interior and degenerates at the horizons, where $N$ vanishes.

In recent years, Dias, Gibbons, Santos and Way~\cite{DGSW2023} numerically constructed static configurations with two small-mass black holes inside a cosmological horizon. In our companion paper~\cite{AX-Existence}, we rigorously proved the existence of such static solutions that converge to de Sitter space as the two additional masses tend to zero. Here, we uncover a similar \emph{non-rididity phenomena} for the whole class of Schwarzschild--de Sitter spaces. 
We start from any compact subextremal Schwarzschild--de Sitter region and construct static vacuum solutions with two additional black-hole horizons. As the areas of the added horizons tend to zero, the solutions converge smoothly to the reference solution away from the two insertion points, including at the original black-hole and cosmological horizons.

The Schwarzschild--de Sitter region is described as follows. A black hole mass $M$ in the subextremal range
\be\label{eq:mass-range}
0<3M\sqrt{\Lambda}<1,
\ee
being fixed, we define the metric
\be\label{eq:F}
h_M\coloneqq\frac{\dd r^2}{F_M(r)}+r^2\dd\omega^2,\qquad
F_M(r)\coloneqq1-\frac{2M}{r}-\frac{\Lambda r^2}{3},\qquad
N_M\coloneqq\sqrt{F_M(r)}.
\ee
The two positive simple zeros $r_b(M)<r_c(M)$ of $F_M$ are the radii of the black-hole and cosmological horizons, respectively.

We seek a static solution with these two original horizons, plus two additional black-hole horizons. We prescribe the area of the central horizon to be its reference area and prescribe the same area for each added horizon.
We introduce the radius at which the lapse attains its maximum and the value of the lapse there:
\be
r_*\coloneqq(3M/\Lambda)^{1/3}, \qquad
N_*\coloneqq N_M(r_*)
\ee
and we choose antipodal points $p_\pm$ on the sphere $r=r_*$, where the function $N_M$ attains its maximum. The rotations fixing these points define the symmetry axis. The antipodal map $\iota(r,\omega)\coloneqq(r,-\omega)$ exchanges the insertion points. In the axially symmetric class, invariance under $\iota$ is equivalent to invariance under equatorial reflection, since the two maps differ by rotation through $\pi$ about the axis. We also impose reversal of the axial angle to remove mixed metric components with the circular direction. Choose an equatorial point $q_0$ on the reference sphere for the normalization of the lapse.

\subsection{Main result}

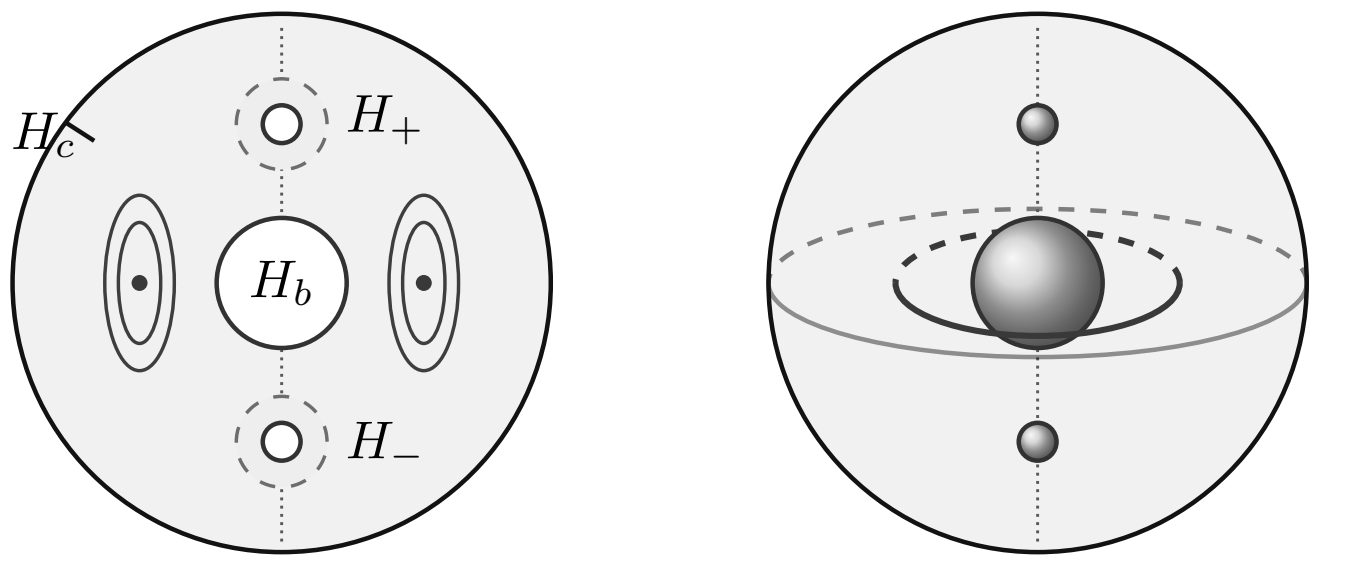
\begin{figure}[!t]\centering
\begin{tikzpicture}[x=1cm,y=1cm,>=Stealth,font=\normalsize,
  cosmological/.style={draw=blue!65!black,line width=.85pt},
  blackhole/.style={draw=red!65!black,line width=.85pt},
  symmetry/.style={draw=black!65,densely dotted,line width=.55pt},
  maximal/.style={draw=red!75!black,line width=1.15pt},
  levelline/.style={draw=teal!70!black,line width=.65pt}]
\path[use as bounding box] (0,0) rectangle (9.4,4.45);

\begin{scope}[shift={(2.20,2.52)}]
\filldraw[cosmological,fill=blue!6] (0,0) circle (1.78);
\draw[symmetry] (0,-1.71)--(0,1.71);
\foreach \z in {-1.05,1.05}{
  \filldraw[fill=orange!16,draw=orange!72!black,dashed,line width=.65pt] (0,\z) circle (.30);
  \filldraw[blackhole,fill=white] (0,\z) circle (.125);
}
\filldraw[blackhole,fill=white] (0,0) circle (.43);
\foreach \x in {-.94,.94}{
  \draw[levelline] (\x,0) ellipse (.14 and .40);
  \draw[levelline] (\x,0) ellipse (.23 and .58);
  \fill[red!75!black] (\x,0) circle (1.5pt);
}
\node at (0,0) {$H_b$};
\node[anchor=west] at (.30,1.08) {$H_+$};
\node[anchor=west] at (.30,-1.08) {$H_-$};
\node[anchor=east] at (-1.23,.98) {$H_c$};
\draw[cosmological] (-1.24,.94)--(-1.43,1.062);
\end{scope}
\node[align=center,text=black,font=\small] at (2.20,.31) {\textbf{(a)} Meridian section\\Axis and nearby levels};

\begin{scope}[shift={(7.20,2.52)}]
\fill[blue!6] (0,0) circle (1.78);
\draw[cosmological] (0,0) circle (1.78);
\draw[cosmological,dashed,opacity=.52] (1.78,0) arc[start angle=0,end angle=180,x radius=1.78,y radius=.49];
\draw[symmetry] (0,-1.71)--(0,1.71);
\draw[maximal,dashed] (.94,0) arc[start angle=0,end angle=180,x radius=.94,y radius=.35];
\foreach \z/\rad in {-1.05/.125,0/.43,1.05/.125}{
  \shade[ball color=red!30!white] (0,\z) circle (\rad);
  \draw[blackhole] (0,\z) circle (\rad);
}
\draw[cosmological,opacity=.45] (-1.78,0) arc[start angle=180,end angle=360,x radius=1.78,y radius=.49];
\draw[maximal] (-.94,0) arc[start angle=180,end angle=360,x radius=.94,y radius=.35];
\end{scope}
\node[align=center,text=black,font=\small] at (7.20,.31) {\textbf{(b)} A ball with three holes\\Three inner boundary spheres};

\end{tikzpicture}
\caption{The four horizons in meridian section and in space.}\label{fig:curved-section}
\end{figure}

We prescribe the central horizon area at its background value and allow the equal areas of the two added horizons to be arbitrarily small. An illustration is provided in \autoref{fig:curved-section}.

\begin{theorem}[Schwarzschild--de Sitter space is not rigid]\label{prob:existence}
Fix a compact subextremal Schwarzschild--de Sitter region with cosmological constant $\Lambda>0$ and mass $M$. Its spatial metric $h_M$ and lapse $N_M$ satisfy the static vacuum Einstein equations~\eqref{eq:static}. There is $\mu_0(M,\Lambda)>0$ such that, for every $0<\mu<\mu_0$, there exists a static pair $(h_\mu,N_\mu)$ with the following properties.
\begin{enumerate}
\item \emph{The static region and its horizons.} The pair solves~\eqref{eq:static} and is smooth on a compact three-ball with three disjoint open balls removed. The four boundary components are the central black-hole horizon, two additional black-hole horizons, and the cosmological horizon. The lapse is positive in the interior, vanishes at all four horizons, and has $|\nabla N_\mu|_{h_\mu}>0$ there.
\item \emph{Symmetry and regularity.} The pair is invariant under rotations about the axis through the two added horizons, equatorial reflection exchanging them, and reversal of the axial angle $\varphi\mapsto-\varphi$. The axis and its intersections with all four horizons are regular for angular period $2\pi$.
\item \emph{Prescribed areas.} The central horizon has area $4\pi r_b(M)^2$, and each added horizon has area $16\pi\mu^2$.
\item \emph{The limits on the background scale and the Schwarzschild scale.} After identifying the regions away from $p_+,p_-$ and normalizing the lapse by a positive constant,
\be\label{eq:outer-limit}
(h_\mu,N_\mu)\longrightarrow(h_M,N_M)
\quad\hbox{in }C^\infty_{\rm loc}\quad\hbox{away from }p_+,p_-.
\ee
This convergence holds up to the two original horizons. Near each added horizon, the spatial metric rescaled by $\mu^{-2}$ converges smoothly up to the horizon to the Schwarzschild metric of mass one.
\end{enumerate}
\end{theorem}

Theorem~\ref{qual:unique-maximal-circle}~below shows that, for sufficiently small $\mu$, the lapse attains its maximum on one equatorial circle. The Hessian is negative definite in its normal plane. The radial eigenvalue stays bounded away from zero, and the magnitude of the polar eigenvalue is asymptotic to a positive constant times $\mu$. The nearby regular level sets are tori (as shown in Figure~\ref{fig:maximal-lapse-foliation} below).

\paragraph{Quantitative estimates.}

The estimates on the reference geometries determine an admissible range for the added masses. We display the dependence of this range on the fixed geometric and analytic constants.

\begin{proposition}[A sufficient upper bound on the added masses]\label{thm:explicit-mass-threshold}
Use the background of Theorem~\ref{prob:existence} and put $\ell\coloneqq\sqrt{3/\Lambda}$. Let $\mathbf c$ denote certain \emph{normalized functional constants}, determined by the constants in the Einstein linearized inverse estimates on the model and the Poincar\'e, Korn, Hardy and Schauder estimates (specified in the proof). Then there exists an explicit positive function $\varepsilon_{\rm suff}$ of these quantities such that the upper bound in Theorem~\ref{prob:existence} may be chosen to be
\be\label{eq:main-explicit-threshold}
 \mu_0(M,\Lambda)=\ell\,\varepsilon_{\rm suff}(\mathbf c).
\ee
The quantities in $\mathbf c$ are determined by the ratio $M/\ell$, a fixed compact subextremal interval containing that ratio in cosmological units, and the prescribed model norms\footnote{Section~\ref{sec:functional-mass-bound} lists the constants and the order in which they are chosen. Proposition~\ref{q:functional-mass-threshold}~below gives the formula~\eqref{q:master-threshold} for $\varepsilon_{\rm suff}(\mathbf c)$, using the estimates in Appendix~\ref{sec:quantitative}.}.
\end{proposition}

An explicit upper bound for the masses allowed by the construction is derived in the course of our analysis. More precisely, the small parameter is measured relative to the fixed central geometry and the cosmological length $\ell$. For fixed $M$, the construction has $\mu/M\to0$ and $\mu/\ell\to0$. Let $H_b,H_c$ be the two original Schwarzschild--de Sitter horizons and define
\be\label{eq:insertion-clearance}
d_*\coloneqq\min\bigl(d_{h_M}(p_+,H_b),d_{h_M}(p_+,H_c),\tfrac12d_{h_M}(p_+,p_-),\ell\bigr)>0.
\ee
For any fixed exponents satisfying~\eqref{eq:free-scale-conditions} below, the nonlinear matching uses the radius $R_\mu\coloneqq\ell(\mu/\ell)^{a_2}$, so its geometric hierarchy is
\be\label{eq:physical-scales}
\mu\ll R_\mu\ll d_*,\qquad
\frac\mu\ell\ll\min\bigl(\frac M\ell,\left(\frac{d_*}{\ell}\right)^{1/a_2}\bigr).
\ee
For a prescribed separation factor $0<\eta<1$, the explicit condition
\be\label{eq:geometric-mass-bound}
0<\mu\le\min\bigl(\eta M,\ell(\eta d_*/\ell)^{1/a_2},\ell\eta^{1/(1-a_2)}\bigr)
\ee
gives $\mu/M\le\eta$, $R_\mu/d_*\le\eta$ and $\mu/R_\mu\le\eta$, by the three displayed bounds. This condition quantifies the separation of scales. The analytical threshold also incorporates the inverse, contraction and force estimates proved below. By scaling, it can be chosen in the form $\mu_0(M,\Lambda)=\ell\,\varepsilon_0(M/\ell)$. Sections~\ref{sec:spectral-models} to~\ref{sec:joined-quantitative} below establish the spectral, weighted and local elliptic estimates underlying the construction. Appendix~\ref{sec:quantitative} combines their constants into an explicit upper bound on the admissible small masses.

\paragraph{An illustration.}

\begin{figure}[htbp]
  \centering
  \resizebox{\linewidth}{!}{\begin{tikzpicture}[x=.81cm,y=.81cm,>=Latex,
  every node/.style={font=\small},
  horizon/.style={draw=black!75,line width=.8pt},
  radial/.style={draw=black!55,line width=.55pt}]
\definecolor{sdSblue}{RGB}{33,79,114}
\definecolor{sdSgold}{RGB}{153,101,27}
\definecolor{sdSink}{RGB}{68,68,68}

\foreach \x/\R/\col in {
  .75/.38/sdSink,
  2.90/.67/sdSblue,
  5.60/1.00/sdSblue,
  8.70/1.28/sdSgold,
  11.80/1.00/sdSblue,
  14.50/.67/sdSblue,
  16.65/.38/sdSink}{
  \begin{scope}[shift={(\x,0)}]
    \fill[\col!5] (0,0) circle[radius=\R];
    \draw[\col,line width=.8pt] (0,0) circle[radius=\R];
    \draw[\col!55,line width=.4pt]
      (0,0) ellipse[x radius={.31*\R},y radius=\R];
    \draw[\col!55,densely dashed,line width=.4pt]
      (\R,0) arc[start angle=0,end angle=180,
      x radius=\R,y radius={.24*\R}];
    \draw[\col!65,line width=.45pt]
      (-\R,0) arc[start angle=180,end angle=360,
      x radius=\R,y radius={.24*\R}];
  \end{scope}
}

\node[align=center] at (.75,2.55)
  {Black-hole horizon\\$N_M=0$};
\node[align=center,text=sdSgold] at (8.70,2.55)
  {Maximal-lapse sphere\\$N_M=N_*$};
\node[align=center] at (16.65,2.55)
  {Cosmological horizon\\$N_M=0$};
\draw[black!45,line width=.45pt] (.75,2.02)--(.75,.48);
\draw[sdSgold!65,line width=.45pt] (8.70,2.02)--(8.70,1.39);
\draw[black!45,line width=.45pt] (16.65,2.02)--(16.65,.48);

\draw[->,sdSblue,line width=.7pt] (2.2,1.72)--(6.5,1.72)
  node[midway,above=3pt,text=black] {$N'_M(r)>0$};
\draw[->,sdSblue,line width=.7pt] (10.9,1.72)--(15.2,1.72)
  node[midway,above=3pt,text=black] {$N'_M(r)<0$};

\draw[radial,->] (.25,-1.92)--(17.30,-1.92)
  node[right,text=black] {$r$};
\foreach \x/\lab in {.75/r_b,8.70/r_*,16.65/r_c}{
  \draw[radial] (\x,-1.84)--(\x,-2.00);
  \node[below=3pt] at (\x,-2.00) {$\lab$};
}
\node at (8.70,-3.22)
  {$N_M(r)=\sqrt{1-\dfrac{2M}{r}-\dfrac{\Lambda r^2}{3}},
    \qquad r_*=\left(\dfrac{3M}{\Lambda}\right)^{1/3}$};
\node[font=\footnotesize\itshape,text=black] at (8.70,-4.10)
  {Schematic sphere sizes indicate the lapse, not the areal radius.};
\end{tikzpicture}}
  \caption{Spherical sections of the Schwarzschild--de Sitter region,
  ordered by the areal radius $r$. The drawing sizes indicate the increase
  and subsequent decrease of the lapse $N_M$. Small spheres mark the two
  horizons where $N_M=0$, and the largest sphere marks the maximum at
  $r=r_*$. The actual sphere areas are $4\pi r^2$ and increase
  throughout the static region.}
  \label{fig:sds-background-lapse}
\end{figure}
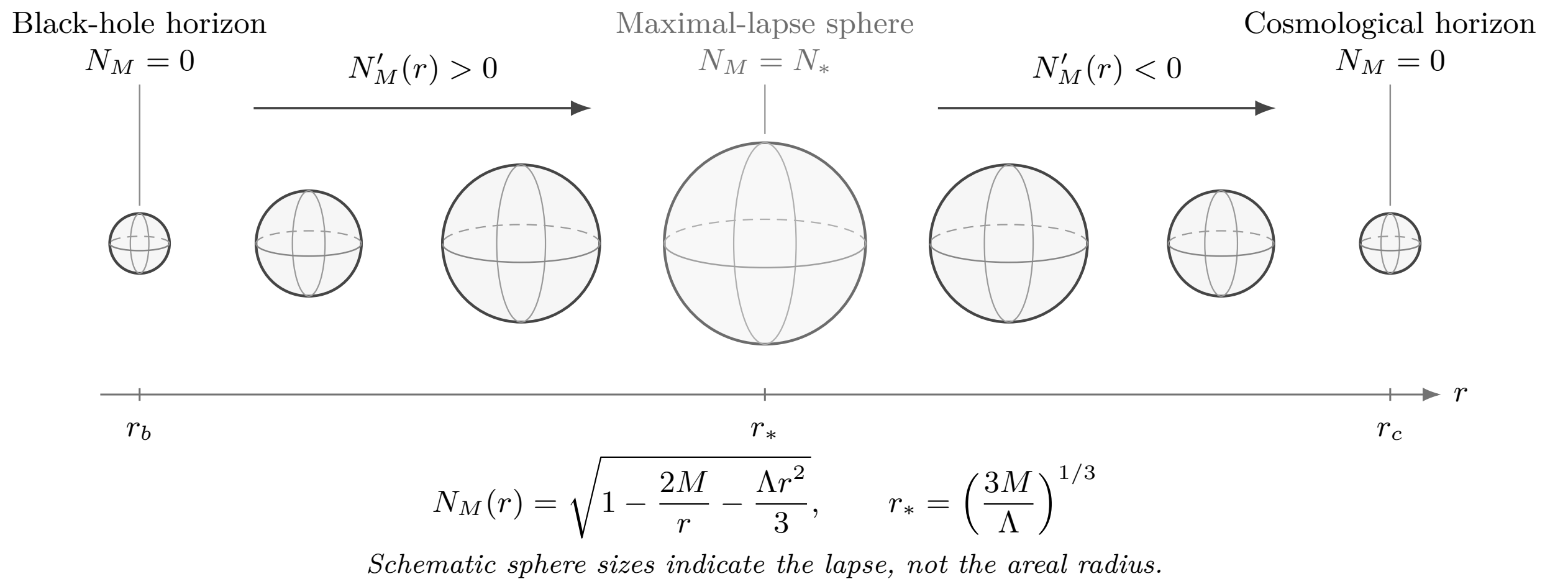

The prescribed central area fixes the mass of the reference region, while the central horizon metric is part of the unknown solution. For $M$ in a compact subinterval of the subextremal range, the weighted estimates are uniform as $\mu$ tends to zero, including in regions where the curvature is of order $\mu^{-2}$. The constants may diverge at the endpoints of the subextremal range.
Figure~\ref{fig:sds-background-lapse} shows how the lapse varies along the spherical sections of the reference region. The two parts of Figure~\ref{fig:curved-section} show the axis and the four boundary spheres. The dashed annuli are the regions where the metric approximations are joined.

\subsection{Related results}

Beyond our companion paper~\cite{AX-Existence}, our result and the methodology of proof proposed in the present paper can be traced back to the following related results.

\paragraph{Black-hole configurations with positive cosmological constant.}

Dias, Santos, and Way~\cite{DSWSpinning} construct numerical stationary
binaries with aligned or opposite spins. The numerical methods for these
stationary problems are discussed in~\cite{DSWMethods}. In the static setting,
Ozuch~\cite{OzuchPositive2026} constructs positive Einstein four-manifolds
that yield axisymmetric spacetimes with arbitrarily many horizons. His
construction varies a cone angle and combines local elliptic continuation
with geometric compactness, using weighted estimates and analysis at
corners~\cite[Sections~1.4, 4 and 5]{OzuchPositive2026}. Broder and
Lafuente~\cite{BroderLafuente2026} also construct torus-invariant positive
Einstein metrics on closed four-manifolds. In Ozuch's construction, smoothness
for a single circle period requires equal surface gravities at the
horizons~\cite[Proposition~6.4]{OzuchPositive2026}. In our family, the
surface gravities of the added horizons grow as $\mu^{-1}$, while those of
the original horizons converge to finite positive limits. The ratio of an
added horizon's surface gravity to that of an original horizon therefore
diverges as $\mu\to0$. This ratio is invariant under constant
renormalization of the lapse.

\paragraph{Gluing and elliptic estimates.}

Hintz~\cite{AX-HintzDeSitterGluing} glues Schwarzschild--de Sitter or
Kerr--de Sitter metrics near points of future conformal infinity to
construct dynamical vacuum spacetimes, subject to balance conditions on
the black-hole parameters. He also inserts small black holes into initial
data satisfying the Einstein constraints~\cite{HintzSmallBlackHoles} and
characterizes solvability of the linearized equations with spatially
compact support by charges associated with Killing
fields~\cite{HintzLinearizedSources}. Here we solve the full static
equations on a compact region, preserving regularity at both original
horizons and determining the positions at which the added black holes
can be in equilibrium. The use of model geometries, elliptic inverse operators,
and finitely many parameters has antecedents in gluing constructions for
constant mean curvature surfaces~\cite{AX-MPPConnectedSums,AX-BKEmbeddedCMC},
Einstein metrics~\cite{AX-BamlerDehn}, and the Yamabe
equation~\cite{AX-EPVYamabe}. For the Einstein constraints,
Carlotto and Schoen~\cite{CSLocalization} localize solutions, while
Le Floch and LeFloch construct asymptotically Euclidean solutions with
prescribed exterior geometry~\cite{BLFPLFShielding,BLFPLFLocalization},
derive stability estimates from weighted Poincar\'e, Korn, and Hardy
inequalities~\cite{BLFPLFStability}, and construct approximate data with
decaying errors in the constraints~\cite{BLFPLFSeeds}. 
For other techniques geometric regularity and continuation, we refer
to~\cite{LS-DeTurckKazdan,AX-KobayashiNomizu}.  

\paragraph{Equilibrium and static vacuum geometries in other settings.}

For $\Lambda=0$, Bunting and Masood-ul-Alam~\cite{AX-BuntingMasood} prove
uniqueness in the asymptotically Euclidean static setting. Other obstructions to multiple-body equilibrium depend on the static or stationary hypotheses, the symmetries, and the asymptotic
conditions~\cite{AX-BeigSchoenStatic,AX-LiTianHorizons,AX-HKWXKerrUniqueness}.
In the stationary axisymmetric problem, Weinstein formulates the equations as a harmonic map system and relates the force between two black holes to the conical deficit on the intervening axis
segment~\cite{AX-WeinsteinEquilibrium,AX-WeinsteinTwoBody}. He determines
its sign in a limiting regime~\cite{AX-WeinsteinForce}, and Li and
Tian~\cite{LiTianHarmonic1992} study the regularity of the associated singular
harmonic maps. Related reductions and coordinate constructions appear
in~\cite{CLSZ2007,CharmousisGregory2004,Charmousis2007,ER2002}.
In our construction the axis is regular throughout, and the coefficient
used to impose equilibrium multiplies a source in the elliptic equation
near each added horizon. The periodic Myers--Korotkin--Nicolai solutions
have a regular axis and Kasner asymptotics. Reiris and
Peraza~\cite[Introduction]{AX-ReirisPeraza} obtain any prescribed finite
number of spherical horizons by quotienting by an axial translation,
possibly combined with a rotation, and also describe arrays with
alternating masses. Further periodic examples in higher dimensions are
constructed in~\cite{KRWY2024}. Other static vacuum problems include
extensions with Bartnik boundary data \cite{AnHuang2024,AlexakisAnEllithyHuang2024,AnHuang2026},
black holes supported by expanding bubbles~\cite{AstorinoEmparanVigano2022},
and, for $\Lambda<0$. Signed masses and degenerate
horizons concern different geometric classes
\cite{AX-KWYSignedMasses,LS-KhuriWoolgar}.

\subsection{Strategy and overview of the proof}\label{sec:proof-outline}

\paragraph{The elliptic equation and the location of the added horizons.} In the interior, where $N>0$, introduce the Riemannian metric $g=h+N^2\dd\tau^2$, where $\tau$ is an auxiliary coordinate. The Einstein equation $\Ric_g=\Lambda g$ is then equivalent to~\eqref{eq:static}. Each horizon has its own smooth Cartesian coordinates in the normal plane, with an angular coordinate normalized by the surface gravity. This allows the two original surface gravities to differ. We fix the coordinates (or gauge) by the so-called \emph{DeTurck vector field}, denoted by $W(g,\overline G)$, which is defined as the contraction of the difference between the connections of $g$ and a reference metric $\overline G$. The modified equation to be solved then has the local schematic form
\be
 \mathcal E_{\overline G}(g)\coloneqq2(\Ric_g-\Lambda g)-\mathcal L_{W(g,\overline G)}g,
 \qquad
 (\mathcal E_{\overline G}(g))_{ij}=-g^{ab}\del_a\del_b g_{ij}+\text{lower-order terms}.
\ee
The operator is nonlinear elliptic in these charts, \emph{including at horizon poles}. We use a contraction argument, after establishing the existence of an inverse to the \emph{linearized Einstein static operator} and controlling it \emph{uniformly as the horizons shrink} and an asymptotic expansion, as a function of the insertion radius,
of the coefficient of the auxiliary source in the equation for the corrected metric. The insertion radius (for the location of the two additional black holes) is then chosen so that both local sources vanish.

In our de Sitter construction of~\cite{AX-Existence}, the two black holes approach one another as their masses tend to zero. The leading force balances their mutual attraction against the cosmological term. Here the two centers approach distinct antipodal points, and the leading force is governed by the radial derivative of the Schwarzschild--de Sitter lapse. The test-particle calculation suggested in~\cite{DGSW2023} gives $r_*=(3M/\Lambda)^{1/3}$, derived in Section~\ref{sec:background}. We put the two trial centers at a common radius $r$ and let the elliptic construction \emph{determine $r$ in a fixed interval $I$ about $r_*$}.

We fix the charts, cutoffs and estimates uniformly on $I$ and on a compact subextremal mass interval for the final area adjustment. In cosmological units $\ell=\sqrt{3/\Lambda}$, write $\mu$ for the trial small mass and $\rho$ for the radius in the spatial normal coordinates at an insertion. The horizon scale is $\mu$. Write $\mu^{a_2}$ for the radius
at which we match the two metrics, and $\mu^{a_1}$ and $\mu^{a_3}$
for the radii at which we cut off the solutions obtained from the
background and the Schwarzschild inverse operators, respectively.
Their order is determined by
$0<a_3<a_2<a_1<1$, which gives
$\mu\ll\mu^{a_1}\ll\mu^{a_2}\ll\mu^{a_3}\ll1$.
Keep the weight and the three radii exponents fixed subject to
\be\label{eq:free-scale-conditions}
 -\tfrac12<\gamma<0,\qquad
 \max(-\gamma,(2-\gamma)^{-1})<a_2<\frac{1}{2+\gamma},\qquad
 0<a_3<a_2<a_1<1.
\ee
These intervals are nonempty. Put
\be\label{eq:error-gains}
 \begin{split}
 \eta_{\rm geom}&\coloneqq\min(a_2,1-a_2),\\
 \eta_{\rm src}&\coloneqq\min(a_2(2-\gamma)-1,-a_2\gamma,
                         1-a_2(2+\gamma),-\gamma),\\
 \delta&\coloneqq\min(a_2,1-a_2,1+\gamma,
                         1-a_2(1-\gamma),1+2\gamma).
 \end{split}
\ee
All these gains are positive under~\eqref{eq:free-scale-conditions}.
Choose, without specializing it,
\be\label{eq:eta-conditions}
 0<\eta<\min(\eta_{\rm src},\delta,1+\gamma,\tfrac12).
\ee
The gain $\eta_{\rm src}$ bounds the source errors at the matching
radius, $\delta$ bounds the error after division by the leading
auxiliary-source coefficient, and $\eta$ is used for the final lapse
expansion. Constants may depend on these fixed exponents.
 Table~\ref{tab:construction-lengths} lists the radii used to combine the inverse operators. (Cf.~also Figure~\ref{fig:cutoff-supports} for the cutoffs.)

\begin{table}[!htbp]\centering
{
\begin{tabular}{@{}l p{11cm}@{}}
\hline
Length & Use in the proof\\
\hline
$1$ & Background region, original horizons and placement intervals.\\
$\mu$ & Schwarzschild coordinates near the new horizons.\\
$\mu^{a_1}$ & Cutoff on the solution from the background inverse.\\
$\mu^{a_2}$ & Matching annulus for the two metric approximations.\\
$\mu^{a_3}$ & Cutoff on the solution from the Schwarzschild inverse.\\
$\rho_{\rm cut}$ fixed as $\mu\to0$ & Radius of the cuts in the convergence proof (Section~\ref{sec:metric-convergence}).\\
\hline
\end{tabular}
}
\caption{Length scales in cosmological units.}\label{tab:construction-lengths}
\end{table}

\paragraph{Inversion on the two models and on the joined region.} At a reference metric $G$, write $S_G\coloneqq\Ric_G-\Lambda G$. The differential $L_G$ of $\mathcal E_G$ at $G$ has the form (cf.~Section~\ref{sec:nonlinear})
\be
 L_Gv=P_Gv+S_G\circ v+v\circ S_G,\qquad
 P_Gv=\nabla_G^*\nabla_Gv-2\operatorname{Riem}_Gv.
\ee
The products in this formula contract one tensor index.  On the Schwarzschild--de Sitter background, $S_G=0$ and the differential reduces to $P_G$. The Schwarzschild model uses the corresponding Ricci-flat differential $P_S$. On the Schwarzschild--de Sitter region the two original surface gravities are fixed during this linear solve. The map from the mass and lapse factor to the two gravities has a nonsingular derivative, so these conditions eliminate both variations. A scalar spectral gap and a vector energy identity eliminate the residual coordinate variations. This gives Proposition~\ref{thm:outer-tensor-inverse}~below. On the unit Schwarzschild metric $G_S$ the decaying kernel in the symmetry class consists of axial displacements of the center, generated by a tensor $T_S$. We use the horizon and axis charts and the kernel classification from~\cite[Section~2.2, Proposition~4.7 and Appendix~B]{AX-Existence}. An integral condition removes $T_S$ from the domain, and a scalar multiple of a compactly supported tensor $\chi_0T_S$ restores solvability, as in Proposition~\ref{q:full-Schwarzschild-model-inverse}~below. Reflection exchanges the two collars in their outward radial frames, leaving one scalar unknown for the two sources.

Let $G_{\mu,r}$ be the simpler joined metric, equal to Schwarzschild near the new horizons and to the background away from the matching annuli. The smooth scale $\omega_\mu$ in~\eqref{join:eq:explicit-weight} is comparable to $\mu+\rho$ near the insertions and to one elsewhere. At a fixed weight $\gamma$ satisfying~\eqref{eq:free-scale-conditions} the norm $X_\mu$ bounds the first four derivatives of a correction by $|\nabla^j v|\le C\omega_\mu^{\gamma-j}\|v\|_{X_\mu}$, including $j=0$. The norm $Y_\mu$ uses two derivatives and weight $\gamma-2$. For the reflected pair of rescaled sources $\psi_\mu$ and integral condition $\ell_\mu(v)=0$, the desired inverse satisfies
\be\label{intro:global-inverse}
 P_{G_{\mu,r}}v-\lambda\psi_\mu=f,\qquad
 \ell_\mu(v)=0,\qquad
 \|v\|_{X_\mu}+|\lambda|\le C\|f\|_{Y_\mu}.
\ee
We split the source with a cutoff at radius $\mu^{a_2}$ and apply the linearized inverse operators associated with the model, with tensor outputs $U_K,U_S$ and scalar output $\lambda$. For output cutoffs $\eta_K,\eta_S$, let $J_\mu$ place the Schwarzschild tensor in the two insertion charts with the normalization of $X_\mu$. The approximate inverse is
\be
 \mathcal R_\mu f=\bigl(\eta_KU_K+J_\mu(\eta_SU_S),\lambda\bigr).
\ee
The exterior cutoff is zero and the interior cutoff is one on the support of the integral condition and the scalar source. This preserves both when the two solutions are added. The cutoff $\eta_K$ varies at radius $\mu^{a_1}$ and $\eta_S$
at radius $\mu^{a_3}$. Choose a weight separation $\sigma>0$ such
that $\gamma-\sigma$ and $\gamma+\sigma$ lie in $(-1,0)$.
The terms obtained by differentiating the two cutoffs are bounded
by constants times $\mu^{\sigma(a_1-a_2)}$ and
$\mu^{\sigma(a_2-a_3)}$, respectively. The differences between the
coefficients on the joined region and those on the models contribute
$\mu^{1-a_1}$ and $\mu^{a_3}$. All four exponents are positive.
We write
\begin{equation}
 \delta_{\rm cut}\coloneqq\min\bigl(1-a_1,\,a_3,\,
           \sigma(a_1-a_2),\,\sigma(a_2-a_3)\bigr)>0
\end{equation}
for the exponent that bounds the sum of these errors. Denote the operator in~\eqref{intro:global-inverse} by $\mathcal A_{\mu,r}$ and the error by
\be
E_\mu\coloneqq\mathcal A_{\mu,r}\mathcal R_\mu-I.
\ee
Our estimates give the control
\be
 \|E_\mu\|_{Y_\mu\to Y_\mu} \le C\mu^\delta_{\rm cut},\qquad
 \mathcal A_{\mu,r}^{-1}=\mathcal R_\mu(I+E_\mu)^{-1}.
\ee
The Neumann series gives a right inverse for the augmented operator. For each fixed $\mu>0$ in the range under consideration, this operator has Fredholm index zero. Surjectivity therefore implies injectivity.
Proposition~\ref{join:thm:joined}~below and Proposition~\ref{q:joined-power} below give the bounds for the cutoffs and the exact rescaling in $J_\mu$.

Several functional inequalities are used in different steps.

\bei

\item A Poincar\'e inequality gives the positive scalar spectral margin $C_{P,M}^{-2}-2\Lambda$ on the \emph{axially symmetric functions on the fixed background that are invariant under equatorial reflection.} The decomposition into exact and co-closed one-forms and Korn's inequality then give the estimates for coordinate variations in Corollary~\ref{q:functional-Poincare-Korn} below.

\item A Hardy estimate (in~\eqref{hv:hardy-estimate} below) bounds $\|\omega_\mu^{-1}V\|_2$ by the $H^1$ norm of a vector field $V$. It controls derivatives of cutoffs of size $\omega_\mu^{-1}$ and the weighted Ricci term in the vector equation. Near the added horizons, the estimate~\eqref{hv:cap} on a rescaled compact Schwarzschild region controls $\mu^{-1}\|V\|_2$ by $\|\nabla V\|_2$ on that region.

\item Sobolev and Schauder estimates on fixed-size rescaled charts give the local norm comparisons. Appendix~\ref{sec:functional-inequalities} below explains our choice of the dilation factors and uniform bounds.
\eei

\paragraph{Matching the Schwarzschild correction with the Green tensor.}

To obtain the uniform inverse estimate, we use the simpler reference metric. To determine the leading term of the auxiliary source coefficient, we match the first-order corrections on the Schwarzschild and Schwarzschild--de Sitter models.
We write $a(r)\coloneqq\del_{\nu_M}N_M(p_+(r))/N_M(p_+(r))$, where $\nu_M$ is the outward unit radial vector on the background. The Schwarzschild correction $A_S$ is regular at the horizon and has affine growth at infinity.

On the Schwarzschild--de Sitter side, we use a symmetric two-tensor
$H_r$ to describe the first-order perturbation of the background metric
caused by the two small black holes. For a given insertion radius $r$, we represent the two holes by point sources at their proposed centers $p_+(r)$ and $p_-(r)$. We call $H_r$ the Green tensor for these two
sources: it solves the linearized Einstein--DeTurck equation in the
sense of distributions on the entire background region, with a
tensor-valued Dirac source at each center. At the two original horizons, we require $H_r$ to be smooth in the normalized Cartesian charts associated with the prescribed surface gravities. These conditions
ensure that the perturbation preserves both surface gravities.

Near either insertion point, let $\rho$ denote the distance to that point in the background spatial metric. In spatial normal coordinates and with the locally normalized auxiliary coordinate $t$, the leading
term of $H_r$ is $2(\delta-\dd t^2)/\rho$, where $\delta$ is the
Euclidean spatial metric in these coordinates. The remainder is bounded
as $\rho\to0$. Thus the leading term of $\mu H_r$ reproduces the mass
term in the expansion of the Schwarzschild metric on the matching
annulus.
We use
\be
 G_S+\mu a(r)A_S \quad\text{in unit Schwarzschild coordinates},\qquad
 G_M+\mu H_r \quad\text{on the background}.
\ee
Both the leading mass term and the term from the first derivative of the lapse agree on the matching annulus. Lemma~\ref{f:jets} below bounds the difference and its scaled derivatives by the four powers associated with the background Taylor remainder, the bounded Green remainder, the quadratic mass term and the decaying Schwarzschild correction. The resulting differential $L_{\overline G}$ includes the two terms involving $S_{\overline G}$. Proposition~\ref{q:evaluated-actual-differential}~below bounds $L_{\overline G}-P_{G_{\mu,r}}$ by $\Ocal(\mu^{\eta_{\rm geom}})$ from $X_\mu$ to $Y_\mu$, so the augmented inverse remains uniform for the actual differential.

\paragraph{The nonlinear correction and the choice of insertion radius.}

Next, we write $g=\overline G+v$ and subtract the leading displacement source $\lambda_\mu^0\psi_\mu$ from $\mathcal E_{\overline G}(\overline G)$, denoting the resulting error by $f_{\mu,r}$. The equation becomes
\be
 L_{\overline G}v-(\lambda_\mu-\lambda_\mu^0)\psi_\mu=-f_{\mu,r}-Q(v),\qquad
 \ell_\mu(v)=0,
\ee
where $Q$ is a nonlinear remainder. Proposition~\ref{f:source} below gives $\|f_{\mu,r}\|_{Y_\mu} \le C\mu$ with one derivative in $r$. The change of the inverse metric in the principal part gives an error term of size $|v|\,|\nabla^2v|$. The terms associated with products of connection coefficients are bounded by $C(|\nabla v|^2+\omega_\mu^{-2}|v|^2)$, while a multiplication by the source weight $\omega_\mu^{2-\gamma}$ leaves the factor $\omega_\mu^\gamma\le\mu^\gamma$. Overall, the corresponding difference estimate is
\be
 \|Q(v)-Q(w)\|_{Y_\mu} \le
 C\mu^\gamma\bigl(\|v\|_{X_\mu}+\|w\|_{X_\mu}\bigr)\|v-w\|_{X_\mu}.
\ee
The source and inverse bounds determine a radius $K\mu$ for the correction ball, with $K$ fixed before $\mu$. On this ball the contraction factor and the relative pointwise size of $v$ are both $\Ocal(\mu^{1+\gamma})$. This gives a positive metric and a solution of the reduced equation. Differentiating the contraction in the moving charts gives $\|v\|_{X_\mu}+\|\del_r v\|_{X_\mu} \le C\mu$ in Proposition~\ref{f:family} below.

Let $Z_\mu$ be a reflected pair of cutoffs of the Schwarzschild displacement tensor. Pairing the correction equation with $Z_\mu$ and integrating its linear term by parts gives
\be
 (\lambda_\mu-\lambda_\mu^0)\langle\psi_\mu,Z_\mu\rangle
 =\langle f_{\mu,r},Z_\mu\rangle+\langle v,L_{\overline G}^*Z_\mu\rangle
   +\langle Q(v),Z_\mu\rangle.
\ee
The pairings use the volume measure of the reference metric per unit auxiliary coordinate. The denominator is of order $\mu^{1+\gamma}$, so the norm of the source alone \emph{would not provide a sufficiently accurate expansion} for the force. On the unit Schwarzschild model the flux of the affine correction gives $-16\pi$ for $P_S$ and another $-16\pi$ from the divergence term in the linearized Ricci tensor. After subtraction of this leading contribution, the remaining model source has zero integral against the displacement tensor. Lemma~\ref{f:test}~below uses this cancellation to control the three pairings with one derivative in $r$. Proposition~\ref{f:force}~below yields
\be\label{intro:force-estimate}
 \lambda_\mu(r)=-\frac{32\pi}{m_0}\mu^{1-\gamma}a(r)+\Ocal_{C^1(I)}(\mu^{1-\gamma+\delta}),\qquad
 m_0\coloneqq\int\chi_0|T_S|^2\dd\nu_S>0 .
\ee
Here $\dd\nu_S$ is Schwarzschild volume per unit auxiliary coordinate. Since $a(r_*)=0$ and $a'(r_*)=-\Lambda/N_*$, division by $\mu^{1-\gamma}$ shows that the $\Ocal_{C^1(I)}(\mu^\delta)$ error preserves the slope and the opposite endpoint signs on a fixed small interval. The zero $r_\mu$ satisfies $|r_\mu-r_*|\le C\mu^\delta$. Reflection makes the two source coefficients equal, so both vanish at $r_\mu$. The contracted Bianchi identity now gives
\be
 (\nabla_g^*\nabla_g-\Ric_g)W^\flat=0 .
\ee
The vector inverse in Proposition~\ref{q:vector-recovery-power}~below gives $W=0$. We obtain the static equations, with regularity in the horizon and axis charts as in Proposition~\ref{reg:bootstrap}~below.

\paragraph{Prescribing the central and added areas.}

The construction uses trial masses, while the theorem prescribes horizon areas. Let $M$ vary near the target mass $M_0$ and write the trial small mass as $u=s\varepsilon$, with $\varepsilon$ prescribed and $s$ in a fixed interval about one. Set $A_K(M)\coloneqq4\pi r_b(M)^2$, and let $A_b(M,u)$ and $A_s(M,u)$ be the actual central and added horizon areas. The area equations are $\mathcal F_{\varepsilon,1}=\mathcal F_{\varepsilon,2}=0$, with
\be
 \begin{aligned}
 \mathcal F_{\varepsilon,1}&=A_b(M,s\varepsilon)-A_K(M_0)
     =A_K(M)-A_K(M_0)+\Ocal(\varepsilon),\\
 \mathcal F_{\varepsilon,2}&=\frac{A_s(M,s\varepsilon)}{16\pi\varepsilon^2}-1
     =s^2-1+\Ocal(\varepsilon^{1+\gamma}).
 \end{aligned}
\ee
The monotonicity of $A_K$ gives the sign of the first component on the two mass faces. On the two faces in $s$, the sign of the second component is that of $s^2-1$. Proposition~\ref{n:retune}~below applies Brouwer's theorem to the projection of $(M,s)-t\mathcal F_\varepsilon(M,s)$ onto the rectangle for a fixed $t>0$. The signs force a fixed point to be interior and to solve both equations. At such a zero $M-M_0=\Ocal(\varepsilon)$ and $u/\varepsilon=1+\Ocal(\varepsilon^{1+\gamma})$. The first-order profiles and local elliptic estimates give the two smooth limits in Theorem~\ref{prob:existence}. The quantitative conditions used in these steps give the bound~\eqref{q:master-threshold}.

The Green tensor also determines the first angular variation of the lapse. Its Bianchi term vanishes at the limiting insertion radius. The improved source estimate then makes the nonlinear correction smaller than the first-order contribution from the Green tensor on the part of the maximal sphere away from the axis. Section~\ref{sec:critical-maximal} obtains its angular profile from a radial Green function and the spherical heat equation. The profile has a strict maximum at the equator with negative second derivative. Its first-order contribution, with a remainder in $C^2$, gives a unique circle of maxima and the transverse Hessian estimates.

\section{The reference geometry}\label{sec:background}

\subsection{The sphere of maximal background lapse}

Let $x$ be signed proper radial distance, increasing with $r$ and zero at $r_*$, and let $\theta\in[0,\pi]$ be the polar angle measured from the positive symmetry axis. Thus $\partial_x=N_M\partial_r$. Recall that $N_*=N_M(r_*)$. The identity $r_*^3=3M/\Lambda$ gives
\be\label{eq:maximum-value}
N_*^2=1-\Lambda r_*^2=1-(9\Lambda M^2)^{1/3}>0.
\ee

\begin{lemma}[The Hessian of the lapse on the maximal sphere]\label{lem:normal}
Use the background~\eqref{eq:F}. Its full critical set is $S_*\coloneqq\{r=r_*\}$, and at every point of this sphere
\be\label{eq:hessian}
\Hess_{h_M}N_M=-\Lambda N_*\,\dd x\otimes\dd x.
\ee
In particular its two tangential Hessian eigenvalues vanish, while its normal eigenvalue is strictly negative.
\end{lemma}
For every subextremal mass $M$, the background lapse attains its maximum on the sphere $r=r_*$. Its second derivative in the unit normal direction is strictly negative at each point of that sphere. Its two tangential directions correspond to rotating the insertion point along the sphere. Once the axis and the reflected pair are fixed, varying the radius is the remaining way to move both centers.
\begin{proof}
For $M>0$, the function $F_M$ tends to $-\infty$ at both ends of $(0,+\infty)$ and has a positive maximum by~\eqref{eq:maximum-value}, so it has two simple positive zeros. The derivative $F_M'=2M/r^2-2\Lambda r/3$ has its unique zero at $r_*$. Direct differentiation gives
\be
(N_M)_x=F_M'/2,\qquad (N_M)_{xx}=N_M F_M''/2.
\ee
Since $F_M''(r_*)=-2\Lambda$, the normal coefficient is the one in~\eqref{eq:hessian}. In the warped metric $\dd x^2+r(x)^2\dd\omega^2$, the tangential Hessian is $(N_M)_xr_x/r$ times the tangential metric. It vanishes at $r_*$. The mixed Hessian is zero by spherical symmetry.\qedhere
\end{proof}

\subsection{Radial equilibrium of a static test particle}

The energy of a static point particle of proper mass $\mu$ relative to the Killing field $\partial_t$ is $E\coloneqq\mu N_M$. This follows by pairing that field with the particle's unit velocity $N_M^{-1}\partial_t$ in the metric $-N_M^2\dd t^2+h_M$. Define its energy force by $-\nabla_{h_M}E$.

\begin{lemma}[The radial force on a static test particle]\label{lem:particle}
The outward proper radial component of this force is
\be\label{eq:particle-force}
\mathcal F_{\rm part}(r)\coloneqq\mu\left(\frac{\Lambda r}{3}-\frac M{r^2}\right).
\ee
Its zero is $r_*$, and
\be
\partial_r\mathcal F_{\rm part}(r_*)=\mu\Lambda,\qquad
\partial_x\mathcal F_{\rm part}(r_*)=\mu\Lambda N_*.
\ee
\end{lemma}
The actual compatibility coefficient of the corrected tensor equation is computed in Proposition~\ref{f:force} below. Its leading term is a nonzero multiple of $\partial_xN_M/N_M$, and its remainder is small enough to select a unique nearby insertion radius. The test-particle calculation identifies the same background radius and its radial nondegeneracy.
The reflected pair has identical outward radial coefficients and opposite signed axial coefficients. Reflection therefore reduces the paired displacement equations to one common radial equation, provided the reference, projections and inverse preserve that symmetry. A single added hole has an odd dipole sector which requires its own outer analysis.
\begin{proof}
Use $-\mu(N_M)_x=-\mu F_M'/2$ and differentiate. At $r_*$, $2M/r_*^3=2\Lambda/3$. This gives both derivatives.\qedhere
\end{proof}

\subsection{Geometry at the horizons and matching surfaces}\label{sec:interfaces}

Near a horizon of surface gravity $\kappa>0$, we use spatial distance $s$ and the Cartesian normal coordinates $s\cos(\kappa\tau)$, $s\sin(\kappa\tau)$ for the associated metric $G=h+N^2\dd\tau^2$. Regularity means smoothness of its tensor components in these coordinates and in tangential coordinates on the horizon. We call them normalized horizon charts. A collar denotes a sufficiently small product neighborhood of the horizon. Each horizon uses its own value of $\kappa$ in these charts. The first reference metric agrees with Schwarzschild--de Sitter near the original horizons and with Schwarzschild near each new horizon. The Green tensor constructed on the background preserves both original surface gravities. The angular lifts, the axis conditions and the tensor projection of~\cite[Section~2.2 and Sections~3 to~5]{AX-Existence} provide the relevant local definitions.

Let $\Sigma$ be an oriented hypersurface inside the positive-lapse region and let $\nu$ be its unit normal. Write its induced metric as $\gamma$, its second fundamental form as $A(X,Y)\coloneqq h(\nabla_X\nu,Y)$, its mean curvature as $H\coloneqq\operatorname{tr}_\gamma A$, and its lapse traces as $u\coloneqq N|_\Sigma$, $v\coloneqq\nu N|_\Sigma$. Its scalar curvature is $R_\gamma=2K_\gamma$.

\begin{lemma}[Constraints for the four interface data]\label{prop:constraints}
On an interface with $u>0$, the data of a static solution satisfy
\bse\label{eq:constraints}
\be
R_\gamma-H^2+|A|_\gamma^2-2\Lambda
=\frac2u(\Delta_\gamma u+Hv),
\ee
\be
u(\operatorname{div}_\gamma A-\dd H)
=\dd v-A(\nabla_\gamma u,\cdot).
\ee
\ese
Here the factor on the left of the second equation is the scalar $u$.
\end{lemma}
These constraints identify the geometric quantities that must agree after the collars are placed in a common domain. At a regular horizon, with the normal directed into the positive-lapse region, $u=0$, $A=0$ and $v$ is a positive constant. The positive-lapse formula~\eqref{eq:constraints} is used on artificial interfaces outside that horizon. The final horizon metric and its surface gravity are selected together with the global solution.
\begin{proof}
Tracing~\eqref{eq:static} gives $\Scal_h=2\Lambda$. The lapse decomposition is $\Delta_hN=N_{\nu\nu}+Hv+\Delta_\gamma u$. Consequently $\Ric_h(\nu,\nu)=-(Hv+\Delta_\gamma u)/u$. Substitution in the Gauss equation proves the first identity. The contracted Codazzi equation is $\Ric_h(\nu,\cdot)=\operatorname{div}A-\dd H$, and the mixed Hessian is $\Hess_hN(\nu,\cdot)=\dd v-A(\nabla u,\cdot)$. Their substitution in the static equation proves the second identity.\qedhere
\end{proof}

\begin{lemma}[Local expansion at a regular static horizon]\label{reg:static-horizon-expansion}
Let $H$ be a nondegenerate regular horizon of a smooth static pair, with its smooth normalized local lift. Let $\gamma_H$ be its induced metric, $K_H$ its Gauss curvature, and $\kappa_H=|\nabla N|_H>0$. In Gaussian spatial distance $s$ directed into the positive-lapse region,
\be\label{reg:static-horizon-jets}
h=\dd s^2+\gamma_H+\frac{s^2}{2}(K_H-\Lambda)\gamma_H+\Ocal(s^4),\qquad
N=\kappa_Hs-\frac{\kappa_HK_H}{6}s^3+\Ocal(s^5).
\ee
The opening $\kappa_H$ is constant on $H$, and the remainder coefficients are smooth in the tangential variables, including the axial poles. Thus the horizon curvature can vary smoothly along the horizon.
\end{lemma}
\begin{proof}
The static equation at $N=0$ gives $\operatorname{Hess}N=0$. Its tangential and mixed components give $A_H=0$ and $\dd_H\kappa_H=0$. Smooth rotation invariance of the normalized lift gives $\gamma(-s)=\gamma(s)$ for the tangential metric and $N(-s)=-N(s)$ for the signed lapse, where $s$ is the signed Gaussian distance from the horizon. Write $\gamma(s)=\gamma_H+s^2\gamma_2+\Ocal(s^4)$ and $N=\kappa_Hs+n_3s^3+\Ocal(s^5)$. At $s=0$ the Ricci components are $\Ric_{ss}=-\operatorname{tr}_{\gamma_H}\gamma_2$ and $\Ric_{ab}=K_H(\gamma_H)_{ab}-(\gamma_2)_{ab}$. The coefficient of $s$ in the tangential static equation gives $2\gamma_2=(K_H-\Lambda)\gamma_H$. The normal equation then gives $6n_3=-\kappa_HK_H$, proving the formulas. The calculations use smooth Cartesian coordinates at a pole, so the tangential coefficients have their ordinary smooth interpretation there.
\end{proof}

\section{Einstein equations linearized about Schwarzschild--de Sitter}\label{sec:linear-Kottler}\label{sec:spectral-models}

\subsection{The spectral gap}

We use here standard interior elliptic estimates and compactness in Sobolev spaces (cf.\ the textbook~\cite[Chapters~5--8]{GT2001}). The tensor systems below have scalar second-order principal part, so their local estimates follow componentwise, absorbing the lower-order couplings by interpolation.
On the static interior, let $G_M\coloneqq h_M+N_M^2\dd\tau^2$ and write $\Delta\coloneqq\operatorname{div}_{G_M}\nabla_{G_M}$. For a function $f$ independent of $\tau$, we have
\be\label{eq:scalar-operator}
-\Delta f=-\frac1{r^2}\partial_r(r^2F_M\partial_rf)
             -\frac1{r^2}\Delta_{\mathbb S^2}f.
\ee
The measure per unit auxiliary parameter is $r^2\dd r\dd\omega$. Regularity means smoothness in the local angular lifts at both horizons. In particular, a radial coefficient and its radial derivatives are bounded in the area coordinate $r$. We require axial symmetry and evenness under $\omega\mapsto-\omega$.

We keep the length scale $\ell=\sqrt{3/\Lambda}$ explicit in this section and write $F\coloneqq F_M$, $F_*\coloneqq F_M(r_*)=N_*^2$. Define the scalar multiplier exponent and spectral margin by
\be\label{q:scalar-constants}
\alpha_*\coloneqq\min\bigl(\frac14,\frac{F_*}{8\Lambda r_c^2}\bigr),\qquad d_s\coloneqq\min\bigl(\alpha_*\Lambda,\frac6{r_c^2}-2\Lambda\bigr)>0.
\ee
The number $d_s$ has units of inverse length squared.

\begin{proposition}[An explicit scalar spectral gap]\label{prop:scalar-gap}\label{q:scalar-gap}
Fix any $M$ satisfying~\eqref{eq:mass-range}. Every nonconstant regular axially and reflection-symmetric scalar eigenfunction of $-\Delta$ has eigenvalue at least $2\Lambda+d_s$. In particular, every regular solution of $-\Delta f=2\Lambda f$ in this class is zero. On the space of regular axially symmetric functions invariant under equatorial reflection, including constants, one has
\be\label{q:scalar-inverse}
\| (-\Delta-2\Lambda)^{-1}f\|_2\le d_s^{-1}\|f\|_2.
\ee
\end{proposition}
In Corollary~\ref{cor:harmonic-gauge} below, the divergence of a vector field in the kernel of the operator on vector fields solves $-\Delta f=2\Lambda f$. The scalar spectral gap forces this divergence to vanish. For static variations of spherical harmonic degree $l\ge2$, the proof of Proposition~\ref{prop:physical-kernel}~below derives the radial equations after choosing coordinates. The perturbation equations of Kodama and Ishibashi~\cite{KI2003} provide the general setting for this decomposition. A tensor inverse additionally requires a choice of gauge, the finite-dimensional mass and lapse parameters, and the permitted variations of horizon surface gravities. These pieces must be combined on the punctured region used in the gluing construction.
\begin{proof}
Decompose into zonal spherical harmonics of degree $l$. Equatorial reflection multiplies $P_l(\cos\theta)$ by $(-1)^l$, so only even degrees $l$ occur. For a nonzero eigenmode $f_l$ of degree $l\ge2$ and eigenvalue $\sigma$, integration of the equation gives
\be
\int_{r_b}^{r_c}\bigl(r^2F_M|f_l'|^2+l(l+1)|f_l|^2\bigr)\dd r
=\sigma\int_{r_b}^{r_c}r^2|f_l|^2\dd r.
\ee
The boundary term vanishes by regularity. Since $F_M(r_c)=0$ implies $r_c^2<3/\Lambda$, this identity gives $\sigma\ge l(l+1)/r_c^2\ge6/r_c^2>2\Lambda$.

For $l=0$, more generally suppose $-(r^2F_Mf')'=\sigma r^2f$ and put $w\coloneqq f'$. Differentiate and use $F_M''+2F_M'/r=-2\Lambda$. This yields
\bse\label{eq:radial-identities}
\be\label{eq:radial-differentiated}
(r^2F_M^2w')'+((\sigma-2\Lambda)r^2F_M-2F_M^2)w=0.
\ee
Multiply by $w$ and integrate. The term $r^2F_M^2ww'$ vanishes at both horizons, so
\be\label{eq:radial-gap}
(\sigma-2\Lambda)\int_{r_b}^{r_c}r^2F_Mw^2\dd r
=\int_{r_b}^{r_c}(r^2F_M^2(w')^2+2F_M^2w^2)\dd r.
\ee
\ese
Every nonconstant regular radial eigenfunction therefore has $\sigma>2\Lambda$. At $\sigma=2\Lambda$, the identity forces $w=0$, and the original equation then forces $f=0$.
For a quantitative bound, use~\eqref{eq:radial-gap}. Its Rayleigh quotient above $2\Lambda$ has numerator $\int(r^2F^2(w')^2+2F^2w^2)\dd r$ and denominator $\int r^2Fw^2\dd r$. For the associated operator
\bse\label{eq:radial-quantitative-gap}
\be
\mathcal D_0 w\coloneqq-\frac{(r^2F^2w')'}{r^2F}+\frac{2F}{r^2}w,
\qquad \phi\coloneqq F^{-\alpha_*},
\ee
direct differentiation gives
\be\label{q:multiplier}
\frac{\mathcal D_0\phi}{\phi}=\alpha_*(1-\alpha_*)\frac{(F')^2}{F}-2\alpha_*\Lambda+\frac{2F}{r^2}.
\ee
The inequality $F''\le-2\Lambda/3$ and $F'(r_*)=0$ imply $(F')^2\ge(4\Lambda/3)(F_*-F)$. To see this, integrate $(F'^2)'=2F'F''$ between $r$ and $r_*$, separately on the two sides of the maximum. Where $F\le F_*/4$, the first term of~\eqref{q:multiplier} is at least $4\alpha_*(1-\alpha_*)\Lambda$, and the whole expression is at least $\alpha_*\Lambda$. Where $F\ge F_*/4$, its last two terms are at least $F_*/(2r_c^2)-2\alpha_*\Lambda\ge F_*/(4r_c^2)\ge\alpha_*\Lambda$.

The identity obtained by writing $w=\phi(w/\phi)$ now gives
\be
\int(r^2F^2(w')^2+2F^2w^2)\dd r\ge\alpha_*\Lambda\int r^2Fw^2\dd r.
\ee
\ese
Its endpoint term is $-\alpha_*r^2FF'w^2$, which vanishes at both horizons because $w$ is regular. The remaining square is integrable despite the divergence of $\phi$ at both horizons. This proves the radial margin. For the nonradial modes allowed by equatorial reflection, namely those of even degree $l\ge2$, the angular energy gives the margin $6/r_c^2-2\Lambda$.

The regular scalar domain has compact resolvent. Indeed choose a positive smooth function $a_0(r)$ equal to $\kappa_j^{-1}$ near horizon $j$, where $\kappa_j$ is its surface gravity. The auxiliary metric $h_M+(a_0N_M)^2\dd\vartheta^2$, with $\vartheta\in\mathbb R/(2\pi\mathbb Z)$, fills both horizons smoothly. Its volume measure multiplied by $a_0^{-1}$ is the invariant density. The scalar operator is a smooth weighted Laplacian on its invariant functions. Rellich compactness and the Dirichlet form therefore give a complete regular eigenbasis.

The eigenbasis proves the estimate on nonconstant functions. The constant mode has shifted eigenvalue $-2\Lambda$, whose modulus is larger than $d_s$; this also proves~\eqref{q:scalar-inverse}.\qedhere
\end{proof}

\subsection{Energy estimates}

A vector field $V$ generates the metric variation $\mathcal L_VG_M$ at first order under pullback by its flow. We use half this variation and its negative divergence in the energy estimates. Define
\be\label{eq:vector-definitions}
\mathcal B V\coloneqq\tfrac12\mathcal L_VG_M,\qquad
\mathcal A V\coloneqq\delta_{G_M}\mathcal B V,
\ee
where $\delta$ is minus the covariant divergence. We use vector fields invariant under the two circular symmetries and their reversals, orthogonal to both circular orbits, and smooth in the normalized local horizon and axis coordinates. Their $H^1$ completion is denoted by $\mathcal V_M$. Norms use the measure $N_M\dd V_{h_M}$, per unit auxiliary parameter. Polar reflection may also be imposed.

\begin{proposition}[Control by the symmetric gradient on a fixed background]\label{prop:vector-energy}
Fix $M$ as in~\eqref{eq:mass-range}. There is $C_M<+\infty$ such that
\be\label{eq:vector-energy}
\|V\|_{H^1} \le C_M\|\mathcal B V\|_{L^2}
\qquad(V\in\mathcal V_M).
\ee
The operator~\eqref{eq:vector-definitions} therefore has a unique weak inverse from $\mathcal V_M^*$ to $\mathcal V_M$. These bounds are uniform when $M$ ranges in a compact subinterval of~\eqref{eq:mass-range}.
\end{proposition}
The smooth auxiliary metric in this proof provides an atlas and compactness. The actual associated Riemannian Einstein metric keeps its own horizon normalizations. The proof does not require equality between the surface gravities of the central and cosmological horizons.

\begin{proof}
Let $a_0(r)>0$ be smooth and equal to $\kappa_j^{-1}$ near horizon $j$, with $\kappa_j\coloneqq|\nabla N_M|$ there. The auxiliary metric
\be
\widetilde G\coloneqq h_M+(a_0N_M)^2\dd\vartheta^2,
\qquad\vartheta\in\mathbb R/(2\pi\mathbb Z),
\ee
fills both horizons smoothly and gives a finite atlas on a compact four-manifold. It is used for estimates. Near each horizon $a_0$ is constant, so normalized covariant derivatives of invariant vector fields agree with those in the local lift of $G_M$. In the remaining compact region the two derivatives and symmetric gradients differ by bounded zeroth-order terms. Their norms are consequently equivalent up to such terms.

On this smooth auxiliary metric, integration by parts gives
\be
2\|\mathcal B_{\widetilde G}V\|_2^2
=\|\widetilde\nabla V\|_2^2+
\|\operatorname{div}_{\widetilde G}V\|_2^2
-\int\Ric_{\widetilde G}(V,V)\dd V_{\widetilde G}.
\ee
The bounded Ricci term and the norm comparisons yield
\be\label{eq:vector-preliminary}
\|V\|_{H^1} \le C_M(\|\mathcal B V\|_2+\|V\|_2).
\ee
The same finite atlas gives compactness of $H^1\hookrightarrow L^2$ in this invariant class.
There is no nonzero Killing field in the class. At a point fixed by both the horizon and axial rotations, invariance forces its value to be zero. Its skew first derivative commutes with the independent rotations in the two normal planes. It is thus a rotation in each plane, and orthogonality to both circular directions forces both coefficients to vanish. Killing transport from this point gives zero on the connected interior. A weak field with $\mathcal B V=0$ is smooth by local elliptic regularity, so the same argument applies to it.

A sequence contradicting~\eqref{eq:vector-energy}, normalized in $H^1$ and with symmetric gradients tending to zero, has an $L^2$-convergent subsequence. Estimate~\eqref{eq:vector-preliminary} makes it converge in $H^1$ to a nonzero Killing field. This contradiction proves~\eqref{eq:vector-energy}. The bilinear form $\int\langle\mathcal BV,\mathcal BW\rangle$ is coercive and yields the stated weak inverse. The local atlas and bounds depend smoothly on $M$ on a compact mass interval. The same compactness argument at a limiting background proves uniformity there.\qedhere
\end{proof}

\begin{corollary}[Absence of regular harmonic gauge fields invariant under equatorial reflection]\label{cor:harmonic-gauge}
Use polar reflection and the nonnegative Hodge Laplacian $\Delta_H=\nabla^*\nabla+\Ric_{G_M}$ on one-forms. A regular vector field in the stated class satisfying $(\Delta_H-2\Lambda)V^\flat=0$ is zero.
\end{corollary}
\begin{proof}
Hodge divergence commutes with the Hodge Laplacian. Thus $\delta V^\flat$ solves the scalar equation excluded by Proposition~\ref{prop:scalar-gap}, and is zero. For the Bianchi operator $\beta T=\delta T+\tfrac12\dd\operatorname{tr}T$, the Einstein identity gives
\be
\beta(\mathcal L_VG_M)=(\Delta_H-2\Lambda)V^\flat=0.
\ee
Since $\operatorname{div}V=0$, it follows that $\delta\mathcal B V=0$. Integration against $V$ gives $\|\mathcal B V\|_2^2=0$, and~\eqref{eq:vector-energy} finishes the proof. The integrations may be performed outside small horizon tubes. In a regular normalized lift, the normal component of $V$ is $O(s)$, mixed components have their smooth parity, and tube area is $O(s)$. The boundary terms tend to zero. The same local smoothness justifies the Hodge identity across the filled fixed sets.\qedhere
\end{proof}

\subsection{Solutions of the homogeneous linearized equations}

We consider variations $(k,n)$ of the spatial metric and lapse that satisfy the homogeneous linearized static equations. The allowed surface-gravity variations are constant on each horizon. Throughout this subsection, we restrict attention to variations $(k,n)$
that preserve the symmetries specified in Section~\ref{sec:introduction}.
Both the metric variation $k$ and the lapse variation $n$ are invariant
under rotations about the symmetry axis, reflection across the equatorial
plane, and reversal of the azimuthal angle $\varphi\mapsto-\varphi$.
In the expansion in scalar spherical harmonics, axial symmetry and
equatorial reflection leave only the modes $P_\ell(\cos\theta)$ with
even $\ell$. Reversal of the azimuthal angle forces the mixed components
$k_{r\varphi}$ and $k_{\theta\varphi}$ to vanish.

The central horizon area and the marked lapse give two scalar normalizations. We use auxiliary mass parameters to prescribe the areas in Theorem~\ref{prob:existence}. The central area removes the background mass variation: implicit differentiation of $F_M(r_b)=0$ gives
\be\label{eq:area-parameter}
\frac{\dd r_b}{\dd M}=\frac{2}{r_bF_M'(r_b)}>0,\qquad
\frac{\dd}{\dd M}(4\pi r_b^2)=\frac{16\pi r_b}{1-\Lambda r_b^2}>0.
\ee
For lapse normalization, use the equatorial point $q_0$ on $S_*$ and prescribe $N(q_0)=N_M(q_0)$. Since $\dd N_M(q_0)=0$, this linearized functional is unchanged by a horizon-preserving coordinate variation. The map from the background mass and the constant rescaling of the lapse to the central horizon area and the lapse at $q_0$ therefore has an invertible differential. The two added areas and the corrected scalar map are prescribed in Section~\ref{sec:areas} below.

\begin{proposition}[Classification of the homogeneous static variations]\label{prop:physical-kernel}
For any fixed $M$ in~\eqref{eq:mass-range}, a regular homogeneous linearized static pair in this class is a variation of mass and lapse factor, plus a smooth horizon-preserving coordinate variation. Fixing the central horizon area and the lapse at the marked equatorial point removes the two scalar parameter variations.
\end{proposition}
The quotient statement in Proposition~\ref{prop:physical-kernel} allows the two horizon surface gravities to vary. We first specify the domain of the tensor operator by fixing both horizon surface gravities. The independent mass and variations of the lapse normalization will then account for the normalization of the central horizon area and the lapse.
\begin{proof}
For a spherical harmonic $Y_l\coloneqq P_l(\cos\theta)$ with $l\ge2$, first put the perturbation in Gaussian gauge near each horizon. The required normal and tangential coordinate vectors have forms $sA(s^2,\theta)$ and $s^2B(s^2,\theta)$. They preserve both boundaries. An angular vector removes the trace-free angular metric coefficient $r^2G_l(\nabla^2Y_l)^{\rm TF}$. If the mixed coefficient is $K_l\dd Y_l$, the additional radial vector
\be
X_l^r\coloneqq-F_M(K_l-\tfrac12r^2G_l')Y_l
\ee
removes that term. To check regularity, use a normal coordinate $s$ at either horizon. The relation $r-r_j=s^2c_j(s^2)$ holds with $c_j(0)\ne0$. Smooth invariance in the normal plane gives $k_{sa}=s$ times a smooth function of $s^2$, so the mixed coefficient $K_l$ in $k_{ra}=K_l\dd Y_l$ is smooth in $r-r_j$; the angular coefficient $G_l$ is smooth as well. Hence $X_l^r=O(r-r_j)$ and $X_l^s=X_l^r/(\dd r/\dd s)$ is $s$ times a smooth function of $s^2$. The vector extends smoothly through the normal lift and is tangent to the horizon. The same argument applies at both ends. In this gauge write
\be
k_{rr}=F_M^{-1}H_{2,l}Y_l,\qquad
k_{ab}=r^2B_lY_l\omega_{ab},\qquad
n=-\tfrac12N_MH_{0,l}Y_l.
\ee
The angular trace-free, mixed and lapse equations give respectively $H_{0,l}=H_{2,l}=H_l$ and
\be\label{eq:physical-H}
B_l'=H_l'+\frac{F_M'}{F_M}H_l,\qquad
(r^2F_MH_l')'-
\left(l(l+1)+2\Lambda r^2+\frac{r^2(F_M')^2}{F_M}\right)H_l=0.
\ee
For completeness, the trace-free angular equation is
\be
 -\tfrac12(H_{2,l}-H_{0,l})(\nabla_a\nabla_bY_l)^{\rm TF}=0.
\ee
Since this tensor harmonic is nonzero for $l\ge2$, the two coefficients
agree.  The mixed equation then reads
\be
 -\tfrac12\left(B_l'-\frac{H_l}{r}\right)
 =-\tfrac12H_l'+\frac{H_l}{2r}
   -\frac{F_M'}{2F_M}H_l,
\ee
which gives the first identity in~\eqref{eq:physical-H}.  In the linearized
lapse equation, the radial flux after removal of the spherical density is
\be
 \frac{r^2F_M'}2+
 \frac{r^2}{2}\bigl(F_M'(B_l-H_l)-F_MH_l'\bigr)Y_l.
\ee
The angular term contributes $l(l+1)H_l/(2r^2)$ after division by $N_M$.
The variations of the volume factor and of the term $\Lambda N$ also contribute. Using $(r^2F_M')'=-2\Lambda r^2$, the coefficient of $Y_l$ in $N_M^{-1}\delta(\Delta_hN+\Lambda N)$ is
\be
\frac1{2r^2}\Bigl(\bigl(r^2F_M'(B_l-H_l)-r^2F_MH_l'\bigr)'
+2\Lambda r^2B_l+l(l+1)H_l\Bigr).
\ee
Inserting $B_l'-H_l'=F_M'H_l/F_M$ cancels $B_l$ and gives the second identity in~\eqref{eq:physical-H}. These equations are regular at $r_*$ and use no division by $F_M'$.
Regularity gives coefficients smooth in $r$ at each simple horizon. To justify the boundary condition for the linearized equations, let $X$ be tangent to a horizon and $\nu$ its unit normal. Because $N=0$ on the horizon, $\Hess N(X,\nu)=X(\nu N)$ there. Linearizing this identity together with the mixed static equation shows that the first variation of $\kappa=|\nabla N|$ is constant along each connected horizon. Its degree-$l$ component therefore vanishes for $l\ge2$. In the gauge above, the variation of the normal-plane opening is
\be
 \frac{\delta\kappa}{\kappa}=\lim_{r\to r_j}
 \left(\frac nN-\frac12 k(\nu,\nu)\right)
 =-H_l(r_j)Y_l,\qquad j=b,c.
\ee
The same boundary condition follows directly from the radial equation in~\eqref{eq:physical-H}. Multiply that equation by $F_M$ and let $r$ tend to $r_j$. Smoothness of $H_l$ in $r$ makes the differentiated flux bounded, so its product with $F_M$ tends to zero. The remaining term gives $r_j^2F_M'(r_j)^2H_l(r_j)=0$. Since the horizon is simple, $H_l(r_j)=0$. Smoothness then gives $H_l=O(F_M)$ at both horizons. Multiplying~\eqref{eq:physical-H} by $H_l$ gives
\be
\int_{r_b}^{r_c}\Bigl(r^2F_M(H_l')^2+
\left(l(l+1)+2\Lambda r^2+\frac{r^2(F_M')^2}{F_M}\right)H_l^2\Bigr)\dd r=0.
\ee
The boundary term vanishes because $H_l=O(F_M)$ and $H_l'$ is bounded. Thus $H_l=0$ and the first equation in~\eqref{eq:physical-H} gives $B_l'=0$. It remains to use the linearized scalar-curvature constraint. With $B_l$ constant, the radial part of the spatial metric is unchanged and the angular metric has first variation $r^2B_lY_lg_{\mathbb S^2}$. The conformal scalar-curvature formula on the unit sphere gives
\be
 \delta\Scal_h=\frac{l(l+1)-2}{r^2}B_lY_l.
\ee
The radial contributions are unchanged because $B_l$ is constant in $r$. Since the static equations impose $\Scal_h=2\Lambda$, the linearized constraint makes $B_l=0$ for $l\ge2$. These modes are coordinate variations. Smooth harmonic coefficients have rapid decay, while the coordinate construction uses fixed derivative orders and polynomial harmonic factors; its vector series therefore converges smoothly on the whole region, including the horizon collars.

Reflection removes degree one. In degree zero the spherical equations in area coordinates are $(rF_M)'=1-\Lambda r^2$ and $(N^2/F_M)'=0$. Their linearizations give a constant mass variation $\delta M$ and a constant variation of the lapse normalization, modulo a radial coordinate change. To check the two boundaries, let $\delta R$ be the variation of the area radius in the original fixed-boundary coordinates. Differentiating $F_M(r_j(M))=0$ at either horizon gives
\be
 \delta R|_{H_j}=r_j'(M)\,\delta M,\qquad
 r_j'(M)=\frac{2}{r_jF_M'(r_j)},\quad j=b,c.
\ee
Choose a smooth family of radial identifications which maps each background horizon to the corresponding horizon of the varied mass, and which is smooth in the normalized horizon coordinates. In area coordinates its derivative with respect to mass takes the values $r_j'(M)$ at the two endpoints. Use these identifications to represent the mass variation on the fixed manifold. After subtracting that represented variation, the remaining area-radius variation vanishes at both horizons. The radial vector which removes it is therefore smooth in the normalized horizon lifts and preserves both boundaries. The positive area derivative in~\eqref{eq:area-parameter} removes the mass variation. The marked lapse condition then fixes the remaining factor. Since $\nabla N_M(q_0)=0$, coordinate variations leave this linearized condition unchanged.\qedhere
\end{proof}

\subsection{Invertibility with prescribed surface gravities on the horizon}

Write $G\coloneqq G_M$ and $\kappa_j=|\nabla N_M|_{H_j}$ for $j=b,c$. Choose a smooth positive radial function $a_0$ equal to $\kappa_j^{-1}$ near horizon $j$. The auxiliary metric $\widetilde G=h_M+(a_0N_M)^2\dd\vartheta^2$, with angular period $2\pi$, fills both horizons and defines a smooth compact four-manifold $X$ diffeomorphic to $\mathbb S^2\times\mathbb S^2$. Let $\mathcal K$ denote the compact group generated by its two rotations, their separate reversals and polar reflection. Throughout this section, we require the static tensors to be invariant under $\mathcal K$.

On covectors in the interior, define $\Phi$ to be the identity on spatial covectors and to send $\dd\tau$ to $a_0\dd\vartheta$, extending this map tensorially. Thus
\be\label{eq:tensor-identification}
\Phi(k+v_{\tau\tau}\dd\tau^2)=k+a_0^2v_{\tau\tau}\dd\vartheta^2.
\ee
This map preserves pointwise tensor norms: the spatial covectors are unchanged and both $\dd\tau$ and $a_0\dd\vartheta$ have length $N_M^{-1}$ in their respective metrics. Moreover $\dd V_{\widetilde G}=a_0N_M\dd V_{h_M}\dd\vartheta$. Thus, for invariant tensors $v,w$,
\be
 \frac1{2\pi}\int_X\langle\Phi v,\Phi w\rangle_{\widetilde G}a_0^{-1}\dd V_{\widetilde G}
 =\int\langle v,w\rangle_G N_M\dd V_{h_M}.
\ee
This identity uses no equality of the two horizon surface gravities. Define the Hilbert target and operator domain through the smooth auxiliary manifold by
\bse\label{eq:tensor-domain}
\be
\mathcal H\coloneqq\Phi^{-1}L^2(X;\operatorname{Sym}^2T^*X,a_0^{-1}\dd V_{\widetilde G})^{\mathcal K},
\ee
\be
\mathcal D\coloneqq\Phi^{-1}H^2(X;\operatorname{Sym}^2T^*X,\widetilde G)^{\mathcal K}.
\ee
\ese
The inner product includes the factor $1/(2\pi)$, so it is exactly integration with $N_M\dd V_{h_M}$ per unit original auxiliary parameter. The norms in~\eqref{eq:tensor-domain} are those transported by $\Phi$. The smooth invariant core specifies all Cartesian parities at the two horizons and the axis.

Write a tensor in the smooth core as $v=k+2N_Mn\dd\tau^2$, where $k$ and $n$ are the variations of the spatial metric and lapse. Let $s$ be distance to a horizon. Smoothness in its fixed normalized normal plane gives
\be
k_{ss}=A+O(s^2),\qquad n=\tfrac12\kappa_jsA+O(s^3),\qquad
\dot\kappa_j/\kappa_j=\lim_{s\to0}(n/(\kappa_js)-k_{ss}/2)=0.
\ee
These equalities are consequences of full Cartesian regularity. A single equality between leading coefficients would not specify the tensor domain.

Define the curvature action by $(\operatorname{Riem}_Gv)_{ij}=R_{ikjl}v^{kl}$, and put
\be\label{eq:tensor-operator}
P_G=\nabla_G^*\nabla_G-2\operatorname{Riem}_G,\qquad
\beta_Gv=\delta_Gv+\tfrac12\dd\operatorname{tr}_Gv.
\ee
The sign convention is
\be\label{eq:tensor-linearization}
2(D(\Ric-\Lambda g)_G[v]+\mathcal B\beta_Gv)=P_Gv.
\ee
Here $\mathcal B$ also acts on one-forms by raising the index with $G$.

\begin{proposition}[The inverse on Schwarzschild--de Sitter with prescribed surface gravities]\label{thm:outer-tensor-inverse}
Fix the mass $M$ of the Schwarzschild--de Sitter background in the subextremal range~\eqref{eq:mass-range}. The linearized Einstein operator $P_G$ with the Bianchi gauge term in~\eqref{eq:tensor-linearization} is self-adjoint on the invariant $L^2$ tensor space $\mathcal H$, with domain $\mathcal D$ as in~\eqref{eq:tensor-domain}. This domain imposes regularity in the normalized charts with both horizon surface gravities fixed. The operator is bijective from $\mathcal D$ to $\mathcal H$, and its inverse satisfies, for every source $f\in\mathcal H$,
\be\label{eq:outer-tensor-inverse}
\|P_G^{-1}f\|_{\mathcal D} \le C_M\|f\|_{\mathcal H}.
\ee
For every integer $m\ge0$ and $0<\alpha<1$, the same inverse maps invariant $C^{m,\alpha}$ tensor sources boundedly into $C^{m+2,\alpha}$ in the normalized horizon charts. The constants in these estimates are uniform on compact subintervals of~\eqref{eq:mass-range} when the functions $a_0,a_0^{-1}$ and their required derivatives are uniformly bounded in the smooth identifications of the background regions.
\end{proposition}
\begin{proof}
\emph{Ellipticity and the adjoint domain.}
On invariant tensors, the conjugated operator $\Phi P_G\Phi^{-1}$ differs from $P_{\widetilde G}$ by an operator of order at most one. Indeed their horizontal principal symbols agree, while vertical derivatives of invariant component functions vanish. Where $a_0$ is constant, the change $\tau=a_0\vartheta$ is an actual local coordinate change on a cover, and the two operators agree. Their first-order difference therefore extends smoothly by zero across both horizon neighborhoods.
The smooth identification used for the uniform estimates can be chosen explicitly. Write $r_b(M)<r_c(M)$ for the two horizon radii and put $d(M)=r_c(M)-r_b(M)$. On a compact subextremal mass interval, both radii and $d^{-1}$ depend smoothly on $M$. With $x=(r-r_b)/d$, factor
\be
 F_M(r_b+dx)=d^2x(1-x)q_M(x),\qquad 0\le x\le1,
\ee
where $q_M$ is positive and smooth jointly in $(M,x)$. The coordinate $x=\sin^2(\chi/2)$ gives
\be
 \widetilde G_M=\frac{\dd\chi^2}{q_M(x)}+r^2\dd\omega^2
 +a_0(M,x)^2d(M)^2x(1-x)q_M(x)\dd\vartheta^2.
\ee
Choose $a_0(M,x)$ smoothly in $(M,x)$, equal to $\kappa_b(M)^{-1}$ near $x=0$ and to $\kappa_c(M)^{-1}$ near $x=1$. Since $\kappa_b=dq_M(0)/2$ and $\kappa_c=dq_M(1)/2$, the polar coefficients have the required normalization at both endpoints. In Cartesian coordinates on the two normal discs, this expression and all its coefficient derivatives depend smoothly on $M$. It therefore defines one fixed smooth compact manifold and fixed invariant tensor spaces for the whole mass interval. The conjugated operators and their smooth elliptic extensions can be chosen to depend continuously on $M$ as bounded maps from $H^2$ to $L^2$, and from $C^{m+2,\alpha}$ to $C^{m,\alpha}$, in these fixed spaces.

Extend the first-order difference to all tensors by its horizontal component formula away from the horizons and by zero on the horizon collars. Adding this extension to $P_{\widetilde G}$ and averaging under $\mathcal K$ gives a smooth strongly elliptic operator $\mathcal Q$ on all tensors over $X$. It commutes with $\mathcal K$ and agrees with $\Phi P_G\Phi^{-1}$ on invariant tensors.

The finite smooth atlas gives
\be\label{eq:tensor-elliptic-estimate}
\|u\|_{H^2(X)} \le C_M(\|\mathcal Q u\|_{L^2(a_0^{-1}\dd V_{\widetilde G})}+\|u\|_{L^2(a_0^{-1}\dd V_{\widetilde G})}).
\ee
For two smooth invariant tensors, integration of $P_G$ outside small horizon tubes proves symmetry. The tensors and their first derivatives are bounded in each normalized lift, and tube measure is $O(s)$, so both boundary terms vanish. Density extends symmetry to the domain~\eqref{eq:tensor-domain}.

To identify the adjoint domain, let $y,f$ be invariant $L^2$ sections such that $\langle\mathcal Q u,y\rangle=\langle u,f\rangle$ for every invariant $H^2$ section. For an arbitrary smooth section $\psi$, average it under $\mathcal K$. The averaging projection is orthogonal for the density $a_0^{-1}\dd V_{\widetilde G}$ and commutes with $\mathcal Q$. Hence the same identity holds with $u=\psi$. It states $\mathcal Q^\dagger y=f$ distributionally on the whole smooth manifold $X$. Elliptic regularity gives $y\in H^2(X)$, with the required invariance. Thus the adjoint domain is exactly the displayed domain. Estimate~\eqref{eq:tensor-elliptic-estimate} and Rellich compactness give finite-dimensional kernel and closed range. Self-adjointness then gives range equal to the orthogonal complement of the kernel, and index zero.

\emph{The homogeneous equation.}
A kernel element is smooth by elliptic regularity. The linearized Bianchi identity gives
\be\label{eq:tensor-bianchi}
\beta_GP_Gv=(\Delta_H-2\Lambda)\beta_Gv.
\ee
Corollary~\ref{cor:harmonic-gauge} gives $\beta_Gv=0$. Equation~\eqref{eq:tensor-linearization} therefore makes $v$ a solution of the homogeneous linearized static equations. Proposition~\ref{prop:physical-kernel} expresses it as the sum of a mass variation, a lapse-normalization variation, and $\mathcal L_ZG$, where $Z$ is a smooth meridional vector field that is invariant under equatorial reflection and generates coordinate changes preserving both horizons.

To remove the two parameters, allow a factor $c>0$ in $N=c\sqrt{F_M}$ and put $\xi\coloneqq r_b/r_c$. The root identities and their derivatives give
\bse\label{eq:opening-jacobian}
\be
M=\frac{\sqrt3}{2\sqrt\Lambda}\frac{\xi(1+\xi)}{(1+\xi+\xi^2)^{3/2}},\qquad
\frac{\kappa_b}{\kappa_c}=\frac{1+2\xi}{\xi(\xi+2)},
\ee
\be
\frac{\dd M}{\dd \xi}=\frac{\sqrt3}{4\sqrt\Lambda}\frac{(2\xi+1)(2-\xi-\xi^2)}{(1+\xi+\xi^2)^{5/2}}>0,\qquad
\frac{\dd}{\dd \xi}\log\frac{\kappa_b}{\kappa_c}=-\frac{2(1+\xi+\xi^2)}{\xi(1+2\xi)(\xi+2)}<0.
\ee
\ese
Consequently the determinant of $\partial(\kappa_b,\kappa_c)/\partial(M,\log c)$ is $\kappa_b\kappa_c\,\partial_M\log(\kappa_b/\kappa_c)<0$. A horizon-preserving coordinate variation leaves the constant surface gravity unchanged. Since the domain fixes both openings, this nonsingular matrix kills both parameter coefficients. The remaining element is $v=\mathcal L_ZG$, and $\beta_Gv=(\Delta_H-2\Lambda)Z^\flat=0$. Corollary~\ref{cor:harmonic-gauge} gives $Z=0$.
The zero kernel and the proved range formula yield bijectivity and~\eqref{eq:outer-tensor-inverse}. Schauder regularity for $\mathcal Q$ gives the H\"older statement. On a compact mass interval the smooth atlases, weighted densities and elliptic estimates are uniform. A sequence violating the inverse bound would converge, by the elliptic estimate and compactness, to a nonzero kernel element at a mass in the fixed subextremal interval. The kernel argument excludes this and proves the stated uniformity.\qedhere
\end{proof}

\begin{corollary}[Solving the linearized Einstein equation with a divergence-free source]\label{cor:outer-compatible-source}
Let $\mathcal C f\coloneqq f-\tfrac12(\operatorname{tr}_Gf)G$ be trace reversal in dimension four. For a regular invariant source $S$ with $\delta_GS=0$, there is a unique solution $v$ of the following system that preserves the horizon surface gravities.
\be
D(\Ric-\tfrac12\Scal\,g+\Lambda g)_G[v]=S,\qquad \beta_Gv=0,
\qquad v=2P_G^{-1}\mathcal C S.
\ee
The solution obeys the corresponding Sobolev or H\"older bound.
\end{corollary}
\begin{proof}
The identities $\beta_G\mathcal C S=\delta_GS$ and~\eqref{eq:tensor-bianchi} give $(\Delta_H-2\Lambda)\beta_Gv=0$. For rough sources this is a distributional identity in the filled normalized lifts, where elliptic regularity first makes $\beta_Gv$ smooth. Corollary~\ref{cor:harmonic-gauge} then gives $\beta_Gv=0$. Apply~\eqref{eq:tensor-linearization} and $\mathcal C^2=I$ to obtain the displayed linearized Einstein equation. Uniqueness follows from Proposition~\ref{thm:outer-tensor-inverse}.\qedhere
\end{proof}

\subsection{Bounds for coordinate variations and a tensor on which the quadratic form is negative}

We use the scalar spectral gap $d_s$ from Proposition~\ref{prop:scalar-gap} to bound the $H^1$ norm of the vector fields that generate infinitesimal coordinate changes. We then exhibit a symmetric two-tensor for which the quadratic form of the linearized Einstein operator $P_G$ is negative on the same Schwarzschild--de Sitter background. To bound the inverse of $P_G$, the relevant spectral quantity is the distance from zero to its spectrum.
For the Schwarzschild comparisons, use the unit-mass lift $G_S=F_S\dd\tau^2+F_S^{-1}\dd r^2+r^2\dd\omega^2$, where $F_S=1-2/r$, $r\ge2$, and $\tau$ has regular period $8\pi$, and write $P_S=\nabla_{G_S}^*\nabla_{G_S}-2\operatorname{Riem}_{G_S}$. All scalar and vector norms use the invariant measure $N_M\dd V_{h_M}$ and the regular domains specified above.

We define the quadratic form
\be
q_{G_M}(V)\coloneqq\int(|\nabla V|^2-\Lambda|V|^2)N_M\dd V_{h_M}
\ee
 and use the norm $\|V\|_{H^1_\ell}^2\coloneqq\|\nabla V\|_2^2+\ell^{-2}\|V\|_2^2$, where $\ell=\sqrt{3/\Lambda}$ is the cosmological length.

\begin{lemma}[Explicit coercivity constants for vector fields and their symmetric gradients]\label{q:vector-gap}
Consider vector fields $V$ on the Schwarzschild--de Sitter background,
invariant under translations of the auxiliary angle $\tau$, rotations
of the azimuthal angle $\varphi$, and reflection across the equatorial
plane; assume that $V$ is orthogonal to both circular orbits and extends
smoothly in the normalized local coordinates at both horizons and along
the axis. Then one has
\be\label{q:vector-constants}
q_{G_M}(V)\ge d_s\|V\|_2^2\ge0,\qquad
q_{G_M}(V)\ge c_{\rm vec}\|V\|_{H^1_\ell}^2,\qquad
c_{\rm vec}\coloneqq\frac{d_s}{d_s+\Lambda+\ell^{-2}}.
\ee
Consequently $\|V\|_{H^1_\ell} \le(2/c_{\rm vec})^{1/2}\|\mathcal BV\|_2$.
\end{lemma}
\begin{proof}
Decompose the dual one-form into an exact part and a co-closed part. The scalar eigenbasis gives this regular orthogonal decomposition: solve for the mean-zero potential using the scalar operator $-\Delta$ in~\eqref{eq:scalar-operator}. The gradient part has the margin in Proposition~\ref{prop:scalar-gap}. For a co-closed mode of even degree $l\ge2$ allowed by equatorial reflection, normalize $\int_{\mathbb S^2}Y_l^2=1$ and write
\be
Y=u(r)Y_l\dd r+v(r)\dd Y_l,\qquad L=l(l+1),\qquad w=r^2Fu,\qquad w'=Lv.
\ee
The degree-zero co-closed part vanishes by regularity. Put $D_lw\coloneqq-Fw''+Lw/r^2$. Direct integration of the form and its exterior derivative gives
\be
\|Y\|_2^2=L^{-1}\langle w,D_lw\rangle_{\dd r/F},\qquad
\|\dd Y\|_2^2=L^{-1}\|D_lw\|_{\dd r/F}^2.
\ee
Regularity gives $w=O(F)$ and removes the endpoint terms. Integration by parts gives $\langle w,D_lw\rangle_{\dd r/F}\ge(L/r_c^2)\|w\|_{\dd r/F}^2$. Cauchy--Schwarz then gives $\|\dd Y\|_2^2\ge(L/r_c^2)\|Y\|_2^2$. The Hodge identity therefore bounds its harmonic-vector form below by $(6/r_c^2-2\Lambda)\|Y\|_2^2$. The exact and co-closed parts are orthogonal also for this form, since the Hodge Laplacian commutes with exterior differentiation and divergence. This proves the first bound. The identity $\|\nabla V\|_2^2=q_{G_M}(V)+\Lambda\|V\|_2^2$ proves the second. Finally $2\|\mathcal BV\|_2^2=q_{G_M}(V)+\|\operatorname{div}V\|_2^2$ gives the stated symmetric-gradient constant. Density extends these bounds to the corresponding $H^1$ closure.
\end{proof}

\begin{corollary}[Poincar\'e and Korn constants on the fixed Schwarzschild--de Sitter region]\label{q:functional-Poincare-Korn}
Use the invariant measure $\dd\nu_M\coloneqq N_M\dd V_{h_M}$ and the regular scalar domain of axially symmetric functions invariant under equatorial reflection. Define the optimal Poincar\'e constant by
\be\label{q:optimal-Poincare}
 C_{P,M}^2\coloneqq\sup_{f\not\equiv\overline f}
 \frac{\|f-\overline f\|_{L^2(\nu_M)}^2}{\|\dd f\|_{L^2(\nu_M)}^2},\qquad
 \overline f\coloneqq\frac{\int f\dd\nu_M}{\int\dd\nu_M}.
\ee
Then $d_{P,M}\coloneqq C_{P,M}^{-2}-2\Lambda>0$. Let $C_{\rm Korn,cc,M}$ be the optimal Korn constant in the same meridional reflection class restricted to $\delta_GV^\flat=0$, with domain norm $H^1_\ell$ and target norm $\|\mathcal BV\|_2$. This constant is finite by Lemma~\ref{q:vector-gap}. Define
\bse\label{q:functional-vector-gap}
\be
 d_{V,M}\coloneqq\min(d_{P,M},6/r_c(M)^2-2\Lambda),
\ee
\be
c_{V,M}\coloneqq\max\Bigl(\frac{d_{V,M}}{d_{V,M}+\Lambda+\ell^{-2}},\min\Bigl(\frac{d_{P,M}}{d_{P,M}+\Lambda+\ell^{-2}},\frac2{C_{\rm Korn,cc,M}^2}\Bigr)\Bigr).
\ee
\ese
For regular meridional vector fields invariant under equatorial reflection, one has
\be\label{q:functional-Korn}
 q_{G_M}(V)\ge c_{V,M}\|V\|_{H^1_\ell}^2,\qquad
 \|V\|_{H^1_\ell} \le C_{\rm Korn,M}\|\mathcal BV\|_2,\qquad
 C_{\rm Korn,M} \le\sqrt{2/c_{V,M}}.
\ee
The optimal Korn constant here is the supremum of the displayed norm quotient over nonzero fields in this class, and $\mathcal BV=\tfrac12\mathcal L_VG_M$. All constants are uniform on a compact subextremal mass interval after taking their appropriate supremum or infimum.
\end{corollary}
\begin{proof}
The regular scalar compact resolvent gives $C_{P,M}^{-2}=\lambda_1(M)$, the first nonzero eigenvalue of $-\Delta$ on the axially symmetric scalar space invariant under equatorial reflection. Proposition~\ref{prop:scalar-gap} gives $\lambda_1>2\Lambda$. The exact/co-closed decomposition in Lemma~\ref{q:vector-gap}, with the exact eigenvalue $\lambda_1$ on its gradient part, gives $q_{G_M}(V)\ge d_{V,M}\|V\|_2^2$. The identities
\be
 \|\nabla V\|_2^2=q_{G_M}(V)+\Lambda\|V\|_2^2,\qquad
 2\|\mathcal BV\|_2^2=q_{G_M}(V)+\|\operatorname{div}V\|_2^2
\ee
give~\eqref{q:functional-Korn}. The field class excludes both circular Killing fields, as required for a finite Korn constant. For the second entry in the maximum defining $c_{V,M}$, we use the exact/co-closed decomposition. Its two parts are orthogonal also in $H^1_\ell$, because the Hodge Laplacian preserves the exact part and $\Ric_G=\Lambda G$. The gradient part has coercivity constant $d_{P,M}/(d_{P,M}+\Lambda+\ell^{-2})$. On the co-closed part the identity $q_G(V)=2\|\mathcal BV\|_2^2$ gives the constant $2/C_{\rm Korn,cc,M}^2$. Their minimum is therefore a further valid lower bound. The uniform spectral and geometric bounds on compact mass intervals prove the last statement.\qedhere
\end{proof}

The tensor estimate requires additional information. For example, the smooth tensor
\be\label{q:negative-tensor}
T_-\coloneqq r^{-4} \left(F\dd\tau^2+F^{-1}\dd r^2-r^2\dd\omega^2\right)
\ee
is trace-free and divergence-free and belongs to the domain preserving the horizon surface gravities. Its normal-plane coefficients agree at each horizon. In an orthonormal frame, direct differentiation gives $|T_-|^2=4r^{-8}$, $|\nabla T_-|^2=80Fr^{-10}$ and $\langle\operatorname{Riem}T_-,T_-\rangle=(-4\Lambda/3+16M/r^3)r^{-8}$. Thus
\be\label{q:negative-quotient}
\frac{\langle P_GT_-,T_-\rangle}{\|T_-\|_2^2}=-\Lambda-\frac57\frac{r_b^{-7}-r_c^{-7}}{r_b^{-5}-r_c^{-5}}<0.
\ee
Indeed the numerator divided by $4\pi$ is $\int_{r_b}^{r_c}(80r^{-8}-192Mr^{-9}-24\Lambda r^{-6})\dd r=[4/(7r^7)+4\Lambda/(5r^5)]_{r_b}^{r_c}$, using $F=0$ at the endpoints. The relevant bound for the full tensor inverse is therefore a bound for the resolvent across both signs of the spectrum, rather than a positive coercivity constant.

\section{Solving the equations linearized about Schwarzschild}\label{sec:Schwarzschild-models}

\subsection{The Schwarzschild metric and its infinitesimal displacement}\label{sec:Schwarzschild-conventions}

\paragraph{The Schwarzschild model, its displacement tensor, and the auxiliary source.}

We use the unit-mass Riemannian Schwarzschild metric $G_S=F_S\dd\tau^2+F_S^{-1}\dd r^2+r^2\dd\omega^2$, where $F_S=1-2/r$ and $r\ge2$. In this section we write $F=F_S$. Its smooth horizon angle is $\tau/4$ and its regular auxiliary period is $8\pi$. The isotropic radius is $s\ge1/2$, with $r=s+1+(4s)^{-1}$. Let $\dd\nu_S$ mean invariant Schwarzschild volume per unit auxiliary coordinate; equivalently integrate over one regular period and divide by $8\pi$.
We use the Schwarzschild classification of~\cite[Proposition~4.7 and Appendix~B]{AX-Existence}: in every weight $-1<\gamma<0$, the regular static axial kernel of $P_S$ on the unit-mass lift with regular period $8\pi$ is spanned by
\be\label{join:eq:T_S}
T_S\coloneqq2\operatorname{Hess}_{G_S}\bigl((R-1)\cos\theta\bigr).
\ee
Here $R$ is the areal radius. This nonzero displacement tensor satisfies $|\nabla^jT_S|\le C_j(1+|y|)^{-2-j}$ in isotropic coordinates. On each individual Schwarzschild model, both parities under polar reflection are allowed. The global equatorial reflection exchanges the two inserted collars.

We define $\eta(t)\coloneqq e^{-1/t}$ when $t>0$ and $\eta(t)\coloneqq0$ when $t\le0$. We also define the cutoff profile
\be\label{q:joined-cutoff-profile}
\vartheta(t)\coloneqq\frac{\eta(t)}{\eta(t)+\eta(1-t)}.
\ee
We choose the smooth model cutoff
\be
\label{q:explicit-centering-cutoff}
\chi_0(r)\coloneqq
\frac{\eta(4-r)}{\eta(4-r)+\eta(r-3)}.
\ee
Thus $\chi_0=1$ for $r\le3$ and $\chi_0=0$ for $r\ge4$.

On $r>2$, use the coframe $e^0=\sqrt F\,\dd\tau$, $e^1=F^{-1/2}\dd r$ and $e^A=r\bar e^A$, where $\bar e^A$ is an orthonormal coframe of the unit sphere. Set $Y_1=\cos\theta$ and $\xi_1\coloneqq\sum_A(D_AY_1)e^A$, with $D_A$ the derivative in the unit-sphere frame dual to $\bar e^A$. The displacement tensor is exactly
\be\label{q:Schwarzschild-displacement-frame}
T_S=\frac2{r^2}\Bigl(Y_1\bigl((e^0)^2+(e^1)^2-\sum_A(e^A)^2\bigr)+\sqrt F(e^1\otimes\xi_1+\xi_1\otimes e^1)\Bigr).
\ee
The scalar function $(r-1)Y_1$ is harmonic, since $(r^2F)'-2(r-1)=0$. The Ricci-flat Hessian commutation identity gives $P_ST_S=0$, zero trace and zero divergence. The tensor is regular at the Schwarzschild horizon, as is also clear from its Hessian definition. Using $\int Y_1^2=4\pi/3$ and $\int|DY_1|^2=8\pi/3$ gives, per unit auxiliary period,
\be\label{q:Schwarzschild-Gram-exact}
\|T_S\|_2^2=16\pi,\qquad \int_{r\ge R}|T_S|^2\dd\nu_S=\frac{64\pi}{3} \left(\frac2R-\frac1{R^2}\right)\quad(R\ge2).
\ee
Indeed the radial density is $(64\pi/3)(1+F)r^{-2}\dd r$. For the fixed cutoff above, we define
\be
\label{q:unit-Gram-definition}
m_0=\int\chi_0|T_S|^2\dd\nu_S>0.
\ee
Integrating over the spheres gives the exact normalization
\be\label{q:Schwarzschild-compact-Gram}
m_0=\frac{64\pi}{3}\int_2^4\chi_0(r)\frac{1+F(r)}{r^2}\,\dd r.
\ee
This integral is positive because $\chi_0=1$ on $[2,3]$. It depends only on the cutoff and the Schwarzschild metric of mass one.

For $-1<d<0$, the augmented unit-model equation $P_ST-\lambda_S\chi_0T_S=f$ has the coefficient
\be\label{q:Schwarzschild-column-coefficient}
\lambda_S=-\frac1{m_0}\int\langle f,T_S\rangle\dd\nu_S,\qquad |\lambda_S|\le C_{\lambda,S}(d)\sup_{r\ge2}r^{2-d}|f|.
\ee
Equation~\eqref{join:eq:column} below gives the corresponding rescaling to a collar of small mass.
To justify the identity, pair the equation with $T_S$ and integrate by parts. The horizon term vanishes in the regular lift, and the outer boundary terms are $O(r^{d-1})$ for $T=O(r^d)$ with its first weighted derivative. The equation $P_ST_S=0$ then gives the identity with the indicated sign. The coefficient estimate follows by taking the absolute value in the integral and using the weighted supremum of $f$. The resulting constant is
\begin{equation}
 C_{\lambda,S}(d)\coloneqq\frac1{m_0}\int r^{d-2}|T_S|\,\dd\nu_S.
\end{equation}
The integral is finite since $T_S$ is regular at the horizon and decays as $O(r^{-2})$ at infinity, while $d<1$.
This constant is optimal for the stated weighted supremum norm.
Indeed, take smooth radial cutoffs $\zeta_j$ supported away from the
horizon and infinity, with $0\le\zeta_j\le1$ and $\zeta_j\to1$
in the interior, and use the sources
\begin{equation}
 f_{j,\epsilon}
 =\zeta_j r^{d-2}\frac{T_S}{\sqrt{|T_S|^2+\epsilon^2}}.
\end{equation}
They have the required symmetries and weighted supremum norm at most
one. The absolute value of the coefficient in
the equation~\eqref{q:Schwarzschild-column-coefficient} tends to
$C_{\lambda,S}(d)$ as $\epsilon\downarrow0$ and then $j\to\infty$,
by dominated convergence.

\subsection{Weighted norms and local elliptic estimates}\label{sec:local-estimates}

\paragraph{Weighted norms, coordinate charts, and local elliptic estimates.} Choose positive chart constants $c_S,c_K$ sufficiently small for the local elliptic estimates below. On Schwarzschild and Schwarzschild--de Sitter, respectively, write
\be\label{q:geometric-chart-radii}
 \widetilde\rho_S(r)\coloneqq c_Sr,\qquad \widetilde\rho_K\coloneqq c_K.
\ee
At large Schwarzschild radius the invariant fields are represented on local covers in the auxiliary direction; their spatial components and covariant derivatives are unchanged. Near a horizon we use its smooth normalized lift. For a base weight $-1<\gamma<0$, define
\bse\label{q:model-Holder-norms}
\be
\|T\|_{X_S}\coloneqq\sup_p r(p)^{-\gamma}\|T\|_{C^{4,\alpha}_{\nabla,r(p)}(B_{\widetilde\rho_S(p)/64})},\qquad
\|f\|_{Y_S}\coloneqq\sup_p r(p)^{2-\gamma}\|f\|_{C^{2,\alpha}_{\nabla,r(p)}(B_{8\widetilde\rho_S(p)})},
\ee
\be
\|T\|_{X_K}\coloneqq\sup_p\|T\|_{C^{4,\alpha}_{\nabla,1}(B_{\widetilde\rho_K/64})},\qquad
\|f\|_{Y_K}\coloneqq\sup_p\|f\|_{C^{2,\alpha}_{\nabla,1}(B_{8\widetilde\rho_K})}.
\ee
\ese
Here the scaled norm at length $a$ is the sum of $a^j\sup|\nabla^j T|$ through the stated order and $a^{j+\alpha}[\nabla^jT]_{\alpha}$ at the highest order, using parallel transport in the regular chart. All balls are centered at the point of the supremum. The model charts are rotated with their centers, so spherical rotations act isometrically. Fixed changes in ball ratios give equivalent norms by finite coverings and the local estimates.

We fix a compact subextremal mass interval and compact placement intervals in the positive-lapse interior. Choose $0<s_K\le1$ so that their insertion points have disjoint normal balls of radius $2s_K$. The original horizons and axis are covered by the normalized smooth charts of Section~\ref{sec:interfaces}. Compactness gives positive ellipticity bounds, finite coefficient seminorms of every fixed order, and finite coordinate comparison norms, uniform over these parameters.

On unit Schwarzschild the compact region is covered by its regular period-$8\pi$ charts. On the end use isotropic Cartesian coordinates, divide lengths by the central areal radius, and use a local cover of the auxiliary coordinate. The metric coefficients then have uniform ellipticity and uniformly bounded scaled derivatives on balls whose spatial radii are a sufficiently small fixed fraction of that radius. The explicit coefficients $1-2/r$ and the isotropic conformal factor tend to their flat values with all scaled derivatives. The compact region and the end therefore admit common positive chart constants. This proves the existence of $c_S,c_K$ in~\eqref{q:geometric-chart-radii}; they are fixed before the inserted mass.

In a regular coordinate ball write a model operator on tensor components as
\be
 P=-a^{ij}\partial_i\partial_j+b^i\partial_i+c,
\ee
with scalar positive definite principal matrix. Let $S_{\rm E}$ be an interior Schauder constant for constant coefficients, uniform for its fixed ellipticity interval, and let $C_{\rm prod}$ bound the component product norms. On a ball of radius $2a$ choose the radius so that
\be\label{q:local-coefficient-condition}
 S_{\rm E}C_{\rm prod}\bigl(\|a^{ij}-a^{ij}(0)\|_{C^{0,\alpha}_{a}}
 +a\|b\|_{C^{0,\alpha}_{a}}+a^2\|c\|_{C^{0,\alpha}_{a}}\bigr)\le\theta_{\rm loc}<1.
\ee
For the uniformly smooth scaled model coefficients this follows by decreasing the fixed chart fraction. The estimate for constant coefficients applied componentwise, with the difference on the right, then absorbs its order-two term. A cutoff between nested balls and the interpolation estimate for order one give
\be\label{q:local-Schauder}
 \|T\|_{C^{m+2,\alpha}_{a}(B_a)} \le S_m\bigl(a^2\|PT\|_{C^{m,\alpha}_{a}(B_{2a})}+\|T\|_{C^0(B_{2a})}\bigr).
\ee
For $m=0$ this is the preceding absorption argument. For $m>0$, differentiate the component equation, apply the order-two estimate on nested balls, and bound the commutators by the already fixed higher coefficient seminorms. Induction and interpolation prove the formula, with $S_m$ depending on these seminorms, ellipticity, $\alpha$, the ball ratios and $(1-\theta_{\rm loc})^{-1}$. The scalar estimate for constant coefficients and interpolation used here are the interior Schauder estimates of~\cite[Chapter~6]{GT2001}. In a four-dimensional lifted ball, local interpolation and the same estimate also permit replacement of the $C^0$ term by $a^{-2}\|T\|_{L^2(B_{2a})}$, with full lifted volume on this ball. This is the local $L^2$ alternative; it follows by absorbing a small multiple of the left side in the scaled interpolation inequality.

All these estimates hold on complete smooth lifted balls at horizons and their poles. Norms formed with covariant derivatives and parallel transport are equivalent to the component norms: the transport equation and successive differentiation express each in terms of the other, with constants bounded by the fixed connection jets. At the Schwarzschild scale $r$ the chart fraction is fixed, so the conversion to the $r$-scaled norms in~\eqref{q:model-Holder-norms} only changes $S_m$ by fixed powers of that fraction. In particular, six output derivatives are controlled by four source derivatives with a uniform constant. The estimates apply also to one-forms and to the scalar operator.

\begin{lemma}[Bounds for the derivatives of an inverse metric]\label{q:matrix-jet-inverse}
In a fixed positive definite metric neighborhood, the $C^{k,\alpha}$ norm of the inverse metric is bounded by a function of the ellipticity lower bound and the $C^{k,\alpha}$ norm of the metric. The inverse map and the coefficient maps formed from it are locally Lipschitz in these norms, with the same fixed dependence.
\end{lemma}
\begin{proof}
The zeroth-order inverse is bounded by the reciprocal ellipticity lower bound. Differentiating $g g^{-1}=I$ gives, for every nonzero multi-index $\beta$,
\be
 D^\beta(g^{-1})=-g^{-1}\sum_{0<\zeta\le\beta}\binom\beta\zeta(D^\zeta g)D^{\beta-\zeta}(g^{-1}).
\ee
Induction and the H\"older product inequality prove the asserted bounds. For differences use $g^{-1}-\widetilde g^{-1}=g^{-1}(\widetilde g-g)\widetilde g^{-1}$ and the same differentiated products. The coordinate formulas for the connection, curvature and DeTurck differential then give the required finite coefficient bounds.\qedhere
\end{proof}

\subsection{An integral condition that removes the infinitesimal displacement}\label{f:range}

We derive the weighted source inverse used to construct the affine Schwarzschild profile in Lemma~\ref{f:profile} below. In the invariant class we define weighted Sobolev norms by integrating $|D^ju|^2(1+s)^{-2d+2j-3}\dd\nu_S$ for $0\le j\le2$. The asymptotic dimension here is three because all fields are static. Choose $-1<d<0$. The integral condition on this unit model is $m_0^{-1}\int\chi_0\langle T,T_S\rangle\dd\nu_S=0$. We write $C^{j,\alpha}_\gamma$ for the scaled H\"older space with weight $1+s$: an order-$i$ derivative carries the factor $(1+s)^{i-\gamma}$, and the highest H\"older seminorm carries $(1+s)^{j+\alpha-\gamma}$ on fixed small model balls of radius proportional to $1+s$. We set $Y_\gamma=C^{k,\alpha}_{\gamma-2}$.

The Euclidean Newton operator for $-\Delta$ gives an inverse in this weight range. Its pointwise weighted bound follows from the exact identity
\be\label{f:weighted-newton-exact}
 \frac1{4\pi}\int_{\mathbb R^3}\frac{|y|^{d-2}}{|x-y|}\,\dd y
 =\frac{|x|^d}{-d(d+1)},\qquad -1<d<0,\quad x\ne0.
\ee
Indeed, radial integration splits the left-hand side into
$|x|^{-1}\int_0^{|x|}t^d\dd t+\int_{|x|}^\infty t^{d-1}\dd t$.
The first integral is finite because $d>-1$, and the second because $d<0$. Thus a source bounded by $C(1+s)^{d-2}$ has a Newton potential bounded by $C' (1+s)^d$, after treating the fixed unit ball separately. The corresponding weighted $L^2$ estimate also follows directly from~\eqref{f:weighted-newton-exact}. On the end, conjugate the Newton kernel by the square roots of the target and source densities. Up to the factor $(4\pi)^{-1}$, the resulting kernel is
\be
 K_d(x,y)=|x|^{-d-3/2}|x-y|^{-1}|y|^{d-1/2}.
\ee
Set $h(x)=|x|^{-3/2}$ and $d'=-1-d$. Applying~\eqref{f:weighted-newton-exact} with $d$ and with $d'$ gives
\be
 \frac1{4\pi}\int K_d(x,y)h(y)\dd y=\frac{h(x)}{-d(d+1)},\qquad
 \frac1{4\pi}\int K_d(x,y)h(x)\dd x=\frac{h(y)}{-d'(d'+1)}.
\ee
Both constants are finite. By Schur's test\footnote{Let $K(x,y)$ be an integral kernel. If there
are a positive weight $w$ and finite constants $A,B$ such that
$\int |K(x,y)|w(y)\,\mathrm dy \le A w(x)$ and
$\int |K(x,y)|w(x)\,\mathrm dx \le B w(y)$, then the integral operator
with kernel $K$ is bounded on $L^2$, with norm at most $\sqrt{AB}$.
For a map between weighted $L^2$ spaces, one applies the test to
the kernel obtained by absorbing the input and output weights.}, the Newton potential operator on the end is bounded
from the weighted space $L^2_{d-2}$ to $L^2_d$.
A cutoff handles the fixed compact region, where $1+s$ replaces $s$. For the derivative estimate, let $A_R=\{R<s<2R\}$ and $A_R'=\{R/2<s<4R\}$. Scaling the interior estimate for $-\Delta u=f$ gives, for $j=1,2$,
\be
 R^{j-d-3/2}\|D^ju\|_{L^2(A_R)}
 \le C\left(R^{-d-3/2}\|u\|_{L^2(A_R')}
 +R^{1/2-d}\|f\|_{L^2(A_R')}\right).
\ee
Taking $R=2^k$ and summing the squares controls both derivative terms in $H^2_d$: the enlarged annuli have bounded overlap and their powers are precisely the weights of $u$ and $f$. The same construction acts componentwise on the static tensor end. Cover the compact Schwarzschild cap by its regular period-$8\pi$ charts. The shifted Dirichlet form on a larger compact cap gives an order-minus-two local inverse, by coercivity and ordinary regularity in the normalized horizon charts.

Patch the inverse constructed near the horizon with the Euclidean Newton inverse using nested cutoffs. Commutators have order one and compact support, hence are compact by local Rellich compactness after applying the inverse, which gains two derivatives. The contribution from shifting the operator near the horizon is compact for the same reason. On the asymptotic end of the static region, the isotropic Schwarzschild coefficients and their derivatives give, componentwise in the Cartesian spatial and auxiliary frame,
\be\label{f:end-operator-difference}
 |(P_S+\Delta_{\mathbb R^3})v|
 \le C(s^{-1}|D^2v|+s^{-2}|Dv|+s^{-3}|v|).
\ee
The same estimates hold after the differentiated products needed in the scaled H\"older norms. For example, in the Sobolev norms above, multiplying the three terms by the target density $(1+s)^{1-2d}$ and integrating over $s\ge R$ gives
\be
 \|(P_S+\Delta_{\mathbb R^3})v\|_{L^2_{d-2}(s\ge R)}
 \le CR^{-1}\|v\|_{H^2_d(s\ge R)},\qquad R\ge2.
\ee
Indeed the three resulting densities are bounded by $CR^{-2}$ times the respective source densities for $D^2v$, $Dv$ and $v$. Thus the end error is arbitrarily small, with the tensor connection and curvature terms included. For the two-sided statement, take a partition $\chi_{\rm c}+\chi_{\rm e}=1$ and output cutoffs $\eta_{\rm c},\eta_{\rm e}$ equal to one on the corresponding source supports. The operator
\be
 B=\eta_{\rm c}A_{\rm c}\chi_{\rm c}+\eta_{\rm e}A_{\rm e}\chi_{\rm e}
\ee
uses the inverse of the shifted operator near the horizon $A_{\rm c}$ and the Euclidean Newton inverse $A_{\rm e}$. In $P_SB-I$, derivatives of the output cutoffs give first-order commutators applied after applying an inverse that gains two derivatives. In $BP_S-I$, the corresponding terms are $A_{\rm c}[P_S,\chi_{\rm c}]$ and $A_{\rm e}[P_S,\chi_{\rm e}]$. Their coefficients have fixed compact support, and they are compact from the weighted $H^2$ domain to itself because $H^2$ embeds compactly into $H^1$ on a compact annulus. The contribution from shifting the operator near the horizon is treated in the same way, using the compact embedding into $L^2$. The remaining end terms have norm $O(R^{-1})$, up to the fixed cutoff and inverse bounds. Choose $R$ sufficiently large that the two end errors have norm less than one. More precisely, write $P_SB=I+K_Y+E_Y$ on the source space and $BP_S=I+K_H+E_H$ on the domain, where the operators $K_Y,K_H$ are compact and $\|E_Y\|,\|E_H\|<1$. Then $B(I+E_Y)^{-1}$ is a right inverse modulo a compact operator, while $(I+E_H)^{-1}B$ is a left inverse modulo a compact operator. The Fredholm criterion therefore gives finite-dimensional kernel and cokernel and closed image. This argument uses smallness, rather than compactness, for the second-order coefficient error on the end.

The classification recalled in Section~\ref{sec:Schwarzschild-conventions} applies to the Sobolev kernel as follows.
\bei

\item For a homogeneous solution $v\in H^2_d$, scaled interior elliptic estimates and Sobolev embedding give $|D^jv|\le C_j s^{d-j}$ on the end, with the analogous scaled H\"older estimates. Regularity on the filled horizon and the axis follows from the local invariant elliptic equation. Thus $v$ belongs to the weighted H\"older kernel in that classification. Conversely, the decay of $T_S$ and its derivatives in Section~\ref{sec:Schwarzschild-conventions} gives, for each $0\le j\le2$,
\be
 \int_{s\ge2}|D^jT_S|^2(1+s)^{-2d+2j-3}\dd\nu_S
 \le C\int_2^\infty s^{-2d-5}\dd s<+\infty.
\ee
The compact part is finite by regularity.

\item The Sobolev kernel is therefore exactly the span of $T_S$.
To determine the cokernel, we consider the annihilator of the range
of $P_S$, that is, the continuous linear functionals on $L^2_{d-2}$
that vanish on every tensor of the form $P_Sv$ with $v\in H^2_d$.
Under the unweighted integral pairing, each such functional is
represented by a tensor $z$ in $L^2_{d'}$, where
$d'\coloneqq-1-d\in(-1,0)$. The vanishing condition reads
\begin{equation}
  \int \langle z,P_Sv\rangle\,\dd\nu_S=0
  \qquad\text{for every }v\in H^2_d.
\end{equation}
By formal self-adjointness, this identity gives the weak equation
$P_Sz=0$ against smooth invariant test tensors.
\eei

  Indeed the target density is $(1+s)^{1-2d}$ and its reciprocal is $(1+s)^{2d-1}=(1+s)^{-2d'-3}$. For a homogeneous dual field $z$, scaled local elliptic estimates on $A_R\subset A_R'$ give
\be
 \sup_{A_R}R^{-d'}\bigl(|z|+R|Dz|\bigr)
 \le C\left(\int_{A_R'}|z|^2(1+s)^{-2d'-3}\dd\nu_S\right)^{1/2},\qquad R\ge2.
\ee
The constant is uniform under rescaling of the Schwarzschild end. Thus the dual weighted $L^2$ condition gives the required pointwise growth and first-derivative bounds; further local estimates give the corresponding higher derivatives. At the horizon, use the period-$8\pi$ lift and a small invariant neighborhood of the Schwarzschild horizon. In the lifted neighborhood, the weak equation $P_Sz=0$ is already
known for smooth invariant test tensors with compact support there.

For an arbitrary test tensor in that neighborhood, average it under the two circle actions and their angle reversals. The averaged test is admissible, while $z$ and $P_S$ are invariant; consequently the same distributional identity holds for the original test. Thus $P_Sz=0$ on the full lifted neighborhood, including the Schwarzschild horizon. The dual weight is bounded above and below there, so $z$ is locally $L^2$ and elliptic regularity gives its smoothness across the horizon and the axis. The same classification gives exactly the span of $T_S$. The integration-by-parts pairing has no boundary contribution at infinity: for $v=O(r^d)$ with $\nabla v=O(r^{d-1})$, the classified dual field satisfies $T_S=O(r^{-2})$ and $\nabla T_S=O(r^{-3})$, so both terms on a radius-$r$ sphere are $O(r^{d-1})$. Here it is essential to use the actual decay of $T_S$; the generic complementary-weight estimates alone would only give an $O(1)$ boundary bound. Thus the closed range is precisely orthogonality to $T_S$. The integral functional is nonzero on $T_S$; restricting by it gives a bounded inverse onto that range.

To obtain the stated H\"older inverse for $f\in Y_\gamma$, choose $d=\gamma/2\in(\gamma,0)$. The source belongs to the intermediate weighted Sobolev target: its squared norm on the end is bounded by a constant times
\be
 \|f\|_{Y_\gamma}^2\int_1^\infty R^{2(\gamma-d)-1}\dd R
 =\frac{\|f\|_{Y_\gamma}^2}{2(d-\gamma)}.
\ee
The compact part is bounded by the same source norm. After imposing the stated orthogonality, solve in the weighted Sobolev spaces just constructed. Elliptic regularity gives $v=\Ocal(s^d)$ with scaled derivatives. On the end rewrite the equation as a Euclidean Poisson equation. Its source is $f+\Ocal(s^{d-3})$, which belongs to weight $\gamma-2$ since $d-\gamma<1$. Multiply that source by a smooth cutoff equal to one outside a fixed ball and zero on a smaller ball, and let $w$ be its componentwise Newton potential. The weighted Newton estimate gives $w=O(s^\gamma)$. Each component of $v-w$ is harmonic outside the larger ball and tends to zero at infinity. The exterior maximum principle, applied with the barrier $C/s$, gives $v-w=O(s^{-1})$; its constant is controlled by the values on the fixed bounding sphere. Since $\gamma>-1$, this term also belongs to the desired weight. Local Schauder estimates and the bounded inverse under the integral condition give the full weighted H\"older bound. This proves precisely the range assertion used for the affine profile.

\begin{proposition}[An inverse for the Einstein equations linearized about Schwarzschild]\label{q:full-Schwarzschild-model-inverse}
Let $X_S^0$ be the kernel of $T\mapsto\int\chi_0\langle T,T_S\rangle\dd\nu_S$. For every regular static polar source $f\in Y_S$, there is a unique pair $(T,\lambda_S)\in X_S^0\times\mathbb R$ satisfying
\be
 P_ST-\lambda_S\chi_0T_S=f,\qquad
 \|T\|_{X_S}+|\lambda_S|\le C_S\|f\|_{Y_S}.
\ee
The constant $C_S$ is a finite bound for the norm of this inverse in precisely these spaces.
\end{proposition}
\begin{proof}
Integration against $T_S$ gives the coefficient in~\eqref{q:Schwarzschild-column-coefficient}. Thus $f+\lambda_S\chi_0T_S$ is orthogonal to $T_S$. The closed-range construction just proved gives its unique inverse under the integral condition. The pairing is bounded on $Y_S$, since $|T_S|=O(r^{-2})$, $|f|=O(r^{\gamma-2})$ and the radial measure is $r^2\dd r$. If $C_{\rm pair}$ is the resulting pairing bound divided by $m_0$ and $C_{\rm col}=\|\chi_0T_S\|_{Y_S}$, then
\be
 |\lambda_S|\le C_{\rm pair}\|f\|_{Y_S},\qquad
 \|f+\lambda_S\chi_0T_S\|_{Y_S}
 \le(1+C_{\rm col}C_{\rm pair})\|f\|_{Y_S}.
\ee
Applying the bounded inverse under the integral condition on the orthogonal range proves the asserted bound for $T$ and $\lambda_S$. Conversely the same integral determines the coefficient of any solution, and the integral condition removes its homogeneous displacement. This proves both uniqueness and the estimate for the sum norm. Every constant refers to the unit model and is independent of the inserted mass.\qedhere
\end{proof}

\section{The linearized equations on Schwarzschild--de Sitter with two insertion points}\label{sec:model-inverse operators}

\subsection{Local inversion near an insertion point}

\paragraph{Estimates on the unpunctured background.}

Proposition~\ref{thm:outer-tensor-inverse} gives the inverse in the regular tensor domain with prescribed horizon surface gravities. Its $L^2$ bound and the local estimates of Section~\ref{sec:local-estimates} give a finite constant $C_{K,\mathrm{all}}$ such that
\be
 \|P_K^{-1}f\|_{X_K} \le C_{K,\mathrm{all}}\|f\|_{Y_K}.
\ee
Here the norms are precisely those of~\eqref{q:model-Holder-norms}. For higher differentiability, differentiate in the finite normalized atlas and apply the same estimates on nested balls. The resulting bounds depend only on the fixed coefficient seminorms and the $L^2$ inverse. They are uniform on a compact subextremal mass interval: in fixed smooth identifications the operator and its domain depend continuously on the mass, and the inverse identity gives a bounded inverse on a neighborhood of each mass. A finite cover of the compact interval proves the \emph{uniform bound.}

 We now fix a compact interval of strictly subextremal central masses and a compact interval of axial insertion positions in the positive-lapse interior. All coordinate and coefficient bounds in this section are \emph{uniform} on these two sets. The two insertion charts are exchanged by reflection. For the fixed Schwarzschild--de Sitter background, write $(h_K,N_K)\coloneqq(h_M,N_M)$, $G_K\coloneqq G_M$ and $P_K\coloneqq P_{G_M}$. We work in cosmological units and use spatial normal coordinates $z$ centered at an insertion point $p$, with $\rho=|z|$, and the normalized auxiliary coordinate $t=N_K(p)\tau$. In these coordinates the invariant tensor operator has the form
\be
\label{green:operator}
P_p=-a_p^{ij}(z)\del_i\del_j+B_p^i(z)\del_i+C_p(z),\qquad
a_p(0)=I,\qquad a_p(z)-I=\Ocal(\rho^2).
\ee
Here $\del_i$ differentiates with respect to $z_i$, and $B_p^i,C_p$ act on the coordinate tensor components. The coefficients and their placement derivatives have bounded derivatives on fixed chart balls. In particular, the bounds used below are coefficient bounds for a known metric, independently of its Green tensor.

\paragraph{The estimate for the Newton potential.}

We fix $0<c<1/8$, $0<\alpha<1$ and an integer $m\ge0$. We use maximum coordinate component norms and the scaled H\"older norms of Section~\ref{sec:local-estimates}. On $\mathbb R^3\setminus\{0\}$ we define
\bse
\label{green:spaces}
\be
\|v\|_{\mathcal X_{\rm b}^{m+2,\alpha}}
\coloneqq\sup_{z\ne0}\|v\|_{C^{m+2,\alpha}_{\rho}(B_{c\rho}(z))},
\ee
\be
\|f\|_{\mathcal Y_{-1}^{m,\alpha}}
\coloneqq\sup_{z\ne0}\rho\,\|f\|_{C^{m,\alpha}_{\rho}(B_{2c\rho}(z))}.
\ee
\ese
The space $\mathcal X_{\rm b}^{m+2,\alpha}$ also requires $v(z)\to0$ as $|z|\to+\infty$. Its norm permits bounded nonsmooth behavior at the origin. For compactly supported sources, let $\mathcal N f$ be convolution, componentwise, with $1/(4\pi|z|)$.

\begin{lemma}[The Newton potential of a source bounded by the inverse radius]\label{green:Newton}
Fix $R>0$. For every source supported in $B_{2R}$, the Newton operator satisfies
\be
\label{green:Newton-bound}
\|\mathcal N f\|_{\mathcal X_{\rm b}^{m+2,\alpha}}
\le C_{\rm N}R\|f\|_{\mathcal Y_{-1}^{m,\alpha}},
\ee
where $C_{\rm N}$ depends only on $m,\alpha,c$ and the component norm. The constant is fixed by the Euclidean interior estimate and the prescribed ball ratios.
\end{lemma}
\begin{proof}
Writing $F=\|f\|_{\mathcal Y_{-1}^{m,\alpha}}$, we have $|f(z)|\le F/|z|$. Spherical integration gives the exact scalar comparison
\be
\label{green:Newton-comparison}
\frac1{4\pi}\int_{|y|\le2R}\frac{\dd y}{|x-y||y|}
=
\begin{cases}
2R-|x|/2,&|x|\le2R,\\
2R^2/|x|,&|x|\ge2R.
\end{cases}
\ee
Consequently $|\mathcal N f|\le2RF$, and its limit at infinity is zero. On each ball $B_{2c\rho}(z)$, the constant-coefficient interior estimate, after scaling by $\rho$, bounds the output norm by
$S_{\rm E}(\|\mathcal N f\|_{C^0}+\rho^2\|f\|_{C^{m,\alpha}_\rho})$.
For $\rho\le4R$, the second term is at most $4RF$ by~\eqref{green:spaces}. For $\rho>4R$, the ball used to measure the source is disjoint from $B_{2R}$, since $c<1/8$. Thus~\eqref{green:Newton-bound} follows. The Euclidean constant $S_{\rm E}$ is the frozen interior Schauder constant of Section~\ref{sec:local-estimates}; it depends only on the derivative order, H\"older exponent and prescribed ball ratios. The proof applies to H\"older sources by local approximation and the same uniform bounds.\qedhere
\end{proof}

\paragraph{Choosing the radius to control the variation of the coefficients.}
We now take $m=4$ and $0<\alpha<1$, as in the bounds for the inverse on the full background. We extend~\eqref{green:operator} from $B_R$ to all of $\mathbb R^3$, equal to $-\Delta$ outside $B_{2R}$, using a fixed smooth cutoff in $\rho/R$. We write the extension as $-\Delta+\mathcal E_p$. Let $J_a,J_B,J_C$ bound, respectively,
$\rho^{-2}(a_p-I)$, $B_p$ and $C_p$ in the scaled $C^{4,\alpha}$ norms, also after one placement derivative. The bounds for $a_p-I$ follow from the spatial normal-coordinate expansion. We choose a product constant $C_{\rm P}$ which includes the H\"older product inequality, the finitely many derivatives of this fixed cutoff, and the finite covering comparison between balls of radii $c\rho$ and $2c\rho$. We set
\be
\label{green:coefficient-constant}
e_0(R)\coloneqq C_{\rm P}(J_B+2R(J_a+J_C)),
\ee
and let $e_1(R)$ be the same expression with the corresponding placement-derivative coefficient bounds. The product rule gives
\be
\label{green:coefficient-bound}
\|\mathcal E_p v\|_{\mathcal Y_{-1}^{4,\alpha}} \le e_0(R)\|v\|_{\mathcal X_{\rm b}^{6,\alpha}},
\qquad
\|\del_p\mathcal E_p v\|_{\mathcal Y_{-1}^{4,\alpha}} \le e_1(R)\|v\|_{\mathcal X_{\rm b}^{6,\alpha}}.
\ee
Here $\del_p$ denotes differentiation in the axial placement parameter in moving normal charts. For example, the principal error contributes
$\rho\,\Ocal(\rho^2)\rho^{-2}|v|_{\mathcal X_{\rm b}}$, the first-order error contributes
$\rho\,\Ocal(1)\rho^{-1}|v|_{\mathcal X_{\rm b}}$, and the potential error contributes
$\rho\,\Ocal(1)|v|_{\mathcal X_{\rm b}}$. Differentiating four times with the scaled norm gives the same three bounds. The cutoff derivatives are bounded after multiplication by the corresponding powers of $\rho$ on $R\le\rho\le2R$.

We choose $0<\theta<1$ and a radius $R$ with disjoint closed balls $\overline B_{2R}(p_\pm)$ contained in the normal charts, such that
\be
\label{green:radius-condition}
R\le1,\qquad C_{\rm N}R e_0(R)\le\theta<1.
\ee
Let $R_{\rm chart}>0$ ensure the stated chart and separation conditions. An explicit sufficient choice, once the coefficient bounds are fixed, is
\be\label{q:Green-radius-choice}
R\coloneqq\min\bigl(R_{\rm chart},1,
\frac{\theta}{2C_{\rm N}C_{\rm P}(1+J_B)},
\left(\frac{\theta}{4C_{\rm N}C_{\rm P}(1+J_a+J_C)}\right)^{1/2}\bigr).
\ee
The linear and quadratic contributions to $C_{\rm N}Re_0(R)$ are then at most $\theta/2$ each. We define
\be
\label{green:local-inverse-constant}
L_{\rm b}\coloneqq\frac{C_{\rm N}R}{1-\theta}.
\ee

\begin{lemma}[A bounded inverse near the point and its parameter derivative]\label{green:local-inverse}
For every family $f_p$ supported in $B_R$ with
$\|f_p\|_{\mathcal Y_{-1}^{4,\alpha}} \le F_0$ and
$\|\del_p f_p\|_{\mathcal Y_{-1}^{4,\alpha}} \le F_1$, there is a solution $v_p$ of
$(-\Delta+\mathcal E_p)v_p=f_p$, vanishing at infinity, with
\be
\label{green:local-and-derivative}
\|v_p\|_{\mathcal X_{\rm b}^{6,\alpha}} \le U_0\coloneqq L_{\rm b}F_0,\qquad
\|\del_pv_p\|_{\mathcal X_{\rm b}^{6,\alpha}} \le U_1\coloneqq L_{\rm b}(F_1+e_1(R)U_0).
\ee
For a $C^1$ source family in the displayed weighted norm, the solution is $C^1$ in the placement parameter. Its restriction to the original chart has no additional delta source at the origin.
\end{lemma}
\begin{proof}
On the closed space $\mathcal X_{\rm b}^{6,\alpha}$, the map
$v\mapsto\mathcal N f_p-\mathcal N\mathcal E_pv$ has contraction factor at most $\theta$ by Lemma~\ref{green:Newton} and~\eqref{green:coefficient-bound}. Its unique fixed point gives the first estimate in~\eqref{green:local-and-derivative}. The background and the normal charts depend smoothly on the placement parameter on the fixed compact parameter set. In these moving charts, $a_p(0)=I$ and $D_za_p(0)=0$ for every $p$; the same identities hold after a placement derivative. Uniform bounds for one further placement derivative and for the required spatial derivatives therefore show, by Taylor's formula and the product estimates above, that $p\mapsto\mathcal E_p$ is $C^1$ in operator norm from $\mathcal X_{\rm b}^{6,\alpha}$ to $\mathcal Y_{-1}^{4,\alpha}$. This differentiates the coefficients in fixed moving coordinates, rather than translating an arbitrary H\"older field. The uniformly convergent inverse series may consequently be differentiated in operator norm, giving
$(I+\mathcal N\mathcal E_p)\del_pv_p=\mathcal N(\del_pf_p-(\del_p\mathcal E_p)v_p)$,
which proves the derivative estimate and the $C^1$ assertion. The bounds $v_p=\Ocal(1)$ and $Dv_p=\Ocal(\rho^{-1})$ make the two boundary terms in Green's formula tend to zero: their orders in the spatial dimension three are $\rho^2$ and $\rho$. Thus the distributional equation has no point mass in addition to $f_p$. All constructions commute with the stabilizing rotations, and reflection exchanges the two charts.\qedhere
\end{proof}

\subsection{The Green tensor and its dependence on the parameters}

In a chart with $z_3$ along the outward axial direction, we put
$a(p)\coloneqq\del_{\nu_K}N_K(p)/N_K(p)$ and $D_p\coloneqq\sum_i\dd z_i^2-\dd t^2$. With $n=z/\rho$, we define the bounded angular tensor $S_{1,p}$ by
\be
\label{green:angular-correction}
(S_{1,p})_{ij}=-a(p)n_3\delta_{ij},\qquad
(S_{1,p})_{tt}=-3a(p)n_3,\qquad
(S_{1,p})_{it}=0.
\ee
We choose a radial cutoff $\chi_S$ equal to one on $B_R$ and zero outside $B_{2R}$, and set $S_p=\chi_S(2D_p/\rho+S_{1,p})$. Its source is
\be
\label{green:source}
P_KS_p=8\pi D_p\delta_{h,p}+f_p^{\,S},\qquad
|D^jf_p^{\,S}|=\Ocal(\rho^{-1-j})\quad(0\le j\le4).
\ee
The same bounds hold after one derivative with respect to the insertion radius, taken in the moving coordinates. At the center of the chart the nonzero connection coefficients are $\Gamma^t_{it}=a(p)\delta_{i3}$ and $\Gamma^i_{tt}=-a(p)\delta_{i3}$. With $u_0=\rho^{-1}$, their rough-Laplacian terms on $2D_p/\rho$ are $-2a(p)\del_3u_0\delta_{ij}$ and $-6a(p)\del_3u_0$. Thus the first lapse jet gives the $\rho^{-2}$ coefficients $a(p)(2n_3\delta_{ij},6n_3)$; applying $-\Delta$ to~\eqref{green:angular-correction} cancels them, since $\Delta_{\mathbb S^2}n_3=-2n_3$. The remaining principal-coefficient error is $\Ocal(\rho^2)D^2(\rho^{-1})$, the first-order coefficient error is $\Ocal(\rho)D(\rho^{-1})$, and the potential is bounded times $\rho^{-1}$. These are all $\Ocal(\rho^{-1})$, with the differentiated bounds. The cutoff annulus is separated from the pole. If $\eta=1$ on $B_{R/2}$ and $\eta=0$ outside $B_R$, sufficient choices are $F_0\ge\sup_p\|\eta f_p^{\,S}\|_{\mathcal Y_{-1}^{4,\alpha}}$ and $F_1\ge\sup_p\|\del_p(\eta f_p^{\,S})\|_{\mathcal Y_{-1}^{4,\alpha}}$. These are finite seminorms of the explicit coefficients and tensors, rather than Green bounds.

Applying Lemma~\ref{green:local-inverse} with the negative of this source gives $v_p$ with $P_Kv_p=-f_p^{\,S}$ on $B_{R/2}$. We choose $\beta=1$ on $B_{R/4}$, supported in $B_{R/2}$, and form the paired tensor $T_p=S_p+\beta v_p$ in the two charts. Its error
\be
\label{green:regular-source}
g_p\coloneqq P_KT_p-8\pi\sum_\pm D_{p_\pm}\delta_{h,p_\pm}
\ee
vanishes on the two balls $B_{R/4}$ and is a regular tensor supported in the two chart annuli. In moving coordinates it equals
$(1-\beta)f_p^{\,S}+[P_p,\beta]v_p$.
For $j=4,5$, let $A_j$ bound $(1-\beta)f_p^{\,S}$ in the ordinary $C^{j,\alpha}$ annular norm and let $C_j$ bound $[P_p,\beta]$ from $\mathcal X_{\rm b}^{6,\alpha}$ to that norm. Let $A_{4,1},C_{4,1}$ bound their placement derivatives. The commutator is first order, so $j=5$ uses derivatives of $v_p$ only through order six. Every one of these constants is a finite product of coefficient and cutoff seminorms on $R/4\le\rho\le2R$.

Let $C_{\rm push,0}$ bound the paired coordinate pushforward on $C^{5,\alpha}$ and $C^{4,\alpha}$, and let $C_{\rm push,1}$ bound its placement derivative from $C^{5,\alpha}$ to $C^{4,\alpha}$. The latter loses one spatial derivative, which has already been allowed in the choice $j=5$. We may therefore take
\bse
\label{green:regular-source-bounds}
\be
F_{\rm ann,0}\ge C_{\rm push,0}(A_5+C_5U_0),
\ee
\be
F_{\rm ann,1}\ge C_{\rm push,1}(A_5+C_5U_0)+C_{\rm push,0}(A_{4,1}+C_{4,1}U_0+C_4U_1),
\ee
\ese
so that $\|g_p\|_{C^{5,\alpha}} \le F_{\rm ann,0}$ and
$\|\del_pg_p\|_{C^{4,\alpha}} \le F_{\rm ann,1}$ in fixed coordinates.

The inverse on the full background gives a smooth correction $w_p=-P_K^{-1}g_p$. We set $K_4\coloneqq C_{\rm atlas}C_{K,\mathrm{all}}$, where $C_{\rm atlas}$ bounds the product of the regular atlas norm comparisons $C^{3,\alpha}\to Y_K$ and $X_K\to C^{4,\alpha}$. In the finite atlas of Section~\ref{sec:local-estimates}, let $D_j\ge1$ bound the differentiated interior estimates
\be
\label{green:intact-elliptic}
\|w\|_{C^{j+2,\alpha}} \le D_j(\|P_Kw\|_{C^{j,\alpha}}+\|w\|_{C^{j+1,\alpha}}),\qquad j=3,4,5.
\ee
These constants follow by differentiating the coordinate equation, applying the local estimate on nested balls, and bounding the resulting commutators by the coefficient seminorms through order $j$. Their finite atlas maximum includes the coordinate changes and inverse powers of the fixed chart radii. They are independent of the placement and of the inserted mass. We choose
\be
\label{green:intact-higher-inverse}
K_{j+2}\ge D_j(1+K_{j+1})\quad(j=3,4,5).
\ee
Then $\|w_p\|_{C^{7,\alpha}} \le K_7F_{\rm ann,0}$ and $\|\del_pw_p\|_{C^{6,\alpha}} \le K_7F_{\rm ann,1}$. The second formula differentiates a regular source on the fixed Schwarzschild--de Sitter region, so it introduces no derivative of a delta distribution. The recurrence uses only the inverse on the full background, which has already been constructed.

\begin{proposition}[Bounds for the Green tensor after subtraction of its leading singularity]\label{green:global}
Choose the radius and constants satisfying~\eqref{green:radius-condition}--\eqref{green:intact-higher-inverse}. The tensor
$H_p=T_p+w_p$ solves the equation with a Dirac source at each insertion point and with both original horizon surface gravities fixed. It is smooth on the axis away from the two points and in the original horizon lifts. In the moving normal balls $\rho<R/4$ it satisfies, for $0\le j\le6$,
\be
\label{green:final-bound}
|D^j(H_p-2D_p/\rho)|\le C_{H,0}\rho^{-j},\qquad
|D^j\del_p(H_p-2D_p/\rho)|\le C_{H,1}\rho^{-j}.
\ee
On $\rho\ge R/4$, the $C^{4,\alpha}$ norms of $H_p$ and its fixed-coordinate placement derivative are bounded by $C_{H,\rm out}$.
All three Green constants are uniform on the fixed compact parameter sets.
\end{proposition}
\begin{proof}
{
Let $A_{\rm ang,0},A_{\rm ang,1}$ bound the scaled angular correction and its placement derivative. Let $T_{\rm cov}$ include the fixed coordinate-to-covariant comparisons, the pullbacks on $C^{6,\alpha}$, and their derivatives from $C^{7,\alpha}$ to $C^{6,\alpha}$. Sufficient choices are
\bse
\label{green:CH-conditions}
\be
C_{H,0}\ge T_{\rm cov}(A_{\rm ang,0}+U_0+K_7F_{\rm ann,0}),
\ee
\be
C_{H,1}\ge T_{\rm cov}(A_{\rm ang,0}+A_{\rm ang,1}+U_0+U_1+K_7(F_{\rm ann,0}+F_{\rm ann,1})).
\ee
\ese
On $\rho\ge R/4$, let $A_{\rm out,0},A_{\rm out,1}$ bound $S_p$ and its moving-chart derivative in $C^{5,\alpha}$ and $C^{4,\alpha}$. Let $V_{\rm out,0},V_{\rm out,1}$ bound multiplication by $\beta$ and its derivative in these annular norms. A common $C^{4,\alpha}$ bound for $H_p$ and its fixed-coordinate placement derivative is obtained by imposing
\be\label{green:outside-constant}
\begin{split}
C_{H,\rm out}\ge{}&(C_{\rm push,0}+C_{\rm push,1})(A_{\rm out,0}+V_{\rm out,0}U_0)\\
&+C_{\rm push,0}(A_{\rm out,1}+V_{\rm out,1}U_0+V_{\rm out,0}U_1)+K_7(F_{\rm ann,0}+F_{\rm ann,1}).
\end{split}
\ee
}
Equations~\eqref{green:regular-source} and $P_Kw_p=-g_p$ give the point-source equation. Lemma~\ref{green:local-inverse} excludes any additional point source. The inverse on the full background imposes the conditions that preserve both horizon surface gravities. Near the pole the remainder is $S_{1,p}+v_p+w_p$, so~\eqref{green:final-bound} follows from~\eqref{green:local-and-derivative} and~\eqref{green:intact-higher-inverse}. To see the loss of one spatial derivative in the moving chart, let $\Phi_p$ denote its spatial coordinate map and $V_p=(\partial_p\Phi_p)\circ\Phi_p^{-1}$. Differentiation gives
\be
 \partial_p(\Phi_p^*w_p)=\Phi_p^*(\partial_pw_p+\mathcal L_{V_p}w_p).
\ee
The normalization of the auxiliary component contributes only the derivative of a smooth algebraic frame factor. On the fixed compact chart, the displayed identity gives a $C^{6,\alpha}$ bound using $\|\partial_pw_p\|_{C^{6,\alpha}}$ and $\|w_p\|_{C^{7,\alpha}}$. These are the two bounds in~\eqref{green:intact-higher-inverse}, and explain the use of the $C^{5,\alpha}$ bound for $g_p$. To justify differentiability in the same H\"older exponent, apply~\eqref{q:local-Schauder} to the construction at the radius $R$ already chosen. On each scaled annulus, $(-\Delta+\mathcal E_p)v_p=f_p$ has uniformly smooth scaled coefficients and a source with all the required weighted derivatives. The bounded $C^0$ norm therefore gives all higher scaled spatial derivatives of $v_p$. Its placement derivative satisfies the same operator equation with source $\partial_pf_p-(\partial_p\mathcal E_p)v_p$, so the same argument applies to it. On the fixed cutoff annuli this gives $g_p\in C^{6,\alpha}$ and its fixed-coordinate placement derivative in $C^{5,\alpha}$. The intact inverse and local regularity then give $w_p\in C^{8,\alpha}$ and $\partial_pw_p\in C^{7,\alpha}$, uniformly on the fixed compact placement interval. The same radius $R$ works at every derivative order. The difference quotient for the pullback then converges in $C^{6,\alpha}$ by the fundamental theorem of calculus and the extra spatial derivative. These higher bounds justify the differentiation; the displayed quantitative estimate still uses the $C^{7,\alpha}$ and $C^{6,\alpha}$ norms above. The definition of $T_{\rm cov}$ and~\eqref{green:CH-conditions} include this term. On the remaining region the pushforward bounds $C_{\rm push,0},C_{\rm push,1}$ and the annular product rule give~\eqref{green:outside-constant}. Further scaled interior estimates give the corresponding qualitative bounds at every fixed derivative order: the explicit source has all the differentiated weighted bounds, and the annular source is smooth. The same argument applies to its placement derivative.

The construction also gives continuity with respect to the central mass. To see this, use the fixed horizon identifications of Proposition~\ref{thm:outer-tensor-inverse} and the moving normal coordinates at the insertion points. On a compact subextremal parameter set, the coordinate maps, the coefficients of $P_p$, and the explicit source $f_p^{\,S}$ depend smoothly on the mass. In the weighted spaces of~\eqref{green:spaces}, their difference is controlled by the difference of their coefficient seminorms. In particular, the local contraction operators and their sources are continuous in operator and source norm, with a common contraction factor less than one. The fixed-point identity then gives continuity of $v_p$ in $\mathcal X_{\rm b}^{6,\alpha}$. On the annulus, the expression for $g_p$ gives continuity in a fixed $C^{4,\alpha'}$ norm, for every $0<\alpha'<\alpha$; the smaller exponent also accounts for transporting the annulus to fixed coordinates. The intact inverse varies continuously in these fixed spaces by the inverse identity, so $w_p$ is continuous in $C^{6,\alpha'}$. Thus, after the indicated identifications, $H_p-2D_p/\rho$ is continuous in the local weighted norm of order four with exponent $\alpha'$, and $H_p$ is continuous in $C^{4,\alpha'}$ on compact sets away from the insertion points. This argument concerns strictly positive distance from the endpoints of the subextremal mass interval.

We will also need a quantitative mass estimate on fixed compact sets that stay a positive distance from the moving poles. For every such compact set $K$, the same construction gives
\be\label{green:mass-Lipschitz}
 \|H_{M,p(M)}-H_{\widetilde M,p(\widetilde M)}\|_{C^4(K)}
 \le C_K|M-\widetilde M|
\ee
when $p(M)$ is the maximal-radius placement, with the fixed radial identifications of the intact regions. Here is the parameter argument, including the horizon domains. In the fixed auxiliary manifold used in~\eqref{eq:tensor-domain}, choose the horizon radial coordinates and the factors $a_0(M)$ smoothly in $M$. This identifies all tensor domains with prescribed horizon surface gravities with a single invariant bundle. The conjugated operator $\widehat P_M$ then has coefficients smooth in $M$, including on the complete horizon charts, and is $C^1$ as a map $C^{6,\alpha}\to C^{4,\alpha}$. Its inverse satisfies
\be
 \widehat P_M^{-1}-\widehat P_{\widetilde M}^{-1}
 =\widehat P_M^{-1}(\widehat P_{\widetilde M}-\widehat P_M)\widehat P_{\widetilde M}^{-1}.
\ee
The inverse bounds and the coefficient difference therefore give a Lipschitz bound in these fixed spaces. In each moving normal ball, the leading tensor $2D/\rho$ has components independent of the mass in the normalized frame. The coefficient error and the explicit corrected source are $C^1$ in $M$ in $\mathcal X_{\rm b}^{6,\alpha}\to\mathcal Y_{-1}^{4,\alpha}$ and in $\mathcal Y_{-1}^{4,\alpha}$, respectively: their first mass derivatives obey the same vanishing orders at the center. The common local contraction bound makes its fixed point $C^1$ in that norm by differentiating the inverse identity. Consequently the annular residual is Lipschitz in $C^{4,\alpha'}$ after transport to fixed coordinates. The extra spatial derivatives already justified above control that transport, also for the mass derivative. Applying the displayed intact inverse identity to this residual gives~\eqref{green:mass-Lipschitz}. On a compact mass bracket the constants are uniform. This argument differentiates the smooth model operators and their prescribed Green sources, not the area-prescribing choice of $M$ as a function of the small mass.
\qedhere
\end{proof}

\subsection{The linearized equation on the background with two points removed}

Let $p_\pm$ be the reflected insertion points. Choose $\rho_w>0$ so that the balls $B_{2\rho_w}(p_\pm)$ are disjoint normal balls and $2\rho_w<\min(s_K,1)$. We take $\omega_\bullet=\rho$ on each $B_{\rho_w}$, $\omega_\bullet=1$ outside $B_{2\rho_w}$ and the fixed interpolation
\be\label{q:fixed-puncture-weight}
 \omega_\bullet\coloneqq\rho+(1-\rho)\vartheta(\rho/\rho_w-1)\qquad(\rho_w\le\rho\le2\rho_w).
\ee
It satisfies $\rho\le\omega_\bullet\le1$ on the transition and a finite uniform bound for $\|\nabla\omega_\bullet\|_\infty$ follows from the fixed profile and $\rho_w$. Its transition seminorms define constants $W_j$ by
\be\label{q:puncture-weight-derivatives}
\sup_{B_{2\rho_w}\setminus B_{\rho_w}}|\nabla^j\omega_\bullet|\le W_j\quad(1\le j\le{5}).
\ee
Choose $c_\bullet>0$ so that the balls below lie in the regular model charts, and so that $8c_\bullet\|\nabla\omega_\bullet\|_\infty<\theta_w<1$. Thus the weight on each ball is comparable to its value at the center with factors $1\pm\theta_w$. Put
\bse\label{q:punctured-Kottler-norms}
\be
\|T\|_{X_{K,\bullet}}\coloneqq\sup_x\omega_\bullet(x)^{-\gamma}\|T\|_{C^{4,\alpha}_{\nabla,\omega_\bullet(x)}(B_{c_\bullet\omega_\bullet(x)}(x))},
\ee
\be
\|f\|_{Y_{K,\bullet}}\coloneqq\sup_x\omega_\bullet(x)^{2-\gamma}\|f\|_{C^{2,\alpha}_{\nabla,\omega_\bullet(x)}(B_{8c_\bullet\omega_\bullet(x)}(x))}.
\ee
\ese
We keep $\gamma$ in the range~\eqref{eq:free-scale-conditions}. The positive Newton kernel satisfies
\be\label{q:Newton}
\frac1{4\pi}\int_{\mathbb R^3}\frac{|y|^{\gamma-2}}{|x-y|}\dd y=\frac{|x|^\gamma}{-\gamma(\gamma+1)}.
\ee
Integration of the inner and outer shells proves this identity for $-1<\gamma<0$. Its constant is $16/3$ at the chosen model weight.

\begin{proposition}[Solving the equation on the background with two points removed]\label{q:punctured-Kottler-inverse}
The static tensor operator on the punctured Schwarzschild--de Sitter region has a unique inverse from $Y_{K,\bullet}$ to $X_{K,\bullet}$ in the regular static polar class invariant under equatorial reflection. Let $L_{\rm loc}$ be the local inverse bound below. Let $M_{\rm src},M_{\rm out}$ bound the combined source and output cutoff maps at the two punctures, let $M_{\rm far}$ bound the remaining source in the unweighted norm, and let $M_{\rm comm}$ bound the output commutator from the local weighted space to that norm. Let $C_{\rm emb}$ bound the restriction of a regular $X_K$ tensor to $X_{K,\bullet}$. A sufficient inverse constant is
\be\label{q:punctured-Kottler-bound}
C_K\ge M_{\rm out}L_{\rm loc}M_{\rm src}+C_{\rm emb}C_{K,\mathrm{all}}(M_{\rm far}+M_{\rm comm}L_{\rm loc}M_{\rm src}).
\ee
\end{proposition}
\begin{proof}
\bse
The multiplication and restriction constants are the operator norms of the fixed cutoff and chart maps.
First divide chart lengths by $s_K$ and freeze the spatial principal matrix of $s_K^2P_G$ at the puncture. After the fixed linear normalization its principal part is $-\Delta$. The Newton identity and the constant-coefficient local estimate give
\be\label{q:frozen-Newton-Holder}
\|\mathcal N f\|_{\mathcal X_\gamma^{\rm E}} \le C_{{\rm N},\gamma}\|f\|_{\mathcal Y_\gamma^{\rm E}}.
\ee
The constant $C_{{\rm N},\gamma}$ depends only on $\gamma$, the frozen ellipticity, the local Schauder constant and the fixed coordinate and ball comparisons. Indeed~\eqref{q:Newton} gives the weighted $C^0$ estimate, and~\eqref{q:local-Schauder} gives the scaled derivative and H\"older bounds on each punctured ball. All these choices concern a fixed model problem.
Choose $\delta>0$. Extend the coefficients smoothly, keeping the original operator on $|x|<\delta/2$ and the frozen operator on $|x|\ge\delta$. Write the extension as $-\Delta+\mathcal E$. The coefficient and cutoff product bounds give a constant $C_{\rm E}$, independent of $\delta$, with
\be\label{q:local-Kottler-coefficient-error}
\|\mathcal EU\|_{\mathcal Y_\gamma^{\rm E}} \le C_{\rm E}\delta\|U\|_{\mathcal X_\gamma^{\rm E}}.
\ee
The principal difference vanishes at the origin; the first-order term gains one weighted radius and the potential gains two. Scaled differentiation of the cutoff preserves these powers. Choose $0<\theta_{\rm loc}<1$ and $\delta$ subject to
\be\label{q:punctured-Kottler-radii}
r_\bullet\coloneqq\delta s_K<\rho_w,\qquad C_{{\rm N},\gamma}C_{\rm E}\delta\le\theta_{\rm loc}.
\ee
Under this condition, the operator
\be\label{q:exact-local-Kottler-inverse}
\mathcal R_{\rm loc}\coloneqq\mathcal N(I+\mathcal E\mathcal N)^{-1},\qquad L_{\rm loc}\coloneqq\frac{C_{{\rm N},\gamma}}{1-\theta_{\rm loc}}
\ee
is a local exact inverse with norm at most $L_{\rm loc}$. The construction preserves the circular symmetries and pairs the charts by reflection.

\ese

\bse
Choose source and output cutoffs equal to one on $B_{r_\bullet/4}$ and supported in $B_{r_\bullet/2}$, so their supports lie in the original-coefficient region. Apply the local inverse operators to the source cutoffs and multiply the outputs. This gives
\be\label{q:punctured-local-parametrix}
\|V\|_{X_{K,\bullet}} \le M_{\rm out}L_{\rm loc}M_{\rm src}\|f\|_{Y_{K,\bullet}},\qquad H\coloneqq f-P_GV=0\quad\hbox{on }B_{r_\bullet/4}(p_\pm).
\ee
To check the residual, let $\chi_{\rm src}$ and $\chi_{\rm out}$ denote the source and output cutoffs at one puncture, and put $U=\mathcal R_{\rm loc}(\chi_{\rm src}f)$. On the support of $\chi_{\rm out}$, the extended and original operators agree. The product rule therefore gives
\be
 f-P_G(\chi_{\rm out}U)
 =(1-\chi_{\rm out}\chi_{\rm src})f-[P_G,\chi_{\rm out}]U.
\ee
For the two disjoint punctures, the first factor is replaced by $1-\sum_{\pm}\chi_{{\rm out},\pm}\chi_{{\rm src},\pm}$, and the commutators are summed. Both resulting terms vanish near each puncture: the first because its pair of cutoffs is one there, and the second because the output cutoffs are constant there. Their supports stay a fixed positive distance from the punctures, where the weighted and unweighted norms are comparable. The residual is regular and obeys
\be\label{q:punctured-residual-bound}
\|H\|_{Y_K} \le(M_{\rm far}+M_{\rm comm}L_{\rm loc}M_{\rm src})\|f\|_{Y_{K,\bullet}}.
\ee
In particular the commutator includes the local inverse factor. The map norm $M_{\rm comm}$ includes the conversion from scaled to unscaled derivatives on the fixed annuli; it is independent of the inserted mass. The tensor $T\coloneqq V+P_G^{-1}H$ solves the equation, and the bound for the inverse on the full background gives~\eqref{q:punctured-Kottler-bound}.

For uniqueness, a homogeneous tensor in $X_{K,\bullet}$ satisfies $T=\Ocal(\rho^\gamma)$, $\nabla T=\Ocal(\rho^{\gamma-1})$. Its boundary terms in Green's formula tend to zero as $\rho^{\gamma+2}$ and $\rho^{\gamma+1}$. It therefore extends first as a distributional solution and then, by elliptic regularity, as a smooth solution across the two points. Injectivity on the full background gives $T=0$.
\ese
\end{proof}

\begin{lemma}[Inverse operators of the model for a range of decay exponents]\label{q:neighboring-weights}
For $-1<d<0$, let $X_S(d),Y_S(d)$ and $X_{K,\bullet}(d),Y_{K,\bullet}(d)$ be the model spaces of~\eqref{q:model-Holder-norms} and~\eqref{q:punctured-Kottler-norms}, with the same balls and derivative scales and with weight $d$ in place of $\gamma$. The space $X_S^0(d)$ is the kernel of $T\mapsto\int\chi_0\langle T,T_S\rangle\dd\nu_S$. The maps
\bse\label{q:neighboring-model-maps}\be
 \mathcal A_S(T,\lambda)=P_ST-\lambda\chi_0T_S:\ X_S^0(d)\times\mathbb R\longrightarrow Y_S(d),\ee\be
 P_K:X_{K,\bullet}(d)\longrightarrow Y_{K,\bullet}(d)
\ee\ese
are bounded isomorphisms. Their inverse norms $C_S(d),C_K(d)$ are finite; the latter is uniform on the fixed compact mass and placement set. The integral functional and compactly supported source are the same for every $d$.
\end{lemma}
\begin{proof}
The weighted range construction in Section~\ref{sec:Schwarzschild-models} applies throughout $-1<d<0$. Its kernel and its unweighted-dual cokernel, of weight $-1-d$, are both spanned by $T_S$. The positive pairing $m_0=\int\chi_0|T_S|^2\dd\nu_S$ makes the compactly supported source transverse to the range, and the same integral condition removes the kernel. This proves the first isomorphism. The areal and isotropic weights, and the prescribed fixed-radius-ratio balls, have uniform model comparison constants, so the conclusion holds in precisely the displayed norms.

For Schwarzschild--de Sitter, repeat the local construction of Proposition~\ref{q:punctured-Kottler-inverse} with $d$. The Euclidean Newton constant is $[-d(d+1)]^{-1}<+\infty$. A fixed sufficiently small coordinate radius makes the coefficient error absorbable. We use the inverse on the full background for the smooth remaining source. A homogeneous tensor of size $O(\rho^d)$ has vanishing distributional boundary terms: the terms involving the tensor and its first derivative have orders $\rho^{d+2}$ and $\rho^{d+1}$. It therefore extends smoothly across the two punctures, and the intact inverse gives zero. All radii and map bounds can be chosen uniformly on the compact parameter set. The bounded inverse theorem gives the stated finite norms.\qedhere
\end{proof}

\section{Solving the tensor equation with two shrinking horizons}\label{sec:insertions}\label{sec:joined-quantitative}

\subsection{A reference metric with four regular horizons}

Work in cosmological length units, so $\ell=\sqrt{3/\Lambda}=1$. The proof is unchanged for arbitrary fixed $\Lambda>0$ after restoring these units. Use the fixed Schwarzschild--de Sitter fields $(h_K,N_K)=(h_M,N_M)$, with central mass $M$. Write $G_K\coloneqq G_M$ and $P_K\coloneqq P_{G_K}$. Choose antipodal points $p_\pm=p_\pm(r)$ on a sphere of radius $r$, where $r$ ranges in a fixed compact interval $I$ about the maximal sphere and inside the static region. Put $b\coloneqq N_K(p_\pm)>0$. The choices below are uniform for $r\in I$. Choose disjoint normal-coordinate balls of radius $4r_0$, contained in the positive-lapse interior. Their Cartesian coordinates $z_\pm$ are compatible with rotations and with the reflection $\iota$ exchanging the two points. Write $\rho_\pm\coloneqq|z_\pm|$.

For the preliminary linear estimate, allow $R_\mu=\mu^a$ with fixed $0<a<1$; the nonlinear construction uses $a=a_2$ from~\eqref{eq:free-scale-conditions}. In each ball remove $\rho_\pm<\mu/2$. On the new manifold define the Schwarzschild fields
\be\label{join:eq:Schw}
h_{S,\mu}\coloneqq(1+\mu/(2\rho))^4\dd z^2,
\qquad N_{S,\mu}\coloneqq b\frac{1-\mu/(2\rho)}{1+\mu/(2\rho)}.
\ee
Take a fixed smooth cutoff $0\le\chi\le1$ equal to one on $[0,1]$ and zero on $[2,+\infty)$. For the effective construction, take $\chi(t)=1-\vartheta(t-1)$ with the explicit profile~\eqref{q:joined-cutoff-profile}. Set, inside either normal ball,
\be\label{join:eq:reference}
h_\mu\coloneqq\chi(\rho/R_\mu)h_{S,\mu}+
(1-\chi(\rho/R_\mu))h_K,
\qquad
N_\mu\coloneqq\chi(\rho/R_\mu)N_{S,\mu}+
(1-\chi(\rho/R_\mu))N_K.
\ee
Outside the two balls use $(h_K,N_K)$. This is positive in the interior and is exactly Schwarzschild near the two new horizons and exactly Schwarzschild--de Sitter near the two original horizons. Convexity of positive metrics gives spatial positivity on the cutoff annuli. The lapse is positive there. The axis is regular because each metric is smooth and invariant in the same spatial Cartesian coordinates. Both new horizons have opening $b/(4\mu)$.
Write $G_\mu\coloneqq h_\mu+N_\mu^2\dd\tau^2$ in the static interior. Use the regular local angle at each horizon. Each reference has all four regular local fillings, though a common global auxiliary period is not required. The operator is
\be\label{join:eq:P}
P_\mu\coloneqq\nabla_{G_\mu}^*\nabla_{G_\mu}-2\mathcal R_{G_\mu},
\qquad (\mathcal R T)_{ij}=R_{ikjl}T^{kl}.
\ee
The convention agrees with~\eqref{eq:tensor-operator}.

All tensors below are static, axisymmetric, invariant under reversal of both circular directions, orthogonal in the two symmetry directions, and invariant under $\iota$. They have only the two circular diagonal components and the symmetric meridional block. They are regular tensors in the normalized local lifts. The reference and the reduced operator preserve these symmetries, so the inverse and the fixed-point iteration act on this same class. Equatorial reflection pairs the two inserted collars and their displacement modes. In particular the domain fixes the reference opening at every horizon. The reference fixes the horizon surface gravities during inversion at each placement. They can vary with the parameters and with the final normalization of the lapse.
The Schwarzschild--de Sitter inverse operator and the Schwarzschild model inverse operator with the integral condition established above provide our two model operators for the construction.

\subsection{Weighted H\"older spaces and the displacement source}

Choose $k\ge0$, $0<\alpha<1$ and $-1<\gamma<0$. We specify a positive scale which is smooth in the normalized horizon charts. Recall the punctured background weight $\omega_\bullet$ from~\eqref{q:fixed-puncture-weight}, and its fixed radius $\rho_w$. For $0<\mu\le\rho_w$, let $\chi_w(\rho)\coloneqq1-\vartheta(\rho/\rho_w-1)$, equal to one for $\rho\le\rho_w$ and zero for $\rho\ge2\rho_w$. On the manifold with the two inserted horizons, define
\be\label{join:eq:explicit-weight}
 \sigma_\mu\coloneqq\omega_\bullet+
    \frac{\mu^2}{4}\sum_{\pm}\frac{\chi_w(\rho_\pm)}{\rho_\pm},
 \qquad \omega_\mu\coloneqq\sqrt{\mu^2+\sigma_\mu^2}.
\ee
Each summand is extended by zero outside its coordinate ball. The balls are disjoint, and both choices are invariant under reflection. Near either new horizon, $\sigma_\mu=R_\mu^{S}-\mu$, where
\be\label{join:eq:areal-scale-radius}
 R_\mu^{S}\coloneqq\rho+\mu+\frac{\mu^2}{4\rho}
\ee
is the areal radius of the inserted Schwarzschild model. The superscript distinguishes this function from the matching radius $R_\mu$.

\begin{lemma}[A smooth scale at the horizons]\label{join:lem:regular-scale}
The function $\omega_\mu$ is positive and smooth in every regular lift. It satisfies
\be\label{join:eq:scale-comparison}
 \frac{\mu+\omega_\bullet}{\sqrt2} \le\omega_\mu\le\mu+\omega_\bullet,
 \qquad \omega_\mu\ge\mu.
\ee
For each integer $j\ge1$ there is a constant $C_j$ independent of sufficiently small $\mu$ and of placement in the fixed compact parameter set such that
\be\label{join:eq:scale-jets}
 \omega_\mu^{j-1}|\nabla_{G_\mu}^j\omega_\mu|\le C_j.
\ee
Write $r_S(s)\coloneqq s+1+(4s)^{-1}\ge2$ for the unit Schwarzschild areal radius. On the inner coordinates $\rho=\mu s\le\rho_w$, the scale has the exact form
\be\label{join:eq:unit-scale}
 \frac{\omega_\mu}{\mu}=w_S(s)\coloneqq\sqrt{1+(r_S(s)-1)^2},
 \qquad w_S\le r_S\le\sqrt2\,w_S,
\ee
\end{lemma}
\begin{proof}
In the inner region, $R_\mu^S$ is a smooth invariant function of the Cartesian coordinates of the normal disc at the horizon. In signed Gaussian distance $d$ from that horizon, its expansion contains only even powers of $d$ and begins with $2\mu+d^2/(8\mu)$. Thus~\eqref{join:eq:explicit-weight} defines a smooth function there. On the remaining part of the insertion balls the cutoffs and coordinates are smooth, and outside these balls $\omega_\mu=\sqrt{1+\mu^2}$, including both original horizons. This proves regularity at all four horizons and their poles.

Put $c\coloneqq\chi_w(\rho)\mu^2/(4\rho)$ on one ball. Since $\rho\ge\mu/2$ and $\omega_\bullet\ge\rho$ there, we have $0\le c\le\mu/2$ and $\omega_\bullet\ge\mu/2$. Consequently
\be
 (\mu+\omega_\bullet)^2-\omega_\mu^2
 =2\omega_\bullet(\mu-c)-c^2\ge\frac{\mu^2}{4}\ge0.
\ee
The lower comparison follows from $\omega_\mu\ge\sqrt{\mu^2+\omega_\bullet^2}\ge(\mu+\omega_\bullet)/\sqrt2$. Outside the balls it is the same elementary inequality with $c=0$. The last comparison in~\eqref{join:eq:unit-scale} follows from
the identities $r_S^2-w_S^2=2r_S-2\ge0$ and $2w_S^2-r_S^2=(r_S-2)^2\ge0$.

On each fixed compact part of the unit Schwarzschild model, $w_S$ and all its covariant derivatives are bounded in the regular charts, including at the horizon. At large $s$, the explicit expressions for $r_S$ and $G_S$ give $|\nabla^j w_S|\le C_j(1+s)^{1-j}$. Constant rescaling by $\mu$ proves~\eqref{join:eq:scale-jets} in the exact model region. Where $\rho$ is beyond that region and below $\rho_w$, the formula for $\omega_\mu$ gives $|D^j\omega_\mu|\le C_j\rho^{1-j}$. In spatial Cartesian coordinates and the local unwrapped coordinate $b\tau$, the interpolated metric and its inverse have uniformly bounded components, and $\rho^j|D^jG_\mu|\le C_j$. Indeed the model differences are $O(\mu/\rho+\rho)$ in these scaled norms, and every differentiated transition cutoff contributes $R_\mu^{-1}\asymp\rho^{-1}$. The condition $\mu/R_\mu\to0$ ensures the uniform inverse-metric bound. The connection formula and induction on $j$ now give the same estimates for covariant derivatives. On the fixed annuli $\rho_w\le\rho\le2\rho_w$, the cutoffs and background coefficients have uniform bounds and the added term is $O(\mu^2)$ with all derivatives. This proves~\eqref{join:eq:scale-jets} globally.

In particular $|\nabla\omega_\mu|\le C_1$. On sufficiently small balls of radius proportional to $\omega_\mu$, the scale is comparable to its value at the center. The bound for derivative $j+1$ and parallel transport on these charts give
\be\label{join:eq:scale-Holder}
 [\nabla^j\omega_\mu]_{\alpha;B_{c\omega_\mu(x)}(x)}
 \le C_{j,\alpha}\omega_\mu(x)^{1-j-\alpha}.
\ee
The chain rule gives the corresponding bounds $|\nabla^j\omega_\mu^a|\le C_{j,a}\omega_\mu^{a-j}$ for every fixed real exponent $a$. All constants depend only on the fixed models, parameter sets and cutoffs.\qedhere
\end{proof}

On fixed compact subsets away from the insertion points, the scale $\omega_\mu$ converges smoothly to $\omega_\bullet$. On each fixed rescaled inner region, $\omega_\mu/\mu=w_S$. At an intermediate scale $a_\mu$ with $\mu/a_\mu\to0$ and $a_\mu\to0$, its expression in $z=a_\mu y$ satisfies
\be\label{join:eq:intermediate-scale}
 \frac{\omega_\mu(a_\mu y)}{a_\mu}
 =\sqrt{\left(\frac{\mu}{a_\mu}\right)^2+
 \left(|y|+\frac{\mu^2}{4a_\mu^2|y|}\right)^2}
 \longrightarrow |y|
\ee
in every $C^j$ norm on compact annuli. Thus the three limiting scales used below are respectively $\omega_\bullet$, $w_S$ and the Euclidean radius; no derivative of an isotropic radius at a horizon is used.

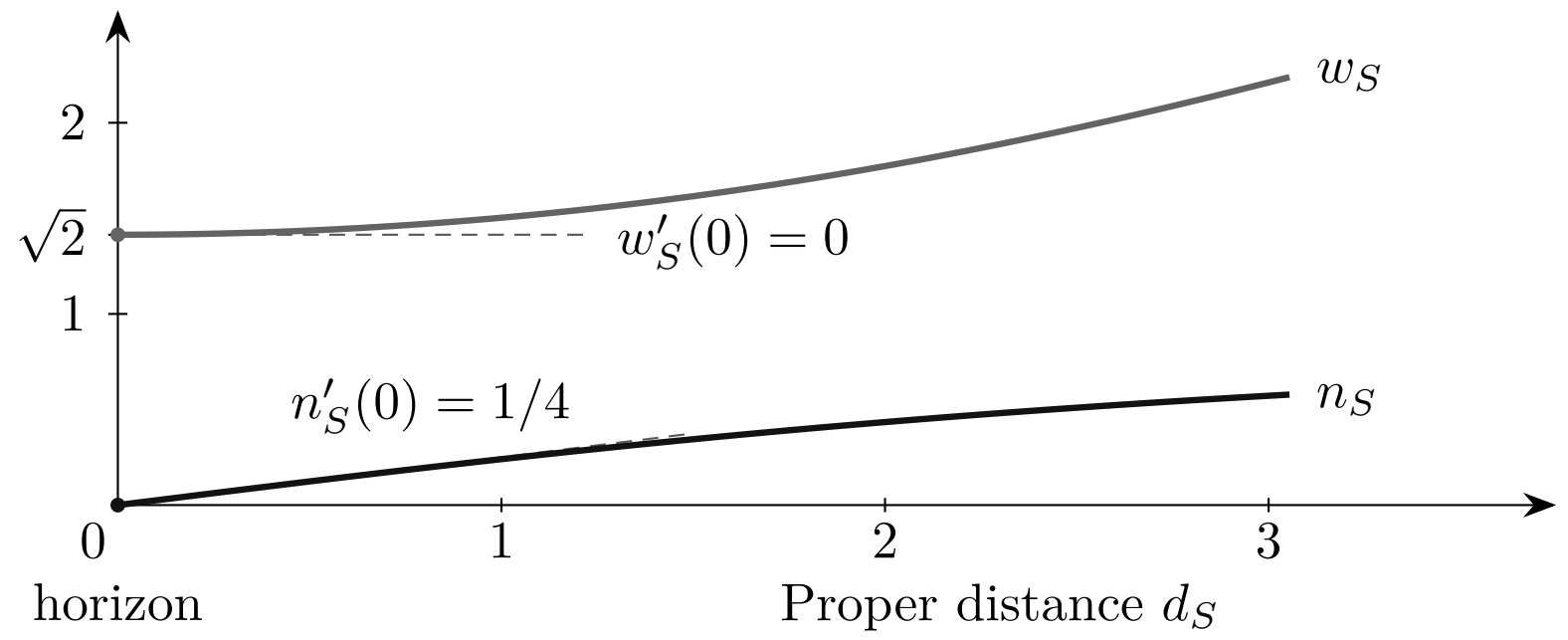
\begin{figure}[!htbp]\centering
\begin{tikzpicture}[x=2.55cm,y=1.27cm,font=\normalsize]
\path[use as bounding box] (-.42,-.72) rectangle (3.99,2.67);
\draw[-{Stealth[length=2.2mm]}] (0,0)--(3.75,0);
\draw[-{Stealth[length=2.2mm]}] (0,0)--(0,2.59);
\foreach \xx in {1,2,3}{
  \draw (\xx,-.037)--(\xx,.037);
  \node[anchor=north,inner sep=2pt] at (\xx,-.04) {$\xx$};
}
\foreach \yy/\lab in {1/1,1.41421356/{\sqrt2},2/2}{
  \draw (-.025,\yy)--(.025,\yy);
  \node[anchor=east,inner sep=4pt] at (-.025,\yy) {$\lab$};
}
\node[anchor=north east,inner sep=2pt] at (0,-.04) {$0$};
\node[anchor=north,inner sep=2pt] at (0,-.35) {horizon};
\node[anchor=north,inner sep=2pt] at (2.3,-.35) {Proper distance $d_S$};
\draw[black!65,densely dashed] (0,0)--(1.5,.375);
\draw[black!65,densely dashed] (0,1.41421356)--(1.23,1.41421356);
\draw[blue!65!black,very thick] plot[domain=0:1.32,samples=100,variable=\p]
  ({\p+(exp(\p)-exp(-\p))/2},{(exp(\p)-1)/(exp(\p)+1)});
\draw[orange!65!black,very thick] plot[domain=0:1.32,samples=100,variable=\p]
  ({\p+(exp(\p)-exp(-\p))/2},{sqrt(1+((exp(\p)+exp(-\p))/2)^2)});
\fill[blue!65!black] (0,0) circle (1.4pt);
\fill[orange!65!black] (0,1.41421356) circle (1.4pt);
\node[anchor=west,inner sep=2pt] at (1.27,1.38) {$w_S'(0)=0$};
\node[anchor=west,inner sep=2pt] at (.42,.52) {$n_S'(0)=1/4$};
\node[anchor=west,inner sep=3pt] at (3.08,.55) {$n_S$};
\node[anchor=west,inner sep=3pt] at (3.08,2.25) {$w_S$};
\end{tikzpicture}
\caption{The normalized lapse $n_S$ vanishes linearly at the Schwarzschild horizon. The scale $w_S$ is positive there and has zero derivative.}\label{fig:inner-levels}
\end{figure}

Figure~\ref{fig:inner-levels} compares the normalized lapse and the scale $w_S$ as functions of the proper distance $d_S$ to the horizon in the unit-mass Schwarzschild metric. Write $r_S=1+\cosh q$ with $q\ge0$. The relation $\dd d_S/\dd r_S=(1-2/r_S)^{-1/2}$ gives $\dd d_S/\dd q=1+\cosh q$. The two profiles are therefore given by
\be\label{join:eq:proper-distance-profiles}
d_S=q+\sinh q,\qquad n_S(d_S)=\tanh(q/2),\qquad w_S(d_S)=\sqrt{1+\cosh^2 q}.
\ee
At $d_S=0$, the lapse has slope $1/4$, and the scale has value $\sqrt2$ and slope zero. The scale sets the radius of the balls used for the weighted norms below. In the mass-$\mu$ model, the proper distance and scale are $\mu d_S$ and $\mu w_S$, respectively.

Recall the model chart factor
\be\label{q:joined-identification-scale}
c_S=\widetilde\rho_S(r)/r,
\ee
with the model chart radius fixed in~\eqref{q:geometric-chart-radii}. We choose the construction ball size subject to
\be\label{q:evaluated-construction-ball}
0<c\coloneqq c_J,\qquad 8c_J\le\min(1,s_K,\widetilde\rho_K,c_S,c_\bullet,[2(1+C_1)]^{-1}).
\ee
For the qualitative estimates we take any sufficiently small fixed $c>0$; for the quantitative estimates we take $c=c_J$. The last restriction in~\eqref{q:evaluated-construction-ball} ensures that the scale varies by at most a factor between $1/2$ and $3/2$ on each enlarged ball of radius $8c_J\omega_\mu(x)$. On balls of radius $c\omega_\mu$, taken in the normalized lifts near the horizons and in the regular interior charts elsewhere, define
\be\label{join:eq:norms}
\|v\|_{X_\mu}\coloneqq\sup_x\omega_\mu(x)^{-\gamma}
\|v\|_{C^{k+2,\alpha}_{\omega_\mu(x)}},\qquad
\|f\|_{Y_\mu}\coloneqq\sup_x\omega_\mu(x)^{2-\gamma}
\|f\|_{C^{k,\alpha}_{\omega_\mu(x)}}.
\ee
Tensor sizes use $G_\mu$. Covariant derivatives of order $j$ in the scaled norms carry the factor $\omega_\mu^j$. Static fields may be unwrapped in the auxiliary direction on interior balls; no injectivity-radius bound for a shrinking auxiliary circle is used.

On the plus collar use
\be\label{join:eq:rescaling}
\Theta_{+,\mu}(y,\sigma)\coloneqq(z_+=\mu y,\tau=\mu\sigma/b),
\qquad \Theta_{+,\mu}^*G_\mu=\mu^2G_S,
\qquad \widehat v_+\coloneqq\mu^{-2}\Theta_{+,\mu}^*v.
\ee
This equality is exact for $|y|<R_\mu/\mu$. A hat on any covariant two-tensor, including a source tensor, denotes $\mu^{-2}\Theta_{+,\mu}^*$ of that tensor. Obtain the minus rescaling by reflection.

We use the cutoff $\chi_0$ and positive normalization integral $m_0$ defined in Section~\ref{sec:Schwarzschild-conventions}. The support of $\chi_0$ is denoted by $K_0$. We define the integral functional by
\be\label{join:eq:centering}
\ell_\mu(v)\coloneqq m_0^{-1}\int\langle\mu^{-\gamma}\widehat v_+,\chi_0 T_S\rangle\dd\nu_S.
\ee
The minus functional is identical on the reflected space. Let $X_\mu^0\coloneqq\ker\ell_\mu$. Define a source $\psi_{+,\mu}$ by
\be\label{join:eq:column}
\mu^{-2}\Theta_{+,\mu}^*\psi_{+,\mu}
=\mu^{\gamma-2}\chi_0 T_S,
\qquad \psi_\mu\coloneqq\psi_{+,\mu}+\iota_*\psi_{+,\mu},
\ee
extended by zero outside the two collars. These fields are regular. On their fixed normalized supports, $\omega_\mu\asymp\mu$ and the scaled $C^{k,\alpha}$ norm of the rescaled tensor multiplying the auxiliary coefficient is $O(\mu^{\gamma-2})$. Multiplication by the source weight therefore gives
\be
 \|\psi_\mu\|_{Y_\mu} \le C\mu^{2-\gamma}\mu^{\gamma-2} \le C,
\ee
with a constant independent of $\mu$; the fixed reflected copy only changes that constant. The same normalization bounds $\mu^{-\gamma}\widehat v_+$ on the support of $\chi_0$ by $C\|v\|_{X_\mu}$, so $\ell_\mu$ is uniformly bounded. It is uniformly nonzero: let $v_\mu$ be the paired compact tensor with normalized plus pullback $\mu^\gamma\chi_0T_S$. Its $X_\mu$ norm is bounded independently of $\mu$, while
\be
 \ell_\mu(v_\mu)=m_0^{-1}\int\chi_0^2|T_S|^2\dd\nu_S>0
\ee
is independent of $\mu$. Thus the norm of $\ell_\mu$ is also bounded below by a fixed positive constant.

\subsection{An inverse with a uniform bound as the horizons shrink}

\paragraph{Fredholm index and the estimate under the integral condition.} At each positive mass, compactness reduces solvability to a kernel estimate. We first identify the operator domain for this argument.

\begin{lemma}[Fredholm index at fixed positive mass]\label{join:lem:Fredholm}
For every fixed $\mu>0$ for which the reference metric $G_\mu$ is defined and positive with its stated regular local fillings, $P_\mu:X_\mu\to Y_\mu$ is Fredholm of index zero. The reference may have a nonzero Einstein defect on its cutoff annuli.
\end{lemma}
\begin{proof}
Use invariant integration per unit $\tau$, with measure $N_\mu\dd V_{h_\mu}$. Its expression in the normalized angle at a horizon of opening $\kappa_j$ is $\kappa_j/(2\pi)$ times the ordinary smooth lifted integral over a regular period. The factor is constant on that horizon.

To justify compactness, choose a positive smooth function $a_0$, invariant under the imposed symmetries and equal to $\kappa_j^{-1}$ near each horizon. A positive partition of unity gives such a function after averaging over the rotations and reflection; paired horizons have equal reference surface gravities. The metric $\widetilde G_\mu=h_\mu+(a_0N_\mu)^2\dd\vartheta^2$, with period $2\pi$ for $\vartheta$, fills all four horizons smoothly. Rescaling the circular tensor component by $a_0^2$ identifies the invariant regular tensor space with a closed symmetry subspace of smooth tensors on this compact auxiliary manifold. Near the horizons $a_0$ is constant and all normalized derivatives agree. On the remaining compact set, the derivatives differ only by bounded zeroth-order terms. Thus the corresponding $H^1,H^2,L^2$ norms are equivalent for this fixed $\mu$, and $H^1\hookrightarrow L^2$ is compact.

For $c_\mu>2\|\mathcal R_{G_\mu}\|_\infty+1$, the symmetric form
\be
\mathfrak a_c(v,z)\coloneqq\int(\langle\nabla v,\nabla z\rangle
-2\langle\mathcal Rv,z\rangle+c_\mu\langle v,z\rangle)
N_\mu\dd V_{h_\mu}
\ee
is coercive on this invariant $H^1$ space. Hilbert space representation solves $(P_\mu+c_\mu)v=f$ weakly. Local tests averaged over the local circle actions, their reversals and the global reflection recover the unrestricted weak equation in each normalized horizon chart. The fixed factor $\kappa_j/(2\pi)$ cancels there. Local elliptic regularity gives $H^2$, and then the usual higher Sobolev and H\"older regularity. In the static interior this is ordinary elliptic regularity on the three-dimensional spatial manifold. Near the horizons and poles it is the smooth four-dimensional local equation.

Consequently the shifted inverse is compact and selfadjoint on invariant $L^2$. Integration by parts has no limiting horizon term: remove normal tubes of radius $s$, whose area is $O(s)$ in a regular lift; smooth invariant tensor coefficients and their first covariant derivatives are bounded. Density gives the form identities. Compactness gives closed range and finite-dimensional kernel for $P_\mu$, and selfadjointness identifies kernel and cokernel. It has index zero. Elliptic regularity identifies the same kernel and cokernel in the stated H\"older spaces. For fixed positive $\mu$ the weight is positive on the compact completed domain, so it changes the norms but not the index or domain.\qedhere
\end{proof}

The geometry has uniform scaled chart bounds. On $C\mu<\rho<r_0$, direct differentiation of~\eqref{join:eq:reference} gives the following estimates, uniformly for $r\in I$:
\be\label{join:eq:neck}
\rho^j|D^j(G_\mu-G_{\rm flat})|\le C_j(\mu/\rho+\rho),
\qquad
\rho^2|\operatorname{Rm}_{G_\mu}|\le C(\mu/\rho+\rho).
\ee
The flat comparison is $\dd z^2+b^2\dd\tau^2$. The linear term in $\rho$ allows the gradient of $N_K$ at the insertion point to be nonzero; in spatial normal coordinates the corresponding spatial metric error is $O(\rho^2)$. The same estimates hold on the cutoff annuli because their derivatives are at scale $R_\mu\asymp\rho$. Near the new horizons the rescaled metric is exactly $G_S$. Near the original horizons it is exactly $G_K$. Scaled local Schauder estimates and interpolation therefore give
\be\label{join:eq:Schauder}
\|v\|_{X_\mu} \le C\bigl(\|P_\mu v\|_{Y_\mu}
+\sup\omega_\mu^{-\gamma}|v|\bigr)
\ee
with a constant independent of small $\mu$.

\begin{proposition}[Uniform estimate under the integral condition]\label{join:prop:centered}
There are $C,\mu_0>0$ such that
\be\label{join:eq:centered}
\|v\|_{X_\mu} \le C(\|P_\mu v\|_{Y_\mu}+|\ell_\mu(v)|)
\qquad(0<\mu<\mu_0).
\ee
\end{proposition}
\begin{proof}
If the estimate fails for every sufficiently small threshold, choose $\mu_i\downarrow0$ and take $\|v_i\|_{X_{\mu_i}}=1$ while both quantities on the right tend to zero. Formula~\eqref{join:eq:Schauder} gives points where $\omega_{\mu_i}^{-\gamma}|v_i|\ge c>0$. Pass to a subsequence.

If these points remain away from the two insertion points, compactness in the scaled charts gives a limiting solution on the background region $v_K$ of $P_Kv_K=0$ off the two punctures, with $|v_K|+\rho|\nabla v_K|\le C\rho^\gamma$. This extends distributionally over each puncture: the boundary terms in Green's formula have orders $\rho^{\gamma+1}$ and $\rho^{\gamma+2}$ per unit auxiliary coordinate, and tend to zero since $\gamma>-1$. Elliptic regularity gives a regular extension. The regularity at the original horizons passes in their fixed local lifts. Proposition~\ref{thm:outer-tensor-inverse} gives zero, contradicting the positive value at the chosen points.

If $\rho_i/\mu_i$ stays bounded near one insertion, use~\eqref{join:eq:rescaling} and set $V_i=\mu_i^{-\gamma}\widehat v_i$. The operator equation becomes $P_SV_i=\mu_i^{2-\gamma}\widehat{P_{\mu_i}v_i}$. Its right side tends to zero on each fixed collar. A limit is regular at the Schwarzschild horizon and has growth at infinity at most $(1+|y|)^\gamma$. The Schwarzschild classification gives a limit $dT_S$. The integral condition~\eqref{join:eq:centering} gives $d=0$, contradicting the normalized positive value.

In the remaining case $\rho_i\to0$ and $\rho_i/\mu_i\to\infty$. Rescale lengths by $\rho_i$, the auxiliary coordinate by $\rho_i/b$, and tensor amplitudes by $\rho_i^{-\gamma}$. Estimate~\eqref{join:eq:neck} gives a flat static limit on $\mathbb R^3\setminus\{0\}$. Every Cartesian component is harmonic, with bound $C|y|^\gamma$. For a compactly supported smooth test function, integration over the complement of a ball of radius $\varepsilon$ gives boundary terms bounded by $C(\varepsilon^{\gamma+1}+\varepsilon^{\gamma+2})$, using the scaled derivative bound. These terms vanish because $\gamma>-1$, so the component is distributionally harmonic across zero and hence smooth there. It tends to zero at infinity because $\gamma<0$; the scalar maximum principle then makes it zero. This again contradicts the positive value at a point of unit radius. These three cases exhaust all weighted concentration points.\qedhere
\end{proof}

\paragraph{The auxiliary source coefficient and the inverse of the augmented operator.}
Let $\chi_R$ be one on $|y|\le R$, zero for $|y|\ge2R$, with the usual derivative bounds, where $R$ is larger than the fixed support of $\chi_0$. Define $Z_{\mu,R}$ by the normalized pullback $\chi_R T_S$ on both collars, with the reflected sign convention of~\eqref{join:eq:column}. For each fixed $R$ and small enough $\mu$, its support lies in the exact Schwarzschild region.

\begin{lemma}[A uniform estimate for the added coefficient]\label{join:lem:coefficient}
If $v\in X_\mu^0$ and
\be\label{join:eq:augmented}
P_\mu v-\lambda\psi_\mu=f,
\ee
then, for all sufficiently large fixed $R$ and all sufficiently small $\mu$ depending on $R$,
\be\label{join:eq:lambda}
|\lambda|\le C(\|f\|_{Y_\mu}
+R^{\gamma-1}\|v\|_{X_\mu}),
\ee
where $C$ is independent of $R$ and $\mu$.
\end{lemma}
\begin{proof}
The measure per unit of the original coordinate $\tau$ is $b\mu^3\dd\nu_S$. The tensor source normalization in~\eqref{join:eq:column} therefore gives
\be\label{join:eq:gram}
\langle\psi_\mu,Z_{\mu,R}\rangle_{L^2(G_\mu)}
=2b\mu^{\gamma+1}m_0.
\ee
The two collars give the factor two. Selfadjoint integration against $Z_{\mu,R}$ transforms the first term of~\eqref{join:eq:augmented} to $\langle v,P_\mu Z_{\mu,R}\rangle$. The normalized right side satisfies
\be
\mu^{2-\gamma}|\widehat f|\le
C\|f\|_{Y_\mu}(1+|y|)^{\gamma-2}.
\ee
Pairing this with $\chi_R T_S=O((1+|y|)^{-2})$ is bounded independently of $R$: its far radial integral is bounded by $C\int_1^\infty r^{\gamma-2}\dd r<+\infty$.

Since $P_S T_S=0$, the tensor $P_S(\chi_R T_S)$ is supported on $R<|y|<2R$ and has size $O(R^{-4})$. The normalized correction $V\coloneqq\mu^{-\gamma}\widehat v$ has size at most $C\|v\|_{X_\mu}(1+|y|)^\gamma$. The error integral is therefore at most $CR^{\gamma-1}\|v\|_{X_\mu}$. Returning to the coordinates before the rescaling by $\mu$ gives the factor $2b\mu^{\gamma+1}$. Divide by the positive Gram factor~\eqref{join:eq:gram} to obtain~\eqref{join:eq:lambda}. The exact Schwarzschild collar makes this calculation free of a metric approximation remainder on the support of $Z_{\mu,R}$.\qedhere
\end{proof}

\begin{proposition}[An inverse for the tensor equation with a bound independent of the small mass]\label{join:thm:joined}
For all sufficiently small $\mu>0$, the operator
\be\label{join:eq:map}
\mathcal A_\mu:X_\mu^0\times\mathbb R\longrightarrow Y_\mu,
\qquad (v,\lambda)\longmapsto P_\mu v-\lambda\psi_\mu
\ee
is a Banach space isomorphism, and
\be\label{join:eq:inverse}
\|v\|_{X_\mu}+|\lambda|\le C_M\|f\|_{Y_\mu}.
\ee
The constant is independent of the two small horizon radii. The central Schwarzschild--de Sitter mass is fixed arbitrarily in its subextremal interval.
\end{proposition}
\begin{proof}
Let $C_0$ be the constant in~\eqref{join:eq:centered} and $C_\psi$ a uniform upper bound for $\|\psi_\mu\|_{Y_\mu}$. Then
$\|v\|_{X_\mu} \le C_0(\|f\|_{Y_\mu}+C_\psi|\lambda|)$.
Choose $R$ so large that the coefficient of $\|v\|$ after inserting~\eqref{join:eq:lambda} is at most $1/2$. This is possible since $\gamma-1<0$. Then choose $\mu$ small enough that $2R\mu<R_\mu$. Absorption proves~\eqref{join:eq:inverse}.

Lemma~\ref{join:lem:Fredholm} gives index zero before imposing the integral condition. The nonzero continuous functional $\ell_\mu$ restricts its domain by codimension one, reducing the index to $-1$. Adding the unused scalar variable gives index zero; the auxiliary source term $-\lambda\psi_\mu$ is a finite-rank perturbation and preserves that index. Estimate~\eqref{join:eq:inverse} makes the kernel zero. Index zero gives zero cokernel, so the map is onto and has the asserted bounded inverse. This is a range statement in $Y_\mu$, not only an estimate on an existing range.\qedhere
\end{proof}

\subsection{Cutoffs for combining the two inverse operators}\label{q:joined-section}

The compactness argument above gives a bound independent of the added mass. To quantify the range of masses for which the inverse exists, we combine the two model inverse operators and estimate the error introduced by the cutoffs. We use cosmological units and write $u$ for the small mass $\mu$ of Section~\ref{sec:insertions}, so that $G_u=G_\mu$ in these units. We set $k=2$, $0<\alpha<1$ and $-\tfrac12<\gamma<0$. Lemma~\ref{q:neighboring-weights} gives the model inverse norms $C_S(d),C_K(d)$ in the prescribed balls. We write $C_S=C_S(\gamma)$ and $C_K=C_K(\gamma)$ for any chosen upper bounds. Choose a normal radius $r_0$ and a model radius $R_0$ with
\be\label{q:evaluated-joined-radii}
 2r_0<\rho_w,\qquad R_0\ge\max(4,\sup_{\operatorname{supp}\chi_0}|y|).
\ee
Write $P_u\coloneqq P_{G_u}$ and $\mathcal A_u(v,\lambda)\coloneqq P_uv-\lambda\psi_u$. Let $C_{\rm geo}>0$ bound the relative model metric differences, through the scaled order required by $X_u\to Y_u$, on the respective output supports:
\be\label{q:joined-metric-coefficients}
 \|G_u-G_K\|_{C^{4,\alpha}_{\rho},\mathrm{rel}} \le C_{\rm geo}(u/\rho+\rho\,\mathbf1_{\rho\le2u^{a_2}}),\qquad
 \|u^{-2}\Theta_u^*G_u-G_S\|_{C^{4,\alpha}_{s},\mathrm{rel}} \le C_{\rm geo}(\rho+(u/\rho)\mathbf1_{\rho\ge u^{a_2}}).
\ee
The pointwise comparisons used below are
$C(u/\rho+\rho\,\mathbf1_{\rho\le2u^{a_2}})$ for the exterior model
and $C(\rho+(u/\rho)\mathbf1_{\rho\ge u^{a_2}})$ for the rescaled
interior model. The exterior reference is exact beyond the matching
annulus, and the interior reference is exact below it. Scaled
differentiation of the cutoff preserves these bounds.
On the exterior output support their supremum is bounded by
$C(u^{1-\beta_{\rm out}}+u^{a_2})$, and on the interior output
support by $C(u^{\beta_{\rm in}}+u^{1-a_2})$. The separation of
the three radii bounds these by the two geometric powers used below.
Fix a positive relative metric margin $0<\rho_{\rm id}<1$.

Let $C_{\rm cov},C_{\rm wt},C_{\rm mult},C_{\rm cover}$ bound the coordinate, weight, multiplication and covering maps in their actual model and construction norms. On the inner output support, $\omega_u/u=w_S$ and $w_S\le r_S\le\sqrt2 w_S$ by~\eqref{join:eq:unit-scale}. On the outer output support $\omega_u\asymp\omega_\bullet$. Lemma~\ref{join:lem:regular-scale} gives the uniform scaled derivatives and H\"older bounds for these weight comparisons and their fixed powers. Thus the weight and covering map constants are finite for the smooth scale~\eqref{join:eq:explicit-weight}. More precisely, for each integer $m\ge0$ and $u\le\rho\le\rho_w/2$,
\be\label{q:outer-smooth-weight}
 \left(\frac{\omega_u}{\rho}\right)^2
 =1+\frac32\left(\frac u\rho\right)^2+
       \frac1{16} \left(\frac u\rho\right)^4,
 \qquad
 \left\|\frac{\omega_u}{\rho}-1\right\|_{C^{m,\alpha}_\rho}
 \le C_m\left(\frac u\rho\right)^2.
\ee
The norm in the last expression is taken on a ball of radius $c_J\rho$, contained in $u/2\le|z|\le\rho_w$ by~\eqref{q:evaluated-construction-ball}. Differentiating the first expression proves the estimate, with one additional derivative for its H\"older part. In particular the comparison tends to one whenever $u/\rho\to0$. The comparison on the remaining fixed annuli is uniformly smooth as well. All comparison and covering constants below refer to this scale and the ball size~\eqref{q:evaluated-construction-ball}. Choose
\be\label{q:evaluated-identification-constants}
 M_{\rm pow}\ge\max(1,C_{\rm cov}C_{\rm wt}C_{\rm mult}C_{\rm cover}).
\ee
All product constants include the finite component sums and both reflected collars. The norm balls meeting a support boundary are included in the fixed covering comparisons. In particular their radius enlargements are fixed independently of $u$; the condition $2r_0<\rho_w$ leaves room for those balls.

The four error powers identified in the introduction only require
the two cutoff scales to lie on opposite sides of the matching radius,
with both between the horizon and background scales. In this section
we write $\beta_{\rm out}=a_1$, $\beta_{\rm in}=a_3$ and
$\sigma_{\rm w}=\sigma$. Choose $0<\sigma_{\rm w}<\min(1+\gamma,-\gamma)$ to keep
the two shifted weights in the range of the model inverses.
Keep the separated output exponents free and define
\be\label{q:power-overlap-exponents}
 \beta_{\rm in}\coloneqq a_3,\qquad \beta_{\rm out}\coloneqq a_1,\qquad
 \delta_{\rm pow}\coloneqq\min(\beta_{\rm in},1-\beta_{\rm out},
 \sigma_{\rm w}(\beta_{\rm out}-a_2),
 \sigma_{\rm w}(a_2-\beta_{\rm in}))>0.
\ee
Let $\alpha_u$ be a radial cutoff equal to one for $\rho\le u^{a_2}$
and zero for $\rho\ge2u^{a_2}$.
Let $\eta_K$ vanish for $\rho\le u^{\beta_{\rm out}}/2$ and equal one for $\rho\ge u^{\beta_{\rm out}}$. Let $\eta_S$ equal one for $\rho\le u^{\beta_{\rm in}}$ and vanish for $\rho\ge2u^{\beta_{\rm in}}$. Use the fixed smooth profile~\eqref{q:joined-cutoff-profile}, with affine rescalings of the radial variable. Each positive-order derivative through order six is bounded by a fixed multiple of $\rho^{-j}$ on its transition annulus. The outer cutoff equals one away from the insertion balls, and the inner cutoffs are paired by reflection.

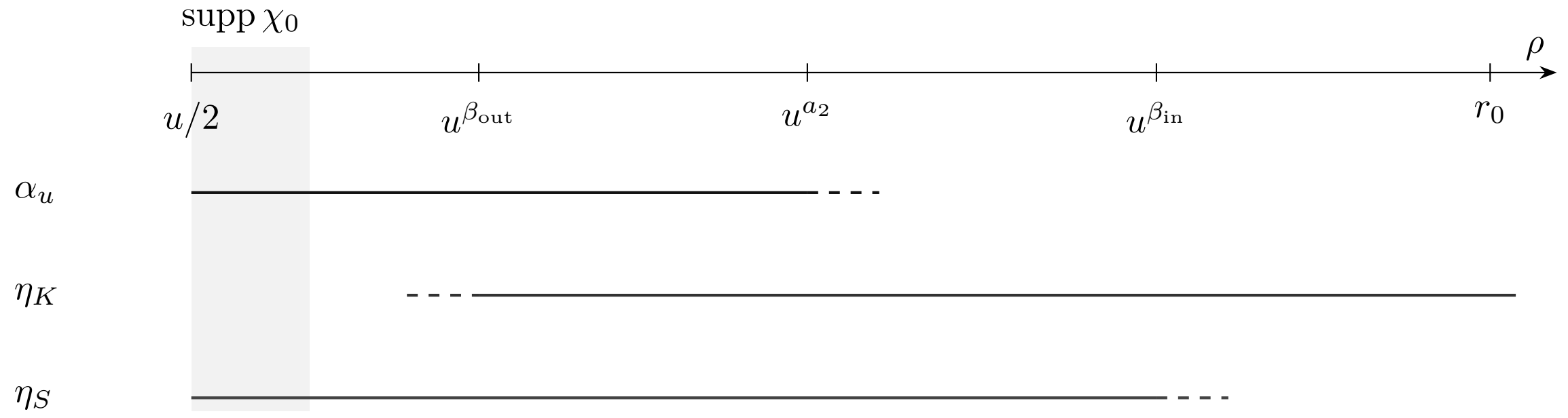
\begin{figure}[!htbp]\centering
\begin{tikzpicture}[x=1cm,y=1cm,>=Stealth,font=\normalsize]
\path[use as bounding box] (0,0) rectangle (15.6,4.45);
\fill[gray!10] (2,.35) rectangle (3.15,3.9);
\node[anchor=west] at (1.78,4.15) {$\operatorname{supp}\chi_0$};
\draw[->] (2,3.65)--(15.3,3.65) node[above left] {$\rho$};
\foreach \x/\lab in {2/{u/2},4.8/{u^{\beta_{\rm out}}},8/{u^{a_2}},11.4/{u^{\beta_{\rm in}}},14.65/{r_0}}{
 \draw (\x,3.56)--(\x,3.74);
 \node[below] at (\x,3.49) {$\lab$};
}
\node[anchor=west] at (.15,2.48) {$\alpha_u$};
\node[anchor=west] at (.15,1.48) {$\eta_K$};
\node[anchor=west] at (.15,.48) {$\eta_S$};
\draw[blue!65!black,thick] (2,2.48)--(8,2.48);
\draw[blue!65!black,thick,dashed] (8,2.48)--(8.7,2.48);
\draw[red!65!black,thick,dashed] (4.1,1.48)--(4.8,1.48);
\draw[red!65!black,thick] (4.8,1.48)--(14.9,1.48);
\draw[green!45!black,thick] (2,.48)--(11.4,.48);
\draw[green!45!black,thick,dashed] (11.4,.48)--(12.1,.48);
\end{tikzpicture}
\caption{Radial cutoffs for combining the inverse operators, shown on a schematic logarithmic scale. Solid segments mark where the cutoff equals one, dashed segments where it varies.}\label{fig:cutoff-supports}
\end{figure}

Choose $u_{\rm pow,id}>0$ so that, whenever $u\le u_{\rm pow,id}$,
\bse\label{q:power-support-conditions}\be
 u\le1/2,\qquad2u^{a_2} \le u^{\beta_{\rm in}},\qquad2u^{\beta_{\rm in}} \le r_0,\ee\be
R_0u\le\tfrac14u^{\beta_{\rm out}},\qquad 2C_{\rm geo}u^{\min(\beta_{\rm in},1-\beta_{\rm out})} \le\rho_{\rm id}.
\ee\ese
Every condition in~\eqref{q:power-support-conditions} is a power restriction with constants fixed before $u$. In particular one may take the minimum of
\be\label{q:identification-threshold}
 1/2,\quad2^{-1/(a_2-\beta_{\rm in})},\quad(r_0/2)^{1/\beta_{\rm in}},\quad(4R_0)^{-1/(1-\beta_{\rm out})},\quad(\rho_{\rm id}/(2C_{\rm geo}))^{1/\min(\beta_{\rm in},1-\beta_{\rm out})}.
\ee
The last condition bounds both geometric errors. These support conditions imply $u\le u^{\beta_{\rm in}} \le r_0/2<\rho_w$, so the restriction used in the definition of the smooth scale is already included. These choices give the exact product identities
\be\label{q:power-product-identities}
 \eta_S\alpha_u=\alpha_u,\qquad\eta_K(1-\alpha_u)=1-\alpha_u.
\ee
The outer output vanishes on the supports of the integral functional and the displacement sources, while the inner output equals one there.
The different supports in Figure~\ref{fig:cutoff-supports} make the error small: the source cutoffs vary on $u^{a_2} \le\rho\le2u^{a_2}$, and the two output cutoffs vary on disjoint annuli on either side. In particular the output cutoffs are constant on the supports that must be reproduced exactly.

\subsection{An error estimate for the combined inverse}

\begin{proposition}[A sufficient mass bound for the tensor inverse]\label{q:joined-power}\label{q:joined}
There are finite constants $B_{\rm pow},D_{\rm pow}>0$, determined by the stated fixed cutoffs, coefficient and comparison maps and the four fixed-model constants
\be\label{q:power-model-constants}
 C_S(\gamma),\quad C_K(\gamma),\quad C_S(\gamma-\sigma_{\rm w}),\quad C_K(\gamma+\sigma_{\rm w}),
\ee
such that the augmented inverse has norm at most
\be\label{q:joined-threshold}
 A_J\coloneqq\frac{B_{\rm pow}}{1-\theta_{J,\rm pow}},\qquad
 u_J\coloneqq\min(u_{\rm pow,id},(\theta_{J,\rm pow}/D_{\rm pow})^{1/\delta_{\rm pow}})
\ee
for $0<u\le u_J$, where $0<\theta_{J,\rm pow}<1$ is fixed. The sufficient condition uses positive powers of fixed ratios.
\end{proposition}
The constants $C_{\rm wt}$, $C_{\rm cover}$ and $M_{\rm pow}$ are chosen in the norms defined with the smooth scale used in the construction; their values need not equal those for another comparable weight function. The integral functional, compactly supported source and the normalization integral are unchanged, since their definitions use the unit-model integrals~\eqref{join:eq:centering}--\eqref{join:eq:column}. Consequently the bound for the inverse of the tensor operator remains uniform, and the bound for the error of the approximate inverse has the same power of $u$. The same weight comparisons apply to one-forms in the vector inverse below. In the nonlinear estimates, $\omega_u\ge u$ gives the factor $u^\gamma$, while $\omega_u\asymp u+\rho$ gives the stated source and force-integral powers. These conclusions require only the explicit reference geometry and the existing support restrictions.
\begin{proof}
\emph{1. Splitting the source and combining the two solutions.} Apply the inverse at weight $\gamma$ on the two models to the source pieces
$f_K=(1-\alpha_u)f$ and $F_S=u^{2-\gamma}\widehat{\alpha_uf}$. Write the outputs as $U_K$ and $(U_S,\lambda)$. The power in $F_S$ follows from the tensor rescaling: on an exact inner collar, $\widehat{P_uv}=u^{-2}P_S\widehat v$, and the normalized field is $U_S=u^{-\gamma}\widehat v$. Thus its source is $u^{2-\gamma}\widehat f$. In the same normalization the prescribed column is exactly $\chi_0T_S$, so no additional power of $u$ enters its scalar coefficient. Multiply the tensor outputs by $\eta_K$ and $\eta_S$, respectively, return the inner tensor with normalized pullback $u^\gamma\eta_SU_S$, and keep the scalar $\lambda$. This defines $\mathcal R_{u,\rm pow}$. The comparison and multiplication maps have uniformly bounded norms. Thus one may take $B_{\rm pow}\ge M_{\rm pow}^2(C_S(\gamma)+C_K(\gamma))$, where $M_{\rm pow}$ bounds their combined norms.

The output satisfies the integral condition, because the outer term vanishes on the support of the integral functional and the inner cutoff is one there. The scalar source term is reproduced exactly. In fact the inner equation is $P_SU_S-\lambda\chi_0T_S=F_S$, with the displacement tensor $T_S$ of Section~\ref{sec:Schwarzschild-conventions}. Multiplication by $\eta_S$ leaves its compactly supported source unchanged. The product identities~\eqref{q:power-product-identities} show that applying the augmented operator gives the original source plus the two output commutators and the two model-operator differences.

\par\addvspace{0.2cm}\noindent
\emph{2. Estimates with nearby weights.} The outer source vanishes below $u^{a_2}$. Its norm with a slightly different weight satisfies
\be\label{q:power-outer-source}
 \|f_K\|_{Y_{K,\bullet}(\gamma+\sigma_{\rm w})} \le C u^{-a_2\sigma_{\rm w}}\|f\|_{Y_u}.
\ee
To justify this for the full local H\"older norms, observe that a ball used to measure the source that meets $\rho\ge u^{a_2}$ has center at least a fixed multiple of $u^{a_2}$. The weights on the ball are comparable by the prescribed relation between ball radius and distance to the insertion point. All differentiated source-cutoff products therefore obey the same estimate. The solution at weight $\gamma+\sigma_{\rm w}$ lies in the base space $X_{K,\bullet}(\gamma)$; uniqueness identifies it with $U_K$. On the outer output transition $\rho\asymp u^{\beta_{\rm out}}$, conversion to the base source weight gives the factor $\rho^{\sigma_{\rm w}}$. On that annulus the neighboring-weight estimate gives $|D^jU_K|\le C u^{-a_2\sigma_{\rm w}}\rho^{\gamma+\sigma_{\rm w}-j}\|f\|_{Y_u}$ for $j\le2$. The product rule for the connection Laplacian gives
\be
 [P_u,\eta_K]U_K=-2\nabla_{\nabla\eta_K}U_K-(\Delta_{G_u}\eta_K)U_K,
\ee
where $\Delta_{G_u}=\operatorname{div}_{G_u}\nabla$; the curvature term commutes with the scalar cutoff. On a ball meeting the transition, its radius and $\rho$ are comparable by the prescribed relation between ball radius and distance to the insertion point. The cutoff derivatives through order four obey $|D^j\eta_K|\le C\rho^{-j}$, so the same product rule applied to the scaled $C^{2,\alpha}$ norm gives a bound by $C\rho^{-2}$ times the scaled $C^{4,\alpha}$ norm of $U_K$ on the corresponding model balls. The first- and second-derivative cutoff terms are therefore both bounded by $C u^{-a_2\sigma_{\rm w}}\rho^{\gamma+\sigma_{\rm w}-2}\|f\|_{Y_u}$. Multiplication by the source weight $\rho^{2-\gamma}$ leaves $u^{-a_2\sigma_{\rm w}}\rho^{\sigma_{\rm w}}$; the scaled H\"older seminorm is controlled by the same product estimate. Consequently
\be\label{q:power-outer-commutator}
 \|[P_u,\eta_K]U_K\|_{Y_u} \le C C_K(\gamma+\sigma_{\rm w})u^{\sigma_{\rm w}(\beta_{\rm out}-a_2)}\|f\|_{Y_u}.
\ee
This estimates the full punctured Schwarzschild--de Sitter solution, including its regular global correction.

The inner source is supported in $s\le2u^{a_2-1}$ and satisfies
\be\label{q:power-inner-source}
 \|F_S\|_{Y_S(\gamma-\sigma_{\rm w})} \le C u^{-(1-a_2)\sigma_{\rm w}}\|f\|_{Y_u}.
\ee
Here a ball used to measure the source that meets this support has center at most a fixed multiple of $u^{a_2-1}$; the same scaled product argument applies. The solution satisfying the integral condition at weight $\gamma-\sigma_{\rm w}$ lies in $X_S(\gamma)$, so uniqueness identifies the entire pair with $(U_S,\lambda)$. On the inner output transition $s\asymp u^{\beta_{\rm in}-1}$, conversion to the base weight gives $s^{-\sigma_{\rm w}}$. The two terms of the commutator obey the same scaled estimate as the second derivative of $U_S$: after multiplication by the base source weight, their common factor is $u^{-(1-a_2)\sigma_{\rm w}}s^{-\sigma_{\rm w}}$. The scaled H\"older product estimate gives the same factor. Since $s\asymp u^{\beta_{\rm in}-1}$, this yields, after the exact constant rescaling,
\be\label{q:power-inner-commutator}
 \|[P_u,\eta_S](u^\gamma U_S)\|_{Y_u} \le C C_S(\gamma-\sigma_{\rm w})u^{\sigma_{\rm w}(a_2-\beta_{\rm in})}\|f\|_{Y_u}.
\ee
The formula denotes the identified inner tensor and commutator. The global scalar estimate follows from the inverse at $\gamma$, namely $|\lambda|\le M_{\rm pow}C_S(\gamma)\|f\|_{Y_u}$; the larger neighboring-weight scalar estimate is not used.

\par\addvspace{0.2cm}\noindent
\emph{3. Bounding the error and obtaining the inverse.} On the outer output support $u/\rho\le2u^{1-\beta_{\rm out}}$; on the inner output support $\rho\le2u^{\beta_{\rm in}}$. The coefficient estimates~\eqref{q:joined-metric-coefficients} and the base-weight inverse bounds therefore estimate both model differences by
$C(C_S(\gamma)+C_K(\gamma))u^{\min(\beta_{\rm in},1-\beta_{\rm out})}\|f\|_{Y_u}$. The four error powers are bounded below by
$\delta_{\rm pow}$ in~\eqref{q:power-overlap-exponents}. Hence
\be\label{q:parametrix-error}
 \|\mathcal A_u\mathcal R_{u,\rm pow}-I\|\le D_{\rm pow}u^{\delta_{\rm pow}}.
\ee
For example, sufficient constants satisfy
\be\label{q:joined-constants}
\begin{split}
 B_{\rm pow}&\ge M_{\rm pow}^2(C_S(\gamma)+C_K(\gamma)),\\
 D_{\rm pow}&\ge C_{K,\rm cut}C_K(\gamma+\sigma_{\rm w})+C_{S,\rm cut}C_S(\gamma-\sigma_{\rm w})
 +C_{\rm diff,pow}(C_S(\gamma)+C_K(\gamma)).
\end{split}
\ee
Here $C_{K,\rm cut},C_{S,\rm cut}$ are the fixed product bounds in the two displayed commutator estimates, including the source and weight comparisons, and $C_{\rm diff,pow}$ bounds the two relative operator differences. The explicit references, fixed profile and finite chart maps determine them before $u$. For $u\le u_J$, \eqref{q:parametrix-error} bounds the error by $\theta_{J,\rm pow}$. The Neumann series gives $\|(\mathcal A_u\mathcal R_{u,\rm pow})^{-1}\|\le (1-\theta_{J,\rm pow})^{-1}$, and composition with $\mathcal R_{u,\rm pow}$ gives the right inverse in~\eqref{q:joined-threshold}. The fixed-scale index is zero by Lemma~\ref{join:lem:Fredholm}. Restricting to the codimension-one domain defined by the integral condition lowers the index by one, and adjoining the scalar source term raises it by one. The augmented index is therefore zero, so this right inverse is the unique inverse.

The exponents remain free: their strict separation makes every error power positive.\qedhere
\end{proof}

\section{The approximate metric and the equation for the insertion radius}\label{sec:nonlinear}

\subsection{The Schwarzschild correction determined by the first derivative of the background lapse}

The metric of Section~\ref{sec:insertions} is accurate enough for inversion. To compute the leading coefficient of the auxiliary source for the corrected metric, we match the first nonconstant terms of the two model expansions. Fix a compact interval $I$ about $r_*=(3M/\Lambda)^{1/3}$ strictly inside the Schwarzschild--de Sitter static region. For $r\in I$, choose the reflected points $p_\pm(r)$ on the axis and smooth equivariant spatial normal coordinates, with $z_3$ directed along the outward unit Schwarzschild--de Sitter radial normal $\nu_K$. Define
\be\label{f:b}
b(r)\coloneqq N_K(p_+(r)),\qquad a(r)\coloneqq\frac{\partial_{\nu_K}N_K(p_+(r))}{b(r)}.
\ee
All constants in this section are uniform for $r\in I$. We now choose the weight in the smaller range
\be\label{f:gamma}
-\tfrac12<\gamma<0.
\ee
Use the operators $P,\beta,\mathcal B$ defined in Section~\ref{sec:linear-Kottler}. For positive block metrics $g,G$ with the same horizon surface gravities, define the DeTurck vector field by $W(g,G)^k\coloneqq g^{ij}(\Gamma(g)^k_{ij}-\Gamma(G)^k_{ij})$. On a Ricci-flat reference, $2D\Ric[v]=Pv-2\mathcal B\beta v$. The true differential of $2(\Ric-\Lambda g)-\mathcal L_{W(g,G)}g$ at its own reference is
\be\label{f:L}
L_Gv=P_Gv+(\Ric_G-\Lambda G)\circ v+v\circ(\Ric_G-\Lambda G),
\ee
Here one index is raised in each of the last two products. To verify the formula, the connection variation is
\be
 2\dot\Gamma^k_{ij}=\nabla_i v_j{}^k+\nabla_j v_i{}^k-\nabla^k v_{ij}.
\ee
Substitute it into $\dot\Ric_{ij}=\nabla_k\dot\Gamma^k_{ij}-\nabla_j\dot\Gamma^k_{ik}$ and commute the covariant derivatives. The curvature terms give $-2\mathcal R_Gv+\Ric_G\circ v+v\circ\Ric_G$, while the divergence and trace terms give $-2\mathcal B_G\beta_Gv$. Hence
\be
 2D\Ric_G[v]=P_Gv+\Ric_G\circ v+v\circ\Ric_G-2\mathcal B_G\beta_Gv.
\ee
At $g=G$, the derivative of the DeTurck vector field is $DW_G[v]=-(\beta_Gv)^\sharp$. The differentiated Lie derivative therefore cancels the last term, and differentiating $-2\Lambda g$ gives~\eqref{f:L}.

For the moving insertion point, the inner map is $\Theta_{\mu,r}(y,\sigma)\coloneqq(z=\mu y,\tau=\mu\sigma/b(r))$; its normalized pullback has the same meaning as in~\eqref{join:eq:rescaling}. Proposition~\ref{thm:outer-tensor-inverse} gives the inverse of the tensor operator on the Schwarzschild--de Sitter background. The inner displacement tensor $T_S$, cutoff $\chi_0$, positive number $m_0$ and compact paired column $\psi_\mu$ are those of Section~\ref{sec:insertions}. In particular $\delta_ST_S=0$, $\operatorname{tr}_ST_S=0$ and $|D^jT_S|\le C_j(1+|y|)^{-2-j}$. The hat convention, integral condition and weighted spaces also remain those of Section~\ref{sec:insertions}. The two individual profiles are paired by reflection; each profile is allowed its own degree-one angular component. We take $k\ge2$ in the scaled H\"older norms. The choice $k=2$, $0<\alpha<1$ agrees with the quantitative model estimates: the differential maps $C^{4,\alpha}$ to $C^{2,\alpha}$, and the quadratic estimate below uses the same two derivatives.

The global inverse is uniform for $r\in I$. If this failed, a sequence violating the bound would have $r_i\to r_0\in I$. Away from the insertion points the limit is the operator on the unpunctured Schwarzschild--de Sitter region. After rescaling near an added horizon it is the operator on Schwarzschild of mass one, and at intermediate scales it is the flat operator: the coefficient error is now $O(\mu/\rho+\rho)$, which still vanishes in that limit. The fixed-scale index argument is unchanged. For every fixed large test radius, the coefficient test remains supported in the exact Schwarzschild part of the simpler reference when $\mu$ is sufficiently small, uniformly in $r$. The integral condition, column and index proof of Proposition~\ref{join:thm:joined} therefore applies on this interval. In moving normalized charts its differential and first parameter derivative have uniformly bounded scaled coefficients. Differentiating the inverse identity gives a uniform $C^1$ parameter bound.

Write $s\coloneqq|y|$, $n\coloneqq y/s$, $n_3\coloneqq n\cdot e_3$. In the asymptotic Cartesian coordinates, the Schwarzschild metric and displacement have
\be\label{f:tails}
 \begin{split}
 G_S&=\dd\sigma^2+\delta+2s^{-1}(\delta-\dd\sigma^2)+\Ocal(s^{-2}),\\
 (T_S)_{ij}&=2s^{-2}(\delta_{i3}n_j+\delta_{j3}n_i-\delta_{ij}n_3)+\Ocal(s^{-3}),\\
 (T_S)_{\sigma\sigma}&=2s^{-2}n_3+\Ocal(s^{-3}).
 \end{split}
\ee
These expansions hold with all differentiated remainder estimates. The leading affine exterior datum is $T\coloneqq2y_3\dd\sigma^2$. Define the bounded angular tensor $B$ by
\be\label{f:B}
 B_{\sigma\sigma}\coloneqq-4n_3,\qquad
 B_{ij}\coloneqq n_3n_in_j-n_3\delta_{ij}+\delta_{i3}n_j+\delta_{j3}n_i,
 \qquad B_{i\sigma}\coloneqq0.
\ee

\begin{lemma}[A regular Schwarzschild correction and its source]\label{f:profile}
There is a regular static axisymmetric tensor $A_S$, satisfying the same integral condition as the global inverse, such that
\be\label{f:profile-eq}
 P_S A_S=c_P\chi_0T_S,\qquad
 c_P\coloneqq-16\pi/m_0,\qquad
 A_S=T+B+\Ocal(s^\gamma).
\ee
All differentiated remainders of order $j$ are $\Ocal(s^{\gamma-j})$. The tensor is invariant under the circular reversals and has only the two circular diagonal entries and the meridional block.
\end{lemma}
\begin{proof}
Let $u\coloneqq s^{-1}$ and let the auxiliary index be $0$. The first-order Schwarzschild Christoffel symbols are
$\Gamma^i_{jk}=\delta_{ik}u_j+\delta_{ij}u_k-\delta_{jk}u_i$,
$\Gamma^0_{i0}=-u_i$ and $\Gamma^i_{00}=u_i$.
Their full contracted sum $\sum_\alpha\Gamma^i_{\alpha\alpha}$ vanishes at this order away from zero. For spatial indices, the divergence is $\partial_\alpha\Gamma^k_{\alpha j}=2u_{kj}$; it does not contribute to the spatial components of the rough Laplacian on $T$, since $T$ has no spatial component. Substitution in the rough Laplacian and curvature contraction gives
\be\label{f:PT}
 (P_ST)_{00}=8s^{-2}n_3+\Ocal(s^{-3}),\qquad
 (P_ST)_{ij}=s^{-2}(4n_3\delta_{ij}-12n_3n_in_j)+\Ocal(s^{-3}).
\ee
For the $00$ component, the spatial part of the covariant Laplacian contributes $(4+2+4)u_3=10u_3$: the three terms come from differentiating the covariant derivative, contracting the spatial connection, and the two tensor slots. Its auxiliary-direction part contributes $-2u_3$ at this order. Thus the rough Laplacian contributes $-8u_3$; the curvature contraction has no term of order $s^{-2}$ in that component. For the spatial components, the rough Laplacian has no term at that order, while the curvature contraction contributes $-4y_3u_{ij}$. The formulas
\be
 u_i=-s^{-2}n_i,\qquad u_{ij}=s^{-3}(3n_in_j-\delta_{ij})
\ee
then give~\eqref{f:PT}. All omitted terms are $O(s^{-3})$ with the stated differentiated bounds. The sphere identities
\be
 \Delta_{\mathbb S^2}n_3=-2n_3,\qquad
 \Delta_{\mathbb S^2}(n_3n_in_j)
 =2(\delta_{i3}n_j+\delta_{j3}n_i+\delta_{ij}n_3)-12n_3n_in_j
\ee
show that $\Delta_{\mathbb S^2}B$ is precisely the coefficient in~\eqref{f:PT}. Hence $P_S(T+B)=\Ocal(s^{-3})$, with differentiated bounds.

Let $r_S=s+1+(4s)^{-1}$ be the unit Schwarzschild areal radius. Extend $\chi_0$ constantly to the left of its defining interval, and set $\zeta_S\coloneqq1-\chi_0(r_S-1)$. Then $A_{S,0}\coloneqq\zeta_S(T+B)$ is zero for $r_S\le4$, so it is regular at the Schwarzschild horizon and satisfies the integral condition. It equals $T+B$ for $r_S\ge5$, in particular for $s\ge4$. Its source $f_0\coloneqq P_S A_{S,0}$ belongs to the weighted space $Y_\gamma\coloneqq C^{k,\alpha}_{\gamma-2}(G_S)$, defined by the scaled norms of~\eqref{join:eq:norms} with weight $1+s$ for every $\gamma>-1$. Indeed $s^{2-\gamma}|D^jf_0|s^j=O(s^{-1-\gamma})$ on the end, and the differentiated bounds give the same estimate for the scaled H\"older seminorm; the cutoff contribution is smooth and compactly supported. The weighted Schwarzschild range is exactly $\{f:\int\langle f,T_S\rangle\dd\nu_S=0\}$, with a bounded inverse under the integral condition; the proof of this range assertion is given in Section~\ref{f:range}.
Let $\nu$ be the outward Schwarzschild unit normal to $|y|=s$. The Green identity and~\eqref{f:tails} give
\be\label{f:P-flux}
 \int\langle f_0,T_S\rangle\dd\nu_S
 =\lim_{s\to\infty}\int_{|y|=s}
 (\langle A_{S,0},\nabla_\nu T_S\rangle-\langle\nabla_\nu A_{S,0},T_S\rangle)
 =-16\pi.
\ee
Indeed the only leading component is $T_{00}=2sn_3$. Since $(T_S)_{00}=2s^{-2}n_3+O(s^{-3})$, the leading boundary integrand is $-12s^{-2}n_3^2$ and its integral is $-12\int_{\mathbb S^2}n_3^2\dd\omega=-16\pi$.
The bounded $B$ and all decaying terms contribute zero. The asymptotic lapse and spatial area factors tend to one, and their errors also vanish in this limit. Thus $-f_0+c_P\chi_0T_S$ is orthogonal to $T_S$. Its inverse under the integral condition added to $A_{S,0}$ proves~\eqref{f:profile-eq}. Averaging preserves the stated symmetry class.
\end{proof}

\begin{lemma}[Bounds for the Schwarzschild correction from its explicit leading term]\label{q:affine-profile-bound}
Use the areal-radius norms $X_S,Y_S$ in~\eqref{q:model-Holder-norms}. Let $Y_S^{(4)}$ be the source norm obtained from $Y_S$ by taking four derivatives, with the same areal weights and balls used to measure the source. Define
\be\label{q:affine-seed-model-norms}
 F_{S,2}\coloneqq\|P_SA_{S,0}\|_{Y_S},\quad F_{S,4}\coloneqq\|P_SA_{S,0}\|_{Y_S^{(4)}},\quad
 C_{\rm col,4}\coloneqq\|\chi_0T_S\|_{Y_S^{(4)}}.
\ee
For the output, use the weight $1+s$ and balls of radius $c_S(1+s)/128$. Let $S_{\rm seed}$ be the corresponding $C^{6,1/2}$ growth norm of $A_{S,0}$ with weight exponent one, and let $S_{S,6}$ satisfy
\be\label{q:affine-Schauder-norm-conversion}
 \|w\|_{C^{6,1/2}_{\gamma}(1+s)} \le S_{S,6}\bigl(\|w\|_{X_S}+\|P_Sw\|_{Y_S^{(4)}}\bigr).
\ee
This constant includes the areal-to-isotropic weight and ball comparisons. Then a sufficient pointwise profile constant through order six is
\be\label{q:affine-profile-constant}
 C_A\coloneqq S_{\rm seed}+S_{S,6}(C_SF_{S,2}+F_{S,4}+|c_P|C_{\rm col,4}).
\ee
All constants are finite and involve the unit Schwarzschild model and the explicit seed.
\end{lemma}
\begin{proof}
The initial approximation is smooth and vanishes near the Schwarzschild horizon. On the end it equals $T+B$; the differentiated cancellation~\eqref{f:PT} gives $P_SA_{S,0}=\Ocal(s^{-3})$. This proves finiteness of its source norms at $-\tfrac12<\gamma<0$. The tensor multiplying the auxiliary coefficient is smooth and compactly supported. The scaled chart bounds and differentiated local estimates give~\eqref{q:affine-Schauder-norm-conversion}: the source balls contain the required local input balls, the smaller output balls are as specified, and $1+s\le r\le4(1+s)/3$ bounds all weight conversions. Thus $S_{S,6}$ is finite in these precise source and target norms.

The centered augmented inverse applied to $-P_SA_{S,0}$ gives the decaying correction $w=A_S-A_{S,0}$, with $\|w\|_{X_S} \le C_SF_{S,2}$. Its source equation is $P_Sw=-P_SA_{S,0}+c_P\chi_0T_S$. The differentiated estimate therefore gives the last term of~\eqref{q:affine-profile-constant}. Adding the seed proves the growth bound for $A_S$; on $s\ge4$ the seed equals $T+B$, so the same estimate gives the decaying remainder bound.\qedhere
\end{proof}

\begin{proposition}[The linearized Ricci tensor and its integral against $T_S$]\label{f:physical-profile}
Define
\be\label{f:J}
 J\coloneqq P_S A_S-2\mathcal B_S\beta_S A_S,\qquad
 c_E\coloneqq-32\pi/m_0,\qquad J_0\coloneqq J-c_E\chi_0T_S.
\ee
The following pointwise and integral statements hold:
\be\label{f:J-bounds}
 \begin{gathered}
 \beta_S A_S=\dd y_3+\Ocal(s^{-1}),\qquad
 |D^jJ_0|\le C_j(1+s)^{-2-j},\\
 \int\langle J_0,T_S\rangle\dd\nu_S=0,\qquad
 \left|\int\chi_R\langle J_0,T_S\rangle\dd\nu_S\right|\le C/R.
 \end{gathered}
\ee
For the first remainder its derivatives of order $j$ are $\Ocal(s^{-1-j})$; derivatives of $\dd y_3$ are understood covariantly. Moreover $\beta_SJ=0$ and $\beta_SJ_0=-c_E\beta_S(\chi_0T_S)$.
\end{proposition}
The coefficient $c_P$ comes from $P_S A_S$. At the reference metric itself the term involving the DeTurck vector field vanishes, so the leading source near a small horizon is determined by $J=2D\Ric_{G_S}[A_S]$. The term $-2\mathcal B_S\beta_S A_S$ contributes to the pairing of $J$ with $T_S$ because $\beta_S A_S$ has a nonzero limit at infinity. Equations~\eqref{f:P-flux} and~\eqref{f:gauge-flux} account for both contributions to $c_E$.
\begin{proof}
The flat Bianchi operator sends $T$ to $\dd y_3$. The bounded angular correction, the connection correction applied to the affine term, and the decaying remainder give the displayed asymptote and its differentiated bounds. In particular $\mathcal B_S\beta_S A_S=\Ocal(s^{-2})$, with all differentiated bounds. The horizon contributes no integration boundary in its regular lift. Since $T_S$ is divergence free,
\be\label{f:gauge-flux}
 \int\langle\mathcal B_S\beta_S A_S,T_S\rangle\dd\nu_S
 =\lim_{s\to\infty}\int_{|y|=s}T_S(\nu,(\beta_S A_S)^\sharp)=8\pi.
\ee
The leading spatial expression in~\eqref{f:tails} gives $T_S(\nu,e_3)=2s^{-2}+\Ocal(s^{-3})$: its three angular terms contract to $n_3^2+1-n_3^2=1$. Equation~\eqref{f:P-flux} and the factor two in $J$ therefore give $-16\pi-16\pi=-32\pi$. This proves the exact orthogonality of $J_0$. Its tail is bounded by $C\int_R^\infty s^{-2}\dd s=C/R$, proving the cutoff estimate. Finally the linearized contracted Bianchi identity is $\beta_S(P_S A_S-2\mathcal B_S\beta_S A_S)=0$ on this Ricci-flat reference. This proves the last two assertions.
\end{proof}

\subsection{Matching the point-source solution to the Schwarzschild correction}

Use the normalized auxiliary coordinate $t=b(r)\tau$ in each normal ball, and put $D_p\coloneqq\delta-\dd t^2$ at its center.

\begin{lemma}[The Green tensor with prescribed horizon surface gravities]\label{f:Green}
There is a paired invariant tensor $H_r$, regular in the fixed normalized lifts of the two original horizons, such that
\be\label{f:Green-eq}
 P_KH_r=8\pi\sum_{\pm}D_{p_\pm}\delta_{h,p_\pm}.
\ee
It depends $C^1$ on $r$ in moving normal charts. At either point,
\be\label{f:Green-tail}
 H_r=2\rho^{-1}(\delta-\dd t^2)+\Ocal(1),\qquad
 |D^j\Ocal(1)|\le C_j\rho^{-j},\qquad
 |D^j\beta_KH_r|\le C_j\rho^{-1-j}.
\ee
The same bounds hold after one derivative with respect to the parameter.
\end{lemma}
\begin{proof}
Proposition~\ref{green:global} gives this tensor and its $C^1$ parameter dependence, with constants specified by~\eqref{green:CH-conditions} and~\eqref{green:outside-constant}. In the flat metric the Bianchi operator annihilates $2D_p/\rho$. The connection correction on this term and the derivative of the bounded remainder are $\Ocal(\rho^{-1})$; scaled differentiation proves the Bianchi bounds, also after a placement derivative.
The delta normalization in~\eqref{f:Green-eq} is literal. With invariant integration per unit $\tau$, its pairing against a test tensor is $8\pi b(r)\langle D_p,\varphi(p)\rangle$ at either point, since $\dd V_G/\dd\tau=N_K\dd V_{h_K}$.  
\end{proof}

The coefficient bounds, local radius and regular correction in Proposition~\ref{green:global} give finite constants $C_{H,0},C_{H,1},C_{H,\rm out}$, uniformly on the fixed parameter set. We use the actual coefficient and map norms in~\eqref{green:coefficient-constant}, \eqref{green:regular-source-bounds} and~\eqref{green:CH-conditions}, and define
\be\label{q:profile-and-differential-constants}
 C_H\coloneqq\max(C_{H,0},C_{H,\rm out}),\qquad
 C_{H,1}^{\rm all}\coloneqq\max(C_{H,0},C_{H,1},C_{H,\rm out}).
\ee
These constants include the derivative with respect to the insertion radius and the seventh spatial derivative of the regular global correction required by its moving-chart pullback.

Set $\rho_\mu\coloneqq\mu^{a_2}$. Define an inner metric in normalized coordinates by
\be\label{f:inner}
 \mu^{-2}\Theta_{\mu,r}^*G_{\mathrm{in}}=G_S+\mu a(r)A_S,
\ee
and an outer metric by
\be\label{f:outer}
 G_{\mathrm{out}}\coloneqq G_K+\mu H_r.
\ee
Use the same cutoff $\chi_\mu=\chi(\rho/\rho_\mu)$ as in~\eqref{join:eq:reference} to interpolate these covariant metrics. Denote the resulting global reference by $\overline G_{\mu,r}$. The metric is positive for small $\mu$: the relative inner change is at most $C\mu(1+s)\le C\mu^{a_2}$ and the relative outer change is at most $C\mu/\rho\le C\mu^{1-a_2}$. These statements, including joint horizon--axis regularity, follow in the normalized horizon charts. All metrics and interpolations have the same static block and symmetry class.

\begin{lemma}[All matching derivatives and the four Cauchy data]\label{f:jets}
On $\rho\asymp\rho_\mu$, for every fixed $j\ge0$,
\be\label{f:jets-all}
 |D^j(G_{\mathrm{out}}-G_{\mathrm{in}})|
 \le C_j\rho^{-j}
 (\rho^2+\mu+\mu^2/\rho^2+\mu^{1-\gamma}\rho^\gamma)
 \le C_j\mu^{m_{\rm match}}\rho_\mu^{-j}.
\ee
The same estimates hold for one derivative with respect to the parameter. In particular the differences of the induced spatial metric and lapse are $\Ocal(\mu^{m_{\rm match}})$; the corresponding normal derivatives are $\Ocal(\mu^{m_{\rm match}-a_2})$, where $m_{\rm match}\coloneqq\min(2a_2,1,2-2a_2,1-\gamma+a_2\gamma)$.
\end{lemma}
The original two surface gravities remain exactly those of $G_K$, since $H_r$ is a regular tensor preserving the horizon surface gravities there. Each inserted opening is exactly $b(r)/(4\mu)$: the inner perturbation is a regular tensor in that Schwarzschild angle. Thus the two new surface gravities vary with $r$, and are fixed during the correction at that $r$. The tensor inverse uses the prescribed opening in the local angle at each horizon.
\begin{proof}
The flat metric, the mass term $2\mu\rho^{-1}D_p$, and the affine lapse term $2az_3\dd t^2$ agree on both sides. The remaining background Taylor term is $\Ocal(\rho^2)$; the bounded Green and affine angular terms have size $\Ocal(\mu)$; the next Schwarzschild term is $\Ocal(\mu^2/\rho^2)$; and the decaying affine-profile term has size $\Ocal(\mu^{1-\gamma}\rho^\gamma)$. Their spatially differentiated estimates were established above. For the derivative with respect to the insertion radius, use the moving normal coordinates and the normalized auxiliary coordinate $t=b(r)\tau$ on both sides. The flat term and the mass term have coefficients independent of $r$ in this frame, while the affine terms have the same coefficient $a(r)$ on both sides. Their derivatives therefore cancel as well. The background Taylor remainder and its placement derivative are $O(\rho^2)$ with scaled derivatives, uniformly on the compact placement interval; the corresponding bounds for the Green remainder and the affine profile follow from their constructions above. Thus the same four-term bound holds after one placement derivative. Substituting $\rho=\mu^{a_2}$ gives the four powers defining $m_{\rm match}$. Taking the square root of the positive auxiliary diagonal coefficient gives the corresponding lapse estimates, since this annulus has lapse bounded away from zero. The changes of unit normal and second fundamental form obey the same first-derivative bound. These four data are approximately matched by the stated estimates; the subsequent global elliptic correction, rather than a separate exact Cauchy matching assertion, produces the single exact metric.
\end{proof}

\subsection{Solving the reduced nonlinear equation}\label{f:nonlinear}

For each $(\mu,r)$ use the improved metric as the DeTurck reference and define
\be\label{f:DT}
 \mathcal E_{\mu,r}(g)\coloneqq2(\Ric_g-\Lambda g)-\mathcal L_{W(g,\overline G_{\mu,r})}g.
\ee
In particular $\mathcal E_{\mu,r}(\overline G)=2(\Ric_{\overline G}-\Lambda \overline G)$, so the leading term near a small horizon is determined by the tensor $J$ in~\eqref{f:J}. Define
\be\label{f:lambda0}
 \lambda^0_\mu(r)\coloneqq c_E a(r)\mu^{1-\gamma},\qquad
 f_{\mu,r}\coloneqq\mathcal E_{\mu,r}(\overline G_{\mu,r})-\lambda^0_\mu(r)\psi_\mu.
\ee

All placement derivatives below are taken after transport to fixed normalized charts: the spatial coordinate $y=z/\mu$ and the normalized auxiliary coordinate $\sigma=b(r)\tau/\mu$ are held fixed on each inner collar. The domain and source tensors are transported by the same bundle identification, with the scalar coefficient unchanged. The transition to fixed exterior coordinates takes place on annuli of fixed radius. The construction of these identifications and the operator-norm derivative are detailed in the proof of~\eqref{q:evaluated-parameter-derivative-constants} below. In particular, $\partial_r v$ in the estimates below denotes this transported derivative.

\begin{proposition}[The source remaining after subtraction of the displacement term]\label{f:source}
The source is global and satisfies $\|f_{\mu,r}\|_{Y_\mu}+\|\partial_rf_{\mu,r}\|_{Y_\mu} \le C\mu$. Its inner pointwise form is
\be\label{f:source-inner}
 \widehat f_{\mu,r}=a(r)\mu^{-1}J_0+\mathcal R_{\mu,r},\qquad
 |D_y^j\mathcal R_{\mu,r}|\le C_j(1+s)^{-j}
 \quad(s\le\rho_\mu/\mu).
\ee
The pointwise bounds on $\mathcal R_{\mu,r}$ also hold after one derivative with respect to the insertion radius $r$.
\end{proposition}
\begin{proof}
On the inner region, the normalization of the auxiliary source gives
\be
 \widehat{\lambda^0_\mu(r)\psi_\mu}
 =c_Ea(r)\mu^{1-\gamma}\mu^{\gamma-2}\chi_0T_S
 =a(r)\mu^{-1}c_E\chi_0T_S.
\ee
The linearized Ricci contribution in the rescaled coordinates is $a(r)\mu^{-1}J$. Their difference is therefore $a(r)\mu^{-1}J_0$, with the sign in~\eqref{f:source-inner}. Taylor expansion of the Ricci operator at $G_S$ gives the remaining terms. The quadratic terms of $\mu aA_S$ are bounded by $C\mu^2$ in normalized Ricci components: $A_S=\Ocal(1+s)$, $D A_S=\Ocal(1)$, $D^2A_S=\Ocal((1+s)^{-2})$, and $\operatorname{Rm}_S A_S^2=\Ocal((1+s)^{-1})$. Multiplication by $\mu^{-2}$ in the hatted source gives the asserted $\Ocal(1)$ remainder. The cosmological term is also $\Ocal(1)$. Higher scaled derivatives satisfy the same bounds because $\mu(1+s)\le C\mu^{a_2}$.
The $J_0$ term in $Y_\mu$ is bounded by
$C\mu^{1-\gamma}(1+s)^{-\gamma} \le C\mu^{1-a_2\gamma}$, and the remainder by
$C\rho_\mu^{2-\gamma}=C\mu^{a_2(2-\gamma)}$.

On the outer region, linearization at the Einstein metric gives
\be
 2(\Ric_{G_K+\mu H_r}-\Lambda(G_K+\mu H_r))
 =-2\mu\mathcal B_K\beta_KH_r+\Ocal(\mu^2\rho^{-4}).
\ee
The Bianchi contribution is bounded by $C\mu\rho^{-2}$, since~\eqref{f:Green-tail} is differentiated once in $\mathcal B_K\beta_KH_r$. Multiplication by the source weight gives $C\mu\rho^{-\gamma}$, which is bounded by $C\mu$ on the fixed outer region. Thus the full source norm is at most $C\mu+C\mu^2\rho_\mu^{-2-\gamma} \le C\mu$. On the matching annulus, Lemma~\ref{f:jets} bounds the weighted cutoff errors by
$C(\mu^{a_2(2-\gamma)}+\mu^{1-a_2\gamma}
+\mu^{2-a_2(2+\gamma)}+\mu^{1-\gamma})$.
Each exponent is at least $1+\eta_{\rm src}$, by~\eqref{eq:error-gains}. The argument uses all required differentiated estimates, and parameter differentiation preserves every bound.
\end{proof}

\begin{proposition}[Solutions of the elliptic equation with one remaining coefficient]\label{f:family}
There is a fixed constant $K>0$ such that, for all sufficiently small $\mu>0$ and every $r\in I$, the reduced equation has a unique solution pair $(v_{\mu,r},\lambda_\mu(r))$ with $\ell_\mu(v_{\mu,r})=0$ in the ball $\|v_{\mu,r}\|_{X_\mu}+|\lambda_\mu(r)-\lambda^0_\mu(r)|\le K\mu$. Here $\lambda_\mu^0$ is given by~\eqref{f:lambda0}. The pair satisfies
\be\label{f:family-eq}
 \mathcal E_{\mu,r}(\overline G_{\mu,r}+v_{\mu,r})
 =\lambda_\mu(r)\psi_\mu,\qquad
 \|v_{\mu,r}\|_{X_\mu}+\|\partial_rv_{\mu,r}\|_{X_\mu} \le C\mu.
\ee
The family and its coefficient are $C^1$ on $I$. For positive $\mu$ they are jointly continuous in $(M,\mu,r)$ after the fixed-domain identifications described below, in a H\"older topology with any exponent smaller than $\alpha$.
\end{proposition}
\begin{proof}
Compare $\overline G$ with the joined reference used for the differential inverse, using the same cutoff at $\rho_\mu$. The affine profile contributes $\mu aA_S=\Ocal(\rho)$ and the Green tensor contributes $\mu H_r=\Ocal(\mu/\rho)$ on their respective supports. Proposition~\ref{q:evaluated-actual-differential}~below also bounds the term from interpolating the squares of the lapses on the matching annulus by $\Ocal(\mu^{2\eta_{\rm geom}})$. The same cutoff calculation applies through any fixed order of scaled derivatives. The full difference is therefore $\Ocal(\mu^{\eta_{\rm geom}})$ in the relative scaled $C^{k+2,\alpha}$ norm, and $\|P_{\overline G}-P_{G_\mu}\|_{X_\mu\to Y_\mu}=\Ocal(\mu^{\eta_{\rm geom}})$. This bound also holds for the difference of the augmented operators, which have the same integral condition and scalar source. In addition,
\be
 \sup\omega_\mu^2|\Ric_{\overline G}-\Lambda \overline G|\le C\mu^{\eta_{\rm geom}},
\ee
with its higher scaled and parameter derivatives. Indeed, on the inner region the normalized defect has size at most $C(\mu^{-1}(1+s)^{-2}+1)$, after restoring the compactly supported source to~\eqref{f:source-inner}. Since $\omega_\mu\asymp\mu(1+s)$ there and $s\le\mu^{a_2-1}$, its weighted pointwise norm before rescaling is bounded by
\be
 C\mu^2(1+s)^2(\mu^{-1}(1+s)^{-2}+1)\le C(\mu+\mu^{2a_2}).
\ee
On the outer region the two terms are $O(\mu\rho^{-2})$ and $O(\mu^2\rho^{-4})$; multiplication by $\omega_\mu^2\asymp\rho^2$ gives $O(\mu+\mu^{2-2a_2})$ for $\mu^{a_2} \le\rho\le r_0$. Outside the insertion charts the defect is $O(\mu)$ in fixed regular charts. On the matching annulus, the scaled defect is bounded by the four terms in~\eqref{f:jets-all}, hence by $C\mu^{\eta_{\rm geom}}$. The scaled derivative and placement bounds follow from the corresponding bounds on these three regions. Formula~\eqref{f:L} therefore proves that the augmented actual differential $L_{\overline G}$ is a uniform isomorphism by its Neumann series. The compactly supported source and integral condition are unchanged. For the parameter estimate we transport the augmented equation to fixed normalized charts, with the unit-model integral condition fixed. We differentiate the transported coefficients and compactly supported source there. Their bounded scaled jets give the operator derivative in $X_u^0\times\mathbb R\to Y_u$ used in~\eqref{q:full-augmented-parameter-map} below.

Write the unknown coefficient as $\lambda^0_\mu+\delta\lambda_\mu$. The equation is
$L_{\overline G}v-\delta\lambda_\mu\psi_\mu=-f_{\mu,r}-Q_{\mu,r}(v)$.
The local Ricci--DeTurck formula gives
\be
 \|Q(v)-Q(w)\|_{Y_\mu}
 \le C\mu^\gamma(\|v\|_{X_\mu}+\|w\|_{X_\mu})\|v-w\|_{X_\mu}.
\ee
For example, the definition of $X_\mu$ gives $|D^jv|\le C\omega_\mu^{\gamma-j}\|v\|_{X_\mu}$ for $j\le2$. Therefore, after multiplying by the source weight,
\be
 \omega_\mu^{2-\gamma}|v|\,|D^2(v-w)|
 +\omega_\mu^{2-\gamma}|Dv|\,|D(v-w)|
 \le C\omega_\mu^\gamma\|v\|_{X_\mu}\|v-w\|_{X_\mu}.
\ee
The term $\omega_\mu^{-2}v(v-w)$ has the same weight. The scaled H\"older product estimates give the corresponding source seminorms. The factor left after taking the supremum is $\sup\omega_\mu^\gamma\asymp\mu^\gamma$. {
Let $A$ bound the inverse on $X_\mu^0\times\mathbb R$, let $E\mu$ bound the reference error and let $B_Q\mu^\gamma$ be the quadratic difference bound above. The image of the ball of radius $K\mu$ is bounded by $AE\mu+AB_QK^2\mu^{2+\gamma}$, since $Q(0)=0$. On this ball the Lipschitz constant is at most $2AB_QK\mu^{1+\gamma}$. Thus the ball is invariant and the map is a contraction if
\be\label{f:fixed-point-conditions}
 AE/K+q/2\le1,\qquad 2AB_QK\mu^{1+\gamma} \le q<1.
\ee
} It lies in the positive metrics because $\sup|v|\le CK\mu^{1+\gamma}\to0$.
Differentiating this fixed-point equation gives a contraction for $\partial_rv$ with source $\Ocal(\mu)$: the inverse, its derivative, the source derivative, and the differentiated quadratic coefficients have the uniform bounds just proved. This yields the asserted $C^1$ estimate. This argument establishes the $\Ocal_X(\mu)$ correction despite the noncompact outer gauge source, rather than assuming that source vanishes.

For joint continuity, fix $(M_0,\mu_0,r_0)$ with $\mu_0>0$. In the insertion coordinates, use a radial diffeomorphism equal to $z\mapsto(\mu/\mu_0)z$ near each small horizon and equal to the identity outside a slightly larger ball. For parameters sufficiently close to the fixed ones this is a smooth diffeomorphism, and it commutes with the rotations and reflection. Compose it with the smooth transport of the centers and the fixed identification of the old horizon collars. At each horizon rescale the circular tensor coefficient by the square of the ratio of the normalized surface-gravity factors, with a smooth transition in the interior. This identifies the domain of fields regular in the horizon charts; on the small horizons it also fixes the unit Schwarzschild coordinates and the normalizing integral. The displacement source has the same property, including its explicit powers of $\mu$.

In these fixed charts the improved reference metrics and the tensors multiplying the auxiliary scalar coefficients depend continuously on $(M,\mu,r)$ in every fixed $C^{j,\alpha'}$ norm with $\alpha'<\alpha$: the profiles near the added horizons and the cutoff are smooth for $\mu>0$, and the Green term has the continuity just established in Proposition~\ref{green:global}. The same holds for the coefficients of the reduced equation and for the integral functional. On a sufficiently small parameter neighborhood the correction and scalar coefficient have uniform bounds, and the contraction has a common closed ball and a common factor below one. Given a convergent sequence of parameters, compact embedding gives a subsequence of the transported corrections converging in $C^{k+2,\alpha'}$; the scalar coefficients also have a convergent subsequence. Passing to the equation and the integral condition gives a solution at the limiting parameters. The uniform $C^{k+2,\alpha}$ bound passes to the limit by lower semicontinuity, so uniqueness in the contraction ball identifies this solution with the constructed one. Every convergent subsequence has the same limit, which proves joint continuity. In particular, the simple force zero varies continuously, and restriction to the horizon gives continuous area functions. This proves continuity for positive $\mu$ only; it does not select a continuous solution of the subsequent area equations.\qedhere
\end{proof}

\begin{proposition}[Smoothness of the corrected metric, horizons and axis]\label{reg:bootstrap}
Fix an admissible positive mass $\mu$ and an insertion radius $r$. The solution of Proposition~\ref{f:family} is smooth in the spatial interior and in every normalized local horizon lift. The corresponding spatial metric $h$ and positive lapse $N$ are smooth up to all four horizons, including their eight intersections with the axis. Each horizon is totally geodesic and has a positive constant opening $\kappa_H$. In Gaussian spatial coordinates $(s,y)$, where $s\ge0$ is distance from that horizon,
\be\label{reg:horizon-parity}
h=\dd s^2+\gamma_H+s^2q_H(s^2,y),\qquad
N=\kappa_Hs+s^3b_H(s^2,y),
\ee
with smooth tensor and scalar coefficients $q_H,b_H$. The axial angle has period $2\pi$, and the axis has this angle on each open segment and at every horizon pole. The spatial interior here includes all the interpolation interfaces.
\end{proposition}
\begin{proof}
\emph{1. Smoothness by repeated elliptic estimates at fixed positive mass.} The reference $\overline G_{\mu,r}$ and the tensor $\psi_\mu$ are smooth in the normalized horizon charts: near the new horizons the affine profile is regular, near the old horizons the Green tensor is regular, and their interpolation uses smooth cutoffs. The solution initially belongs to $C^{k+2,\alpha}$ in these charts and is positive by the contraction estimate. In any full Cartesian lifted chart, the coordinate expression of its equation is
\be\label{reg:coordinate-equation}
-g^{ab}\del_a\del_bg_{ij}
=\lambda_\mu(\psi_\mu)_{ij}
+F_{ij}(g^{-1},\del g;\overline G_{\mu,r},\del\overline G_{\mu,r},\del^2\overline G_{\mu,r}),
\ee
where $F_{ij}$ is smooth in its arguments. The cancellation of the other second derivatives is the defining DeTurck cancellation. On each smaller chart the positive matrix $g^{ab}$ is uniformly elliptic for this fixed $\mu$. If $g\in C^{j,\alpha}$ with $j\ge2$, the right side belongs to $C^{j-1,\alpha}$ and the coefficients to $C^{j,\alpha}$. The componentwise interior Schauder estimate therefore gives $g\in C^{j+1,\alpha}$. Iteration proves smoothness. The invariant equation is the full local equation: all its tensor components have the imposed symmetries, and the omitted circular mixed components vanish under their separate reversals. This argument applies on complete lifted balls at horizons and poles, and on ordinary spatial balls with an unwrapped auxiliary coordinate in the interior.

\par\addvspace{0.2cm}\noindent
\emph{2. The horizon and its poles.} Near a horizon use its own normalized angle $\vartheta_H=\kappa_H\tau$. Smooth rotation invariance gives $\gamma(-s,y)=\gamma(s,y)$ for the tangential metric and $N(-s,y)=-N(s,y)$ for the signed lapse in signed Gaussian distance $s$. The normal-plane metric is isotropic at its center. Consequently $N/(\kappa_Hs)$ extends to a positive smooth function invariant under $s\mapsto-s$ and equal to one at $s=0$, giving~\eqref{reg:horizon-parity}. The constant $\kappa_H$ is the prescribed opening of this local lift. The identity $\gamma(-s,y)=\gamma(s,y)$ gives $\partial_s\gamma(0,y)=0$, so the horizon is totally geodesic.

At an intersection with the axis, use Cartesian coordinates $(x_1,x_2,y_1,y_2)$ for the two normal rotation planes. The horizon circle rotates $(x_1,x_2)$ and the axial circle rotates $(y_1,y_2)$. Their smooth commuting actions give a regular lifted four-ball. The spatial region is the half-ball $x_2=0$, $x_1\ge0$, with smooth coordinates $(x_1,y_1,y_2)$. Thus the pole is a regular boundary point of the spatial manifold. Isotropy of the metric in the axial normal plane gives the ratio $2\pi$ of the small orbit circumference to its normal radius. The same argument applies to every open axis segment. The four boundary spheres have two axial fixed points each, giving all eight poles.

\par\addvspace{0.2cm}\noindent
\emph{3. Smoothness across the interpolation regions and at critical points.} The reference, correction and equation are global smooth tensors in the common atlas, so the first argument also applies on every artificial interpolation interface. At an interior point with $\dd N=0$ the lapse remains positive. The lifted metric and its principal matrix are therefore positive there, and~\eqref{reg:coordinate-equation} has the same ellipticity. This proves the asserted regularity at those points. The proof concerns each fixed positive mass. The limiting estimates on fixed and rescaled compact sets follow by the same local bootstrap once the corresponding coefficient bounds are uniform.
\end{proof}

\subsection{The coefficient of the auxiliary source}

Let $L_\mu\coloneqq\rho_\mu/(4\mu)$, and let $Z_\mu$ have normalized pullback $\chi_{L_\mu}T_S$ on the two collars, with the same reflection convention as the auxiliary source. Its support lies strictly inside the inner ansatz. Integrals in this section use $\overline G$ per unit $\tau$. The Gram denominator satisfies
\be\label{f:gram}
 \langle\psi_\mu,Z_\mu\rangle
 =2b(r)\mu^{\gamma+1}(m_0+e_G(\mu,r)),
\ee
where $e_G(\mu,\cdot)=\Ocal_{C^1(I)}(\mu)$. The support restriction~\eqref{q:power-support-conditions} gives $L_\mu\ge R_0$ in the quantitative range, so $\chi_{L_\mu}=1$ on the support of $\chi_0$. Equation~\eqref{join:eq:column} gives $\widehat\psi_{+,\mu}=\mu^{\gamma-2}\chi_0T_S$. The normalized metric is $G_S+\mu aA_S$, so both its inverse metric in the tensor contraction and its volume density differ from those of $G_S$ by $O(\mu)$ there, also after one placement derivative. The conversion from normalized volume to volume per unit original auxiliary parameter is $b\mu^3$. Consequently one collar contributes $b\mu^{\gamma+1}(m_0+O(\mu))$; reflection gives the same contribution from the other collar and accounts for the factor two.

\begin{lemma}[The three estimates in the force calculation]\label{f:test}
For the solution of Proposition~\ref{f:family}, the following estimates hold for the normalized pairings and their first derivatives with respect to $r$:
\be\label{f:test-estimates}
 \begin{split}
 \frac{|\langle f_{\mu,r},Z_\mu\rangle|}{2b\mu^{\gamma+1}}
 &\le C(\mu^{2-\gamma-a_2}+\mu^{1-\gamma+a_2}+\mu^{2-\gamma}(1+|\log\mu|)),\\
 \frac{|\langle v_{\mu,r},L_{\overline G}^*Z_\mu\rangle|}{2b\mu^{\gamma+1}}
 &\le C(\mu^2+\mu^{2-\gamma+a_2(\gamma-1)}),\\
 \frac{|\langle Q(v_{\mu,r}),Z_\mu\rangle|}{2b\mu^{\gamma+1}}
 &\le C\mu^{2+\gamma}.
 \end{split}
\ee
\end{lemma}
\begin{proof}
For the first estimate, use the model measure $b\mu^3\dd\nu_S$ and the unit Schwarzschild contraction in the leading pairing from~\eqref{f:source-inner}. After division, this contribution is the integral of $\chi_{L_\mu}\langle J_0,T_S\rangle$ against $\dd\nu_S$, multiplied by $\mu^{1-\gamma}a(r)$.
The full model integral vanishes and the integral over its cutoff complement is $\Ocal(L_\mu^{-1})$ by~\eqref{f:J-bounds}. In the normalized moving charts, $J_0$, $T_S$ and $L_\mu$ are independent of the placement radius. The placement derivative acts only on $a(r)$, so the same cancellation and tail bound hold for that derivative. This gives $C\mu^{2-\gamma-a_2}$. The bounded remainder has integral at most $C\mu^{2-\gamma}L_\mu=C\mu^{1-\gamma+a_2}$. The metric and volume changes from $G_S$ to $G_S+\mu aA_S$ add at most
$C\mu^{2-\gamma}\bigl(1+\int_1^{2L_\mu}s^{-1}\dd s\bigr)$ in the $J_0$ pairing, since their relative size is $C\mu(1+s)$. The constant term accounts for the compact part of the collar. This proves the first bound.

For the second line, $L_{\overline G}$ is selfadjoint in this reference measure by~\eqref{f:L}. Write $\omega\asymp\mu+\rho$ and use $|v|\le C\mu\omega^\gamma$. The part $P_S(\chi_{L_\mu}T_S)$ has normalized size $\Ocal(L_\mu^{-4})$ on its cutoff annulus. On that annulus, the normalized correction is bounded by $C\mu^{1+\gamma}L_\mu^\gamma$, while the normalized volume is $O(L_\mu^3)$. Restoring the factor $\mu^{-2}$ from the operator and $b\mu^3$ from the volume gives
\be
 |\langle v,P_{S,\mu}Z_\mu\rangle|
 \le Cb\mu^{2+\gamma}L_\mu^{\gamma-1}.
\ee
After division by the Gram scale $2b\mu^{\gamma+1}$ this becomes $C\mu L_\mu^{\gamma-1}=C\mu^{2-\gamma}\rho_\mu^{\gamma-1}=C\mu^{2-\gamma+a_2(\gamma-1)}$.
In the pointwise tensor norm before rescaling, the difference of the two operators, including the zeroth-order Ricci terms in~\eqref{f:L}, satisfies
\be
 |(L_{\overline G}^*-P_{S,\mu})Z_\mu|
 \le C\mu^2(\mu+\rho)^{-3}.
\ee
Indeed the inner metric change is $\Ocal(\rho)$, its first derivative is bounded, its second derivative is $\Ocal((\mu+\rho)^{-1})$, and $|D^jZ_\mu|\le C_j\mu^2(\mu+\rho)^{-2-j}$. The reference Ricci potential has size at most $C\mu(\mu+\rho)^{-2}+C$ and obeys the same resulting bound. The pairing is therefore at most
$C\mu^3\int_\mu^{\rho_\mu}\rho^{\gamma-1}\dd\rho\le C\mu^{3+\gamma}$,
which gives $C\mu^2$ after division. A regular scaled bolt chart gives the identical bound for the omitted endpoint region $\rho\asymp\mu$.

Finally the local quadratic formula gives
$|Q(v)|\le C\mu^2\omega^{2\gamma-2}$.
The pairing with $|Z_\mu|\le C\mu^2\omega^{-2}$ is at most
$C\mu^4\int_\mu^{\rho_\mu}\rho^{2\gamma-2}\dd\rho\le C\mu^{3+2\gamma}$.
Division gives the third line. For derivatives with respect to the insertion radius, use the normalized moving charts and define the density ratio $j_{\mu,r}$ by
\be
 \frac{\dd V_{\overline G}/\dd\tau}{b(r)\mu^3\dd\nu_S}=j_{\mu,r}.
\ee
On the support of the test tensor, the normalized metric is $G_S+\mu a(r)A_S$. The profile bounds and the determinant formula give
\be
 |\partial_r^k(j_{\mu,r}-1)|\le C\mu(1+s),\qquad k=0,1.
\ee
The normalized test tensor $\chi_{L_\mu}T_S$ and its cutoff are independent of $r$. The tensor contraction also varies with the reference metric. In normalized coordinates, differentiating either inverse metric in this contraction gives a relative coefficient bounded by $C\mu(1+s)$, since $\partial_r\widehat{\overline G}=\mu a'(r)A_S$. Its higher scaled spatial derivatives satisfy the same bound. For clarity, write $\mathcal I_r(A,B)$ for the normalized tensor contraction of $A$ and $B$ multiplied by $j_{\mu,r}$, and let $A$ and $B$ be the normalized entries of any of the three pairings. With $\mu$ fixed, reflection and the volume factor $b(r)\mu^3$ give the derivative formula
\begin{equation}
 \partial_r\Bigl(\mu^{2-\gamma}\int\mathcal I_r(A,B)\dd\nu_S\Bigr)
 =\mu^{2-\gamma}\int\Bigl(
 \mathcal I_r(\partial_r A,B)+\mathcal I_r(A,\partial_r B)
                  +(\partial_r\mathcal I_r)(A,B)\Bigr)\dd\nu_S .
\end{equation}
The factor $b(r)$ has canceled before differentiation. The coefficient of the last term is bounded by $C\mu(1+s)$, so it uses the same bounds as the changes of contraction and density above. The Schwarzschild part of the adjoint on the test tensor is independent of $r$. Differentiating the remaining coefficients uses $\mu a'(r)A_S$ in place of $\mu a(r)A_S$ and gives the same pointwise estimate for the operator difference. The derivative of the quadratic remainder is the sum of the explicit coefficient derivative and $D_vQ(v)[\partial_r v]$. Both terms are bounded by $C\mu^2\omega^{2\gamma-2}$, since $v$ and $\partial_r v$ have the same weighted estimate. This gives the stated powers of $\mu$ also for the derivatives of all three normalized pairings.
\end{proof}

\begin{proposition}[The coefficient of the auxiliary source at the two added horizons]\label{f:force}
The actual nonlinear coefficient in~\eqref{f:family-eq} satisfies
\be\label{f:force-formula}
 \lambda_\mu(r)
 =-\frac{32\pi}{m_0}\,\mu^{1-\gamma}
 \frac{\partial_{\nu_K}N_K(p_+(r))}{N_K(p_+(r))}
 +\Ocal_{C^1(I)}(\mu^{1-\gamma+\delta}),
 \qquad
 \delta>0\ \text{as defined in~\eqref{eq:error-gains}}.
\ee
In particular it has a unique zero in a fixed sufficiently small interval about $r_*$, and that zero satisfies $|r_\mu-r_*|\le C\mu^\delta$. At this zero both local obstruction coefficients vanish: reflection identifies the two collar coefficients, and the global displacement source is the sum of their reflected copies with disjoint supports.
\end{proposition}
On the background region, the reference metric includes the Green tensor multiplied by the small mass. This correction is matched near each new horizon to the regular Schwarzschild correction with affine growth. Its essential cancellation is the exact integral of $J_0$, while its nonlinear remainder is the integrable weighted product producing $\mu^{2+\gamma}$. The latter is smaller than the force scale precisely under the stated restriction $\gamma>-1/2$.
\begin{proof}
Testing the exact correction equation against $Z_\mu$ and moving the linearized operator to its formal adjoint gives
\begin{equation}
 \lambda_\mu-\lambda_\mu^0
 =\frac{\langle f_{\mu,r},Z_\mu\rangle
       +\langle v_{\mu,r},L_{\overline G}^*Z_\mu\rangle
       +\langle Q(v_{\mu,r}),Z_\mu\rangle}
        {\langle\psi_\mu,Z_\mu\rangle}.
\end{equation}
After division by $2b(r)\mu^{\gamma+1}$ in numerator and denominator, the denominator is $m_0+e_G(\mu,r)$. Its reciprocal is uniformly bounded in $C^1(I)$ by~\eqref{f:gram} and $m_0>0$. Applying~\eqref{f:test-estimates} and the $C^1$ product bound gives
\be
 \|\lambda_\mu-\lambda^0_\mu\|_{C^1}
 \le C(\mu^{2-\gamma-a_2}+\mu^{1-\gamma+a_2}+\mu^{2-\gamma}(1+|\log\mu|)
 +\mu^2+\mu^{2-\gamma+a_2(\gamma-1)}+\mu^{2+\gamma}).
\ee
After division by $\mu^{1-\gamma}$, the powers in the resulting terms are, respectively,
\be
 1-a_2,\quad a_2,\quad 1\ \text{with a logarithm},\quad 1+\gamma,\quad 1-a_2(1-\gamma),\quad 1+2\gamma.
\ee
They are positive by~\eqref{eq:free-scale-conditions}, and their minimum is the stated $\delta$; the logarithmic term is also bounded by a constant times $\mu^\delta$ on $(0,1/2]$. This proves~\eqref{f:force-formula}. For the Schwarzschild--de Sitter normalization above, $a=F_M'/(2\sqrt{F_M})$, so $a(r_*)=0$ and $a'(r_*)=-\Lambda/\sqrt{F_M(r_*)}\ne0$. The uniform $C^1$ remainder preserves the sign of the derivative and the opposite signs at the endpoints of a smaller fixed interval. The intermediate value theorem and the mean value theorem prove the last assertion. The same proof is uniform when $M$ ranges in a compact subinterval of the subextremal interval and the intact inverse is uniformly bounded there. Replacing $\mu$ by $s\mu_{\rm target}$ with $s$ in a compact positive interval preserves every estimate.
\end{proof}

\section{Vanishing of the DeTurck vector field}\label{sec:gauge}

\subsection{Quadratic forms for vector fields on the reference geometries}

\paragraph{The vector equation on the background.} When the coefficient of the auxiliary source vanishes, the contracted Bianchi identity gives a homogeneous equation for the DeTurck vector field. We prove that this equation has only the zero solution, with estimates independent of the shrinking horizon size. The corrected metric then satisfies the static vacuum Einstein equations. Integrals below are per unit original auxiliary parameter, with measure $N_g\dd V_{h_g}$. All vector fields under consideration are meridional, invariant under the global equatorial reflection that exchanges the two added horizons, and smooth in the normalized local lifts.

Write $G=G_M=h_M+N_M^2\dd\tau^2$ and $\dd m_G=N_M\dd V_{h_M}$. Recall the form $q_G$ from~\eqref{q:functional-Korn} and define on one-forms
\be
 Q_G\coloneqq\nabla_G^*\nabla_G-\Ric_G=\Delta_{H,G}-2\Lambda.
\ee

\begin{lemma}[Positive energy for the vector equation on the fixed background]\label{hv:outer}
There is $c_M>0$ such that $q_G(V)\ge c_M\|V\|_{H^1(G)}^2$ for every regular field in the stated symmetry class, and hence for its $H^1$ closure.
\end{lemma}
\begin{proof}
Corollary~\ref{q:functional-Poincare-Korn} gives $q_G(V)\ge c_{V,M}\|V\|_{H^1_\ell}^2$ on this class and its $H^1$ closure. Since
\be
 \|V\|_{H^1_\ell}^2
 =\|\nabla V\|_2^2+\ell^{-2}\|V\|_2^2
 \ge\min(1,\ell^{-2})\|V\|_{H^1(G)}^2,
\ee
we may take $c_M\coloneqq c_{V,M}\min(1,\ell^{-2})>0$.\qedhere
\end{proof}

\paragraph{The reference metric near the shrinking horizons.}
Use the reference $G_\mu$ of~\eqref{join:eq:reference}, allowing the insertion points to vary in a fixed compact interval about the maximal sphere. The common lapse factor $b$ stays bounded away from zero. Its inner regions are exactly Schwarzschild and its two original horizon collars are unchanged. The normal radius is denoted by $\rho$. With $R_\mu=\mu^{a_2}$, all the bounds below hold uniformly in placement. The improved reference differs from this simpler reference by a scaled $C^2$ quantity tending to zero.

Write $\omega_\mu\asymp\max(\mu,\rho)$ in the two fixed insertion neighborhoods, with a smooth positive continuation comparable to one outside them. The reference has uniformly bounded scaled local geometry, converges smoothly to $G$ off the insertion points, and
\be\label{hv:reference-ric}
\sup\omega_\mu^2|\Ric_{G_\mu}-\Lambda G_\mu|_{G_\mu}
\le C_M(\mu/R_\mu+R_\mu)\longrightarrow0.
\ee
Indeed on the inner exact Schwarzschild region the Ricci tensor is zero, giving at most $\Lambda R_\mu^2$. On the cutoff annulus the metric difference and its scaled first two derivatives are $O_M(\mu/R_\mu+R_\mu)$; the linear term in the background lapse is included in this bound. The scaled curvature formula proves~\eqref{hv:reference-ric}; outside the insertion neighborhoods the error vanishes.

Consider also any positive block metric $g_\mu$ with the same local angular regularity and openings as its reference and
\be\label{hv:perturbation}
\sup\bigl(|g_\mu-G_\mu|+\omega_\mu|\nabla(g_\mu-G_\mu)|
+\omega_\mu^2|\nabla^2(g_\mu-G_\mu)|\bigr)\le\epsilon_\mu\longrightarrow0.
\ee
The tensor norms and derivatives in this display are those of the reference in normalized local lifts. Near either old horizon the metric converges in those lifts; near each new horizon the rescaled fixed compact Schwarzschild neighborhoods converge in $C^2$. The curvature formula yields
\be\label{hv:eta}
\eta_\mu\coloneqq\|\omega_\mu^2(\Ric_{g_\mu}-\Lambda g_\mu)\|_{L^\infty(g_\mu)}
\le C_M(\mu/R_\mu+R_\mu+\epsilon_\mu)\longrightarrow0.
\ee

\begin{lemma}[Sharp Schwarzschild Hardy constants]\label{q:Hardy}
Fix $u>0$ and let $G_{S,u}$ be the mass-$u$ Schwarzschild metric. Write $\rho\ge u/2$ for its isotropic radius and $R\coloneqq\rho+u+u^2/(4\rho)$ for its areal radius. Every regular static vector field supported at finite radius satisfies
\be\label{q:Hardy-constant}
 \int R^{-2}|V|^2\dd\nu_{S,u} \le\int(u+\rho)^{-2}|V|^2\dd\nu_{S,u} \le4\int|\nabla V|^2\dd\nu_{S,u}.
\ee
The constant $4$ is optimal for either weight on the complete Schwarzschild model, including within the axial meridional class. The measure is taken per unit of the original parameter $\tau$.
\end{lemma}
\begin{proof}
We set $\psi\coloneqq1+u/(2\rho)$ and $p\coloneqq\rho^2-u^2/4$. The invariant radial measure is $p\psi^4\dd\rho\dd\omega$, and the radial scalar derivative energy is $p|f'|^2\dd\rho\dd\omega$. For the positive function $\phi\coloneqq(u+\rho)^{-1/2}$, direct differentiation gives
\bse\label{q:sharp-Hardy-identity}
\be
 -\frac{(p\phi')'}{\phi}=\frac{\rho^2/4+u\rho+3u^2/16}{(u+\rho)^2},
\ee
\be
 -\frac{(p\phi')'}{\phi}-\frac{p\psi^4}{4(u+\rho)^2}
 =\frac{2u\rho-u^2/2+5u^4/(16\rho^2)+u^5/(8\rho^3)+u^6/(64\rho^4)}{4(u+\rho)^2}\ge0.
\ee
\ese
The numerator is positive for $\rho\ge u/2$. The ground-state identity therefore proves
\be
 \frac14\int\frac{p\psi^4}{(u+\rho)^2}f^2\dd\rho
 \le\int p|f'|^2\dd\rho.
\ee
Its boundary term vanishes at $\rho=u/2$ because $p=0$ and $\phi'/\phi$ is bounded; compact support removes the other endpoint. We apply this along each angular ray to $f=|V|$, using a smooth approximation at zeros and Kato's inequality. The nonnegative angular and auxiliary derivative energies can be kept on the right. This proves the second inequality. The first follows from $R\ge u+\rho$.

For optimality, we work outside a large radius in isotropic Cartesian coordinates and fix the constant axial vector $Z=\partial_{z_3}$. Choose a nonzero $\chi\in C_c^\infty((1,2))$ and set
$V_L\coloneqq\rho^{-1/2}\chi((\log\rho)/L)Z$. These fields are axial and meridional, regular on the axis, and vanish near the horizon. The Schwarzschild metric is Euclidean up to $\Ocal(u/\rho)$ there, while $\nabla Z=\Ocal(u/\rho^2)$. Their weighted norm and energy, divided by $4\pi$, are respectively
\be
 L\int\chi^2\dd t+o(L),\qquad
 \frac L4\int\chi^2\dd t+\frac1L\int(\chi')^2\dd t+o(L).
\ee
The cross term integrates to zero, and the metric errors tend to zero uniformly on the support $e^L<\rho<e^{2L}$. The same numerator asymptotic holds for the areal-radius weight. The quotient tends to four, proving optimality. For each of these test fields fixed on the unit model, its rescaled support eventually fits in an exact Schwarzschild portion as the inserted mass decreases. Reflection gives the corresponding field at the second insertion.

For comparison with the smooth construction scale, put $\omega_{S,u}\coloneqq\sqrt{u^2+(R-u)^2}$ on the complete model. The inequality $\omega_{S,u} \le R\le\sqrt2\,\omega_{S,u}$ from~\eqref{join:eq:unit-scale} gives
\be\label{q:regular-weight-Hardy}
 \int\omega_{S,u}^{-2}|V|^2\dd\nu_{S,u}
 \le2\int R^{-2}|V|^2\dd\nu_{S,u}
 \le8\int|\nabla V|^2\dd\nu_{S,u}.
\ee
The upper bound $8$ is uniform in $u$ on the exact Schwarzschild model. Although $\omega_{S,u}=\omega_u$ for $\rho\le\rho_w$, the metrics agree only on the exact Schwarzschild portion of the construction collar. The uniform Hardy inequality for the whole joined metric, including the interpolation annuli, is proved in Lemma~\ref{hv:hardy}~below.\qedhere
\end{proof}

\subsection{A Hardy inequality}

\begin{lemma}[A Hardy inequality uniform as the new horizons shrink]\label{hv:hardy}
For the references and perturbations above there is $C_M$ independent of small $\mu$ such that
\be\label{hv:hardy-estimate}
\int\omega_\mu^{-2}|V|^2\dd m_{g_\mu}
\le C_M\int(|\nabla V|^2+|V|^2)\dd m_{g_\mu}.
\ee
Consequently the squared $L^2$ norm of $V$ on either insertion ball of radius $R$ is at most $C_MR^2\|V\|_{H^1(g_\mu)}^2$, for fixed small $R$ and all sufficiently small $\mu$.
\end{lemma}
\begin{proof}
First fix a large number $L$. On the compact Schwarzschild cap up to normal radius $2L\mu$, covariant differentiation controls the $L^2$ norm with scale $\mu$:
\be\label{hv:cap}
\mu^{-2}\int_{\rho<2L\mu}|V|^2\dd m_{g_\mu}
\le C_L\int_{\rho<2L\mu}|\nabla V|^2\dd m_{g_\mu}.
\ee
To prove this on the unit cap, a countersequence with derivative norm tending to zero has an $L^2$ limit which is a parallel field. The Schwarzschild curvature has no common nonzero null vector. Indeed, in an orthonormal frame $e_0,e_1,e_2,e_3$ adapted to the auxiliary circle, the radial direction and the tangent sphere, the curvature operators of the planes $e_0\wedge e_1$ and $e_2\wedge e_3$ are nonzero multiples of the rotations in those respective planes; their kernels are complementary two-planes. A parallel field is annihilated by both operators and therefore vanishes. Thus no nonzero parallel field exists. Compactness gives the estimate with no outer boundary condition. Scaling back gives~\eqref{hv:cap}; the factor converting the normalized circle measure into the measure per unit original parameter cancels on both sides. To pass to the perturbed metric, keep $L$ fixed and take $\mu$ small enough that the cap lies in the exact Schwarzschild region. Compare the two connections on the unit cap. The scaled $C^1$ bound in~\eqref{hv:perturbation} gives
\be
 \|\nabla^{S}V\|_{L^2(S)}
 \le(1+C_L\epsilon_\mu)\|\nabla^{g_\mu}V\|_{L^2(g_\mu)}
 +C_L\epsilon_\mu\|V\|_{L^2(g_\mu)},
\ee
and the two $L^2$ norms of $V$ are uniformly comparable. Inserting this in the model cap estimate and choosing $\mu$ so that $C_L\epsilon_\mu$ is sufficiently small absorbs the last term. Scaling back proves~\eqref{hv:cap} for the perturbed metric, with $L$ fixed before $\mu$.

On the annulus $L\mu<\rho<2L\mu$, extend the spatial Cartesian components to the filled Euclidean ball. The scaled annular extension estimate and~\eqref{hv:cap} bound its Euclidean derivative energy by $C_L$ times the actual derivative energy on the Schwarzschild cap. The possible zeroth-order term $(L\mu)^{-2}\int|V|^2$ in that extension estimate is controlled precisely by~\eqref{hv:cap}. Away from this cap, in the normal Cartesian coordinates,
\be
|D V|\le C|\nabla^{g_\mu}V|
+C_M\rho|V|+C\mu\rho^{-2}|V|+C\epsilon_\mu\rho^{-1}|V|.
\ee
This is the background normal-coordinate connection estimate, the Schwarzschild tail estimate, and the scaled perturbation bound. Full covariant differentiation includes the spatial connection terms needed here. The measure is uniformly comparable to Euclidean three-dimensional measure on this region, per unit original auxiliary parameter.

Apply the Euclidean three-dimensional Hardy inequality to the extended field with a cutoff at a fixed outer radius. It gives, up to the fixed outer $L^2$ term, the integral of $\rho^{-2}|V|^2$. In the preceding derivative estimate the singular coefficient is at most $C(L^{-1}+\epsilon_\mu)\rho^{-1}$ when $\rho\ge L\mu$. Write $I_{\mu,L}$ for the integral of $\rho^{-2}|V|^2$ over the part of the insertion chart with $\rho\ge L\mu$, and $E_\mu$ for $\int(|\nabla V|^2+|V|^2)\dd m_{g_\mu}$. The extension and Hardy estimates give
\be
 I_{\mu,L} \le C_L E_\mu+C_0(L^{-2}+\epsilon_\mu^2)I_{\mu,L}.
\ee
Here $C_0$ is independent of $L$ and $\mu$: it comes from the connection estimate on the exterior annulus, where the metric and measure comparisons are uniform for $L$ above a fixed value. The cap extension contributes to $C_L E_\mu$, not to the coefficient of $I_{\mu,L}$. Choose $L$ first so that $C_0L^{-2}<1/4$, and then choose $\mu$ so small that $C_0\epsilon_\mu^2<1/4$. Absorption bounds $I_{\mu,L}$ by $2C_LE_\mu$, uniformly in those $\mu$. Equation~\eqref{hv:cap} controls the original inner cap as well. The fixed outer cutoffs contribute only a bounded $L^2$ term. Summing over the two disjoint insertion neighborhoods proves~\eqref{hv:hardy-estimate}. Since $\omega_\mu\le CR$ inside radius $R$ once $\mu<R$, the last assertion follows.\qedhere
\end{proof}

\subsection{Coercivity as the horizons shrink}

Let us consider the quadratic form
\be
q_{g_\mu}(V)\coloneqq\int \Big( |\nabla V|^2-\Ric_{g_\mu}(V,V) \Big) \, \dd m_{g_\mu}.
\ee
\begin{proposition}[A uniform lower bound for the quadratic form on vector fields]\label{hv:uniform}
For all sufficiently small $\mu$, there is $c_M'>0$ independent of $\mu$ such that
\be
q_{g_\mu}(V)\ge c_M'\|V\|_{H^1(g_\mu)}^2
\ee
on the space of regular meridional vector fields invariant under equatorial reflection. The statement holds uniformly over families satisfying~\eqref{hv:perturbation} with a common bound $\epsilon_\mu\to0$.
\end{proposition}
\begin{proof}
All norms below use $g_\mu$ and the measure $\dd m_{g_\mu}$, with $\|V\|_{H^1(g_\mu)}^2\coloneqq\|\nabla V\|_2^2+\|V\|_2^2$. Denote the uniform constant in~\eqref{hv:hardy-estimate} by $C_{\rm H,M}$. The definition of $\eta_\mu$ in~\eqref{hv:eta} gives, for every regular field $V$,
\be
 \left|\int(\Ric_{g_\mu}-\Lambda g_\mu)(V,V)\dd m_{g_\mu}\right|
 \le\eta_\mu\int\omega_\mu^{-2}|V|^2\dd m_{g_\mu}
 \le C_{\rm H,M}\eta_\mu\|V\|_{H^1(g_\mu)}^2.
\ee
Hardy's estimate also extends this bound to the corresponding $H^1$ closure. Substitution in the definition of $q_{g_\mu}$ yields
\be\label{hv:quantified-energy}
 q_{g_\mu}(V)\ge\|\nabla V\|_2^2-\Lambda\|V\|_2^2
 -C_{\rm H,M}\eta_\mu\|V\|_{H^1(g_\mu)}^2.
\ee
By~\eqref{hv:eta}, the error is at most $C_M(\mu/R_\mu+R_\mu+\epsilon_\mu)\|V\|_{H^1(g_\mu)}^2$, where $C_M$ is independent of $\mu$, $V$ and the placement within the fixed intervals.

If no positive uniform constant existed, we could choose $\mu_i\to0$ and fields with $\|V_i\|_{H^1(g_{\mu_i})}=1$ and $q_{g_{\mu_i}}(V_i)\le i^{-1}$. Estimate~\eqref{hv:quantified-energy} gives $(1+\Lambda)\|V_i\|_2^2\ge1-i^{-1}-C_{\rm H,M}\eta_{\mu_i}$. Since $\eta_{\mu_i}\to0$, this implies $\|V_i\|_2^2\ge[2(1+\Lambda)]^{-1}$ for all sufficiently large $i$. Local Rellich compactness off the two insertion points, together with the no-concentration estimate in Lemma~\ref{hv:hardy}, gives strong global $L^2$ convergence to a nonzero field $V_0$ on the background, after the usual identifications off the shrinking balls. Its derivative energy is bounded by lower semicontinuity. The Hardy estimate and local convergence give $\rho^{-1}V_0\in L^2$ near each limiting point, by applying the estimate first outside a fixed small ball and then letting its radius tend to zero. Choose a smooth cutoff $\chi_\varepsilon$ which vanishes for $\rho\le\varepsilon$, equals one for $\rho\ge2\varepsilon$, and satisfies $|D\chi_\varepsilon|\le C/\rho$ on its support. Then
\be
 \|(D\chi_\varepsilon)V_0\|_2^2
 \le C\int_{\varepsilon<\rho<2\varepsilon}\rho^{-2}|V_0|^2\dd m_G\longrightarrow0.
\ee
The remaining terms converge by dominated convergence. Thus $\chi_\varepsilon V_0\to V_0$ in $H^1$, and $V_0$ belongs to the full background $H^1$ class across both points. All symmetries and the regular old-horizon closure conditions pass to the limit.
To pass to the energy, first fix a compact set away from the two limiting points. The metrics converge there in $C^1$ in the fixed charts, and weak lower semicontinuity bounds the energy of $V_0$ by the limit inferior of the energies of $V_i$ on that set. Exhausting the punctured background and using positivity gives the same inequality for the full gradient energy. In addition, the Ricci error is controlled independently of concentration:
\be
 \left|\int(\Ric_{g_{\mu_i}}-\Lambda g_{\mu_i})(V_i,V_i)\dd m_{g_{\mu_i}}\right|
 \le C_{\rm H,M}\eta_{\mu_i}\longrightarrow0,
\ee
because $\|V_i\|_{H^1(g_{\mu_i})}=1$. Strong $L^2$ convergence therefore gives
\be
0\ge\liminf_i q_{g_{\mu_i}}(V_i)
\ge q_G(V_0)>0
\ee
by Lemma~\ref{hv:outer}, a contradiction. The proof depends only on the displayed uniform geometric bounds, which proves the last assertion.\qedhere
\end{proof}

Use the DeTurck vector field $W(g,\overline G_{\mu,r})$ defined in Section~\ref{sec:nonlinear}, with $\mathcal B_gW\coloneqq\tfrac12\mathcal L_Wg$ and $\beta_g\coloneqq\delta_g+\tfrac12\dd\operatorname{tr}_g$. For the smooth metrics considered below, both $g$ and $\overline G_{\mu,r}$ are invariant under the circle actions, their separate reversals, and polar reflection, so the same is true of $W$. The two reversals change the signs of the corresponding circular components while leaving $W$ invariant; hence those components vanish. Moreover the two metrics are smooth in the same normalized horizon and axis charts. Their connection difference is a smooth tensor there, so $W$ is a regular meridional field in the vector class used above. At each fixed positive $\mu$, its smoothness and the positive weight also place $W^\flat$ in the weighted vector domain used below. For any metric,
\be
2\beta_g\mathcal B_gW=(\nabla_g^*\nabla_g-\operatorname{Ric}_g)W^\flat.
\ee
\begin{corollary}[Recovery of the static equations after cancellation of the auxiliary source]\label{hv:recovery}
Suppose a smooth metric satisfying~\eqref{hv:perturbation} solves $\Ric_g-\Lambda g-\mathcal B_gW=0$. Then $W=0$ and $\Ric_g=\Lambda g$.
\end{corollary}
\begin{proof}
The contracted Bianchi identity gives $\beta_g(\Ric_g-\Lambda g)=0$. Thus $Q_gW^\flat=0$. Remove a normal tube of radius $\varepsilon$ around each horizon and integrate by parts. In its regular normalized lift, $W$ and $\nabla W$ are bounded, while the tube has area $O(\varepsilon)$; hence the absolute value of its boundary integral is $O(\varepsilon)$. Near a pole, the same estimate holds in a smooth chart for the two independent rotations, so the axis creates no additional boundary term. Letting $\varepsilon\to0$ gives
\be
 0=\int\langle Q_gW^\flat,W^\flat\rangle\dd m_g
 =\int\bigl(|\nabla W|^2-\Ric_g(W,W)\bigr)\dd m_g.
\ee
Proposition~\ref{hv:uniform} gives $W=0$.\qedhere
\end{proof}
For the solution of Proposition~\ref{f:family}, the scaled $C^2$ difference from the simpler joined reference tends to zero, with bound $O_M(\mu^{\eta_{\rm geom}}+\mu^{1+\gamma})$. Proposition~\ref{f:force} selects the radius, and Proposition~\ref{q:vector-recovery-power} below gives the specified mass range in which the reduced solution satisfies the static equations. We prescribe its areas and lapse normalization in Section~\ref{sec:areas} below.

\subsection{An explicit mass condition for recovery of the static equations}

The preceding energy estimates give quantitative $H^1$ coercivity. For recovery of the static equations, injectivity of the harmonic-vector operator is sufficient. We now obtain an alternative sufficient condition from fixed-model weighted inverse operators.

For $-1<d<0$, define $X_S^V(d),Y_S^V(d)$ and $X_{K,\bullet}^V(d),Y_{K,\bullet}^V(d)$ by the same coordinate balls, derivative orders and weights of~\eqref{q:model-Holder-norms} and~\eqref{q:punctured-Kottler-norms}, using covariant one-forms in place of symmetric two-tensors. The one-forms are static, axisymmetric, meridional and regular in the normalized lifts. We impose global polar reflection on Schwarzschild--de Sitter and on the joined region. On an individual Schwarzschild collar both polar parities are allowed, since reflection exchanges the two insertions. Define $X_u^V,Y_u^V$ in the same way from~\eqref{join:eq:norms}, with $k=2$, $0<\alpha<1$ and $-\tfrac12<\gamma<0$.

\begin{lemma}[Solving the vector equation on the two fixed models]\label{q:vector-model-inverses}
For every fixed $d\in(-1,0)$, the maps
\be\label{q:vector-model-maps}
 Q_S:X_S^V(d)\longrightarrow Y_S^V(d),\qquad
 Q_K:X_{K,\bullet}^V(d)\longrightarrow Y_{K,\bullet}^V(d)
\ee
are isomorphisms. Their inverse norms have finite bounds $C_{S,V}(d)$ and $C_{K,V}(d)$, uniform on the fixed compact Schwarzschild--de Sitter mass and placement sets.
\end{lemma}
\begin{proof}
\emph{1. The Schwarzschild model.} On Schwarzschild, $Q_S=\nabla_S^*\nabla_S$ because the model is Ricci flat. The construction that combines inverses near the horizon and on the asymptotic end of Section~\ref{f:range} applies to this Laplace-type operator on one-forms. More explicitly, its static Cartesian principal part at infinity is the componentwise three-dimensional negative Laplacian. The Euclidean Newton inverse is bounded in every weight $d\in(-1,0)$. The coefficient difference has norm $O(1/s)$ from the weighted order-two space to its source space outside radius $s$, so it is absorbed after choosing a fixed sufficiently large radius. A shifted regular inverse on a larger compact Schwarzschild region, with nested output and source cutoffs, gives the cap part. The cutoff commutators have order one and fixed compact support, and the cap shift has order zero; after applying an inverse that gains two derivatives these give compact errors by local Rellich compactness. The same construction on both sides gives a Fredholm operator on weighted Sobolev spaces.

A continuous functional vanishing on the range of $Q_S$ is represented, under integration without a weight, by a one-form that solves $Q_S\xi=0$ weakly in the dual weight $d'=-1-d\in(-1,0)$, exactly as in Section~\ref{f:range}: the reciprocal target density is $(1+s)^{2d-1}=(1+s)^{-2d'-3}$. Local elliptic estimates on scaled annuli give pointwise decay to zero for a homogeneous field in either the domain weight or this dual weight. If $\xi$ is either homogeneous field, the Bochner identity, with $\Delta_{G_S}=\operatorname{div}_{G_S}\nabla$, is
\be\label{q:vector-model-maximum}
 \Delta_{G_S}|\xi|^2=2|\nabla\xi|^2\ge0.
\ee
With auxiliary period $8\pi$, the metric extends smoothly across the horizon in the local Cartesian coordinates, where the horizon is an interior set. The maximum principle on compact regions that increase to the full Schwarzschild model and the zero limit at infinity give $\xi=0$. Both kernel and cokernel therefore vanish.  For a H\"older source, choose a temporary Sobolev weight strictly between $d$ and zero and use the Newton representation on the end, followed by the same local estimates as in Section~\ref{f:range}, to obtain the full weighted H\"older inverse in the stated balls. Boundedness follows from the bounded inverse theorem and the fixed norm comparisons.

\par\addvspace{0.2cm}\noindent
\emph{2. The punctured Schwarzschild--de Sitter model.} On the unpunctured Schwarzschild--de Sitter region, Lemma~\ref{hv:outer} gives coercivity for regular meridional vector fields invariant under equatorial reflection. Hilbert space representation followed by the elliptic estimates in the normalized horizon charts gives its bounded inverse. To puncture the model, apply the local construction of Proposition~\ref{q:punctured-Kottler-inverse} to $Q_K$. Its frozen principal symbol is again the scalar negative Laplacian, so the Newton bound~\eqref{q:Newton} applies at weight $d$. The first-order coefficient error gains one weighted radius and the zeroth-order error gains two. A small fixed coefficient extension is absorbed, giving an exact local inverse. Fixed source and output cutoffs leave a regular annular residual, which is corrected by the complete intact vector inverse. This constructs a bounded right inverse on both punctures. A homogeneous field has size $O(\rho^d)$ with first derivative $O(\rho^{d-1})$; its Green boundary terms have orders $\rho^{d+1}$ and $\rho^{d+2}$ and vanish because $d>-1$. It extends weakly and then smoothly across the punctures. The intact inverse proves uniqueness. The fixed radii, coefficient bounds and intact estimates are uniform on the compact parameter sets, which proves the last assertion.\qedhere
\end{proof}

Use the separated cutoffs and exponents~\eqref{q:power-overlap-exponents}--\eqref{q:power-support-conditions}. The one-form rescaling is
\be\label{q:vector-model-rescaling}
 \widehat\xi\coloneqq u^{-1}\Theta_u^*\xi,\qquad
 Q_S\widehat\xi=u^2\widehat f\quad\hbox{when }Q_{G_u}\xi=f\hbox{ on the exact inner collar}.
\ee
The factor $u^{-1}$ preserves pointwise amplitudes under the constant metric rescaling. To check the power in the equation, naturality gives $\Theta_u^*(Q_{G_u}\xi)=Q_{u^2G_S}(\Theta_u^*\xi)$. Both the connection Laplacian and the Ricci endomorphism scale by $u^{-2}$ under $G_S\mapsto u^2G_S$. Hence $\widehat f=u^{-2}Q_S\widehat\xi$, which proves~\eqref{q:vector-model-rescaling}. Let $B_{J,V},D_{J,V}>0$ bound the assembled vector inverse and its four error terms, using
\be\label{q:vector-assembly-constants}
 C_{S,V}(\gamma),\quad C_{K,V}(\gamma),\quad
 C_{S,V}(\gamma-\sigma_{\rm w}),\quad C_{K,V}(\gamma+\sigma_{\rm w})
\ee
and the fixed coefficient, cutoff, covering and identification maps. In particular $B_{J,V}$ can be taken as $M_V^2(C_{S,V}(\gamma)+C_{K,V}(\gamma))$. Define
\be\label{q:vector-reference-threshold}
 u_{J,V}\coloneqq\min(u_{\rm pow,id},(\theta_{J,V}/D_{J,V})^{1/\delta_{\rm pow}}),\qquad
 A_Q\coloneqq\frac{B_{J,V}}{1-\theta_{J,V}},\qquad0<\theta_{J,V}<1.
\ee

\begin{lemma}[A uniform inverse for the vector equation on the full domain]\label{q:vector-joined-power}
For $0<u\le u_{J,V}$, the map $Q_{G_u}:X_u^V\to Y_u^V$ is an isomorphism with inverse norm at most $A_Q$.
\end{lemma}

\begin{proof}
Apply the two model inverse operators to the outer source $(1-\alpha_u)f$ and the inner source $F_S^V\coloneqq u^{2-\gamma}(u^{-1}\Theta_u^*(\alpha_uf))$, then multiply their outputs by the separated output cutoffs.  The outer source has norm with a slightly different weight at most $Cu^{-a_2\sigma_{\rm w}}\|f\|_{Y_u^V}$, and on its output transition $\rho\asymp u^{\beta_{\rm out}}$ the conversion back to $\gamma$ gives $\rho^{\sigma_{\rm w}}$. Its commutator is bounded by $Cu^{\sigma_{\rm w}(\beta_{\rm out}-a_2)}\|f\|_{Y_u^V}$. The inner source is supported in $s\le2u^{a_2-1}$ and has norm at most $Cu^{-(1-a_2)\sigma_{\rm w}}\|f\|_{Y_u^V}$ at weight $\gamma-\sigma_{\rm w}$. At $s\asymp u^{\beta_{\rm in}-1}$ the output conversion gives $s^{-\sigma_{\rm w}}$, giving $Cu^{\sigma_{\rm w}(a_2-\beta_{\rm in})}\|f\|_{Y_u^V}$. The neighboring-weight solutions equal the solutions in the space with the original weight by uniqueness and the corresponding inclusions of weighted spaces. The source-ball support comparisons are the same as in the proof of Proposition~\ref{q:joined-power}.

The metric-coefficient differences for the Laplace-type vector operator have the same two-term bounds as in~\eqref{q:joined-metric-coefficients}, so their contributions are bounded by $Cu^{1-\beta_{\rm out}}\|f\|_{Y_u^V}$ and $Cu^{\beta_{\rm in}}\|f\|_{Y_u^V}$. All four exponents are at least $\delta_{\rm pow}$. Thus the assembled right approximate inverse has error norm at most $D_{J,V}u^{\delta_{\rm pow}}$ and norm at most $B_{J,V}$. The Neumann series gives a right inverse of norm at most $A_Q$. At fixed positive $u$, the compact regular-fill argument of Lemma~\ref{join:lem:Fredholm} applies to $Q_{G_u}$ on one-forms and gives index zero. Surjectivity therefore implies injectivity.\qedhere
\end{proof}

Choose positive constants $D_{\rm ref},K$ and a threshold $u_{\rm red}$ on which the known reference and correction ball satisfy
\be\label{q:vector-reference-ball-bounds}
 \|\overline G-G_u\|_{C^{4,\alpha}_{\omega_u},\mathrm{rel}} \le D_{\rm ref}u^{\eta_{\rm geom}},\qquad
 \|v\|_{X_u} \le Ku,\qquad0<u\le u_{\rm red}.
\ee
The reference estimate follows from its explicit profiles; the correction ball and its positivity are those of Proposition~\ref{f:family}. Appendix~\ref{sec:quantitative} specifies these constants and takes $u_{\rm red}=\min(u_D,u_{\rm fp})$. Choose a coefficient constant $C_Q>0$ such that, on this entire positive metric ball,
\be\label{q:vector-coefficient-comparison}
 \|Q_g-Q_{G_u}\|_{X_u^V\to Y_u^V} \le C_Q(D_{\rm ref}u^{\eta_{\rm geom}}+Ku^{1+\gamma}).
\ee
Indeed the formula for $Q_g$ uses the metric and its coordinate derivatives through order two, and its source $C^{2,\alpha}$ norm uses coordinate derivatives of the metric through order four. Write $Q_g-Q_{G_u}=\sum_{j=0}^2A_jD^j$ in a scaled regular chart and put $\varepsilon_u=D_{\rm ref}u^{\eta_{\rm geom}}+Ku^{1+\gamma}$. The inverse-metric and connection formulas, applied to the positive metric ball, give
\be
 \sum_{j=0}^2\omega_u^{2-j}\|A_j\|_{C^{2,\alpha}_{\omega_u}}
 \le C\varepsilon_u.
\ee
The $X_u$ bound gives the second power in $\varepsilon_u$, since $-\tfrac12<\gamma<0$. A one-form in $X_u^V$ has scaled derivatives $|D^j\xi|\le C\omega_u^{\gamma-j}\|\xi\|_{X_u^V}$; multiplying by the source weight $\omega_u^{2-\gamma}$ cancels the powers of $\omega_u$. The scaled H\"older product estimate gives the same bound for the source seminorm, including balls in the normalized horizon charts. All order-four derivative, frame, covering and relative metric comparisons are included in $C_Q$. This is a finite coefficient bound on the known reference and correction ball; it uses no inverse for the unknown metric.

\begin{proposition}[Vanishing of the DeTurck vector field for small masses]\label{q:vector-recovery-power}
Fix $0<\theta_V<1$ and define
\be\label{q:gauge-power-threshold}
 u_{g,\rm pow}\coloneqq\min\bigl(u_{\rm red},u_{J,V},
 \left(\frac{\theta_V}{A_QC_Q(D_{\rm ref}+K)}\right)^{1/\eta_{\rm geom}}\bigr).
\ee
For $0<u\le u_{g,\rm pow}$, every solution of the reduced equation with a zero coefficient of the auxiliary source in the stated fixed-point ball has vanishing DeTurck vector field and satisfies the static Einstein equations.
\end{proposition}
The power criterion proves the injectivity needed for recovery. The preceding energy estimates also describe the harmonic-vector form and its $H^1$ control.
\begin{proof}
The difference bound is at most $C_Q(D_{\rm ref}+K)u^{\eta_{\rm geom}}$ because $u\le1$. Thus
$\|Q_{G_u}^{-1}(Q_g-Q_{G_u})\|\le\theta_V<1$, and the Neumann series gives an inverse for $Q_g$. The normalized horizon charts and symmetries give the same one-form domain for $g$ and $G_u$. For fixed positive $u$ the weight is positive on the compact completed domain, so the metric dual $W^\flat$ of the DeTurck vector field belongs to $X_u^V$. At a zero of the scalar obstruction, the reduced equation is $2(\Ric_g-\Lambda g)=\mathcal L_Wg$. Applying $\beta_g$ and using the contracted Bianchi identity gives
\be
 0=\beta_g(\mathcal L_Wg)
   =(\nabla_g^*\nabla_g-\Ric_g)W^\flat
   =Q_gW^\flat.
\ee
This identity holds for the corrected metric without assuming that it is Einstein; the Ricci endomorphism remains that of $g$. Injectivity gives $W=0$, proving the claim.\qedhere
\end{proof}

\section{Prescribing the horizon areas and completing the construction}\label{sec:areas}

\subsection{The two equations for the horizon areas}

We complete the proof of Theorem~\ref{prob:existence}. Fix the target central mass $M_0$ strictly inside the subextremal interval and a target small mass $\varepsilon>0$. Use an auxiliary central mass $M$ near $M_0$ and an auxiliary small mass $u$ near $\varepsilon$. Proposition~\ref{f:family} and Proposition~\ref{f:force}, followed by Proposition~\ref{q:vector-recovery-power}, give a smooth static solution at its force zero $r=r_u(M)$, with correction bounded by $Cu$ in $X_u$. The zero is simple and depends continuously on positive $u$ and nearby $M$. All estimates are uniform for $M$ in a compact interval about $M_0$ and $u=s\varepsilon$ with $s$ in a fixed compact positive interval. The fixed weight satisfies~\eqref{eq:free-scale-conditions}.

Write $A_K(M)\coloneqq4\pi r_b(M)^2$ for the central Schwarzschild--de Sitter area, and let $A_b(M,u)$ and $A_s(M,u)$ be the central and one small horizon area in the recovered static family. The two small areas are equal by reflection.

\begin{lemma}[The geometric area map]\label{n:areas}
The constructed metrics obey
\be\label{n:area-estimates}
 A_b(M,u)=A_K(M)+O(u),\qquad
 A_s(M,u)=16\pi u^2(1+O(u^{1+\gamma})).
\ee
Both error estimates are uniform in the indicated compact parameter sets. Moreover one has 
\be\label{n:area-derivative}
 A_K'(M)=\frac{16\pi r_b(M)}{1-\Lambda r_b(M)^2}>0.
\ee
\end{lemma}
\begin{proof}
At the central horizon the reference is $G_K(M)+uH_r$, with $H_r$ bounded in its regular fixed-opening lifts, uniformly in $M,r$. The weight there is comparable to one, so the correction $v$ and its tangential restriction are $O(u)$. Smooth dependence of the area density on a positive metric proves the first estimate.
On the other hand, on a unit rescaled small horizon, the inner reference is $G_S+u a(r)A_S$; its tangential metric differs from the Schwarzschild metric by $O(u)$. The correction has relative scaled $C^2$ size at most $C u^\gamma\|v\|_{X_u} \le Cu^{1+\gamma}$. Since $\gamma<0$, the latter bound also includes the reference perturbation for $u<1$. The rescaled horizon area is therefore $16\pi(1+O(u^{1+\gamma}))$. Restoring spatial length scale $u$ gives the second formula. Finally differentiate $F_M(r_b(M))=0$ to get $r_b'=2/(1-\Lambda r_b^2)>0$, and hence~\eqref{n:area-derivative}.\qedhere
\end{proof}

\begin{proposition}[Exact central and small horizon areas]\label{n:retune}
For every sufficiently small target $\varepsilon>0$ there are construction parameters $M_\varepsilon,u_\varepsilon$ such that
\be\label{n:exact}
 A_b(M_\varepsilon,u_\varepsilon)=4\pi r_b(M_0)^2,\qquad
 A_s(M_\varepsilon,u_\varepsilon)=16\pi\varepsilon^2.
\ee
They satisfy
\be\label{n:parameters}
 M_\varepsilon-M_0=O(\varepsilon),\qquad
 u_\varepsilon/\varepsilon=1+O(\varepsilon^{1+\gamma}).
\ee
No differentiability at $u=0$ is assumed (or needed).
\end{proposition}

The conclusion is existence of parameters for each target $\varepsilon$. This topological argument by itself does not select a unique parameter pair or a continuous choice as $\varepsilon$ varies. Uniform convergence of every such choice is enough for the stated existence and limiting geometry. Any claim of a smooth one-parameter family would require a derivative estimate for the area map and must be stated separately.

\begin{proof}
We fix the mass interval $J=[M_-,M_+]$ already chosen around $M_0$ and choose an independent scale margin $0<d<1$. On the rectangle $J\times[1-d,1+d]$, we define
\be
 F_1(M,s)\coloneqq A_b(M,s\varepsilon)-A_K(M_0),\qquad
 F_2(M,s)\coloneqq\frac{A_s(M,s\varepsilon)}{16\pi\varepsilon^2}-1.
\ee
These functions are continuous. By Lemma~\ref{n:areas}, $F_1=A_K(M)-A_K(M_0)+O(\varepsilon)$ and $F_2=s^2-1+O(\varepsilon^{1+\gamma})$, uniformly on the rectangle. Strict monotonicity of $A_K$ makes $F_1$ negative on its lower mass face and positive on its upper mass face, for small $\varepsilon$. The second function is negative at $s=1-d$ and positive at $s=1+d$.

For completeness, the face-sign version of Brouwer's theorem~\cite[Chapter~11]{GT2001} gives a simultaneous zero as follows. Let $\pi$ be the coordinatewise projection to the rectangle. The continuous map $(M,s)\mapsto\pi((M,s)-t(F_1,F_2))$, for any fixed $t>0$, has a fixed point. A fixed point cannot lie on a face, because that face's corresponding strict sign moves its coordinate towards the interior. At an interior fixed point the projection can fix that coordinate only if the corresponding function is zero. Thus $F_1=F_2=0$, proving~\eqref{n:exact} with $u_\varepsilon=s_\varepsilon\varepsilon$.

The first zero and~\eqref{n:area-estimates}, using the positive lower bound for $A_K'$ on the compact interval, give $|M_\varepsilon-M_0|\le C\varepsilon$. The second zero gives $s_\varepsilon^2=1+O(\varepsilon^{1+\gamma})$. Since $s_\varepsilon$ stays positive and bounded, it gives the second estimate in~\eqref{n:parameters}.\qedhere
\end{proof}

\subsection{Normalization of the lapse and smooth local limits}\label{sec:bridge-local-limit}

Fix an equatorial point $q_0$ on the maximal sphere of the target background $M_0$, using a smooth identification of the intact regions for nearby $M$. This point remains away from both insertions and horizons. The area-normalized static lapse is positive there and converges to $N_{M_0}(q_0)>0$. Multiply it by the positive constant
\be
 c_\varepsilon\coloneqq N_{M_0}(q_0)/N_{\mathrm{constructed}}(q_0).
\ee
This preserves all static equations, spatial areas, axis regularity and horizon nondegeneracy. It fixes the marked lapse exactly. Since $M_\varepsilon-M_0=O(\varepsilon)$ and the outer metric correction is $O(u_\varepsilon)$, this factor is $1+O(\varepsilon)$ in the normalization where the construction starts with lapse factor one.

On each fixed compact set away from the insertion points, including the two original horizon collars, use the fixed-domain identifications and the normalized Cartesian lifts of Proposition~\ref{reg:bootstrap}. The weight is bounded below on this set. The exterior reference converges in each fixed $C^j$ norm by Proposition~\ref{green:global}, and Proposition~\ref{f:family} gives an $O(u_\varepsilon)$ difference from that reference in $C^{k+2,\alpha}$. On a fixed collar in the unit Schwarzschild coordinates, the normalized lift of the reference differs from $G_S$ by $O(u_\varepsilon)$ in every fixed $C^j$ norm. In these coordinates the weight is comparable to $u_\varepsilon$, so the correction after rescaling has $C^{k+2,\alpha}$ norm $O(u_\varepsilon^{1+\gamma})$. The matching annuli lie outside the chosen sets in both comparisons for small enough $\varepsilon$.

In these lifts, the equation~\eqref{reg:coordinate-equation} has a vanishing auxiliary source once the insertion radius is chosen. In the rescaled lifts the Einstein constant is $u_\varepsilon^2\Lambda$. The initial convergence to a positive model gives uniform ellipticity on each fixed chart, and the references have uniform bounds at every fixed order there. Applying the interior Schauder estimates on smaller lifted charts gives uniform bounds of all orders. Compactness in each lower-order norm and the already determined $C^{k+2}$ limit give smooth convergence, also at the rotation centers. Spatial restriction gives convergence of $h$, including in the half-ball charts at the horizon poles. At an original horizon $H$, write $\vartheta=\kappa_H(M_\varepsilon)\tau$ for the normalized angle and $(x_1,x_2)$ for its Cartesian rotation coordinates. If $g_\varepsilon$ is the lifted metric before the final lapse normalization, then on $x_2=0$, $x_1\geq0$, the lapse is $c_\varepsilon\kappa_H(M_\varepsilon)x_1\sqrt{(g_\varepsilon)_{22}}$, since $\del_\vartheta=x_1\del_{x_2}$ there. The coefficient under the square root is smooth and uniformly positive, and $c_\varepsilon\to1$ and $\kappa_H(M_\varepsilon)\to\kappa_H(M_0)$. This proves smooth convergence of the lapse up to the original horizons. At each new horizon, $u_\varepsilon/\varepsilon\to1$ gives the same unit-mass spatial Schwarzschild limit when we rescale by $\varepsilon^{-2}$.

The positive static metric has the original ball-minus-three-balls topology. Proposition~\ref{reg:bootstrap} gives its smoothness at all four horizons, on the axis, at the eight poles and across the interpolation interfaces. Lemma~\ref{reg:static-horizon-expansion} gives its local static horizon expansion. Regularity in the joint normalized horizon and axial Cartesian charts gives angle $2\pi$ on the whole open axis and at every horizon pole. The square root of the auxiliary circular coefficient is positive in the interior and has the form $\kappa s$ times a positive smooth function of $s^2$ near a horizon. It therefore defines the positive lapse with its regular simple zero. These facts prove the geometric, area and limit assertions of Theorem~\ref{prob:existence}. The sufficient mass condition in terms of the model constants is stated next.

The recovered solutions are real analytic in the analytic structure determined by the Einstein equation. Indeed each smooth local Riemannian metric $G=h+N^2\dd\tau^2$ is Einstein, so~\cite[Theorem~5.2]{LS-DeTurckKazdan} makes its coefficients analytic in harmonic coordinates. Its Killing field and the reflection $\tau\mapsto-\tau$ are analytic: their local expressions follow from Killing transport and from the analytic exponential map, respectively. The fixed spatial hypersurface of this reflection is therefore analytic, and its induced metric is $h$. In the positive-lapse interior, $N=\sqrt{G(\partial_\tau,\partial_\tau)}$ is analytic. At a horizon, use its analytic normal exponential map within the spatial slice to define signed Gaussian distance $s$. The circle action fixes that horizon and has nonzero linear rotation there, so its squared Killing norm is $s^2a(s^2,y)$, where $a$ is analytic and $a(0,y)=\kappa^2>0$. Consequently $N=s\sqrt{a(s^2,y)}$ on the spatial side $s\ge0$, which is analytic up to the horizon. The same Cartesian construction applies at the axis and at its intersections with the horizons. These local analytic structures agree on overlaps, since isometries between analytic Riemannian metrics are analytic. The argument concerns the recovered Einstein metrics; no analyticity of the intermediate cutoff references is required.

Section~\ref{sec:critical-maximal} determines how the maximum of the lapse changes under this construction.

\subsection{A sufficient mass bound in terms of geometric and functional constants}\label{sec:functional-mass-bound}

We collect the constants entering the mass threshold in cosmological units $\ell=1$; restoring units multiplies the mass by $\ell$. Fix $k=2$, $0<\alpha<1$, $-\tfrac12<\gamma<0$, the compact mass interval $J$ and the placement intervals $I_M$ in~\eqref{q:evaluated-mass-interval}--\eqref{q:uniform-location}. The charts, cutoffs and overlap exponents are fixed before the inserted mass. The tensor model constants are
\bse\label{q:functional-model-constants}
\be
 C_S\ge\|\mathcal A_S^{-1}\|_{Y_S\to X_S^0\times\mathbb R},\qquad
 \mathcal A_S(T,\lambda)\coloneqq P_ST-\lambda\chi_0T_S,
\ee
\be
 C_K\ge\sup_{M\in J,\ r\in I_M}
 \|P_{G_M}^{-1}\|_{Y_{K,\bullet}\to X_{K,\bullet}}.
\ee
\ese
Here $X_S^0$ is the domain satisfying the integral condition, with product norm $\|T\|_{X_S}+|\lambda|$. The norms~\eqref{q:model-Holder-norms} and~\eqref{q:punctured-Kottler-norms} specify both fixed model problems. The two inverse estimates give finite bounds $C_S,C_K$ in these weighted norms.

The other constants are defined in the estimates where they are first used:
\begin{itemize}
\item The invariant Poincar\'e and Korn constants are those of Corollary~\ref{q:functional-Poincare-Korn}. Lemma~\ref{q:Hardy} gives the optimal model Hardy constant $4$, with the smooth-scale comparison~\eqref{q:regular-weight-Hardy}. The vector inverse operators use the domains of Lemma~\ref{q:vector-model-inverses}.
\item The chart, cutoff, rescaling, multiplication and covering norms are fixed in~\eqref{q:evaluated-identification-constants}. The affine and Green profile norms are given by~\eqref{q:affine-profile-constant}, \eqref{green:CH-conditions} and~\eqref{green:outside-constant}.
\item The bounds $D_{\rm ref},D_L,E,E_1,B_Q,L_1,B_1$ are defined by~\eqref{q:reference-constant-conditions}--\eqref{q:evaluated-parameter-derivative-constants}. They include the derivative of the augmented operator in fixed normalized coordinates.
\item The force pairings, vector coefficient comparison and horizon area bounds are specified in~\eqref{q:evaluated-force-pairing-constants}, \eqref{q:vector-coefficient-comparison} and~\eqref{q:evaluated-area-constants}.
\end{itemize}
The inverse and source bounds fix $K$. The nonlinear coefficient bounds use the whole ball of radius $Ku$ of positive metrics.

\begin{proposition}[An explicit upper bound for the masses allowed by the construction]\label{q:functional-mass-threshold}
Use the constants from the inverse constructions and Appendix~\ref{sec:quantitative}. The margins $\theta_{J,\rm pow},\theta_{J,V},\theta_V$ and $q$ lie in $(0,1)$. With the thresholds $u_J,u_D,u_{\rm fp},u_F,u_g,u_A$ and the area margin $\Delta_A$ defined there, set, for $0<d<1$,
\be\label{q:master-threshold}
\varepsilon_{\rm suff}(\mathbf c)\coloneqq\min\bigl(\frac{\min(u_J,u_D,u_{\rm fp},u_F,u_g,u_A)}{1+d},\frac{\theta_A\Delta_A}{C_b(1+d)},\left(\frac{\theta_A d(2-d)}{C_s(1+d)^{3+\gamma}}\right)^{1/(1+\gamma)}\bigr)>0.
\ee
Then Theorem~\ref{prob:existence} holds with
\be\label{q:functional-mass-bound}
 \mu_0(M,\Lambda)=\ell\,\varepsilon_{\rm suff}(\mathbf c),\qquad \ell=\sqrt{3/\Lambda},
\ee
where the collection $\mathbf c$ consists of the constants and auxiliary choices fixed above, all evaluated in cosmological units at the target ratio $M/\ell$ and the fixed interval containing that mass.
\end{proposition}
\begin{proof}
Appendix~\ref{sec:quantitative} gives the fixed model bounds and the six restrictions on the auxiliary mass. We recall their order of choice. The model inverse and geometric map norms give $u_J$ and $A_J$. The affine and Green profile bounds give $D_{\rm ref},D_L,u_D$ and $A$. The explicit reference error and the local nonlinear formula then give $E,E_1,B_Q,L_1,B_1$, after which we fix $K$ and $K_1$. The test integrals determine $C_F$ in~\eqref{q:force-constant} and hence $u_F$ in~\eqref{q:force-threshold}. The vector model inverse operators and metric coefficient comparisons determine $u_g$. The horizon estimates determine $u_A,C_s,C_b$. Every constant is finite by the corresponding estimate, and every margin or denominator used in a threshold is positive.

For an auxiliary mass $u=s\varepsilon$, $1-d\le s\le1+d$, the first entry of~\eqref{q:master-threshold} ensures simultaneously the actual differential inverse, invariance and contraction, the force equation, the vanishing of the DeTurck vector field, and positivity of the horizon metrics. The last two entries give the required signs on the two pairs of faces for the area constraints. The finite-dimensional adjustment therefore prescribes the target central and small horizon areas. At this stage only the fixed initial H\"older order is used; smoothness at all horizons, their poles, the axis and every interior point follows at each fixed positive mass from Proposition~\ref{reg:bootstrap}. It introduces no further mass restriction at higher differentiability orders. This completes the proof of Proposition~\ref{q:functional-mass-threshold}.\qedhere
\end{proof}

\section{The equatorial circle of maximal lapse}\label{sec:critical-maximal}

\subsection{Localization of the maxima}

The lapse attains its maximum in the interior, since it is positive there and zero on the horizons. Every maximum away from the axis belongs to a circle of maxima. We first show that all maxima lie on a radial graph away from the poles. We then determine the maximizing angle by computing the first angular term on this graph.
In this section, we adapt the maximum-principle argument for the radial monotonicity
of the heat kernel~\cite{CheegerYau1981,AlonsoOranChamizoMartinezMas2023}
to the sum of two kernels centered at antipodal poles.

\begin{lemma}[Exclusion of fixed insertion neighborhoods]\label{qual:maximum-away}
There is a fixed $\delta>0$ such that, for all sufficiently small masses, no global maximum of the normalized lapse lies in either insertion ball $\rho_\pm\le\delta$.
\end{lemma}
\begin{proof}
Write $u$ for the auxiliary small mass. The area adjustment gives $u\asymp\mu$, and the lapse is normalized by $N(q_0)=N_*$. Its normalization factor and the change of central mass contribute $O(u)$ to the squared background maximal lapse. All constants below are fixed on the original compact parameter sets before $\varepsilon_b$ is chosen.

At the insertion radius where the coefficient of the auxiliary source vanishes, $|a(r_u)|\le Cu^\delta$. On $\rho\le2u^{a_2}$, the relative affine correction is therefore at most $Cu^{a_2+\delta}$, and the correction obtained by solving the nonlinear equation has relative size at most $Cu^{1+\gamma}$. The Schwarzschild lapse has squared deficit at least $cu^{1-a_2}$ from its asymptotic value on $u/2\le\rho\le2u^{a_2}$. On the interpolation annulus, the exterior squared lapse has the same deficit by the negative leading term of $H_{\tau\tau}$. These deficits dominate the errors because $a_2+\delta>1-a_2$, $1+\gamma>1-a_2$ and $1>1-a_2$. The first two inequalities follow from~\eqref{eq:free-scale-conditions} and~\eqref{eq:error-gains}. Interpolation of the covariant metrics interpolates their $\tau\tau$ components, so the deficit persists throughout this region. At the horizon these comparisons are interpreted as $|v(K,K)|\le|v|_G G(K,K)$ for the static Killing field $K$; no division by its vanishing norm is used.

For $2u^{a_2} \le\rho\le\varepsilon_b$, the reference equals $G_K+uH$. The coefficient $H_{\tau\tau}$ of the Green tensor is $-2b^2/\rho+O(1)$, while the correction is bounded by $Cu\rho^\gamma$ in this component. Since $b$ is uniformly positive, comparison with the target maximal value gives
\be
 N^2\le N_*^2-\frac{cu}{\rho}+Cu\rho^\gamma+Cu.
\ee
Choose $\varepsilon_b$ so small that $C\rho^{1+\gamma}+C\rho\le c/2$ for $0<\rho\le\varepsilon_b$. This proves $N^2<N_*^2$ on that annulus. The same strict inequality holds in the inner region for small $u$, as shown above. Since $N(q_0)=N_*$, neither ball contains a global maximum. Reflection gives the conclusion for both insertions.
\qedhere
\end{proof}

\begin{lemma}[Locating the maxima where the radial derivative vanishes]\label{qual:radial-reduction}
There are fixed $\theta_0\in(0,\pi/2)$ and a radial collar of the background maximal sphere such that every global maximum lies on a unique smooth graph $x=X_\mu(\theta)$, with $\theta_0<\theta<\pi-\theta_0$. Here $x$ is signed background radial distance, and $X_\mu=O(\mu)$. The maximal set is the union of the axial orbits corresponding to the maximizers of
\be
 f_\mu(\theta)=N_\mu(X_\mu(\theta),\theta).
\ee
Its maximizing angles form a nonempty compact subset of $(\theta_0,\pi-\theta_0)$.
\end{lemma}

Equatorial reflection gives $\partial_\theta N_\mu(x,\pi/2)=0$. Together with the radial critical equation on the graph of Lemma~\ref{qual:radial-reduction} and axial invariance, this shows that the orbit at $x=X_\mu(\pi/2)$ is a circle of critical points. Its background distance from the maximal sphere is $O(\mu)$ and its radial second variation is strictly negative. The angular calculation below determines its polar second variation.

\begin{proof}
Lemma~\ref{qual:maximum-away} excludes fixed neighborhoods of the insertion points. On their complement the lapse converges uniformly to the background lapse, including the original horizon collars. Since its value at $q_0$ is $N_*$, every global maximum must approach the background maximal sphere. Combining this localization with the excluded polar neighborhoods places all maxima in a fixed angular interval away from the axis.

Choose $\theta_0$ smaller than this angular separation, so that the points of the background maximal sphere at $\theta_0$ and $\pi-\theta_0$ lie strictly inside the excluded neighborhoods. On a fixed radial collar with these angular bounds, the reference and correction estimates give $N_\mu-N_M=O(\mu)$ in $C^2$. The background radial second derivative is strictly negative on a sufficiently small collar, and its radial first derivative has opposite signs at the two ends. The same statements hold for $N_\mu$ for small $\mu$, uniformly in $\theta$. The implicit function theorem and strict radial monotonicity therefore give a unique smooth solution $X_\mu(\theta)$ of $\partial_xN_\mu=0$, with $|X_\mu|\le C\mu$.

Every global maximum is an interior critical point and hence lies on this graph. Conversely, maximizing $f_\mu$ gives the global maximal value, since the graph contains all global maxima. Its two endpoints remain inside the excluded neighborhoods for small $\mu$, so neither is maximizing. Compactness gives the stated conclusion. Reflection preserves the graph and the reduced function.
\qedhere
\end{proof}

\subsection{Angular monotonicity from the Green function}

\paragraph{The point sources at the maximal sphere.} We now compute the first angular variation on the graph on which the radial derivative of the lapse vanishes. All estimates in this subsection are for a fixed strictly subextremal background; their constants need not remain bounded at the endpoints of the subextremal mass interval. Write $F=F_{M_0}$, $r_*=r_*(M_0)$ and $b=\sqrt{F(r_*)}=N_*$. Let $H_*$ be the paired Green tensor of Lemma~\ref{f:Green}, with its poles at the north and south points of the background maximal sphere.

\begin{lemma}[The Green tensor at the maximal sphere]\label{qual:green-Bianchi-zero}
The tensor $H_*$ satisfies $\beta_KH_*=0$ distributionally. The same identity holds for each background in the fixed compact interval of subextremal masses, at its own maximal radius. At the insertion radius where the coefficient of the auxiliary source vanishes, the correction in the auxiliary mass parameter $u$ satisfies
\be\label{qual:improved-correction}
 \|v_{u,r_u}\|_{X_u} \le C u^{1+\eta}.
\ee
In particular, on fixed compact sets away from the two poles, its contribution to the metric and lapse is $o(u)$ in $C^4$.
\end{lemma}
\begin{proof}
Trace reversal sends $D_p$ to $-2\,\dd t^2$. The trace-reversed point source is therefore supported on the static orbit through $p$ and tangent to that orbit in both indices. The divergence of such an invariant point source vanishes precisely when the orbit has zero acceleration. To check this assertion directly, test its divergence against a smooth invariant meridional field $Z$. The only contraction is with
\be
 (\mathcal B_K Z)_{\tau\tau}=N_K\langle\nabla N_K,Z\rangle_{h_K}.
\ee
This is zero at either pole because $\nabla N_K=0$ on the maximal sphere. General test fields reduce to invariant ones by averaging; the derivative in $\tau$ integrates to zero over a local static period. Thus $\beta_K(P_KH_*)=0$ as a distribution. The identity~\eqref{eq:tensor-bianchi} gives $Q_K\beta_KH_*=0$. Elliptic regularity makes this distribution smooth across the poles. It is regular at the original horizons, so Corollary~\ref{cor:harmonic-gauge} gives $\beta_KH_*=0$.

The placement derivative bounds in Lemma~\ref{f:Green}, applied also to the smoothly varying Bianchi operator, now give
\be
 |D^j\mathcal B_K\beta_KH_r|\le C_j|r-r_*|\rho^{-2-j}
\ee
in moving insertion charts, with the corresponding fixed-chart bound away from the poles. In the proof of Proposition~\ref{f:source}, the only term of order $u$ was this exterior Bianchi contribution. Its source norm is now bounded by $Cu|r-r_*|$. The inner remainder, the outer quadratic term and the matching error have source norms bounded by $Cu^{1+\eta_{\rm src}} \le Cu^{1+\eta}$; the inner term proportional to $a(r)$ obeys the same bound on the fixed placement interval. Consequently
\be
 \|f_{u,r_u}\|_{Y_u} \le C(u|r_u-r_*|+u^{1+\eta})\le Cu^{1+\eta},
\ee
because Proposition~\ref{f:force} gives $|r_u-r_*|\le Cu^\delta$ and $\eta<\delta$. The correction equation and its uniform augmented inverse give
\be
 \|v_{u,r_u}\|_{X_u} \le C\|f_{u,r_u}\|_{Y_u}
       +Cu^\gamma\|v_{u,r_u}\|_{X_u}^{2}.
\ee
The previously proved $O(u)$ bound absorbs the second term for small $u$, proving~\eqref{qual:improved-correction}. On a fixed compact set the weight is bounded below, which proves the last assertion.
\qedhere
\end{proof}

\paragraph{A scalar Green function.}
On $(r_b,r_c)$ define
\be\label{qual:angular-radial-operator}
 p(r)\coloneqq r^2F(r),\qquad V(r)\coloneqq2\Lambda r^2+\frac{r^2F'(r)^2}{F(r)},\qquad
 \mathcal L_{\rm rad}\coloneqq-\frac{\dd}{\dd r} \left(p\frac{\dd}{\dd r}\right)+V.
\ee
Work in $L^2((r_b,r_c),\dd r)$. The domain of the quadratic form is the completion of $C_c^\infty((r_b,r_c))$ for the norm with square $\int(p|w'|^2+(V+1)|w|^2)\dd r$. The operator associated with this form acts on those $w$ in the form domain for which there is an $f\in L^2$ satisfying $\int(pw'\overline{\phi'}+Vw\bar\phi)\dd r=\int f\bar\phi\dd r$ for every $\phi$ in the form domain; its value at $w$ is $f$. For a source supported away from the endpoints, the resolvent therefore selects the regular solution $w=O(F)$ near either endpoint. The other local solution has order $F^{-1}$ and infinite form energy. The form satisfies $\mathcal L_{\rm rad}\ge2\Lambda r_b^2>0$. To justify the heat kernel bounds at the interior point used below, set $x(r)\coloneqq\int_{r_b}^r p(s)^{-1/2}\dd s$, which ranges over a finite interval $(0,L)$. The unitary change $w\mapsto p^{1/4}w$ transforms $\mathcal L_{\rm rad}$ into
\be
 -\partial_x^2+W(x),\qquad
 W(x(r))\coloneqq V(r)+\frac{p''(r)}4-\frac{p'(r)^2}{16p(r)}.
\ee
This follows by differentiating $w=p^{-1/4}v(x)$. Near either endpoint, if $d$ is its $x$-distance, then $W=15/(4d^2)+O(1)$. Indeed $p=a(r-r_b)+O((r-r_b)^2)$ and $V=a/(r-r_b)+O(1)$ at the left endpoint, while $x=2\sqrt{(r-r_b)/a}+O((r-r_b)^{3/2})$; the right endpoint is identical with reversed distance. Thus $W$ is bounded below. Scalar parabolic comparison with the free Dirichlet heat equation gives the local diagonal bound $Ct^{-1/2}$ for $0<t\le1$, after undoing the unitary change at a fixed interior point. The form lower bound and the semigroup identity give exponential decay there for $t\ge1$.

Let $k_t(r,s)$ denote its heat kernel. Let $K_t(\theta)$ denote the heat kernel on the unit round sphere with source at the north pole, normalized to have integral one. Define, away from the poles,
\be\label{qual:angular-profile}
 \Phi(\theta)\coloneqq-4\pi b^2\int_0^\infty
       k_t(r_*,r_*)(K_t(\theta)+K_t(\pi-\theta))\dd t.
\ee
This integral is the Green function of $\mathcal L_{\rm rad}-\Delta_{\mathbb S^2}$, evaluated at $r=r_*$, with two polar sources and the indicated coefficient. It is smooth on $0<\theta<\pi$. Near $t=0$, the angular heat kernel and each of its derivatives on a closed subinterval of $(0,\pi)$ have an exponentially small factor, while $k_t(r_*,r_*)=O(t^{-1/2})$. At large $t$ the positive lower bound for $\mathcal L_{\rm rad}$ gives exponential decay. These bounds justify differentiating~\eqref{qual:angular-profile} on every such subinterval.

\begin{lemma}[Strict angular monotonicity of the scalar profile]\label{qual:angular-profile-sign}
The function $\Phi$ is symmetric about $\pi/2$ and satisfies
\be\label{qual:angular-profile-signs}
 \Phi'(\theta)>0\quad(0<\theta<\pi/2),\qquad
 \Phi''(\pi/2)<0.
\ee
\end{lemma}
\begin{proof}
Put $U_t(\theta)\coloneqq K_t(\theta)+K_t(\pi-\theta)$. This is the spherical heat solution with equal unit point sources at the two poles. We use the derivative maximum-principle argument for heat-kernel monotonicity, as in~\cite{AlonsoOranChamizoMartinezMas2023}, and give its antipodal version here. Its derivative $w=\partial_\theta U_t$ satisfies
\be
 \partial_t w=\partial_\theta^2 w+\cot\theta\,\partial_\theta w
              -\frac{w}{\sin^2\theta},\qquad
 w(t,0)=w(t,\pi/2)=0.
\ee
Approximate the initial pair of point masses by positive smooth functions proportional to $\exp(\cos^2\theta/\varepsilon)$, normalized to have total integral two. Their derivatives are nonpositive on $[0,\pi/2]$. The parabolic maximum principle preserves this sign. More explicitly, if $w_+\coloneqq\max(w,0)$, multiplication by $w_+\sin\theta$ and integration give
\be
 \frac12\frac{\dd}{\dd t}\int_0^{\pi/2}w_+^2\sin\theta\,\dd\theta
 =-\int_0^{\pi/2}|\partial_\theta w_+|^2\sin\theta\,\dd\theta
  -\int_0^{\pi/2}\frac{w_+^2}{\sin\theta}\,\dd\theta\le0.
\ee
Smooth spherical symmetry gives $w=O(\theta)$ at the pole and removes the endpoint term. The initial positive part is zero. Passing to the heat kernel gives $w\le0$ for $t>0$.
The degree-two spherical harmonic coefficient of $U_t$ is nonzero at every finite positive time, since it is its nonzero initial coefficient multiplied by $e^{-6t}$. Thus $w$ is not identically zero. The strong maximum principle gives $w<0$ for $t>0$ and $0<\theta<\pi/2$. The boundary point lemma at $\theta=\pi/2$ gives $\partial_\theta w(t,\pi/2)>0$. The coefficients are smooth near this endpoint.

The radial heat kernel is strictly positive at $(r_*,r_*)$ for $t>0$. For completeness, its nonnegative sign follows by applying the scalar parabolic maximum principle to the Dirichlet problems on compact subintervals and letting the compact subintervals increase to $(r_b,r_c)$ in the closed form; strict positivity in the interior follows from the strong maximum principle on the connected interval. Differentiating~\eqref{qual:angular-profile} therefore gives both strict inequalities in~\eqref{qual:angular-profile-signs}.
\qedhere
\end{proof}

\subsection{The first angular variation and the circle of maxima}

\begin{proposition}[The first angular term of the lapse where its radial derivative vanishes]\label{qual:angular-first-term}
Let $f_\mu$ be the lapse on the graph on which the radial derivative of the lapse vanishes in Lemma~\ref{qual:radial-reduction}, with the prescribed small area parameter $\mu$ and normalization $N_\mu(q_0)=N_*$. On every fixed angular interval used in that lemma,
\be\label{qual:angular-expansion}
 f_\mu(\theta)=N_*+\mu(\Phi(\theta)-\Phi(\pi/2))
                     +O_{C^2}(\mu^{1+\eta}).
\ee
This conclusion holds for every area-prescribing choice furnished by Proposition~\ref{n:retune}; no differentiability of that choice at $\mu=0$ is assumed.
\end{proposition}

\begin{proof}
\bse
\emph{1. The angular dependence of the Green tensor for the linearized Einstein equation.}
Lemma~\ref{qual:green-Bianchi-zero} shows that $H_*$ satisfies the linearized Einstein equation with the two point sources
\be
 2D(\Ric-\Lambda G)_{G_K}[H_*]
       =8\pi\sum_\pm D_{p_\pm}\delta_{h,p_\pm}.
\ee
For each degree $l\ge2$, we use the gauge of Proposition~\ref{prop:physical-kernel} in radial distributions. At the exact background, covariance gives
\be
 D(\Ric-\Lambda g)_{G_K}[\mathcal L_XG_K]=0.
\ee
This is an identity of differential operators with smooth coefficients on the interior, so it holds on distributions and the change of gauge preserves the point sources. The spatial block of each source is proportional to $h_K$. Its angular trace-free and radial--angular projections therefore vanish, and the corresponding equations give $H_{0,l}=H_{2,l}=H_l$ and $B_l'-H_l'=F'H_l/F$ in distributions. The $\tau\tau$ component of the equation has the point source
\be
 D(\Delta_hN+\Lambda N)[H_*]=4\pi b\sum_\pm\delta_{h,p_\pm}.
\ee
Indeed $\Ric_{\tau\tau}=-N\Delta_hN$, so the $\tau\tau$ component of the preceding Ricci equation is $-N\,D(\Delta_hN+\Lambda N)=-4\pi b^2\sum\delta_{h,p_\pm}$. To obtain the scalar equation with the correct normalization, use the flux calculation leading to~\eqref{eq:physical-H}. The coefficient of $P_l(\cos\theta)$ in $N_K^{-1}D(\Delta_hN+\Lambda N)[H_*]$ in this gauge is
\begin{equation}
 \frac{1}{2r^2}\Bigl(\bigl(r^2F'(B_l-H_l)-pH_l'\bigr)'
                  +2\Lambda r^2B_l+l(l+1)H_l\Bigr).
\end{equation}
Using $(r^2F')'=-2\Lambda r^2$ and $B_l'-H_l'=F'H_l/F$ gives $(\mathcal L_{\rm rad}+l(l+1))H_l/(2r^2)$. Since $\dd V_{h_K}=r^2 F^{-1/2}\dd r\dd\omega$, each spatial point measure has coordinate density $b\,r_*^{-2}\delta_{r_*}\delta_{\mathbb S^2}$, with $\delta_{\mathbb S^2}$ at the corresponding pole. Multiplication by $2r^2/N_K$ thus yields
\be\label{qual:angular-radial-source}
 (\mathcal L_{\rm rad}+l(l+1))H_l
   =8\pi b\,q_l\,\delta_{r_*},\qquad
 q_l\coloneqq\frac{2l+1}{4\pi}(1+(-1)^l).
\ee
Here $P_l(\cos\theta)$ is used for the angular harmonic, and
$\delta_{\rm north}+\delta_{\rm south}=\sum_{l\ge0}q_lP_l(\cos\theta)$.
Because $p(r_*)>0$, homogeneous solutions on either side extend smoothly to $r_*$ from their own side. Any distribution supported at $r_*$ that is left after subtracting these piecewise smooth solutions is a finite sum of derivatives of the Dirac mass. The principal term $-(pH_l')'$ would send the highest such derivative to one of order two greater, which cannot occur in the equation. Thus $H_l$ is piecewise smooth, and the absence of a delta derivative in~\eqref{qual:angular-radial-source} makes it continuous at $r_*$. Integrating the equation across $r_*$ gives the flux jump
\be
 [pH_l']_{r_*}=-8\pi bq_l,
\ee
where the brackets denote right minus left limits.
\ese

\bse
The coordinate gauge is regular at the horizons and gives $H_l=O(F)$ there. It can also be used across the source radius in distributions. In the spatial normal coordinates of~\eqref{f:Green-tail}, the components of $h_K-\delta$ are $O(\rho^2)$ and their first derivatives are $O(\rho)$. Replacing $2\delta/\rho$ by $2h_K/\rho$ in the spatial block thus changes the remainder by $O(\rho)$ with bounded first derivatives. In spherical coordinates, $2h_K/\rho$ has zero angular trace-free and radial--angular components. The first derivatives of the remaining bounded term are $O(\rho^{-1})$, which controls their angular projections uniformly near the source radius. In particular, for $z=r-r_*$ and a fixed small polar cap, the possibly singular part of the projected first derivative is bounded by
\be
 C_l\int_0^{\theta_1}\frac{\theta\,\dd\theta}{\sqrt{z^2+\theta^2}} \le C_l\theta_1.
\ee
The radial--angular coefficient is also bounded. Hence $X_l^r=-F(K_l-r^2G_l'/2)Y_l$ is locally bounded for each fixed degree. Its lapse contribution $X_l^rN_K'$ tends to zero at $r_*$ because $N_K'(r_*)=0$, including when $X_l^r$ has distinct one-sided limits. Thus the gauge preserves the lapse trace despite radial derivative jumps in its coefficients. The coefficient $H_{*,\tau\tau}$ in the original coordinates is bounded by $C(1+\rho^{-1})$ near either source, where $\rho$ is the distance to that source. On nearby spheres this is bounded by a constant times one plus the reciprocal of the angular distance to the corresponding pole, which is integrable against the spherical measure. Dominated convergence gives its harmonic traces at $r_*$. We use only the individual harmonic gauges to compute these traces. The $C^2$ comparison below is made in the original charts.

On a fixed interval about $r_*$, the form norm controls the local $H^1$ norm since $p$ is bounded below there and $V\ge0$. The one-dimensional trace estimate makes evaluation at $r_*$ continuous on the form domain, so the Dirac source belongs to its dual. The projected solution $H_l$ is continuous with finite one-sided derivatives at $r_*$. At each endpoint, $p=O(F)$, $V=O(F^{-1})$, $H_l=O(F)$ and $H_l'$ is bounded. Cutting off within distance $\eta$ of the endpoints therefore gives an error $O(\eta^2)$ in the square of the form norm. Together with local $H^1$ approximation, this puts $H_l$ in the form domain. The coercive form for $\mathcal L_{\rm rad}+l(l+1)$ then gives the unique solution of~\eqref{qual:angular-radial-source} by applying its resolvent to the right side in the dual of the form domain. Its lapse component on the maximal sphere is $-bH_l(r_*)/2$. To identify the trace, first interpret the heat integral~\eqref{qual:angular-profile} as a distribution on the sphere. Indeed, pairing the sum of the two spherical heat kernels with a smooth test function $\varphi$ is bounded by $2\|\varphi\|_\infty$, and the radial diagonal is integrable in time at both endpoints. The degree-$l$ coefficient of this distribution is
\begin{equation}
 -4\pi b^2q_l\int_0^\infty k_t(r_*,r_*)e^{-tl(l+1)}\dd t .
\end{equation}
By the spectral integral for the resolvent, this is $-bH_l(r_*)/2$ for $l\ge2$. Reflection symmetry removes all odd degrees on both sides. The two traces thus differ only by the constant degree-zero mode, as distributions and as smooth functions away from the poles.

The radial coefficients can be computed without a tensor inversion. Let $u_l,v_l$ be positive solutions of $(\mathcal L_{\rm rad}+l(l+1))w=0$, regular at $r_b$ and $r_c$, respectively, and define
\be
 W_l\coloneqq p(u_l'v_l-u_lv_l'),\qquad
 g_l\coloneqq\frac{u_l(r_*)v_l(r_*)}{W_l}.
\ee
The quantity $W_l$ is constant. The equation $(pw')'=(V+l(l+1))w$ shows that $u_l'>0$ and $v_l'<0$, so $W_l>0$. The Green function formed from $u_l$ on the left and $v_l$ on the right has flux jump minus one, proving that its diagonal value is $g_l$. The global equatorial reflection restricts the spherical expansion of the profile to even degrees. We write the constant term separately.
\be
 \Phi(\theta)=-2b^2\sum_{\substack{l\ge2\\l\ {\rm even}}}(2l+1)g_lP_l(\cos\theta)+\text{constant}.
\ee
This equality is understood in distributions, or after heat regularization; the integral~\eqref{qual:angular-profile} supplies its smooth value and derivatives away from the poles.  
\ese

\bse
\par\addvspace{0.2cm}\noindent
\emph{2. The nonlinear correction and the normalizations.}
Write $u$ for the auxiliary small mass and $M$ for the auxiliary central mass. On the fixed collar away from the poles, Lemma~\ref{qual:green-Bianchi-zero} and the exterior reference give
\be
 N_{\rm constructed}=N_M+
        u\frac{H_{M,r_u,\tau\tau}}{2N_M}+O_{C^4}(u^{1+\eta}).
\ee
The Green tensor changes by $O(u^\delta)$ there when $r_u$ is replaced by $r_*(M)$. The area adjustment gives $M-M_0=O(u)$ and $u/\mu=1+O(\mu^{1+\gamma})$. The lapse normalization multiplies the expression by $1+O(u)$. Changes of $M$ in the background lapse are radial, and hence constant on its maximal sphere. Estimate~\eqref{green:mass-Lipschitz} shows that changes of $M$ in the Green term contribute $O_{C^4}(u^2)$ on this collar. The derivative bounds on the graph follow explicitly from
\be
 X_\mu'=-\frac{N_{x\theta}}{N_{xx}},\qquad
 X_\mu''=-\frac{N_{x\theta\theta}+2N_{xx\theta}X_\mu'+N_{xxx}(X_\mu')^2}{N_{xx}},
\ee
with the derivatives evaluated at $x=X_\mu(\theta)$. The denominator is bounded away from zero, while the background has no angular dependence. The fixed $C^4$ comparison thus gives $\|X_\mu\|_{C^2}=O(u)$. Moving from the background maximal sphere to this graph contributes $O_{C^2}(u^2)$: the background radial derivative vanishes there, its displacement is $O(u)$ in $C^2$, and the Green term already carries a factor $u$.

Consequently the reduced lapse has the form $N_*+u\Phi(\theta)+c_u+O_{C^2}(u^{1+\eta})$, where $c_u$ is independent of $\theta$. Its value on the equatorial critical graph differs from $N_\mu(q_0)=N_*$ by $O(u^2)$, since the equatorial critical point is at radial distance $O(u)$ from $q_0$ and the radial derivative at $q_0$ is $O(u)$. Subtracting this value eliminates $c_u$. Replacing $u$ by $\mu$ preserves the stated remainder and proves~\eqref{qual:angular-expansion}.
\qedhere
\ese
\end{proof}

\begin{theorem}[A unique circle of maxima with negative definite normal Hessian]\label{qual:unique-maximal-circle}
For every fixed strictly subextremal background and all sufficiently small prescribed added masses $\mu>0$, the global maximal lapse set of the constructed solution is one equatorial axial circle. The Hessian is negative definite in its two transverse directions. More precisely, if $\lambda_{\theta,\mu}$ denotes the polar eigenvalue of the spatial Hessian along that circle, then
\be\label{qual:angular-Hessian-expansion}
 \lambda_{\theta,\mu}=\frac{\mu}{r_*^2}\Phi''(\pi/2)+O(\mu^{1+\eta})<0.
\ee
The radial eigenvalue is $-\Lambda N_*+O(\mu)$, and the axial eigenvalue is zero. In particular the transverse nondegeneracy is not uniform as $\mu\to0$.
\end{theorem}
\begin{proof}
All global maxima lie on the graph where the radial derivative of the lapse vanishes, in the region away from the poles. On a sufficiently small fixed interval about $\pi/2$, Lemma~\ref{qual:angular-profile-sign} and~\eqref{qual:angular-expansion} give $f_\mu''<0$ for small $\mu$. Reflection gives $f_\mu'(\pi/2)=0$. On the compact remainder of $[\theta_0,\pi/2)$, the strict positive lower bound for $\Phi'$ and the $C^1$ expansion give $f_\mu'>0$. Reflection supplies the opposite sign on the other half. Hence the equatorial angle is the unique maximizing angle.

At this circle reflection makes the radial--polar Hessian entry zero, while axial invariance at a critical point annihilates the axial direction and its mixed entries. The radial entry is $-\Lambda N_*+O(\mu)$ by the fixed $C^2$ comparison. The polar coordinate vector has squared length $r_*^2+O(\mu)$, and reflection makes the radial graph derivative zero at the equator. Thus its polar Hessian entry is $f_\mu''(\pi/2)/(r_*^2+O(\mu))$, which gives~\eqref{qual:angular-Hessian-expansion}. Both transverse entries are strictly negative, proving the assertion. We use the orthonormal radial, polar and axial frame $(e_x,e_\theta,e_\varphi)$ along this circle and define $\rho_x\coloneqq\Ric_h(e_x,e_x)$. The static equations and $\Scal_h=2\Lambda$ give
\be
 \begin{gathered}
 \Ric_h(e_\varphi,e_\varphi)=\Lambda,\\
 \Hess N=N\,\operatorname{diag}(\rho_x-\Lambda,-\rho_x,0),\\
 \rho_x=-\frac{\mu\,\Phi''(\pi/2)}{r_*^2N_*}+O(\mu^{1+\eta})>0.
 \end{gathered}
\qedhere
\ee
\end{proof}

For each fixed sufficiently small positive mass, the local form at the maximal circle is nondegenerate. We denote its lapse value by $N_{\max,\mu}$. The metric and lapse are analytic, and the axial Killing field is analytic by Killing transport. Its flow gives local analytic quotient coordinates away from the axis. Completing the square in the radial direction leaves a one-variable analytic function with strictly negative second derivative. An analytic change in that variable therefore gives local quotient coordinates $(s,t)$ in which
\be
 N_\mu=N_{\max,\mu}-s^2-t^2.
\ee
The change of coordinates depends on the added mass $\mu$.

Figure~\ref{fig:maximal-lapse-foliation} locates the circle of maxima in the full static region. A plane through the axis meets the circle at two points. Rotating either closed level curve about the symmetry axis gives the corresponding torus.

\begin{figure}[!htbp]\centering
\begin{tikzpicture}[x=1cm,y=1cm, font=\normalsize, >=Stealth,
  outer/.style={draw=blue!60!black,line width=.8pt},
  hole/.style={draw=black!78,line width=.7pt,fill=white},
  level/.style={draw=teal!70!black,line width=.75pt},
  guide/.style={draw=black!42,densely dashed,line width=.45pt}]
  \path[use as bounding box] (0,0) rectangle (14.2,5.65);
  \begin{scope}[shift={(3.25,3.03)}]
    \fill[blue!4] (0,0) circle (2.22);
    \draw[guide] (0,-2.22)--(0,2.47);
    \draw[guide] (-2.22,0)--(2.22,0);
    \draw[outer] (0,0) circle (2.22);
    \draw[hole] (0,0) circle (.49) node {$H_b$};
    \draw[hole] (0,1.34) circle (.17);
    \draw[hole] (0,-1.34) circle (.17);
    \node[anchor=west] at (.20,1.34) {$H_+$};
    \node[anchor=west] at (.20,-1.34) {$H_-$};
    \node[anchor=east,text=black] at (-1.55,1.43) {$H_c$};
    \node[anchor=west,text=black] at (.09,2.47) {Axis};
    \foreach \x in {-1.39,1.39}{
      \foreach \rx/\ry in {.12/.23,.21/.38,.31/.54}{
        \draw[level] (\x,0) ellipse[x radius=\rx,y radius=\ry];
      }
      \fill[red!75!black] (\x,0) circle[radius=1.65pt];
    }
  \end{scope}

  \draw[guide] (5.01,3.53)--(8.45,4.61);
  \draw[guide] (5.01,2.53)--(8.45,1.47);

  \begin{scope}[shift={(10.63,3.03)}]
    \draw[->,black!60,line width=.45pt] (-2.05,0)--(2.29,0)
      node[right] {$e_x$};
    \draw[->,black!60,line width=.45pt] (0,-1.64)--(0,1.86)
      node[above] {$e_\theta$};
    \foreach \rx/\ry in {.35/.56,.67/.99,.99/1.40}{
      \draw[level] (0,0) ellipse[x radius=\rx,y radius=\ry];
    }
    \fill[red!75!black] (0,0) circle[radius=2.0pt];
    \node[anchor=west,text=black] at (1.15,.93) {Lapse levels};
  \end{scope}
  \node[text=black] at (3.25,.25) {\textbf{(a)} Meridian section of the full region};
  \node[text=black] at (10.63,.25) {\textbf{(b)} Enlargement near one maximum};
\end{tikzpicture}
\caption{\texorpdfstring{The circle of maximal lapse and nearby regular levels. The filled dots in the meridian section are the intersections with the same equatorial circle. The two matching curves at each level are sections of the same torus. The enlargement is schematic. Its axes are the radial and polar directions.}{The circle of maximal lapse and nearby regular levels.}}\label{fig:maximal-lapse-foliation}
\end{figure}

The angular calculation also gives a quantitative local picture of the levels. In the background distance coordinate $x$, write
\be\label{qual:scaled-collar-coordinate}
 x=X_\mu(\theta)+\sqrt{\mu}\,\xi,\qquad
 \mathcal Z_\mu(\xi,\theta)\coloneqq
 \frac{N_{\max,\mu}-N_\mu(X_\mu(\theta)+\sqrt{\mu}\,\xi,\theta)}{\mu}.
\ee
The scale $\sqrt{\mu}$ in this formula follows from balancing the quadratic radial Taylor term with the angular perturbation of order $\mu$. It is determined by the expansion. Here $\mathcal Z_\mu$ denotes the rescaled lapse deficit, and $N_{\max,\mu}$ is the global maximal value. For bounded $\xi$ and a fixed closed angular interval away from the poles,
\be\label{qual:scaled-collar-deficit}
 \mathcal Z_\mu(\xi,\theta)
 =\frac{\Lambda N_*}{2}\xi^2+\Phi(\pi/2)-\Phi(\theta)
       +O_{C^2}(\mu^\eta).
\ee
Indeed Taylor expansion in $x$ about the graph on which the radial derivative of the lapse vanishes has zero linear term. The second derivative is $-\Lambda N_*+O_{C^2}(\mu^{1/2})$ in these scaled coordinates, by the fixed $C^4$ estimates and $\|X_\mu\|_{C^2}=O(\mu)$. Its quadratic contribution after division by $\mu$ is therefore $\Lambda N_*\xi^2/2+O_{C^2}(\mu^{1/2})$. The angular contribution follows from~\eqref{qual:angular-expansion}, with $N_{\max,\mu}=f_\mu(\pi/2)$. Since $\eta<1/2$, one has $\mu^{1/2} \le\mu^\eta$ for $0<\mu\le1$, which gives the stated remainder. Thus regular contours of the limiting deficit determine the limiting level curves in this scaled collar. The implicit function theorem gives $C^1$ convergence on compact regular contours; a fixed neighborhood of the equatorial minimum of this deficit has closed contours, whose rotations are tori. This description concerns the collar of the maximal sphere and not the regions near the horizons.

For each fixed sufficiently small positive mass, all levels sufficiently close to the maximal value are tori. Indeed the negative transverse Hessian gives a small disc about the maximum in the meridional quotient, with nearby levels simple closed curves by the Morse lemma. Rotating such a curve about the axis gives a torus; the neighborhood stays away from the axis and has the product coordinates supplied by the construction. By compactness and uniqueness of the maximal circle, every level sufficiently close to the maximum lies in this neighborhood.

\section{Convergence as the additional horizons shrink}\label{sec:metric-convergence}

\subsection{The geometry near and away from the added horizons}

Let $\mathcal M_\mu$ be the compact spatial region of the constructed solution with metric $h_\mu$. Each added horizon has area radius $2\mu$. Measuring lengths in units of $\mu$ gives a sphere of area $16\pi$ at the boundary of the rescaled region. In the moving isotropic coordinates of the construction the model has inner radius $u_\mu/2$, where $u_\mu/\mu\to1$ by Proposition~\ref{n:retune}. At a fixed rescaled distance from either horizon the interpolation takes place farther and farther away. By Section~\ref{sec:bridge-local-limit}, the spatial metric $\mu^{-2}h_\mu$ converges smoothly on each such fixed neighborhood, including the horizon, to the unit-mass Schwarzschild metric.

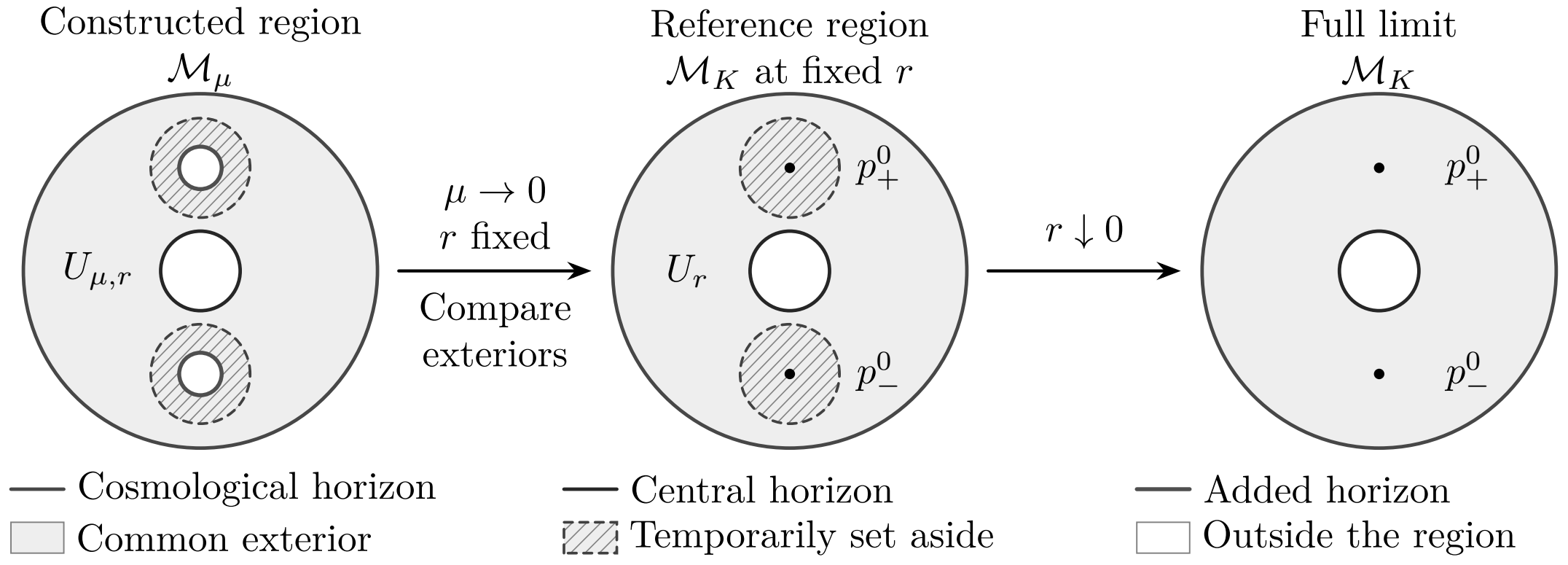
\begin{figure}[!htbp]\centering
\begin{tikzpicture}[x=1cm,y=1cm,font=\normalsize,line cap=round]
\definecolor{meridianExterior}{RGB}{231,240,247}
\definecolor{meridianCosmological}{RGB}{35,80,120}
\definecolor{meridianSmall}{RGB}{160,45,41}
\definecolor{meridianOmitted}{RGB}{249,237,212}
\definecolor{meridianHatch}{RGB}{162,125,75}
\tikzset{outer horizon/.style={draw=meridianCosmological,line width=.90pt},
  central horizon/.style={draw=black!85,line width=.90pt},
  added horizon/.style={draw=meridianSmall,line width=1.03pt},
  comparison circle/.style={draw=black!75,dash pattern=on 2.2pt off 1.8pt,line width=.69pt}}
\path[use as bounding box] (-1.91,-2.74) rectangle (12.95,2.46);

\foreach \shift/\stage/\heading/\subtitle in {
  0/0/{Constructed region}/{$\mathcal M_\mu$},
  5.35/1/{Reference region}/{$\mathcal M_K$ at fixed $r$},
  10.70/2/{Full limit}/{$\mathcal M_K$}}{
  \begin{scope}[xshift=\shift cm]
    \node[align=center] at (0,2.01) {\heading\\\subtitle};
    \begin{scope}[scale=.82]
      \fill[meridianExterior] (0,0) circle (1.96);
      \foreach \pos/\sign in {1.14/{+},-1.14/{-}}{
        \begin{scope}[shift={(0,\pos)}]
          \ifnum\stage<2
            \fill[meridianOmitted] (0,0) circle (.55);
            \begin{scope}
              \clip (0,0) circle (.55);
              \foreach \stripe in {-8,...,8}{
                \draw[meridianHatch,line width=.31pt]
                  ({.128*\stripe-.63},-.65)--({.128*\stripe+.63},.65);
              }
            \end{scope}
            \draw[comparison circle] (0,0) circle (.55);
          \fi
          \ifnum\stage=0
            \fill[white] (0,0) circle (.235);
            \draw[added horizon] (0,0) circle (.235);
          \else
            \fill[black] (0,0) circle[radius=1.65pt];
            \node[anchor=west,inner sep=1.2pt] at (.68,0) {$p_{\sign}^{0}$};
          \fi
        \end{scope}
      }
      \fill[white] (0,0) circle (.44);
      \draw[central horizon] (0,0) circle (.44);
      \draw[outer horizon] (0,0) circle (1.96);
      \ifnum\stage=0
        \node at (-1.13,0) {$U_{\mu,r}$};
      \fi
      \ifnum\stage=1
        \node at (-1.13,0) {$U_r$};
      \fi
    \end{scope}
  \end{scope}
}
\draw[-{Stealth[length=2.2mm]},thick]
  (1.80,0)--(3.55,0)
  node[midway,above=2pt,align=center] {$\mu\to0$\\$r$ fixed}
  node[midway,below=2pt,align=center] {Compare\\exteriors};
\draw[-{Stealth[length=2.2mm]},thick] (7.15,0)--(8.90,0)
  node[midway,above=2pt] {$r\downarrow0$};

\draw[outer horizon] (-1.72,-1.98)--++(.48,0)
  node[anchor=west,inner sep=3pt] {Cosmological horizon};
\draw[central horizon] (3.30,-1.98)--++(.48,0)
  node[anchor=west,inner sep=3pt] {Central horizon};
\draw[added horizon] (8.50,-1.98)--++(.48,0)
  node[anchor=west,inner sep=3pt] {Added horizon};

\filldraw[fill=meridianExterior,draw=black!50,line width=.36pt]
  (-1.72,-2.56) rectangle ++(.48,.28);
\node[anchor=west,inner sep=3pt] at (-1.24,-2.42) {Common exterior};
\fill[meridianOmitted] (3.30,-2.56) rectangle ++(.48,.28);
\begin{scope}
  \clip (3.30,-2.56) rectangle ++(.48,.28);
  \foreach \stripe in {0,...,5}{
    \draw[meridianHatch,line width=.31pt]
      ({3.04+.13*\stripe},-2.62)--({3.43+.13*\stripe},-2.23);
  }
\end{scope}
\draw[comparison circle] (3.30,-2.56) rectangle ++(.48,.28);
\node[anchor=west,inner sep=3pt] at (3.78,-2.42) {Temporarily set aside};
\draw[black!50,line width=.36pt] (8.50,-2.56) rectangle ++(.48,.28);
\node[anchor=west,inner sep=3pt] at (8.98,-2.42) {Outside the region};
\end{tikzpicture}
\caption{Meridional cuts in the constructed and reference regions. The dashed circles are sections of the cutting spheres of radius $r$. The unhatched interiors of the outer circles, with the horizon holes removed, are the exteriors compared at fixed $r$. The full reference region includes the two points $p_\pm^0$.}\label{fig:vanishing-small-masses}
\end{figure}

Write $\mathcal M_K$ for the compact target region with metric $h_{M_0}$ and $p_\pm^0$ for the limiting insertion points. Fix a small $r>0$ and cut at coordinate radius $r$ in each moving insertion chart and in the corresponding limiting chart. The chart maps identify these spheres for sufficiently small $\mu$ at fixed $r$. They extend to the exterior by interpolating the small displacement of the charts inside fixed disjoint neighborhoods of $p_\pm^0$ and using the fixed-horizon identifications away from them. Write $U_{\mu,r}$ and $U_r$ for the closed regions outside these cuts in $\mathcal M_\mu$ and $\mathcal M_K$, respectively. The exterior identification on these regions gives smooth convergence of the metric and lapse, including at the original black-hole and cosmological horizons. In Figure~\ref{fig:vanishing-small-masses}, the first arrow compares the exteriors at fixed $r$. The second restores the two limiting points in the reference region as $r\downarrow0$.

In isotropic coordinates of radius $\rho\ge u_\mu/2$, the Schwarzschild conformal factor $1+u_\mu/(2\rho)$ is between $1$ and $2$. The interpolation with the smooth background gives uniform comparison with the Euclidean metric throughout the cut region. As in the proof of Proposition~\ref{f:family}, the relative size of the additional terms in the approximate metric is $O(u_\mu^{\eta_{\rm geom}})$, and that of the solution correction is $O(u_\mu^{1+\gamma})$. Both tend to zero, so the comparison also holds for $h_\mu$. There is a single inner coordinate ball of radius $u_\mu/2$ deleted from each such region. Joining points by straight segments with a detour on the inner sphere gives an intrinsic diameter at most $Cr$. Volume comparison gives volume at most $Cr^3$ and area at most $Cr^2$ for each cutting sphere, with $C$ independent of $r$ and of sufficiently small $\mu$ at fixed $r$. The same estimates hold for the cut balls on the background. Thus in the original scale the four boundary components give way to the two original horizons, while the neighborhoods of the two added horizons shrink to $p_\pm^0$. Smooth convergence of the volume densities outside the cuts and the bound $Cr^3$ inside also give convergence of the total volumes.

\paragraph{Distances between the compact spatial regions}

Use the length distances of the full regions $\mathcal M_\mu$ and $\mathcal M_K$. Their diameters are uniformly bounded by fixing one comparison radius, using smooth convergence on its connected exterior, and adding the diameters of the two cut regions. For a small radius $r$, let $1+\delta_{\mu,r}$ bound the distortion of lengths under the exterior identification in both directions, with $\delta_{\mu,r}\to0$ at fixed $r$. A minimizing curve entering a cut region can be diverted along its cutting sphere at cost at most $Cr$ by replacing its segment from first entrance to last exit. Applying this in both manifolds and transporting the diverted curves bounds the discrepancy of the full distances on the exterior by $C(r+\delta_{\mu,r})$. Pair the exterior points by the identification and pair each cut region with the corresponding background ball. Each such pair is within $Cr$ on either side of a corresponding pair on the cutting spheres. Hence the distortion of this correspondence is at most $C(r+\delta_{\mu,r})$.

We use the intrinsic flat distance introduced by Sormani and
Wenger~\cite{SormaniWenger2011}. To estimate this distance, we apply
the comparison theorem of Lakzian and
Sormani~\cite{LakzianSormani2013} to the truncated exteriors.

For the intrinsic flat distance, orient the regions consistently on the exterior. Their interiors have the given compact regions as metric and settled completions, since curves can be pushed inward in smooth horizon collars with arbitrarily small length change and the horizon points have positive volume density. The integration currents thus extend across the collars to the compact regions with their usual boundary currents. To apply the comparison of Lakzian and Sormani~\cite[Theorem~4.6]{LakzianSormani2013} to subregions compactly contained in those interiors, trim a collar of width $\eta$ along each original horizon, using corresponding collars in the two exteriors. The volumes and boundary areas of the trimmed exteriors are uniformly bounded, and the discarded volumes are bounded by $C(r^3+\eta)$. The discrepancy of the full distances on the smaller exterior still satisfies the same bound. The connected trimmed exteriors have uniformly bounded intrinsic diameter. One can detour around each cut along its sphere and around the trimmed horizons in their smooth collars. Smooth convergence at fixed $r$ bounds the relative metric distortion by a quantity tending to zero, uniformly for small collar widths. Together with the discrepancy of full distances, this gives width and height terms bounded by $C\sqrt{r+\delta_{\mu,r}}$, using the uniform diameter bound. First let $\eta$ tend to zero at fixed $\mu,r$, and then let $\mu$ tend to zero at fixed $r$. Together these comparisons give
\be
 \limsup_{\mu\downarrow0}d_{\rm GH}(\mathcal M_\mu,\mathcal M_K)\le Cr,\qquad
 \limsup_{\mu\downarrow0}d_{\mathcal F}(\mathcal M_\mu,\mathcal M_K)\le C(\sqrt r+r^3).
\ee
Letting $r\downarrow0$ gives convergence in both distances.

\subsection{Local spacetime extensions and the spatial double}\label{sec:horizon-continuation}

\paragraph{Analytic continuation and comparison with the reference geometries.} The Lorentzian metric extends locally across each horizon into a neighborhood on the future side. To see the local continuation, use signed Gaussian distance $s$ and analytic horizon coordinates $y$. By the parity in Lemma~\ref{reg:static-horizon-expansion} and analyticity of the constructed solutions, the spatial metric is $\dd s^2+\gamma(s^2,y)$ and $N^2=\kappa^2s^2(1+s^2b(s^2,y))$ with $\gamma,b$ analytic. For $s>0$ put $X=s\cosh(\kappa t)$, $T=s\sinh(\kappa t)$ and $z=X^2-T^2$. Differentiation gives $X\dd T-T\dd X=\kappa s^2\dd t$ and $\dd X^2-\dd T^2=\dd s^2-\kappa^2s^2\dd t^2$. Consequently the Lorentzian metric is
\be
 -\dd T^2+\dd X^2+\gamma(z,y)-b(z,y)(X\dd T-T\dd X)^2 .
\ee
It extends analytically to a neighborhood of the horizon section, with the same Lorentzian signature by continuity. The vacuum Einstein equations continue there by analyticity. The static Killing field extends as $\kappa(X\partial_T+T\partial_X)$ and has squared norm $-\kappa^2z(1+zb(z,y))$. It is spacelike on the nearby future side $T>|X|$. Each black-hole horizon and the cosmological horizon have such local extensions.

The analytic coefficients in the normal-plane formula determine the continuation at each horizon. The region $T>|X|$ is the local future continuation across the horizon. This is the interior side at $H_b$ and at the added horizons, and the exterior side at $H_c$. An exact de Sitter exterior would fail to match the curvature at order two. The first normal derivatives of $\gamma(X^2-T^2,y)$ vanish on the bifurcation sphere, so its Gaussian curvature is the ambient sectional curvature of its tangent plane by the Gauss equation. At the cosmological horizon this curvature converges to $r_c(M_0)^{-2}=\Lambda/3+2M_0/r_c(M_0)^3$ by smooth convergence up to the boundary. It is therefore strictly greater than the de Sitter curvature $\Lambda/3$ for sufficiently small added masses.

An exact Schwarzschild--de Sitter region is excluded from the connected analytic continuation by a curvature identity. Regard the Weyl tensor with components $C_{ab}{}^{cd}$ as an endomorphism $\mathcal W$ of two-forms, acting by $(\mathcal W\omega)_{ab}=\tfrac12C_{ab}{}^{cd}\omega_{cd}$. On the static region let $\rho_i$, $i=1,2,3$, be the eigenvalues of $\Ric_h$. The spatial slices are totally geodesic by time reflection. In a frame diagonalizing $\Ric_h$, the spatial sectional curvature of the plane complementary to the $i$th direction is $\Lambda-\rho_i$, since $\Scal_h=2\Lambda$. The Einstein equation gives the same sectional curvature for the timelike plane containing the $i$th direction. Time reflection makes curvature components with one time index vanish, and the Einstein equation makes the time block diagonal in this frame. After subtracting the constant-curvature part, the eigenvalues of $\mathcal W$ are thus $2\Lambda/3-\rho_i$, each twice. Applying the cubic discriminant to these three numbers of sum zero gives
\be
 (\operatorname{tr}\mathcal W^2)^3-12(\operatorname{tr}\mathcal W^3)^2
 =16\prod_{i<j}(\rho_i-\rho_j)^2 .
\ee
For Schwarzschild--de Sitter the two spherical Ricci eigenvalues agree, so the scalar on the left vanishes throughout its analytic continuation. For our solutions the static Hessian equation and Theorem~\ref{qual:unique-maximal-circle} give three distinct Ricci eigenvalues at the maximal circle when the added masses are sufficiently small. The scalar is positive there. It is analytic, and the identity principle excludes an open subset of exact Schwarzschild--de Sitter geometry in the connected analytic continuation. In a smooth gluing in the same horizon charts, an exact Schwarzschild--de Sitter region abutting the horizon would force all derivatives of this scalar to vanish there. The analytic static germ rules this out as well.

\paragraph{Extension across the added horizons and the Schwarzschild limit.}
Take two copies of $\mathcal M_\mu$ and identify the corresponding added
horizon spheres. Keep the spatial metric on both copies and reverse the
sign of the lapse on the second. In signed Gaussian distance $s$ from
either sphere, the metric and lapse take the form
\be
 h_\mu=\dd s^2+\gamma(s^2,y),\qquad
 N_\mu=\kappa s\sqrt{1+s^2b(s^2,y)},
\ee
with analytic coefficients. These formulas extend the pair analytically
across the sphere. The static equations are homogeneous in the lapse, so
they hold on both copies and, by continuity, across the identified
spheres. The resulting connected compact manifold has four boundary
components, the central and cosmological horizons of each copy.

After rescaling the spatial metric by $\mu^{-2}$ about either added
horizon, the convergence of Section~\ref{sec:bridge-local-limit} extends
across the sphere in Gaussian coordinates. Since the matching radius
divided by $\mu$ tends to infinity, the comparison extends to prescribed
compact neighborhoods on both sides. Truncating the limiting Schwarzschild
geometry far enough out on each side places the two truncation boundaries
beyond any prescribed distance from the base point. The same distance
bound holds for small $\mu$ by smooth convergence on the truncated region.
Thus the pointed limit is the complete spatial Schwarzschild manifold of
mass one, with one asymptotically flat end corresponding to each copy.

\begin{figure}[!htbp]
\centering
\begin{tikzpicture}[>=Stealth,font=\small]
\foreach \x/\sign in {0/>,4/<}{
  \shade[ball color=gray!20,opacity=.55] (\x,0) circle (1.45);
  \draw[thick] (\x,0) circle (1.45);
  \draw[densely dashed,gray] (\x,0) ellipse (1.45 and .41);
  \shade[ball color=gray!45] (\x,0) circle (.35);
  \node at (\x,0) {$H_b$};
  \node at (\x,1.68) {$N_\mu\sign0$};
  \node at (\x,-1.66) {$H_c$};
}
\foreach \y/\lab in {.77/+, -.77/-}{
  \shade[top color=black!5,bottom color=black!38]
    (1.09,\y+.22) rectangle (2.91,\y-.22);
  \draw[thick] (1.09,\y+.22)--(2.91,\y+.22)
    (1.09,\y-.22)--(2.91,\y-.22);
  \draw[gray] (1.09,\y) ellipse (.10 and .22)
    (2.91,\y) ellipse (.10 and .22);
  \filldraw[fill=black!12,very thick] (2,\y) ellipse (.13 and .22);
  \node[above] at (2,\y+.22) {$H_{\lab}$};
}
\draw[->,thick] (2.19,.95) .. controls (3.1,2.9) and (6.55,2.75) ..
  node[pos=.63,above] {$h_\mu\mapsto\mu^{-2}h_\mu$} (7.84,.54);
\begin{scope}[shift={(8,0)}]
\shade[top color=black!4,bottom color=black!35]
 (-1.65,1) .. controls (-1.13,.54) and (-.5,.30) .. (0,.30)
 .. controls (.5,.30) and (1.13,.54) .. (1.65,1)
 -- (1.65,-1) .. controls (1.13,-.54) and (.5,-.30) .. (0,-.30)
 .. controls (-.5,-.30) and (-1.13,-.54) .. (-1.65,-1) -- cycle;
\draw[thick] (-1.65,1) .. controls (-1.13,.54) and (-.5,.30) .. (0,.30)
 .. controls (.5,.30) and (1.13,.54) .. (1.65,1);
\draw[thick] (-1.65,-1) .. controls (-1.13,-.54) and (-.5,-.30) .. (0,-.30)
 .. controls (.5,-.30) and (1.13,-.54) .. (1.65,-1);
\draw[very thick] (0,0) ellipse (.14 and .30);
\draw[->] (-1.1,0)--(-1.95,0);
\draw[->] (1.1,0)--(1.95,0);
\node at (0,-.72) {Minimal sphere};
\node[align=center] at (0,-1.42) {Two asymptotically flat ends};
\end{scope}
\node at (2,-2.12) {\textbf{(a)} Compact spatial double};
\node at (8,-2.12) {\textbf{(b)} Spatial Schwarzschild limit};
\end{tikzpicture}
\caption{Schematic perspective of the compact spatial double and
the limit near $H_+$. The dark ellipses represent the two added horizon spheres $H_\pm$, along which the copies are identified. Each copy retains its central boundary $H_b$ and its
cosmological boundary $H_c$. The rescaled limit about $H_+$ has a
minimal sphere separating two asymptotically flat ends.}
\label{fig:spatial-double}
\end{figure}
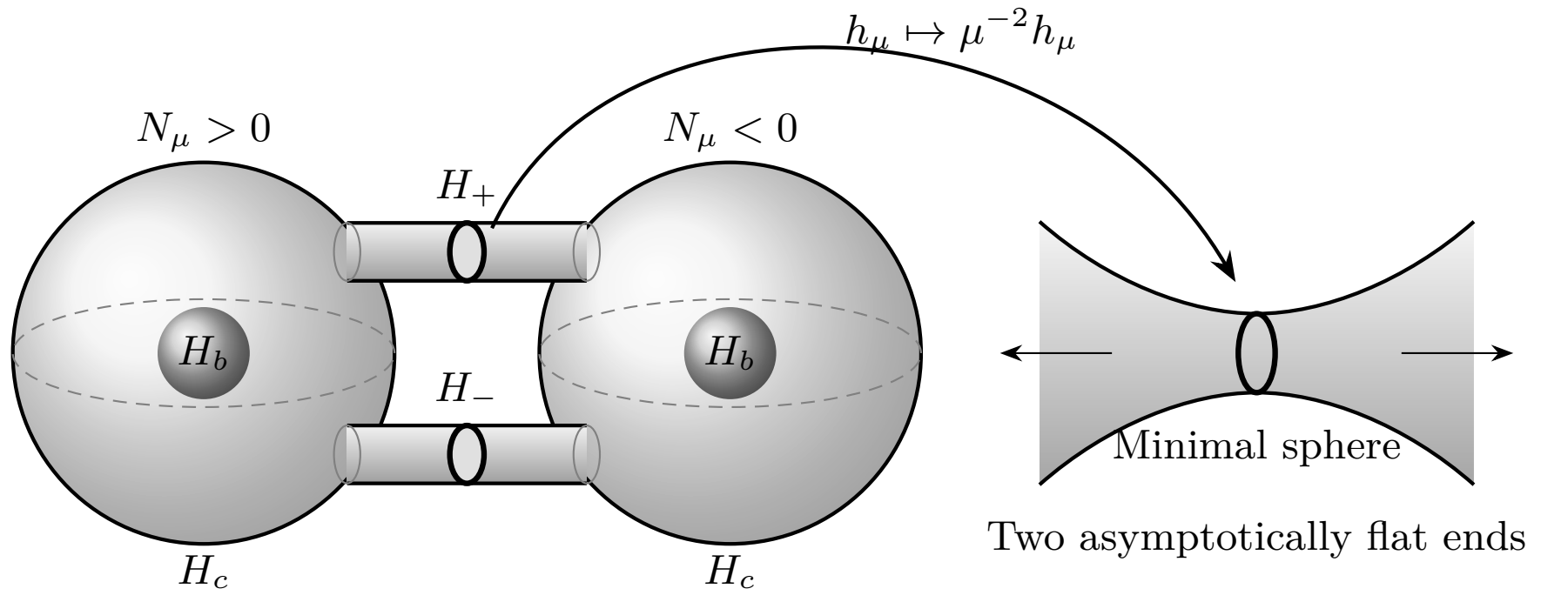

In Figure~\ref{fig:spatial-double}(a), each dark ellipse represents an
entire added horizon sphere. The short tubes depict neighborhoods of the
identified spheres. The arrow singles out $H_+$ as the base horizon for
the rescaling. In panel~(b), the minimal sphere is its limit, and each
branch comes from one of the two copies.

\paragraph*{Acknowledgments.}

The authors are grateful to Bruno Le Floch for many insightful discussions. 
This project began several years ago, when the first author (PLF) was a visiting research fellow at the Courant Institute of Mathematical Sciences, New York University. He is grateful to Jeff Cheeger, Pierre Germain (now at Imperial College London), Fang-Hua Lin, and Esteban Tabak, as well as to the late Robert V. Kohn and Peter D. Lax, for their warm hospitality during his visits to Courant over many years.
The Courant Institute provides an exceptionally stimulating environment for research, especially at the interface between PDEs, differential geometry, and applied mathematics. 

\paragraph*{Funding and technical support}

The first author (PLF) was partially supported by the research project ANR-23-CE40-0010-02: Einstein-PPF, entitled \emph{``Einstein constraints: past, present, and future''}, funded by the Agence Nationale de la Recherche, and by the European MSCA Staff Exchange Project 101131233: Einstein-Waves, entitled \emph{``Einstein gravity and nonlinear waves: physical models, numerical simulations, and data analysis''}. After writing this paper, the authors used ChatGPT 6 to help improving the flow of its presentation and drafting the TikZ code for all of the figures.

\small 

\appendix

\section{Quantitative bounds on the masses of the added black holes}\label{sec:quantitative}

\subsection{Uniform background parameters and profile bounds}

We specify each estimate constant by the inequality in which it is used.
For a bounded linear map, we may take its operator norm or any proved
upper bound. For a family of maps, the bound is taken uniformly on the
stated parameter set. When a comparison is used to obtain an explicit
numerical bound, we give the comparison that implies it. The auxiliary
margins are free parameters subject to the inequalities stated at their
introduction. The resulting upper restriction on the added masses is
sufficient for the construction.

We use the dimensionless variables of Section~\ref{sec:joined-quantitative} and a fixed weight $\gamma$ satisfying~\eqref{eq:free-scale-conditions}. All upper-bound constants used in denominators are chosen strictly positive, enlarging them if necessary; this adds no condition on the geometric data. Fix a target central mass $M_0$ strictly between zero and $M_N\coloneqq1/(3\sqrt\Lambda)$. Choose $h_M>0$ and then $h>0$ subject to
\bse\label{q:evaluated-mass-interval}
\be
M_-\coloneqq M_0-h_M>0,\qquad M_+\coloneqq M_0+h_M<M_N,\qquad r_b(M_+)<r_*(M_-),
\ee
\be\label{q:uniform-location}
0<h<\min(r_*(M_-)-r_b(M_+),\ r_c(M_+)-r_*(M_+)).
\ee
\ese
We put $J\coloneqq[M_-,M_+]$ and $I_M\coloneqq[r_*(M)-h,r_*(M)+h]$. Such choices exist by strict subextremality and continuity of the horizon and placement radii; first shrink the mass interval about $M_0$, then choose $h$ less than the positive minimum clearance on it. Every upper-bound constant below is uniform for $M\in J$ and $r\in I_M$; every radius and positive lower bound is replaced by its minimum on this compact set. Let $C_{P,J}\coloneqq\sup_{M\in J}C_{P,M}$, with the scalar class fixed in~\eqref{q:optimal-Poincare}. We may use any positive spectral margin satisfying
\be\label{q:uniform-scalar}
 0<d_{s,J} \le\min(2\Lambda,C_{P,J}^{-2}-2\Lambda,6/r_c(M_-)^2-2\Lambda).
\ee
Corollary~\ref{q:functional-Poincare-Korn} proves positivity and gives the corresponding vector and Korn estimates. The direct geometric choice
$\min(\Lambda/4,F_*(M_+)/(8r_c(M_-)^2),6/r_c(M_-)^2-2\Lambda)$ remains valid when a sharper Poincar\'e constant is unavailable.
The inequalities defining $J$ keep the horizons nondegenerate. The compact interval $J$ stays a positive distance from the extremal mass.

For $I_M=[r_-,r_+]$, define
\bse\label{q:improved-reference-background-constant}
\be
F_I\coloneqq\min(F(r_-),F(r_+)),\quad U_I\coloneqq M/r_-^2+\Lambda r_+/3,\quad V_I\coloneqq2M/r_-^3+\Lambda/3,
\ee
\be\label{q:parameter-background-constants}
A_I\ge\max(1,U_IF_I^{-1/2}),\qquad
A_{I,1}\ge1+A_I+F_I^{-1/2}+V_IF_I^{-1/2}+U_I^2F_I^{-3/2}.
\ee
\ese
Concavity of $F$ gives $F\ge F_I>0$. These inequalities control $b,b^{-1},a,a'$ in~\eqref{f:b}. We specify the profile constants by the norms which determine them.

For the affine profile we take $C_A$ from Lemma~\ref{q:affine-profile-bound}. For the Green tensor we take $C_H,C_{H,1}^{\rm all}$ from~\eqref{q:profile-and-differential-constants}, with the quantitative construction of Proposition~\ref{green:global}. These are proved norm bounds on the fixed models.

{
For $0\le j\le6$, the model profiles satisfy
\bse\label{q:evaluated-profile-bounds}
\be
|D^jA_S|\le C_A(1+s)^{1-j},\qquad |D^j(A_S-T-B)|\le C_A(1+s)^{\gamma-j}\quad(s\ge4),
\ee
\be
|D^j(H_r-2D_p/\rho)|\le C_H\rho^{-j}\quad(\rho<R/4).
\ee
\ese

The Green bound after one moving-chart derivative uses $C_{H,1}^{\rm all}$; outside these balls the corresponding four-derivative bounds hold. Let $C_{\rm join}$ bound the cutoff, rescaling and chart product maps in the interpolation defining $\overline G$. Let $C_{\rm lead}$ bound the leading Green tensor in these maps. Let $C_{\rm mix}$ bound the term due to interpolating the squares of the lapses instead of squaring their interpolation, with the factor $u^{2\eta_{\rm geom}}$ removed in the relative $C^{4,\alpha}_{\omega_u}$ norm. Its bound is obtained from the fixed lapse profiles and the cutoff as shown below. Choose $C_{\rm Ric}$ so that $\|L_{G_{u,r}}-P_{G_{u,r}}\|_{X_u\to Y_u} \le C_{\rm Ric}u^{\eta_{\rm geom}}$ uniformly for $0<u\le u_J$. By~\eqref{f:L}, this operator is the sum of the two zeroth-order terms with $\Ric_{G_{u,r}}-\Lambda G_{u,r}$, with indices raised using $G_{u,r}$. Let $C_{\rm lin}$ bound the coefficient Lipschitz estimate for the differential~\eqref{f:L} in a relative metric neighborhood of radius $0<\rho_{\rm ref}<1$.
}

\begin{proposition}[Comparison of the reference metrics and their linearized operators]\label{q:evaluated-actual-differential}
Choose $D_{\rm ref},D_L$ such that
\be\label{q:reference-constant-conditions}
D_{\rm ref}\ge C_{\rm join}(A_IC_A+C_H+C_{\rm lead})+C_{\rm mix},\qquad D_L\ge C_{\rm lin}D_{\rm ref}+C_{\rm Ric}.
\ee
For $u\le u_{\rm ref}\coloneqq\min(u_J,(R/8)^{1/a_2},(\rho_{\rm ref}/D_{\rm ref})^{1/\eta_{\rm geom}})$ they give
\bse\label{q:evaluated-reference-difference}
\be
\|\overline G_{u,r}-G_{u,r}\|_{C^{4,\alpha}_{\omega_u},\mathrm{rel}} \le D_{\rm ref}u^{\eta_{\rm geom}},
\ee
\be\label{q:evaluated-differential-difference}
\|L_{\overline G_{u,r}}-P_{G_{u,r}}\|_{X_u\to Y_u} \le D_Lu^{\eta_{\rm geom}}.
\ee
\ese
\end{proposition}
\begin{proof}
Lemma~\ref{q:affine-profile-bound} gives the affine estimates. Proposition~\ref{green:global} gives the Green estimates in the weighted source norm appropriate to $\rho^{-1}$, with the norm choices recalled above.
On the inner support \(s\le2u^{a_2-1}\), the relative affine profile is bounded by a fixed multiple of \(A_IC_Au^{\eta_{\rm geom}}\). On the outer support \(\rho\ge u^{a_2}\), the Green tensor contributes at most a fixed multiple of \((C_{\rm lead}+C_H)u^{\eta_{\rm geom}}\). These estimates apply after multiplying by the common cutoff \(\chi_u\) or \(1-\chi_u\). In the spatial block the simpler reference already interpolates the covariant metrics. In the auxiliary circle component, let \(N_S=N_{S,u}\) and \(N_K=N_M\) be the lapses of the two simpler models in the same coordinates. Subtracting \(G_{u,r}\) from \(\overline G_{u,r}\) leaves, in addition to the affine and Green terms, the tensor
\be
\label{q:reference-interpolation-term}
 \chi_u(1-\chi_u)(N_S-N_K)^2\,\dd\tau^2 .
\ee
Indeed the coefficient here is \(\chi_u N_S^2+(1-\chi_u)N_K^2-(\chi_u N_S+(1-\chi_u)N_K)^2\).

On the support of this term and on the scaled chart balls meeting that support, \(\omega_u\asymp\rho\asymp u^{a_2}\). The explicit formula~\eqref{join:eq:Schw} gives \(N_S-b=O(u/\rho)\) through four scaled derivatives and their H\"older seminorm. Smoothness of \(N_K\) and \(N_K(p_\pm)=b\) give \(N_K-b=O(\rho)\) in the same norm. For the seminorm one may differentiate once more on the enlarged scaled balls. The lapse of \(G_{u,r}\) is bounded away from zero on these balls by the uniform positive lower bound for \(b\) and the bounds for the two lapse profiles. The scaled norms of its reciprocal and of \(\chi_u\) are bounded there by the same profile and cutoff estimates. Thus the scaled product inequality, with the two factors of \(\dd\tau\) measured in the reference frame, bounds~\eqref{q:reference-interpolation-term} by \(C_{\rm mix}u^{2\eta_{\rm geom}}\) in the stated relative norm. The product constant, the norm of the cutoff and its complement, the reciprocal-lapse and frame comparisons, and the squared bound on \(u^{-\eta_{\rm geom}}(N_S-N_K)\) give a choice of \(C_{\rm mix}\) independent of \(u\). Since \(u\le1\) by the support condition~\eqref{q:power-support-conditions}, adding this estimate to the affine and Green bounds proves~\eqref{q:evaluated-reference-difference}.

The coefficients in~\eqref{f:L} are smooth rational functions of the metric components and their coordinate derivatives in the neighborhood of positive metrics of radius $\rho_{\rm ref}$. Their derivatives there give $C_{\rm lin}$ for the difference $L_{\overline G_{u,r}}-L_{G_{u,r}}$. The product estimate in the norms~\eqref{join:eq:norms} bounds the operator norm of $L_{G_{u,r}}-P_{G_{u,r}}$ by a fixed multiple of the supremum, over the scaled balls centered at $x$, of the scaled $C^{2,\alpha}$ norm of $\omega_u(x)^2(\Ric_{G_{u,r}}-\Lambda G_{u,r})$. In the exact Schwarzschild region, $\Ric_{G_{u,r}}=0$ and $\omega_u^2=O(u^{2a_2})$. In the exact Schwarzschild--de Sitter region the tensor in the norm vanishes. On the matching annulus, the scaled $C^{4,\alpha}$ metric estimate follows from~\eqref{join:eq:neck} through five derivatives and gives an $O(u^{\eta_{\rm geom}})$ bound on the scaled curvature term. Balls meeting the annulus have comparable scales. This gives $C_{\rm Ric}$, and adding the two operator estimates proves~\eqref{q:evaluated-differential-difference}. Both constants are determined by the known metric profiles and the cutoff.
\end{proof}
 
\subsection{Conditions for solving the nonlinear equation}

Choose $0<\theta_D<1$. The actual augmented inverse has norm at most $A\coloneqq A_J/(1-\theta_D)$ when
\be\label{q:differential-threshold}
u\le u_D\coloneqq\min(u_{\rm ref},(\theta_D/(A_JD_L))^{1/\eta_{\rm geom}}).
\ee
We describe the remaining constants without assigning large numerical values. For each product of coefficients and derivatives, we use the norm of multiplication between the scaled H\"older spaces in the estimate. A nonlinear coefficient constant is the supremum of the relevant derivatives of the coordinate Ricci--DeTurck formula on the compact set of normalized metric jets specified by the known profile bounds and the chosen positive metric neighborhood. Inverse metrics have eigenvalues bounded by $(1-\rho_{\rm met})^{-1}$ on a relative neighborhood of radius $0<\rho_{\rm met}<1$. These prescriptions give finite constants from explicit functions of finitely many variables and the local Schauder constants.

{
We identify the parameter-dependent spaces using the normalized insertion charts and a smooth transport supported in the fixed compact tubes swept by the insertion points. Near each small horizon the circular tensor coefficient is rescaled by the ratio of the lapse factors; this keeps the normalized angle and the unit-model integral condition fixed. The transition to the unchanged outer identification lies in a fixed region, independently of $u$. The coordinate jets of these spatial maps and algebraic frame rescalings depend smoothly on the placement parameter, uniformly on its compact interval. A geometric pullback derivative is estimated from $C^{m+1,\alpha}$ to $C^{m,\alpha}$; the known reference and Green jets provide this order. After transporting the equation to fixed normalized charts, we differentiate its coefficients there. In these fixed spaces defined by the integral condition we define the full augmented operator by
\be\label{q:full-augmented-parameter-map}
 \mathbf L_{u,r}(v,\lambda)\coloneqq L_{\overline G_{u,r}}v-\lambda\psi_u.
\ee
On either exact inner collar the rescaling~\eqref{join:eq:rescaling} removes the placement dependence of the simpler reference: $u^{-2}\Theta_{u,r}^*G_{u,r}=G_S$. The normalized source tensor and the integral condition are likewise independent of $r$, since
\be
 u^{2-\gamma}\widehat\psi_{+,u,r}=\chi_0T_S,\qquad
 \ell_{u,r}(v)=m_0^{-1}\int\langle u^{-\gamma}\widehat v_+,\chi_0T_S\rangle\dd\nu_S.
\ee
Thus differentiating these normalized objects introduces no inverse power of $u$ on their fixed support. For the improved reference the normalized metric is $G_S+u a(r)A_S$; its placement derivative is $u a'(r)A_S$ in the same coordinates. The profile bounds through order six give, on scaled balls contained in the inner region,
\be
 \|\partial_r(u^{-2}\Theta_{u,r}^*\overline G_{u,r})\|_{C^{4,\alpha}_{1+s}}
 \le Cu(1+s)\le Cu^{a_2},
 \qquad s\le u^{a_2-1}.
\ee
The first inequality includes the scaled H\"older seminorm and the uniform bound on $a'(r)$; the second uses the full extent of the inner region. The derivative bound below also includes the transition to the outer identification.
}

\begin{lemma}[Bounds for the source, the remainder and their derivatives]\label{q:evaluated-source-quadratic}
With the norms and the operator $\mathbf L_{u,r}$ defined above, there are uniform constants $E,E_1,B_Q,S_0,L_1,B_1>0$ such that
\be\label{q:evaluated-source-bound}
\|f_{u,r}\|_{Y_u} \le Eu,\qquad \|\del_rf_{u,r}\|_{Y_u} \le E_1u,
\ee
and, in the ball $\|v\|_X,\|w\|_X\le\rho_{\rm met}u^{-\gamma}/S_0$,
\bse\label{q:quadratic-constant}
\be
\|Q_u(v)-Q_u(w)\|_Y\le B_Qu^\gamma(\|v\|_X+\|w\|_X)\|v-w\|_X,\qquad Q_u(0)=0,
\ee
\be\label{q:evaluated-Czero-embedding}
\|v\|_{C^0,\mathrm{rel}} \le S_0u^\gamma\|v\|_X.
\ee
\ese
In the same ball, and with placement derivatives taken in the fixed normalized charts,
\be\label{q:evaluated-parameter-derivative-constants}
\|\del_r\mathbf L_{u,r}\|\le L_1,\qquad \|\del_rQ_u(v)\|_Y\le B_1u^\gamma\|v\|_X^2.
\ee
\end{lemma}
\begin{proof}
{
Let $E_{\rm in},E_{\rm out},E_{\rm match}$ be the coefficient bounds for the three regional source calculations in the proof below, and let $E_{\rm in,1},E_{\rm out,1},E_{\rm match,1}$ be their placement-derivative bounds. Choose
\be\label{q:evaluated-nonlinear-constants}
E\ge E_{\rm in}+E_{\rm out}+E_{\rm match},\qquad E_1\ge E_{\rm in,1}+E_{\rm out,1}+E_{\rm match,1}.
\ee
Let $C_{\rm prod}$ bound the products of coordinate derivatives measured at the local scale in the H\"older norms and pointwise, and let $C_{\rm nl}$ bound the second and third derivatives of the Ricci--DeTurck formula with respect to the metric components and their coordinate derivatives on the stated neighborhood, and set $S_0\coloneqq(1-\rho_{\rm ref})^{-1}$. The relative correction norm is the operator norm with respect to $\overline G$. Take $B_Q\ge C_{\rm prod}C_{\rm nl}$. Choose $L_1,B_1$ as bounds for the differentiated products below.
}
After subtracting the auxiliary source term, the leading inner term is $a u^{-1}J_0$. Its source weight gives $u^{1-\gamma}(1+s)^{-\gamma} \le C u^{1+\eta}$ on $s\le2u^{a_2-1}$. The inner quadratic and cosmological terms have the same or a higher power. On the outer region the Bianchi term is bounded by $Cu\rho^{-2}$; its weighted size is $Cu\rho^{-\gamma} \le Cu$ on a fixed region. The quadratic term $u^2\rho^{-4}$ gives $u^{1+\eta}$ at the inner endpoint. The matching error~\eqref{f:jets-all}, with two cutoff derivatives, also gives $u^{1+\eta}$. Taking the finite product bounds for these expressions defines $E_{\rm in},E_{\rm out},E_{\rm match}$. The parameter derivative uses $A_{I,1}$ and $C_{H,1}^{\rm all}$ and preserves each power. Summing proves~\eqref{q:evaluated-source-bound}.

Taylor's formula for the coordinate operator writes each quadratic difference as a sum with one factor $v-w$, one factor $v$ or $w$, and at most two derivatives in total before taking the source norm. The weights leave one factor $\omega_u^\gamma\le u^\gamma$. The product and coefficient constants give~\eqref{q:quadratic-constant}. Before taking the weighted source supremum, the local formula gives $|Q_u(v)|\le B_Q\omega_u^{2\gamma-2}\|v\|_X^2$. The derivative with respect to the correction also satisfies a pointwise bound. At fixed $u,r$, Taylor's formula gives $D_vQ_u(v)[w]=\int_0^1 D^2\mathcal E_{u,r}|_{\overline G+tv}[v,w]\,\dd t$. Each product has two perturbation factors and at most two derivatives in total. In the scaled charts their weights give the factor $\omega_u^{2\gamma-2}$. With $B_Q$ bounding these products as above, this gives
\be\label{q:quadratic-pointwise-derivative}
 |D_vQ_u(v)[w]|\le 2B_Q\omega_u^{2\gamma-2}\|v\|_X\|w\|_X.
\ee
The inequality $\omega_u\ge u$ and the comparison $\overline G\ge(1-\rho_{\rm ref})G_u$ give the embedding with this $S_0$. For the derivative with respect to the insertion radius, fix a nearby radius $r_0$ and use the moving normal-coordinate maps on two balls of fixed radius. Extend their spatial transports to the identity across fixed annuli, pairing the two maps by reflection. Normalize the circular tensor component by $b(r)^{-2}$ on the inner balls, and make its transition on the same fixed annuli. This is an identification of tensor bundles, rather than a change to a common angular period. Apply this same identification to the source bundle. If $J_r$ and $K_r$ denote the resulting domain and target maps to the spaces at $r_0$, respectively, the differentiated operator is $K_r\mathbf L_{u,r}(J_r^{-1}\cdot,\cdot)$. The scalar unknown is unchanged. The radius $\rho$, the scale $\omega_u/u=w_S(\rho/u)$, the cutoff $\chi(\rho/u^{a_2})$, the model column and the normalizing integral are fixed on the moving inner balls. In particular, differentiating these cutoffs introduces no factor $u^{-1}$. The derivatives of the circular normalization involve $b'/b$, which is uniformly bounded on the placement interval. Write the local component expression of this transported operator as $\sum_{j=0}^2 A_jD^j$. The differentiated reference and transition bounds above, together with the smooth fixed transport and inverse-metric identities, give
\be
 \omega_u^{2-j}\|\partial_r A_j\|_{C^{2,\alpha}_{\omega_u}} \le C,\qquad j=0,1,2,
\ee
uniformly in $u$ and $r$. Near an inserted horizon this uses the exact normalized expression for the metric; on the matching annulus it uses~\eqref{f:jets-all} and its placement derivative. In the outer region we use the Green bounds in moving normal charts where the distance to an insertion point can tend to zero. We use the fixed-chart bounds at a fixed positive distance from both centers, including on the support of the transition to the identity. The same estimates hold in the normalized horizon charts at the original horizons and the axis. Multiplication by the source weight gives
\be
 \omega_u^{2-\gamma}|(\partial_r A_j)D^jv|
 \le C\|v\|_{X_u},
\ee
and the scaled H\"older product estimates give the same bound for the source seminorm. The normalized source tensor and the integral condition are fixed as shown above. This proves the first bound in~\eqref{q:evaluated-parameter-derivative-constants}. Differentiating the coefficients of the quadratic formula with $v$ fixed gives the local bound $|\partial_r Q_u(v)|\le B_1\omega_u^{2\gamma-2}\|v\|_X^2$, with $B_1$ chosen to bound the differentiated products. The scaled H\"older product estimate gives the second bound in~\eqref{q:evaluated-parameter-derivative-constants}.\qedhere
\end{proof}

Choose $0<q<1$ and $K\ge AE/(1-q/2)$. Equality is permitted because the fixed-point ball is closed. Conditions~\eqref{f:fixed-point-conditions}, together with the relative metric bound, are satisfied if $u\le u_{\rm fp}$, where
\be\label{q:fixed-point-threshold}
u_{\rm fp}\coloneqq\min\bigl(1,\left(\frac q{2AB_QK}\right)^{1/(1+\gamma)},\left(\frac{\rho_{\rm met}}{S_0K}\right)^{1/(1+\gamma)}\bigr).
\ee
For $u\le\min(u_D,u_{\rm fp})$, the map on $\|v\|_X\le Ku$ has contraction factor at most $q$ and maps this ball into itself: division of the image estimate by $Ku$ gives $AE/K+AB_QKu^{1+\gamma} \le AE/K+q/2\le1$. Thus the same smallness condition controls both estimates. The embedding estimate also gives $\|v\|_{C^0,\mathrm{rel}} \le S_0Ku^{1+\gamma} \le\rho_{\rm met}<1$ throughout the iteration ball. Consequently
\be
 (1-\rho_{\rm met})\overline G\le\overline G+v
 \le(1+\rho_{\rm met})\overline G.
\ee
These are inequalities on the full normalized horizon charts, including the horizon fixed sets. Thus every metric used in the iteration lies in the positive neighborhood on which the nonlinear coefficient bounds were chosen. For its derivative, put
\be\label{q:parameter-constant}
A_1\coloneqq A^2L_1,\qquad K_1\ge\frac{A_1E+AE_1+(A_1B_Q+AB_1)K^2}{1-q}.
\ee
For each fixed $u>0$, the transported coefficients and the transported source tensor are $C^1$ in $r$ in the displayed operator norms. In the inner charts this follows from the explicit profile and its bounds; in the outer charts it follows from the differentiability with respect to the insertion radius in Proposition~\ref{green:global}; the fixed smooth transition uses the additional spatial derivative in that proof. The integral condition is fixed in these identifications. All operator norms in this argument are taken in the fixed spaces at $r_0$. The scaled coordinate norms are uniformly equivalent to the covariant norms, so no differentiation of a norm or translation of an arbitrary H\"older tensor is required. For each coefficient of the transported operator, continuity of its placement derivative in the scaled coefficient norm gives
\be
 \omega_u^{2-j}
 \left\|\frac{A_j(r+h)-A_j(r)}{h}-\partial_rA_j(r)\right\|_{C^{2,\alpha}_{\omega_u}}
 \longrightarrow0\qquad(h\longrightarrow0).
\ee
Here $u>0$ is fixed. On the inner balls this follows from the explicit profile, and on the exterior from the moving-chart differentiability of the Green tensor. The transitions have fixed support and smooth coefficients. The scaled product estimate turns this coefficient convergence into convergence in the operator norm from the fixed domain to the fixed target. The uniform coefficient bound proved above bounds the resulting derivative independently of $u$. The resolvent identity
\be
 \mathbf L_{u,r+h}^{-1}-\mathbf L_{u,r}^{-1}
 =-\mathbf L_{u,r+h}^{-1}(\mathbf L_{u,r+h}-\mathbf L_{u,r})\mathbf L_{u,r}^{-1}
\ee
therefore yields $\del_r\mathbf L_{u,r}^{-1}=-\mathbf L_{u,r}^{-1}(\del_r\mathbf L_{u,r})\mathbf L_{u,r}^{-1}$ in operator norm. Its derivative is bounded by $A_1$, uniformly in the stated parameter range. Let $x_u=(v_u,\delta\lambda_u)$ denote the fixed point in the product norm. On the open set of positive metrics, the coordinate formula for the Ricci--DeTurck operator and the H\"older product estimates make the fixed-point equation continuously Fr\'echet differentiable in $x_u$ and in the transported placement parameter. Its derivative with respect to $x_u$ is the identity plus an operator of norm at most $q<1$, and hence is invertible by its Neumann series. The derivative bound extends to the boundary of the closed ball by continuity. Starting from zero, the iterates are continuous in $r$ and converge locally uniformly in the transported norms because $q<1$. The resulting fixed point therefore agrees locally with the solution given by the implicit function theorem, justifying its differentiation. Moving the derivative of the quadratic term in $v_u$ to the left gives
\be
 (1-q)\|\partial_rx_u\|
 \le(A_1E+AE_1)u+(A_1B_Q+AB_1)K^2u^{2+\gamma}.
\ee
The factor $q$ is the same contraction bound as above. Since $u^{2+\gamma} \le u$, the choice of $K_1$ proves $\|\partial_rx_u\|\le K_1u$, and in particular the asserted bound for $\partial_rv_u$. For fixed $A,E,B_Q,q$ and metric margin, the choice $K=AE/(1-q/2)$ maximizes both $K$-dependent entries of~\eqref{q:fixed-point-threshold}. 

\subsection{Estimates for the auxiliary source coefficient and its zero}

For the force we use the exact fixed-model Gram integral $m_0$ in~\eqref{q:unit-Gram-definition}. For scalar functions on $I_M$ we use $\|f\|_{C^1}\coloneqq\sup_{I_M}|f|+\sup_{I_M}|f'|$. This norm is submultiplicative. Let $G_0>0$ satisfy $\|e_G(u,\cdot)\|_{C^1} \le G_0u$ for the Gram error in~\eqref{f:gram}. This is a finite integral on the fixed support of $\chi_0$, computed from $G_S+uaA_S$ and its density. Let $T_1,T_2,T_3$ bound the three pairings of Lemma~\ref{f:test}, including one placement derivative. Their coefficients may be chosen by the following precise product conditions. Let $J_{\rm tail}$ bound the model tail in~\eqref{f:J-bounds} and $J_{\rm rem}$ the inner source remainder. Let $J_{\rm vol}$ bound the coefficient of $u^{2-\gamma}(1+|\log u|)$ in the estimate for the correction to the pairing of $a(r)J_0$ with the test tensor, including the change of volume density and the changes in both inverse metrics used to contract the tensors. This bound includes one placement derivative of the entire correction. Let $C_{\rm test}$ bound the test-cutoff commutator, the inner operator difference on the test tensor, and their first placement derivatives, and let $C_{\rm pair}$ include the angular integral, regular horizon charts and their first parameter derivatives. The Gram denominator appears separately in~\eqref{q:force-constant} below. Then sufficient choices are
\bse\label{q:evaluated-force-pairing-constants}
\be
T_1\ge C_{\rm pair}\bigl(\|a\|_{C^1(I_M)}J_{\rm tail}+J_{\rm rem}+J_{\rm vol}\bigr),\qquad
T_2\ge C_{\rm pair}C_{\rm test}(K+K_1),
\ee
\be
T_3\ge\frac{C_{\rm pair}((B_Q+B_1)K^2+2B_QKK_1)}{1-2\gamma}.
\ee
\ese
For the third pairing, the derivative with respect to the insertion radius of $Q_u(v_u)$ in the fixed charts is $\partial_rQ_u(v_u)+D_vQ_u(v_u)[\partial_r v_u]$, with the first term taken at fixed correction. The pointwise estimates for $Q_u$, $\partial_rQ_u$ and~\eqref{q:quadratic-pointwise-derivative} bound the undifferentiated and differentiated terms by $u^2\omega_u^{2\gamma-2}$ times $B_QK^2$ and $B_1K^2+2B_QKK_1$, respectively. In the normalized pairing the volume factor $b(r)u^3$ leaves $u^{2-\gamma}$ after division by the Gram scale $2b(r)u^{\gamma+1}$ and summation over the two collars. On each collar, $\omega_u\asymp u(1+s)$, the normalized test tensor has size at most $C(1+s)^{-2}$, and the model volume per unit period is bounded above by $C(1+s)^2\dd s\dd\omega$ down to $s=1/2$. The radial integral is bounded by $\int_{1/2}^\infty(1+s)^{2\gamma-2}\dd s\le1/(1-2\gamma)$. Together with the factor $u^{2\gamma}$ from the bounds on the quadratic term this yields $u^{2+\gamma}$. In the $C^1$ estimate, $C_{\rm pair}$ includes both the norm and the first placement derivative of the contraction and density in~\eqref{f:gram}, as well as the uniform weight, test tensor and volume comparisons on the entire rescaled collar. This proves the stated choice of $T_3$.

The three bounds for the source pairing use the affine profile and its source. The norm $\|a\|_{C^1(I_M)}$ is bounded by $A_{I,1}$ in~\eqref{q:parameter-background-constants}. When choosing the constants uniformly in $M$, take this bound uniformly on $J$. The bound $C_{\rm test}$ uses~\eqref{f:L} and the explicit test cutoff. The bound $C_{\rm pair}$ must include the product of these fixed pairing-map norms and $\max(1,(-\gamma)^{-1},4^{1-\gamma})$. The last two factors come from the inner radial integral and $L_u=u^{a_2-1}/4$.  

\begin{proposition}[Force estimates with specified coefficient bounds]\label{q:evaluated-force-pairings}
For $0<u\le\min(u_D,u_{\rm fp})$, the constants just described give the three inequalities~\eqref{f:test-estimates} in $C^1(I_M)$. For $0<\theta_G<1$, put
\be\label{q:force-constant}
C_F\ge\frac{(2+C_{\log})T_1+2T_2+T_3}{(1-\theta_G)m_0},\qquad
C_{\log}\coloneqq\sup_{0<u\le1}u^{1-\delta}(1+|\log u|)<+\infty.
\ee
When $G_0u\le\theta_Gm_0$, one has
\be
\left\|u^{\gamma-1}\lambda_u+\frac{32\pi}{m_0}a\right\|_{C^1(I_M)} \le C_Fu^\delta.
\ee
\end{proposition}
\begin{proof}
The full model integral vanishes. The leading pairing is therefore bounded in $C^1(I_M)$ by a fixed multiple of $\|a\|_{C^1(I_M)}J_{\rm tail}\,u^{2-\gamma-a_2}$, since the model tail is independent of $r$ and $L_u^{-1}=4u^{1-a_2}$. The inner source remainder is bounded by $Cu^{1-\gamma+a_2}$. The change in the density and tensor contractions gives $u^{2-\gamma}(1+|\log u|)$ by the integral in the proof of Lemma~\ref{f:test}. The adjoint test calculation gives $u^2+u^{2-\gamma+a_2(\gamma-1)}$ with coefficients $C_{\rm test}(K+K_1)$. For the quadratic term, the radial integral is bounded by $u^{2\gamma-1}/(1-2\gamma)$; the product rule for one placement derivative gives exactly the numerator in the condition for $T_3$. This proves the asserted sufficient coefficients. Set $a_G\coloneqq m_0^{-1}\sup|e_G|$ and $b_G\coloneqq m_0^{-1}\sup|e_G'|$. The hypothesis gives $a_G+b_G\le\theta_G<1$. Direct differentiation gives
\be\label{q:Gram-C1-reciprocal}
 \|(m_0+e_G)^{-1}\|_{C^1}
 \le\frac1{m_0} \left(\frac1{1-a_G}+\frac{b_G}{(1-a_G)^2}\right)
 \le\frac1{(1-\theta_G)m_0}.
\ee
For the second inequality, $(1-a_G+b_G)(1-a_G-b_G)=(1-a_G)^2-b_G^2\le(1-a_G)^2$. This reciprocal bound is sharp under the stated $C^1$ norm constraint, as the constant function $e_G=-\theta_Gm_0$ shows. Testing the correction equation gives $(m_0+e_G)(\lambda_u-\lambda_u^0)$ as the sum of the three normalized pairings of Lemma~\ref{f:test}. Apply the reciprocal bound in $C^1(I_M)$. Dividing the three test estimates by $u^{1-\gamma}$ gives, respectively, $u^{1-a_2}+u^{a_2}+u(1+|\log u|)$, $u^{1+\gamma}+u^{1-a_2(1-\gamma)}$ and $u^{1+2\gamma}$. The displayed logarithmic supremum proves the result.
\end{proof}

The leading force derivative has the uniform lower bound
\bse\label{q:force-slope}
\be
-a'(r)=\frac{2M/r^3+\Lambda/3}{\sqrt F}+\frac{(F')^2}{4F^{3/2}},
\ee
\be\label{q:uniform-force}
\underline\kappa_{F,J}\coloneqq{\frac{32\pi}{m_0}}\frac{2M_-/(r_*(M_+)+h)^3+\Lambda/3}{N_*(M_-)}>0.
\ee
\ese
The prefactor $32\pi/m_0$ is the exact Gram normalization of the leading force. Choose $0<\theta_F<1$ and set
\be\label{q:force-threshold}
u_F\coloneqq\min\bigl(1,\frac{\theta_Gm_0}{G_0},\left(\frac{\theta_F\underline\kappa_{F,J}\min(1,h)}{C_F}\right)^{1/\delta}\bigr).
\ee
To verify both assertions, write $G_u(r)\coloneqq u^{\gamma-1}\lambda_u(r)$ and $G_{\rm ref}(r)\coloneqq-(32\pi/m_0)a(r)$. The definition of $u_F$ gives $C_Fu^\delta\le\theta_F\underline\kappa_{F,J}\min(1,h)$. Hence $G_u'\ge(1-\theta_F)\underline\kappa_{F,J}>0$. Since $G_{\rm ref}(r_*)=0$ and $G_{\rm ref}'\ge\underline\kappa_{F,J}$, the values of $G_{\rm ref}$ at $r_*\pm h$ have opposite signs and modulus at least $\underline\kappa_{F,J}h$. The same signs hold for $G_u$ by the $C^0$ part of the error estimate. The intermediate value theorem gives a zero, and the derivative bound makes it unique. The same inequalities yield $|r_u-r_*|\le C_Fu^\delta/\underline\kappa_{F,J}$.

\subsection{A mass bound that ensures the DeTurck vector field vanishes}\label{q:gauge-section}

Use the one-form inverse constants of Lemma~\ref{q:vector-model-inverses}, uniformly on the fixed mass and placement sets. Let $M_V$ bound the source and output identification maps for the separated cutoffs. Choose $B_{J,V}\ge M_V^2(C_{S,V}(\gamma)+C_{K,V}(\gamma))$, and let $D_{J,V}$ be the sum of the four coefficient-product bounds in Lemma~\ref{q:vector-joined-power}. The weights in~\eqref{q:vector-assembly-constants} specify its model norms. These choices give $u_{J,V},A_Q$ by~\eqref{q:vector-reference-threshold}. Next, take $D_{\rm ref}$ and $K$ as above, and set $u_{\rm red}\coloneqq\min(u_D,u_{\rm fp})$. A sufficient coefficient constant $C_Q$ is the norm of the product estimate for derivatives of the metric in~\eqref{q:vector-coefficient-comparison}; Lemma~\ref{q:matrix-jet-inverse} controls its inverse-metric derivatives on the same positive neighborhood. The gauge threshold used in the final mass condition is
\be\label{q:gauge-smallness}
 u_g\coloneqq\min\bigl(u_D,u_{\rm fp},u_{J,V},
 \left(\frac{\theta_V}{A_QC_Q(D_{\rm ref}+K)}\right)^{1/\eta_{\rm geom}}\bigr).
\ee
Proposition~\ref{q:vector-recovery-power} proves recovery of the static equations throughout this range. All entries are power restrictions in fixed geometric and functional constants.

\subsection{Prescribing both horizon areas and choosing the mass bound}\label{q:area-section}

Choose any $0<d<1$ for the scale interval $s\in[1-d,1+d]$, and define
\be\label{q:area-margin}
\Delta_A\coloneqq\min(A_K(M_+)-A_K(M_0),A_K(M_0)-A_K(M_-))>0.
\ee
Let $P_s$ be the relative tangential profile bound on a unit small horizon, let $H_b$ bound $H_r$ on the central horizon, and let $S_s,S_b$ bound the restrictions of the correction norm to these horizons after the factors $u^\gamma$ and $1$, respectively, have been removed. With the actual weight~\eqref{join:eq:explicit-weight} and tensor sizes taken in $G_u$, we may choose $S_s=2^{\gamma/2}$ and $S_b=1$. Indeed $\rho=u/2$ and $\omega_u=\sqrt2\,u$ on a small horizon, whereas $\omega_u=\sqrt{1+u^2}\ge1$ on the original horizons. Restricting a covariant tensor to tangential directions does not increase its operator norm. Choose a relative horizon margin $0<\rho_A<1$. Sufficient area constants are
\be\label{q:evaluated-area-constants}
C_s\ge P_s+S_sK,\qquad C_b\ge4\pi r_b(M_+)^2(H_b+S_bK),\qquad P_s\ge\tfrac32 A_IC_A.
\ee
Set
\be\label{q:area-positivity-threshold}
u_A\coloneqq\min(1,(\rho_A/C_s)^{1/(1+\gamma)},\rho_A/(H_b+S_bK)).
\ee
For $u\le u_A$, these bounds give $|A_b-A_K(M)|\le C_bu$ and $|A_s/(16\pi u^2)-1|\le C_su^{1+\gamma}$. Indeed a relative two-dimensional metric change of operator norm $t<1$ has area Jacobian between $1-t$ and $1+t$. The condition $u\le u_A$ makes the relative tangential changes at most $\rho_A$, and the profile and correction bounds give the displayed constants.

Choose $0<\theta_A<1$. The face signs for the two area parameters follow from
\be\label{q:area-threshold}
\varepsilon\le\frac{\theta_A\Delta_A}{C_b(1+d)},\qquad
\varepsilon\le\left(\frac{\theta_A d(2-d)}{C_s(1+d)^{3+\gamma}}\right)^{1/(1+\gamma)}.
\ee
In fact $u=s\varepsilon$, the central area error is at most $C_b(1+d)\varepsilon$, and the normalized small-area error is at most $C_s(1+d)^{3+\gamma}\varepsilon^{1+\gamma}$. The smallest unperturbed face margin is $d(2-d)$. Each error is at most the chosen fraction $\theta_A$ of its face margin.
All constants have now been chosen before the inserted mass. The six auxiliary restrictions and the two area-face conditions give precisely~\eqref{q:master-threshold}. Restoring units gives $\mu_{\rm suff}=\ell\varepsilon_{\rm suff}(\mathbf c)$, as stated in Proposition~\ref{q:functional-mass-threshold}. The constants in this formula are the proved geometric and functional bounds described in Section~\ref{sec:functional-mass-bound}.

\section{Functional inequalities and their scaling}\label{sec:functional-inequalities}

\subsection{Function spaces and weights used in the proof}
We collect here the norms used in the proof. There are two distinct uses
of weights. The Sobolev weights specify decay at the Schwarzschild end,
while the H\"older weights measure both regularity and size on balls whose
radii vary with the geometry. At a horizon or an axis point these balls
are taken in the smooth normalized local coordinates.

For the energy estimates on the compact static region, we use $L^2$ with
measure $\dd\nu=N\dd V_h$, the volume per unit auxiliary coordinate.
For vector fields the norm $H^1_\ell$ has square
$\|\nabla V\|_2^2+\ell^{-2}\|V\|_2^2$, where $\ell$ is the cosmological
length. The Hardy inequality on the region with the added horizons
also controls $\|\omega_\mu^{-1}V\|_2$. On the end of Schwarzschild
of mass one, the weighted Sobolev spaces $H^2_d$ and $L^2_{d-2}$ used
in Section~\ref{f:range} have squared norms given by
\begin{equation}
 \sum_{j=0}^{2}\int |D^jv|^2(1+s)^{-2d+2j-3}\,\dd\nu_S,
 \qquad
 \int |f|^2(1+s)^{1-2d}\,\dd\nu_S,
 \qquad -1<d<0.
\end{equation}
Here $s$ is the isotropic radius. Under integration without a weight,
the dual of the target is represented using $L^2_{-1-d}$. The
Schwarzschild displacement is removed by the integral condition in
Section~\ref{f:range}.

For the elliptic estimates in H\"older spaces, let $a$ be the length
used to rescale a coordinate ball. The local scaled norm multiplies
the term with $j$ derivatives by $a^j$ and the highest H\"older
seminorm by $a^{j+\alpha}$. We then take a supremum over the
centers of the balls. With exponent $\gamma$, the norm of the
unknown contains the factor $a^{-\gamma}$ and that of the right-hand
side contains $a^{2-\gamma}$. The two powers account for the two
derivatives in the equation. The following choices cover the
model problems and the region where the solutions are joined.
\begin{itemize}
\item On the fixed Schwarzschild model we take $a=r$ in
$X_S,Y_S$, where $r$ is the areal radius. In the estimates formulated
in isotropic coordinates we use the comparable scale $1+s$. On the
unpunctured Schwarzschild--de Sitter model we use $a=1$ in
$X_K,Y_K$. The precise ball radii and derivative orders are given
in~\eqref{q:model-Holder-norms}.
\item On Schwarzschild--de Sitter with the insertion points removed,
we take $a=\omega_\bullet$ in $X_{K,\bullet},Y_{K,\bullet}$.
This scale is the distance to the center in its spatial normal
coordinates near each puncture and is one away from the two
coordinate neighborhoods. The local Newton estimates used to
construct the Green tensor also use the two norms
$\mathcal X_{\rm b}^{m+2,\alpha}$ and
$\mathcal Y_{-1}^{m,\alpha}$ in~\eqref{green:spaces}.
\item On the region with the added horizons, we take $a=\omega_\mu$
in $X_\mu,Y_\mu$, as in the equations~\eqref{join:eq:explicit-weight}
and~\eqref{join:eq:norms}. This scale is comparable to $\mu$ near an
added horizon, to the distance from the insertion point in the
intermediate region, and to one away from the insertion
neighborhoods. Where the equation includes an auxiliary scalar
coefficient, the domain norm is the sum of the norm of the tensor
subject to the integral condition and the absolute value of that
coefficient.
\end{itemize}
The quantitative estimates on the models use $-\tfrac12<\gamma<0$.
The proof also uses the same norms with nearby exponents to control
the derivatives of the cutoffs that combine the model inverses.
For the vector equation, the corresponding spaces are defined by
the same weights and local scales with covariant one-forms in
place of symmetric two-tensors. The definitions of the vector
model spaces and their symmetry conditions are given in
Section~\ref{q:gauge-section} and in the construction of the
vector inverse in the body of the article.

\subsection{Functional inequalities on the fixed and rescaled domains}

\paragraph{Poincar\'e and symmetric-gradient inequalities.}

We summarize the Poincar\'e, Korn, Hardy, Sobolev, interpolation and Schauder estimates used in the construction, together with the estimates for extension, products and the Newton potential. The formulas below specify the dependence on the radius of a rescaled domain. For families of domains, the constants also depend on the geometry and coefficient bounds after rescaling.

Put $\dd\nu_M=N_M\dd V_{h_M}$ and $\overline f=(\int f\dd\nu_M)/(\int\dd\nu_M)$. For regular scalar functions invariant under axial rotations and polar reflection, Proposition~\ref{prop:scalar-gap} and Corollary~\ref{q:functional-Poincare-Korn} give
\be
 \|f-\overline f\|_{L^2(\nu_M)} \le C_{P,M}\|\dd f\|_{L^2(\nu_M)},
 \qquad C_{P,M}^{-2}>2\Lambda.
\ee
For regular meridional vector fields invariant under equatorial reflection, as in Section~\ref{sec:linear-Kottler}, the same corollary gives
\be
 \|V\|_{H^1_\ell} \le C_{\rm Korn,M}\|\mathcal BV\|_2,
 \qquad C_{\rm Korn,M} \le\sqrt{2/c_{V,M}},\qquad
 \mathcal BV=\tfrac12\mathcal L_VG_M.
\ee
The scalar mean and symmetries are essential; the vector class excludes the circular Killing fields. On compact subextremal mass intervals these constants are uniform. Under $G_M\mapsto\lambda^2G_M$ and $\ell\mapsto\lambda\ell$, the Poincar\'e constant is multiplied by $\lambda$ and its spectral margin by $\lambda^{-2}$. Both sides of the Korn inequality, with the length $\ell$ in the norm, have the same scaling, so the Korn constant is unchanged. The positive constant $c_{V,M}$ is controlled by the scalar spectral margin and co-closed decomposition. For tensors, the negative direction~\eqref{q:negative-quotient} requires a bounded inverse rather than positive energy: Proposition~\ref{q:full-Schwarzschild-model-inverse} gives the Schwarzschild inverse with its displacement condition, and Lemma~\ref{q:neighboring-weights} gives nearby decay exponents.

\paragraph{Hardy inequalities and annular extension.}
{
For $f\in C_c^\infty(\mathbb R^3)$, and then by density, $\int|x|^{-2}|f|^2\dd x\le4\int|Df|^2\dd x$. The estimate also holds for compactly supported vector-valued functions $V$. Apply the scalar estimate to $|V|$ and use $|D(|V|)|\le|DV|$ almost everywhere. For the mass-$u$ Schwarzschild metric, Lemma~\ref{q:Hardy} gives
\be
 \int R^{-2}|V|^2\dd\nu_{S,u} \le\int(u+\rho)^{-2}|V|^2\dd\nu_{S,u} \le4\int|\nabla V|^2\dd\nu_{S,u}.
\ee
Here $R$ is the areal radius and $\rho$ the isotropic radius. The constant $4$ is sharp on the complete model. Let $C_{\mathrm H,S}^{\omega}$ denote a constant, uniform in $u$,
in the Schwarzschild Hardy inequality with weight $\omega_{S,u}^{-2}$,
where $\omega_{S,u}=\sqrt{u^2+(R-u)^2}$. Since
$R\le\sqrt2\,\omega_{S,u}$, the weighted integral is at most twice
the integral with weight $R^{-2}$. Thus one may take
$C_{\mathrm H,S}^{\omega} \le 2\cdot4=8$. On the construction collars the model constants apply to fields supported where the metric is exactly Schwarzschild. The estimate on the joined domain uses a separate constant.
}
Under the metric and symmetry assumptions of Lemma~\ref{hv:hardy}, the joined geometries satisfy
\be
 \int\omega_\mu^{-2}|V|^2\dd m_{g_\mu}
 \le C_{\rm H,M}\int(|\nabla V|^2+|V|^2)\dd m_{g_\mu},
\ee
uniformly for sufficiently small $\mu$. In particular the squared $L^2$ norm on an insertion ball of radius $R$ is bounded by $C_M R^2\|V\|_{H^1(g_\mu)}^2$ for small $\mu$. The annular extension used in that lemma is a linear map $E_R:H^1(A_R)\to H^1(B_{2R})$, equal to the field on $A_R\coloneqq\{R<|x|<2R\}$, with
\be
 \int_{B_{2R}}|D E_Rv|^2\dd x
 \le C\left(\int_{A_R}|Dv|^2\dd x
       +R^{-2}\int_{A_R}|v|^2\dd x\right).
\ee
Dilation of a unit-annulus extension makes $C$ independent of $R$; averaging preserves the required rotations and reflections. {
For each fixed $L$, the cap inequality~\eqref{hv:cap} controls the last term on the annulus $L\mu<\rho<2L\mu$:
\be
 \mu^{-2}\int_{\rho<2L\mu}|V|^2\dd m_{g_\mu}
 \le C_L\int_{\rho<2L\mu}|\nabla V|^2\dd m_{g_\mu}.
\ee
Compactness and the absence of parallel fields give the bound on the unit Schwarzschild cap, with no outer boundary condition. Both sides scale by the same factor, and the estimate remains valid under the scaled $C^1$ perturbations in Lemma~\ref{hv:hardy}. The estimate on the joined domain also uses the scaled metric comparisons on the matching annulus and the bounds on the fixed exterior from Lemma~\ref{hv:hardy}.
}

 \paragraph{Weighted estimates, dilation and compactness.}
For the three-dimensional Newton operator,
\be
 \frac1{4\pi}\int_{\mathbb R^3}\frac{|y|^{d-2}}{|x-y|}\dd y
 =\frac{|x|^d}{-d(d+1)},\qquad -1<d<0,\quad x\ne0.
\ee
{
The Schur test in Section~\ref{f:range} gives $\|\mathcal N f\|_{L_d^2} \le[-d(d+1)]^{-1}\|f\|_{L_{d-2}^2}$ on the homogeneous Euclidean model. A fixed cutoff handles the compact part of the Schwarzschild model; scaled interior estimates and the bounded overlap of dyadic annuli control the derivatives. Proposition~\ref{q:punctured-Kottler-inverse} uses the weighted H\"older estimates at the insertion points. For $v_R(y)=v(Ry)$, Euclidean dilation gives
\be
 \|D^jv\|_{L^p(B_R)}=R^{3/p-j}\|D^jv_R\|_{L^p(B_1)}
 \qquad(1\le p\le\infty).
\ee
A Poincar\'e constant scales as a length, and the derivative and weighted terms in the Hardy inequality have the same scaling. This calculation also gives the $R^{-2}$ term in the squared extension estimate.
}
On the three-dimensional model annuli, with $A'_R\coloneqq\{R/2<|x|<4R\}$, scaled Sobolev embedding gives
\be
 \|u\|_{L^\infty(A_R)} \le C \, R^{-3/2}
 \sum_{j=0}^2R^j\|D^ju\|_{L^2(A'_R)}.
\ee
The compact embeddings $H^2\Subset H^1\Subset L^2$ on fixed bounded smooth domains are used in the model inverse operators and concentration argument. At horizons we use the normalized charts and their local elliptic estimates. In the norms~\eqref{join:eq:norms}, derivative $j$ carries the factor $\omega_\mu^j$. On an exact Schwarzschild region, $\Theta_\mu^*G_\mu=\mu^2G_S$. With a vector field normalized to have the same pointwise length on the unit model, its squared $L^2$ norm per unit of the original parameter scales by $b\mu^3$. Its derivative energy and its $\omega_\mu^{-2}$-weighted squared $L^2$ norm both scale by $b\mu$. These equal powers explain the uniformity of the model Hardy and cap estimates; the proof of Lemma~\ref{hv:hardy} transfers them to the joined metrics.

 \paragraph{Products, interpolation and Schauder estimates.}
On fixed charts, $\|uv\|_{C^{j,\alpha}} \le C_{\rm prod}\|u\|_{C^{j,\alpha}}\|v\|_{C^{j,\alpha}}$, also for tensor contractions with component sums. In the scaled norms~\eqref{join:eq:norms}, this gives the estimate used in Proposition~\ref{f:family}:
\be
 \|Q(v)-Q(w)\|_{Y_\mu} \le C\mu^\gamma(\|v\|_{X_\mu}+\|w\|_{X_\mu})\|v-w\|_{X_\mu}.
\ee
The factor $\mu^\gamma$ bounds $\sup_x\omega_\mu(x)^\gamma$, since $\gamma<0$ and $\omega_\mu\ge\mu$. On nested smooth balls $B'\Subset B$, interpolation gives, for every $\eta>0$, $\|u\|_{C^{j+1,\alpha}(B')} \le\eta\|u\|_{C^{j+2,\alpha}(B)}+C_\eta\|u\|_{C^0(B)}$. For uniformly elliptic systems with scalar principal symbol,
\be
 \|u\|_{C^{j+2,\alpha}(B')} \le C(\|Pu\|_{C^{j,\alpha}(B)}+\|u\|_{C^0(B)}).
\ee
{
On balls of radii $a$ and $2a$, the scaled $C_a^{m,\alpha}$ norm weights derivatives and seminorms by $a^j$ and $a^{j+\alpha}$, respectively. In a scaled frame, \eqref{q:local-Schauder} reads
\be
 \|T\|_{C_a^{m+2,\alpha}(B_a)} \le
 S_m\bigl(a^2\|PT\|_{C_a^{m,\alpha}(B_{2a})}+\|T\|_{C^0(B_{2a})}\bigr).
\ee
The same estimate holds with $a^{-2}\|T\|_{L^2(B_{2a})}$ in place of the last term in a four-dimensional lifted ball, using its full volume. The power $-2$ follows from scaling the four-dimensional $L^2$ norm. The power $-3/2$ above concerns the three-dimensional annuli. The scaled interpolation and Schauder constants depend on the ratios of the radii, ellipticity and bounds on the coefficients after rescaling the operator by $a^2$. Section~\ref{sec:local-estimates} establishes these uniform bounds on the chart family, including at the horizon poles; Proposition~\ref{reg:bootstrap} uses the estimates at fixed  mass.
}

\section{The nonlinear equations and their linearization}\label{sec:equations-summary}

\paragraph{The static equations for the spatial metric and the lapse}

Let $h$ be a spatial metric, let $N$ be its lapse, and let $\nabla$ and $\Delta_h=\nabla^i\nabla_i$ be the connection and scalar Laplacian of $h$. The nonlinear equations are
\be\label{sum:static}
\mathcal E(h,N)\coloneqq\Hess_hN-N(\Ric_h-\Lambda h)=0,\qquad
\mathcal S(h,N)\coloneqq\Delta_hN+\Lambda N=0.
\ee
Consider $h_t=h+tk$ and $N_t=N+tv$, where $k$ is a symmetric covariant two-tensor and $v$ a scalar. Raise indices with $h$, put $\operatorname{tr}k=h^{ij}k_{ij}$, and define the connection variation and its contraction by
\be\label{sum:connection}
A^p{}_{ij}\coloneqq\tfrac12h^{pq}(\nabla_i k_{jq}+\nabla_j k_{iq}-\nabla_q k_{ij}),\qquad
Z^p\coloneqq\nabla^ik_i{}^p-\tfrac12\nabla^p\operatorname{tr}k.
\ee
The Ricci variation is $\dot{\Ric}_{ij}=\nabla_pA^p{}_{ij}-\nabla_jA^p{}_{ip}$. Differentiating gives the two linearized static equations
\bse\label{sum:static-linearization}
\be
D\mathcal E[k,v]_{ij}=\nabla_i\nabla_jv-A^p{}_{ij}\nabla_pN-v(\Ric_{ij}-\Lambda h_{ij})-N(\dot{\Ric}_{ij}-\Lambda k_{ij}),
\ee
\be
D\mathcal S[k,v]=\Delta_hv+\Lambda v-k^{ij}\nabla_i\nabla_jN-Z^p\nabla_pN.
\ee
\ese
These identities follow from differentiating the inverse metric, the connection and the scalar Hessian. 

\paragraph{The elliptic equation for symmetric tensors}

For the auxiliary metric $g=N^2\dd\tau^2+h$, the static system is equivalent, where $N>0$, to $\Ric_g=\Lambda g$. In particular $\Ric_g{}_{ij}=\Ric_h{}_{ij}-N^{-1}(\Hess_hN)_{ij}$, $\Ric_g{}_{\tau\tau}=-N\Delta_hN$, and the mixed components vanish. Let $G$ be the reference metric. Using its covariant derivative, we express the difference of the two connections and the difference $\Ric_g-\Ric_G$ as follows.
\bse\label{sum:nonlinear-Ricci}
\be
 \mathcal A^a{}_{bc}\coloneqq\tfrac12g^{ad}(\nabla^G_bg_{cd}+\nabla^G_cg_{bd}-\nabla^G_dg_{bc}),
\ee
\be
 (\Ric_g)_{ij}=(\Ric_G)_{ij}+\nabla^G_a\mathcal A^a{}_{ij}-\nabla^G_j\mathcal A^a{}_{ia}+\mathcal A^a{}_{ab}\mathcal A^b{}_{ij}-\mathcal A^a{}_{jb}\mathcal A^b{}_{ia}.
\ee
\ese
These formulas display all nonlinear terms using the reference covariant derivatives. With the conventions of~\eqref{eq:tensor-operator} and~\eqref{f:L}, define
\bse\label{sum:DeTurck}
\be
W(g,G)^a\coloneqq g^{bc}(\Gamma(g)^a{}_{bc}-\Gamma(G)^a{}_{bc}),\qquad
\mathcal F_G(g)\coloneqq2(\Ric_g-\Lambda g)-\mathcal L_{W(g,G)}g,
\ee
\be
D\mathcal F_G|_G[k]=L_Gk
=P_Gk+(\Ric_G-\Lambda G)\circ k+k\circ(\Ric_G-\Lambda G).
\ee
\ese
Here $P_G=\nabla_G^*\nabla_G-2\operatorname{Riem}_G$ and one index is raised in each product. Indeed $DW|_G[k]=-(\beta_Gk)^\sharp$, where $\beta_Gk=\delta_Gk+\frac12\dd\operatorname{tr}_Gk$ and $(\delta_Gk)_j=-\nabla_G^ik_{ij}$. The Lie derivative cancels the gauge term in the Ricci variation. At an Einstein reference $L_G=P_G$; on the improved approximate metric the displayed zeroth-order terms remain part of the differential. The lapse variation corresponds to the tensor variation $k+2Nv\,\dd\tau^2$ in the static block form. The augmented equation additionally carries the scalar obstruction column of Section~\ref{sec:nonlinear}; its coefficient is removed before the auxiliary-vector estimate recovers~\eqref{sum:static}. All horizon formulas for the auxiliary metric are interpreted in the normalized horizon charts.


\begin{thebibliography}{99}
\bibitem{AlexakisAnEllithyHuang2024} Alexakis S., An Z., Ellithy A., and Huang L.-H., Existence of static vacuum extensions for Bartnik boundary data near Schwarzschild spheres, Comm. Anal. Geom. 34 (2026), 423--446.
\bibitem{AlonsoOranChamizoMartinezMas2023} Alonso-Or\'an D., Chamizo F., Mart\'inez \'A.~D., and Mas A., Pointwise monotonicity of heat kernels, Rev. Mat. Complut. 36 (2023), 207--220.
\bibitem{AnHuang2024} An Z. and Huang L.-H., Static vacuum extensions with prescribed Bartnik boundary data near a general static vacuum metric, Ann. PDE 10 (2024), article 6.
\bibitem{AnHuang2026} An Z. and Huang L.-H., Strict stability of the ADM mass and static vacuum extensions, Preprint arXiv:2609.28101, 2026.

\bibitem{AstorinoEmparanVigano2022} Astorino M., Emparan R., and Vigan\`o A., Bubbles of nothing in binary black holes and black rings, and viceversa, J. High Energy Phys. 2022 (2022), no.~7, 007.
\bibitem{AX-BamlerDehn} Bamler R.H., Construction of Einstein metrics by generalized Dehn filling, J. Eur. Math. Soc. 14 (2012), no.~3, 887--909.
\bibitem{AX-BeigSchoenStatic} Beig R. and Schoen R.M., On static $n$-body configurations in relativity, Class. Quantum Grav. 26 (2009), 075014.
\bibitem{AX-BKEmbeddedCMC} Breiner C. and Kapouleas N., Embedded constant mean curvature surfaces in Euclidean three-space, Math. Ann. 360 (2014), 1041--1108.
\bibitem{BroderLafuente2026} Broder K. and Lafuente R.A., $T^2$-invariant Einstein metrics on closed $4$-manifolds, Preprint arXiv:2610.18219, 2026.
\bibitem{AX-BuntingMasood} Bunting G.L. and Masood-ul-Alam A.K.M., Nonexistence of multiple black holes in asymptotically Euclidean static vacuum space-time, Gen. Relativ. Gravit. 19 (1987), 147--154.
\bibitem{CSLocalization} Carlotto A. and Schoen R., Localizing solutions of the Einstein constraint equations, Invent. Math. 205 (2016), no.~3, 559--615.
\bibitem{Charmousis2007} Charmousis C., Ernst equation and spheroidal coordinates with a cosmological constant term, J. Phys. Conf. Ser. 68 (2007), 012007.
\bibitem{CharmousisGregory2004} Charmousis C. and Gregory R., Axisymmetric metrics in arbitrary dimensions, Class. Quantum Grav. 21 (2004), 527--554.
\bibitem{CLSZ2007} Charmousis C., Langlois D., Steer D., and Zegers R., Rotating spacetimes with a cosmological constant, J. High Energy Phys. 2007 (2007), 064.

\bibitem{CheegerYau1981} Cheeger J. and Yau S.-T., A lower bound for the heat kernel, Comm. Pure Appl. Math. 34 (1981), 465--480.

\bibitem{LS-DeTurckKazdan} DeTurck D.M. and Kazdan J.L., Some regularity theorems in Riemannian geometry, Ann. Sci. \'{E}c. Norm. Sup\'er. (4) 14 (1981), 249--260.
\bibitem{DGSW2023} Dias \'O. J.C., Gibbons G.W., Santos J.E., and Way B., Static black binaries in de Sitter space, Phys. Rev. Lett. 131 (2023), 131401.
\bibitem{DSWMethods} Dias \'O. J.C., Santos J.E., and Way B., Numerical methods for finding stationary gravitational solutions, Class. Quantum Grav. 33 (2016), 133001.
\bibitem{DSWSpinning} Dias \'O. J.C., Santos J.E., and Way B., Spinning black binaries in de Sitter space, Phys. Rev. Lett. 133 (2024), 191401.
\bibitem{ER2002} Emparan R. and Reall H.S., Generalized Weyl solutions, Phys. Rev. D 65 (2002), 084025.
\bibitem{AX-EPVYamabe} Esposito P., Pistoia A., and V\'etois J., The effect of linear perturbations on the Yamabe problem, Math. Ann. 358 (2014), no.~1--2, 511--560.
\bibitem{GT2001} Gilbarg D. and Trudinger N.S., Elliptic partial differential equations of second order, Classics in Mathematics, Springer, Berlin--Heidelberg, 2001.
 

\bibitem{AX-HKWXKerrUniqueness} Han Q., Khuri M., Weinstein G., and Xiong J., Kerr black hole uniqueness, Preprint arXiv:2608.23325, 2026.
\bibitem{AX-HintzDeSitterGluing} Hintz P., Black hole gluing in de Sitter space, Comm. Partial Differential Equations 46 (2021), no.~7, 1280--1318.
\bibitem{HintzSmallBlackHoles} Hintz P., Gluing small black holes into initial data sets, Comm. Math. Phys. 405 (2024), article 114.
\bibitem{HintzLinearizedSources} Hintz P., The linearized Einstein equations with sources, Lett. Math. Phys. 114 (2024), article 94.
\bibitem{KRWY2024} Khuri M., Reiris M., Weinstein G., and Yamada S., Gravitational solitons and complete Ricci flat Riemannian manifolds of infinite topological type, Pure Appl. Math. Q. 20 (2024), no.~4, 1895--1921.
\bibitem{AX-KWYSignedMasses} Khuri M., Weinstein G., and Yamada S., Balancing static vacuum black holes with signed masses in 4 and 5 dimensions, Phys. Rev. D 104 (2021), no.~4, 044063.
\bibitem{LS-KhuriWoolgar} Khuri M. and Woolgar E., Nonexistence of degenerate horizons in static vacua and black hole uniqueness, Phys. Lett. B 777 (2018), 235--239.
\bibitem{AX-KobayashiNomizu} Kobayashi S. and Nomizu K., Foundations of differential geometry, Vol.~I, Interscience Publishers, New York--London, 1963.
\bibitem{KI2003} Kodama H. and Ishibashi A., A master equation for gravitational perturbations of maximally symmetric black holes in higher dimensions, Prog. Theor. Phys. 110 (2003), 701--722.
\bibitem{LakzianSormani2013} Lakzian S. and Sormani C., Smooth convergence away from singular sets, Comm. Anal. Geom. 21 (2013), 39--104.
\bibitem{BLFPLFStability} Le Floch B. and LeFloch P.G., Harmonic, radial, and shell stability of the weighted Einstein constraints on the sphere at infinity, Preprint arXiv:2607.28599, 2026.
\bibitem{BLFPLFLocalization} Le Floch B. and LeFloch P.G., Optimal localization for the Einstein constraints, arXiv:2312.17706 (2023, revised 2026).
\bibitem{BLFPLFShielding} Le Floch B. and LeFloch P.G., Optimal shielding for Einstein gravity, Class. Quantum Grav. 41 (2024), 13LT02.
\bibitem{BLFPLFSeeds} Le Floch B. and LeFloch P.G., The seeds and silhouettes of optimal gravitational shielding, Preprint arXiv:2610.39262, 2026.
\bibitem{AX-Existence} LeFloch P.G. and Rozoy L., Existence of static axisymmetric solutions to the Einstein equations, Preprint arXiv:2610.34555, 2026.
\bibitem{LiTianHarmonic1992} Li Y.-Y. and Tian G., Regularity of harmonic maps with prescribed singularities, Comm. Math. Phys. 149 (1992), no.~1, 1--30.
\bibitem{AX-LiTianHorizons} Li Y.-Y. and Tian G., Nonexistence of axially symmetric, stationary solution of Einstein vacuum equation with disconnected symmetric event horizon, Manuscripta Math. 73 (1991), no.~1, 83--89.
\bibitem{AX-MPPConnectedSums} Mazzeo R., Pacard F., and Pollack D., Connected sums of constant mean curvature surfaces in Euclidean $3$ space, J. reine angew. Math. 536 (2001), 115--165.
\bibitem{OzuchPositive2026} Ozuch T., Positive Einstein metrics on infinitely many four-manifolds, Preprint arXiv:2610.08052, 2026.
\bibitem{AX-ReirisPeraza} Reiris M. and Peraza J., A complete classification of $S^1$-symmetric static vacuum black holes, Class. Quantum Grav. 36 (2019), 225012.

\bibitem{SormaniWenger2011} Sormani C. and Wenger S., The intrinsic flat distance between Riemannian manifolds and other integral current spaces, J. Differential Geom. 87 (2011), 117--199.
\bibitem{AX-WeinsteinEquilibrium} Weinstein G., On rotating black holes in equilibrium in general relativity, Comm. Pure Appl. Math. 43 (1990), no.~7, 903--948.
\bibitem{AX-WeinsteinTwoBody} Weinstein G., The stationary axisymmetric two-body problem in general relativity, Comm. Pure Appl. Math. 45 (1992), no.~9, 1183--1203.
\bibitem{AX-WeinsteinForce} Weinstein G., On the force between rotating co-axial black holes, Trans. Amer. Math. Soc. 343 (1994), no.~2, 899--906.
\end{thebibliography}
\end{document}